\documentclass[aps,preprintnumbers,amsmath,amssymb,prd,superscriptaddress,longbibliography,twocolumn]{revtex4-1}
\usepackage{dcolumn}
\usepackage{color}
\usepackage[caption=false]{subfig}
\usepackage{feynmp}
\usepackage{feynmp-auto}
\usepackage{systeme,mathtools}
\usepackage{stmaryrd}
\usepackage{bbm}

\def\comment#1{}

\newcommand{\nc}{\newcommand}
\nc{\beq}{\begin{eqnarray}}
	\nc{\eeq}{\end{eqnarray}}
\nc{\scs}{\scriptstyle}
\nc{\setval}{\fmfset{wiggly_len}{3mm} \fmfset{arrow_len}{1.5mm}
	\fmfset{arrow_ang}{13} \fmfset{dash_len}{1.5mm}\fmfpen{0.125mm}
	\fmfset{dot_size}{2thick}}

\usepackage{bm,latexsym,mathrsfs,enumerate,color}
\usepackage[mathcal]{euscript}
\usepackage[breaklinks=true,unicode=true,urlcolor = blue,colorlinks 
= true,citecolor = blue,linkcolor = blue]{hyperref}
\usepackage{graphicx}
\usepackage{todonotes}
\usepackage{wrapfig}

\renewcommand{\vec}[1]{\bm{#1}}

\def\slashchar#1{\setbox0=\hbox{$#1$}           
	\dimen0=\wd0                                 
	\setbox1=\hbox{/} \dimen1=\wd1               
	\ifdim\dimen0>\dimen1                        
	\rlap{\hbox to \dimen0{\hfil/\hfil}}      
	#1                                        
	\else                                        
	\rlap{\hbox to \dimen1{\hfil$#1$\hfil}}   
	/                                         
	\fi}                                         %

\DeclareMathAlphabet\mathbfcal{OMS}{cmsy}{b}{n}

\begin{document}

\title{Exact Universal 
Chaos, Speed Limit, Acceleration, Planckian Transport Coefficient, ``Collapse'' to equilibrium, 
and Other Bounds in Thermal Quantum Systems}

\author{Zohar Nussinov}
\email{zohar@wustl.edu}
\affiliation{Department of Physics, Washington University, St.\ Louis, MO 
63130, USA}

\author{Saurish Chakrabarty}
\affiliation{Department of Physics, Acharya Prafulla Chandra College, Kolkata 700131, India}

\date{\today}

\begin{abstract}
We introduce ``local uncertainty relations'' in thermal many body systems. Using these relations, we derive basic bounds. These results include the demonstration of universal non-relativistic speed limits (regardless of interaction range), bounds on acceleration or force/stress, acceleration or material stress rates, transport coefficients (including the diffusion constant and viscosity), electromagnetic or other gauge field strengths, correlation functions of arbitrary spatio-temporal derivatives, Lyapunov exponents, and thermalization times. We further derive analogs of the Ioffe-Regel limit. These bounds are relatively tight when compared to various experimental data. In the $\hbar \to 0$ limit, all of our bounds either diverge (e.g., the derived speed and acceleration limit) or vanish (as in, e.g., our viscosity and diffusion constant bounds). Our inequalities hold at all temperatures and, as corollaries, imply general power law bounds on response functions at both asymptotically high and low temperatures. Our results shed light on how apparent nearly instantaneous effective ``collapse'' to energy eigenstates may arise in macroscopic interacting many body quantum systems. We comment on how random off-diagonal matrix elements of local operators (in the eigenbasis of the Hamiltonian) may inhibit their dynamics. 
\end{abstract}

\maketitle

\section{Introduction}
\label{sec:intro}

In classical statistical mechanics, Planck's constant appears, in a sleight of hand, as an amended divisor in phase space integrals. The insertion of a divisor having the units of Planck's constant is mandated by the dimensionless character of both the canonical partition functions and of the number of classical ``microstates'' in the microcanonical ensemble. Its necessary introduction hints at something deep missed by classical physics. In trivially solvable quantum many body systems such as ideal gases and harmonic oscillators, it can be seen how the counting of states in Boltzmann's expression for the entropy is associated with this required classical constant not only carrying the same units as Planck's constant but rather being exactly equal to it. This equivalence implies quantitative experimental consequences for these otherwise seemingly classical thermal systems. Indeed, the more than century old Sackur-Tetrode equation \cite{Sackur,tetrode,experiment} for the entropy of a classical three-dimensional ideal gas enabled early estimates of Planck's constant from thermodynamic measurements of monatomic mercury vapor \cite{experiment}. The emergence of Planck's constant in phase space integrals describing high temperature ``classical'' systems is often colloquially ascribed to the existence of fundamental minimal cells, of a volume set by the momentum-position uncertainty relation, that tesselate phase space. 

Recent years saw the revival of questions concerning the venerable role of Planck's constant in high temperature systems and a flurry of new hypotheses concerning universal ``Planckian" bounds \cite{KSS1,KSS2,Shuryak,Tom,nnbk,benk,bound,jan-book,HLS-book,Matteo2019,jan-planck,KT,planck1,planck3,planck4,juan-martin,melting-speed,sudip,kapitulnik,HM,Lucas1,Ramshaw,sazo} on various physical quantities (including, notably, viscosities, their ratio to the entropy density, heat diffusion, and conductivities) suggested to be nearly saturated in quark-gluon plasmas generated in heavy ion colliders \cite{KSS2,Shuryak,Tom} and numerous other physical systems. These more recent conjectures were largely triggered by Maldacena's celebrated AdS-CFT correspondence \cite{ads1,Witten}. These new surmised bounds \cite{KSS1,KSS2,Shuryak,Tom,nnbk,benk,bound,jan-book,HLS-book,Matteo2019,jan-planck,KT,planck1,planck3,planck4,juan-martin,melting-speed,HM,sudip,kapitulnik,Lucas1,Ramshaw} are intimately related to the aptly called ``{\it Planckian time scales}'' ($\tau_{Planck} = \hbar/(k_{B}T))$ multiplying the frequency in expressions for black-body radiation \cite{Planck1900} (involving the temperature $T$ of the system, Planck's constant, and the Boltzmann constant (albeit their name, both constants were first introduced by Planck \cite{Planck1900})), Einstein's and Debye's subsequent calculations \cite{Einstein,Debye} of phonon contributions to the heat capacity, and many other arenas \footnote{Notwithstanding similarity in name, these ubiquitous thermal time scales set by $\tau_{Planck}$ are, of course, not to be confused with the (possibly experimentally unattainable \cite{aba1}) Planck time ($t_{P} = \sqrt{\hbar G/c^5}$ with $G$ the gravitational constant and $c$ the speed of light) below which non-renormalizable quantum gravity effects appears. Both the thermal ``Planck scale'' $\tau_{Planck}$ (a term coined, in the context of thermal dissipation, by Zaanen \cite{planck3}) and the Planck time $t_P$ are associated with hitherto conjectured shortest possible timescales. In the current work, we rigorously establish the thermal Plackian time $\tau_{Planck}$ to be a minimal possible scale for the variation of local observables in typical thermal systems. As we will explain, our exact inequalities involve effective local heat capacities whose scale is typically set by the Boltzmann constant $k_B$ yet not exactly equal to it (especially at low temperatures where there may be a strong quantum suppression of these heat capacities leading to yet stronger inequalities on the allowed rates of change of local observables). This, in turn, gives rise to bounds with dimensionless prefactors amending $\tau_{Planck}$.}. Eyring's theory for chemical reaction rates and its original suggested extensions \cite{eyring,eyringViscosity} involve such a Planckian time scale multiplying an exponential in the Gibbs free energy activation barrier. Wigner's ``quantum correction'' expansion about the classical equilibrium distribution \cite{Wigner} similarly involves powers of Planck's constant multiplied by those of the inverse temperature. In the path integral formulation of statistical physics, the imaginary time direction is compactified with a circumference set by $1/\tau_{Planck}$ and thus its emergence in quantum statistical mechanics is inevitable (as it directly does in, e.g., Matsubara frequencies \cite{AGD,coleman,das}). Such Planckian time scales appear in myriad other arenas. For instance, the temperature of the Hawking radiation \cite{Hawking} (a black-body radiation) and the effective Davies-Unruh temperature \cite{Un1,Un-2} are, respectively, prescribed by setting, up to a factor of $2 \pi$, the Planckian time $\tau_{Planck}$ equal to the ratio of the speed of light to the surface gravitational acceleration of black holes \cite{Hawking} or to the ratio of the speed of light to the acceleration of observers \cite{Un1,Un-2}. As fundamental entities, thermal Planckian time scales are consistent with measurements of strongly correlated electronic systems, e.g., \cite{Marel} and are pivotal to quantum critical phenomena \cite{Hertz,subir,NFL} as well as various quantum effects \cite{musketeers} at finite temperature. Planckian type time scales are also central to numerous questions concerning correlations, equilibration, and information scrambling \cite{nnbk,juan-martin,bound,op-scram1,op-scram2,ent-growth,typical0,typical}. The Sachdev-Ye-Kitaev model \cite{SY-SYK,K-SYK,Juan-SYK} and its myriad extensions and applications, e.g., \cite{0-SYK,1-SYK,2-SYK,3-SYK,4-SYK,5-SYK} have made some of these concepts more tangible. In the current work, we make many of these earlier suggested bounds rigorous by systematically deriving general inequalities. In certain limits, these inequalities relate to conjectured minimal time scales  \cite{KSS1,KSS2,Tom,nnbk,jan-book,Matteo2019,jan-planck,KT,planck1,planck3,planck4,juan-martin,bound,melting-speed} that are typically given by universal numerical constants multiplying $\tau_{Planck}$ and the thermal de-Broglie wavelength. However, in general, our rigorous bounds differ from these (replacing the Boltzmann constant in $\tau_{Planck}$ by a local heat capacity that is relevant for a given local observable). We will further arrive at several new inequalities. Our bounds will be derived by applying the standard operator form uncertainty relations within an Araki-Lieb or thermofield double type framework \cite{lieb-pure,das} while taking note of the relevant equations of motion and thermal expectation values.  

The oldest and, by far, the most basic variance type uncertainty relations \cite{uncertain,Konrad,Robertson,Mandelshtam-Tamm,lenny,Davidson,Anandan-Aharonov} revolve around the momentum-position and energy-time uncertainty relations (relating quantities that carry units) and their extensions. Similar variance bounds stemming from the Cauchy-Schwarz inequality include Bologiubov's celebrated inequality \cite{Bogol} that has proved instrumental in rigorously establishing the absence of finite temperature symmetry breaking in low dimensional systems \cite{MW} (and particular higher dimensional extensions \cite{naglass,harrisa,boundary-bounds1,topohigh}). In the current work, we will build on elementary textbook type variance uncertainty relations in a systematic way. Our analysis greatly expands on the initial results of \cite{bound} in this direction. Numerous related illuminating (and, at times, much broader) forms of the uncertainty relations exist. The latter provide additional insight and finer detail into the consequences and extensions of the simplest variance type uncertainty relations. These inequalities include the penetrating information theoretic \cite{info0,info1,info2,info3,info4,info5,info6,info7,info8,info9,info10,info11,info12,info13} uncertainty relations that relate Shannon entropies to eigenstate overlaps (and, unlike the variance based relations, do not directly involve quantities having physical units). Other celebrated inequalities include extensions to Markovian systems \cite{thermo1,thermo2}, the information-time \cite{time-information}, energy-temperature \cite{E-T} relations, and many others. In the current work, already by using the simplest variance based uncertainty relations, we will derive bounds (Sections \ref{setup} and \ref{derive:time}) on rates of change (both temporal and spatial) of general observables in thermal systems. To avoid confusion in terminology, we note that while we introduce bounds on the general rates of change of local observables and their correlations, we will not touch on ``quantum speed limits'' relating to how rapidly quantum systems may evolve between distinguishable states \cite{Mandelshtam-Tamm,speed2,speed1,campo}. The derivations in this paper are aimed to be exceedingly transparent and non-technical with all details explicitly spelled out. At their core, our calculations will rely on obtaining ``local'' variants of the variance based uncertainty relations in many body systems. We will explain what the resulting bounds imply in trivial theories with no connected correlations (Section \ref{decoupled}). These relations will carry over to certain interacting Reflection Positive systems (Appendix \ref{sec:RP}) and, more generally, to every theory in which all connected correlation functions of relevant local Hamiltonians are positive semi-definite. Subsequently, we will turn to obtaining bounds for specific observables in general interacting theories and discuss their implications. 

Specifically, we will illustrate how our uncertainty inequalities lead to 

(1) A universal {\it limit on particle speeds} in non-relativistic thermal systems (Section \ref{v-bounds}). Unlike the Lieb-Robinson bounds \cite{Lieb_Robinson,Bruno,Sergey,alioscia,kaden,Else20}, this bound {\it does not require locality (nor sufficiently rapid decay \cite{Else20})} of the interactions, finite Hilbert spaces, or other similar restrictions. 

(2) Upper bounds on the {\it acceleration} (Section \ref{a-bounds}) and on the temporal rate of change of the acceleration or, equivalently, that of the forces or stresses (Section \ref{sec:stress}) and their implications for a lower bound for the diffusion constant and an upper bound on the shear viscosity when the Stokes-Einstein relation holds. 

(3) Bounds on general transport coefficients in local theories (Section \ref{GsKsec}) that illustrate that due to quantum effects, these coefficients may not, generally, vanish. Section \ref{sec:viscosity} details lower bounds on the (bulk and shear) {\it viscosity} in local theories  similar to those suggested in \cite{nnbk} and other more recent (and, for common fluids, typically stronger) bounds \cite{KT}. We discuss bounds on the {\it electrical response} in Section \ref{sec:conduct}. 

(4) Extensions of our inequalities to spatial gradients (Section \ref{sec:space}) illustrate that in thermal semiclassical systems, the {\it ratio between the fields and the potentials} is bounded from above by the inverse de-Broglie wavelength. Analogous considerations apply to {\it field strengths in general gauge theories}.

(5) Direct {\it chaos type} bounds (Section \ref{sec:chaos1}) that do not hinge on the currently heavily studied ``Out of Time Order Correlation'' (OTOC) functions. Our rigorous bounds conform with those conjectured (and further motivated by gravitational analogs) several years ago by Maldacena, Shenker, and Stanford \cite{juan-martin} and relate to earlier conjectures by Sekino and Susskind \cite{op-scram1}. We derive direct semiclassical bounds on the time evolution of the out of time order correlation (OTOC) functions (i.e., not those associated with their deformed regularized variant \cite{juan-martin}) in Section \ref{appOTOC}. In Section \ref{OTOCBound}, we discuss possible qualitative bounds on semiclassical transport coefficients that derive from the Lyapunov exponent bounds. 

(6) A stringent inequality (Section \ref{sec:finitetime}) similar to that proposed by Ioffe and Regel \cite{Ioffe-Regel} in systems with {\it quasiparticles undergoing ballistic motion}.

 (7) Sections \ref{sec:lowT} and \ref{sec:highT} explain how asymptotic low and high temperature {\it corollaries} of our inequalities lead to {\it universal relaxation rate bounds that scale algebraically with the temperature}. These bounds are consistent with those satisfied by Fermi liquids \cite{NFL,AGD,LL,coleman} and linear in $T$ ``bad metal'' resistivities (that include those in the normal phase of high temperature superconductors) \cite{NFL,planck1,planck4}. It has been pointed out, e.g. \cite{sarma,dassarmaPh}, such linear in $T$ resistivities may, e.g., result from electron-phonon coupling.

(8) Lower bounds on {\it thermalization times} (Section \ref{quantum-thermalization}) {\it relating to the average of quantum measurements performed over a given time window}. The latter Planckian time averaged measurements may become {\it eigenstate expectation values} in systems satisfying the Eigenstate Thermalization Hypothesis \cite{eth1,eth2,eth3,eth4,rigol,pol,polkovnikov1,polkovnikov2,srednicki-95}. More generally, these short time averages are equal to a weighted average of eigenstate expectation values amongst eigenstates that share the same energy density. 
We will further connect thermalization times with transport and derive anew viscosity bounds in semiclassical systems. 

(9) Illustrate that if the off-diagonal matrix elements of local operators in the energy eigenbasis are random (as the Eigenstate Thermalization Hypothesis asserts) then the dynamics of these local observables are constrained in random
non-equilibrium states (Section \ref{sec:off}). We will further motivate why, on average, the statistical properties of the off-diagonal matrix elements may appear to be random
and highlight their importance.

Our bounds will hold in {\it an arbitrary number} of spatial dimensions. In most of these derivations (those pertaining to bounds on time derivatives of general observables), heat capacities (either exact or effective) will play a prominent role. 

In Table \ref{real-values}, we provide a synopsis of the bounds that we will derive for several transport coefficients and contrast these with representative experimental values. Many of these systems (e.g., water at high temperature) can readily be thought of bona fide classical systems. Our bounds are {\it relatively tight} when contrasted with various empirical values of systems that may be typically regarded as classical, e.g., water at non-cryogenic temperatures. Since the parameters governing these systems have their origin in quantum effects, it is might not be surprising that the order of magnitude of some of our quantum bounds may be nearly saturated when contrasted with empirical measurements. As we will explain, our transport coefficient bounds are for positive autocorrelation function contributions to the Green-Kubo formula. As such, when significant negative autocorrelation function contributions to the Green-Kubo integral appear (e.g., oscillatory velocity autocorrelation functions of ions in solids), our bounds on these proxies do not imply transport coefficient bounds. 

To help the reader navigate this work, numerous results and additional background have been relegated to the appendices. These include a review of the Araki-Lieb or thermofield double construct and a discussion of how it implies finite temperature uncertainty bounds (Appendix \ref{sec:thermofield}), a discussion of the non semiclassical limit of quadratic moment bounds (Appendix \ref{QUB}), further detail on semiclassical higher moment generalizations of our inequalities (Appendix \ref{triv-triv-long}), an explicit discussion of how our inequalities apply to identical particles (Appendix \ref{sec:identical}), explaining how our bounds vary when different choices are made for the local Hamiltonian endowing the observables with dynamics (Appendix \ref{sec:comments}), a discussion of Fermi quasiparticle systems (Appendix \ref{Fermigas}), further details on the derivation of the velocity bound (Appendix \ref{velocity-bound-App}), Reflection Positive systems (Appendix \ref{sec:RP}) and more general instances of positive semi-definite connected correlation functions of relevant local Hamiltonians. In Appendices \ref{higher_grad} and \ref{sec:2-point} we derive bounds on {\it high order gradients} and discuss inequalities concerning the ratios between pair correlators to their derivatives. Appendix \ref{sec:generalcorrelatorgrad} details bounds on correlators of the gradients of general operators in the semiclassical limit. In order to make our more formal temporal bounds more lucid, we analyze (in Appendix \ref{sec:toy}) as pedagogical textbook type examples a harmonic solid (Appendix \ref{sec:harmonic}) and a simple XY spin model (Appendix \ref{sec:XY}). We further discuss prethermalized systems (Appendix \ref{minprethermal}). We conclude by briefly commenting on the limits of time measurements (Appendix \ref{sec:clock}). 

\section{The Idea in a Nutshell}
\label{central}

\begin{table*}
    \begin{tabular}{|c|c|c|c|c|m{2.1in}|}
    \hline
    Quantity & Bound & \parbox{2.01cm}{{\tiny ~}\\Approximate  bound\\\vspace{1pt}} &
           \parbox{2cm}{Typical value of quantity} & Equation \# &\parbox{2in}{Remarks}\\
    \hline
    $\displaystyle{
    \overline{\tau^{-1}_{Q}}}$
             & $\displaystyle{\frac{2 \sqrt{k_{B} T^2 C_{v, i}}}{\hbar}}$
                     & $\begin{matrix}\overline{\tau^{-1}_{Q}} \lesssim 7.9 \times 10^{13}\mbox{Hz}\\ \mbox{at }T=300~K\end{matrix}$ & \parbox[c][0.59in]{0.9in}{See  below rows for specific quantities} & \ref{central1},\ref{cv1} & 
                     \parbox[c][0.59in]{2.1in}{Upper bound on the average {\em relaxation rate} of general local quantities $Q$. As an ``approximate value'', we set here $C_{v,i}$ to be $k_B$. 
                     }\\
    \hline
    $\displaystyle{\sqrt{\overline{ \langle (v^{H}_{i})^{2} \rangle/(\sigma^{H}_{r_{i}})^{2}}}}$
             & $\displaystyle{\frac{k_{B} T \sqrt{2}}{\hbar}}$
                           & $\sqrt{\overline{(\langle v^{H}_i)^2\rangle}}\lesssim 6.98~km/s$ & \parbox[c][0.59in]{1.1in}{$v_{\mbox{sound}}\sim3.1~km/s$\\ \cite{alSpeed}} & \ref{vbound12} & \parbox[c][0.92in]{2.1in}{Bound on non-relativistic root mean square particle {\em speed}. The values are for Aluminum (lattice const. 4.04{\AA} \cite{aluminum}) just below melting ($660^\circ C)$ substituting position fluctuations $\sigma^{H}_{r_{i}}$ of Lindemann ratio 0.1 \cite{lindemannAl}
                           .
}
                           \\ 
                           \hline
    $\displaystyle{D}$
             & $\displaystyle{\frac{\hbar}{2\pi m}}$
                     & $D \gtrsim 5.6\times10^{-10}m^2/s$  & $1.1\times10^{-9}m^2/s$  & \ref{minDiff+},\ref{DOTOC} & 
                     \parbox[c][1.32in]{1.97in}{Lower bounds on the {\em diffusion constant} when long time oscillatory tails of the velocity autocorrelation function may be ignored. The universal bound on the left is that of Eq. (\ref{DOTOC}). 
                     Similar bounds result from the exact inequality (\ref{minDiff+}). 
                     The values are
                     for water at STP \cite{waterD}.}\\ 
    \hline
    $\displaystyle{\zeta}$
             & $\displaystyle{{\cal O}\left(\frac{{\sf n}\hbar}{\sqrt{d^3({\sf z}+1)}}\right)}$
                     & $\zeta \gtrsim  2.9\times10^{-7}$ Pa-s & $4.9\times10^{-4}$ Pa-s
                     & \ref{upeta:eq} & \parbox[c][0.9in]{2.1in}{Bound on {\em bulk viscosity}. Values are for water at 100$^\circ C$ and atmospheric pressure. The dimensionality and effective coordination number are, respectively, $d=3$ and ${\sf z}=4.5$ \cite{watercoord1,watercoord2}.}\\
    \hline
    $\displaystyle{\eta}$
             & $\displaystyle{{\sf n}h}$
                     & $ \eta \gtrsim 2 \times10^{-5}$ Pa-s & $6.3\times10^{-5}$ Pa-s
                     & \ref{nhb} & \parbox[c][0.59in]{2.1in}{Simplified universal bound on the {\em shear viscosity}. The value quoted is the minimal viscosity of water over all temperatures \cite{jing,NIST,KT}.}\\
    \hline
    $\displaystyle{\eta}$
             & $\displaystyle{\frac{k_BT}{6\pi RD_{\min}}}$
                     & $ \eta \lesssim 2.8\times10^{-3}$Pa-s  & $1.8\times10^{-3}$Pa-s  & \ref{etaup},\ref{etaup'} & \parbox[c][1.05in]{2.1in}{Maximal {\em shear viscosity} when the classical Stokes-Einstein relation is valid.
                     $D_{\min}$ denotes our lower bound on the diffusion constant. The value quoted is for water at STP (computed via Eq. (\ref{etaup'}) for molecular radius $R=1.35$ {\AA}).
                     } \\ 
    \hline
  \end{tabular}
  \caption{A synopsis of some of our bounds and their comparison to experiment for sample quantities (water and Aluminum). The equation numbers of the derived inequalities appears in the column before the last.}
  \label{real-values}
\end{table*}

In the current work, we will 
largely focus on {\it open} non-relativistic systems $\Lambda$ with a general {\it time independent} global Hamiltonian $H_{\Lambda}$ and examine the associated general equilibrium expectation values. For such an open system at a temperature $T$, the variance of $H_{\Lambda}$ in thermal equilibrium is given by $k_{B} T^2 C^{(\Lambda)}_v$ where $C^{(\Lambda)}_v$ is the standard constant volume thermodynamic heat capacity of the full system $\Lambda$. A central inequality that we will repeatedly use links the dynamics of general local observables with effective (often exactly computable) thermodynamics,
\begin{equation}
\label{central1}
\! \! \! \! \! \! \! \! \frac{2 \sqrt{k_{B} T^{2} C_{v, i}}}{\hbar} \geq
 \frac{1}{N_{\Lambda}} \sum_{i=1}^{N_{\Lambda}}
 \frac{\left|\left\langle\frac{d Q^H_{i}}{d t}\right\rangle\right|}{\sigma_{Q^H_{i}}}   
\end{equation}
We will explain the meaning of this relation and derive this and related inequalities in Section \ref{derive:time}. We will then explain, in some depth, the consequences of these inequalities in subsequent Sections. 
In essence, Eq. (\ref{central1}) asserts that the relative rate of change of {\it any local operator} ~$Q^H_{i}$~
in an open thermal system cannot exceed a bound containing the system temperature $T$, Planck's and Boltzmann's constants and, as we will describe in later Sections, an associated {\it effective local heat capacity} $C_{v, i}$. This effective heat capacity is a local analogue of the extensive thermodynamic heat capacity $C^{(\Lambda)}_v$ of the full global system $\Lambda$. The ratio appearing in the sum on the righthand side of Eq. (\ref{central1}) is associated with the relative temporal fluctuations of the general local Heisenberg picture operator $Q^H_i$ with the angular brackets denoting averages taken with respect to the density matrix of the thermal system. $N_{\Lambda}$ is the total number of particles, spins, etc., in the system $\Lambda$. Eq. (\ref{central1}) is a consequence of a {\it local} variant of the time-energy uncertainty relation that we will discuss in Section \ref{setup}. This inequality was derived in \cite{bound} yet its consequences were only relatively briefly examined. The appearance, in Eq. (\ref{central1}), of an effective local heat capacity $C_{v,i}$ controlling the maximal temporal fluctuations is not entirely surprising. In an open system at a fixed temperature, the magnitude of heat exchange with the environment and ensuing fluctuations are set by the heat capacity; a larger heat capacity enables more rapid fluctuations. In Eq. (\ref{central1}), for a local observable $Q^H_i$, the relevant part of the ``environment''  now becomes a local region in the open system $\Lambda$ with fluctuations set by the said effective local heat capacity $C_{v, i}$ (instead of the global extensive thermodynamic heat capacity $C^{(\Lambda)}_v$). In the current work, we will make these notions precise. We will find that for typical local quantities $Q^H_i$, the effective heat capacity will generally be a number of order unity times the Boltzmann constant, 
\begin{equation}
\label{typical11}
C_{v,i}=\mathcal{O}\left(k_{B}\right).
\end{equation}
This will, in turn, imply that the lefthand side of Eq. (\ref{central1}) scales as 
$\mathcal{O}\left(\frac{k_{B} T}{\hbar}\right)$.  
In other words, given an equilibrated system at a given temperature $T$, it is not possible for any local quantity to exhibit appreciable dynamics at times whose scale is shorter than the ``Planckian time scale'' $ \tau_{Planck} = 
\hbar/(k_{B} T)$ noted in the Introduction. To be complete, however, we must note that although Eq. (\ref{typical11}) typically holds, in general, the effective local heat capacity $C_{v,i}$ may differ from $k_{B}$ by dimensionless prefactors that will make our bounds not precisely equal to the Planckian time $\tau_{Planck}$. In particular, due to quantum suppression, at low temperatures, the effective heat capacity $C_{v,i}$ may drop precipitously as $T \to 0$ leading to bounds on the dynamics that are substantially tighter than those suggested by $\tau_{Planck}$. (An exactly solvable example in which such a suppression comes, analytically, to life is afforded by the harmonic solid (Appendix \ref{sec:harmonic}, see Eq. (\ref{harmonic-final}) therein in particular)). At room temperature, $\tau_{Planck} = 2.5 \times 10^{-14}$ seconds. We will illustrate how Eq. (\ref{central1}) and its extensions mandate many of the bounds listed in the Introduction and briefly discuss measurements and equilibration times. As we will illustrate in Section \ref{quantum-thermalization}, averages performed over a time window of width $\tau_{Planck}$ may (in systems that saturate this lower time scale bound) tend to their thermal equilibrium averages. Equilibrium averages correspond, for observables that are smooth functions of the energy, to eigenstate averages \cite{eth1,eth2,eth3,eth4,rigol,pol,polkovnikov1,polkovnikov2,srednicki-95,bound,glass}. Thus, in this sense, various observables cannot ``{\it instantaneously collapse}'' to their equilibrium (or, equivalently Hamiltonian eigenstate) expectation values on time scales shorter than $\tau_{Planck}$. The time scale bounds that we derive have universal high and low temperature limits that are satisfied by various systems (e.g., Fermi liquids and ``bad metal'' systems). 

In theories with local interactions, we will find that Eq. (\ref{central1}) and related inequalities imply a bound on the viscosity,
\begin{equation}
\label{etanhh}
\eta \geq {\cal{O}}( \mathrm{n} h).
\end{equation}
Here, ${\mathrm n}$ is the particle number density. In the current work, we will arrive at Eq. (\ref{etanhh}) multiple times. The more rigorous of the viscosity bounds that we will derive will be smaller than $ \mathrm{n} h$ by constant numerical prefactors as well as factors of involving the number of particles that a given particle interacts with and the spatial dimensionality of the system. Empirically, ${\mathrm n} h$ is indeed a lower bound on the viscosity of all liquids that were recently examined \cite{nnbk,KT,jing}. 
For, e.g., water \cite{jing,NIST,KT}, the minimal viscosity is $6.3 \times 10^{-5}$ Pascal $\times$ second whereas ${\mathrm n} h =2 \times 10^{-5}$ Pascal $\times$ second. For comparison, the viscosity of water at room temperature and atmospheric pressure is $10^{-3}$ Pascal $\times$ second (see Table \ref{real-values} for further comparisons).

In a manner similar to that leading to Eq. (\ref{central1}), we will derive an additional set of spatial dual bounds. In particular, we will demonstrate 
that at sufficiently high temperatures where classical equipartition is valid or whenever classical equipartition provides an upper bound on the kinetic energy,
\begin{equation}
\frac{1}{N_{\Lambda}} \sum_{i=1}^{N_{\Lambda}} \frac{\left|\left\langle\frac{\partial f}{\partial x_{i \ell}}\right\rangle\right|^{2}}{\left\langle f^{2}\right\rangle} \leq \frac{4 m k_{B} T}{\hbar^{2}}=\frac{8 \pi}{\lambda_{T}^{2}}.
\end{equation}
Here,
\begin{equation}
\label{dBeq}
\lambda_{T} \equiv \sqrt{\frac{2 \pi \hbar^{2}}{m k_{B} T}}
\end{equation}
is the non-relativistic thermal de-Broglie wavelength for a particle of mass $m$ and $f$ is an {\it arbitrary function} of the particle coordinates and any other degrees of freedom. At sufficiently low temperatures, when $\lambda_{T}$ becomes comparable to the interatomic spacing, condensation typically onsets in bosonic systems. To provide an order of magnitude scale for these spatial gradient inequalities in a conventional non-cryogenic setting, for an O$_{2}$ molecule at room temperature (moving at $\sim$ 460 meters/second), the thermal de-Broglie wavelength $\lambda_{T}  \sim  1.8 \times 10^{-11}$ meters is far smaller than an atomic scale (further dwarfed by intermolecular distance in air at atmospheric pressure, of $\sim$ 4 nanometers (and the mean free path which is another order of magnitude larger yet)).

\section{Setup for proving thermal temporal bounds}
\label{setup}

To set the stage, we briefly recall basic facts concerning many body systems and their steady state thermal equilibrium properties. Stationary equilibrium averages are associated with a coarse, time-independent, ``macrostate'' of the system. Indeed, in equilibrium, the global system wide average of the expectation values of physically measurable local observables are not delicately tied to the specific intricacies of an individual many body quantum state nor those of a classical ``microstate''  of the system. The macrostate is defined by only a small number of thermodynamic state variables (that are preparation history independent). These stationary equilibrium state variables may be set by extensive global sums and/or their system-wide averages and other related intensive variables. However, the detailed many body quantum states (and their classical microstate approximations) generally evolve in time. A cornerstone of thermodynamics is the observation that even though the many body system harbors a divergent number of microscopic degrees of freedom (with most of these {\it exhibiting, system size independent, dynamic fluctuations}) only a very small number of static (macro)state variables suffice to fully capture all of the global properties of the equilibrated system. In a gas in a closed vessel that is in equilibrium, the momenta of the (divergent number of) colliding gas molecules continuously vary with time. The equilibrated system is not static- the velocities of individual gas molecules are not zero nor constant. In classical phase space, when the system equilibrates, the probability distribution becomes stationary. However, different phase space points evolve according to Hamilton's equations of motion. Similarly, for such a gas, the single body probability density is a sum of delta functions (or, more precisely, localized functions of the position for the finite size molecules) centered about the locations of these particles in a given time slice. Classically, at any time $t$, the full many body probability density of an individual $N_\Lambda$ particle thermal system is set by a product of delta functions in phase space for the position and momentum coordinates for each of the particles. That is, the classical probability density for a specific system $\Lambda$ (not the for the full ensemble of such systems) is $ \rho_{\Lambda}^{\sf classical}(\{{\bf x}_{i}\}; \{{\bf p}_{i}\}) = \prod_{i=1}^{N_{\Lambda}} \delta({\bf x}_{i} - {\bf X}_{i}) \prod_{i=1}^{N_{\Lambda}} \delta({\bf p}_{i} - {\bf P}_{i})$ with the specific $\{{\bf X}_{i}\}$ and $\{{\bf P}_i\}$ denoting the specific positions and momenta of the $N_{\Lambda}$ particles at time $t$ and $\{{\bf x}_{i}\}$ and $\{{\bf p}_{i}\}$ marking the phase space variables. 
The global average of the squares of the individual particle momenta (or in the absence of magnetic fields, equivalently, velocities) is given by the canonical ensemble average (which, of course, would not be finite if the equilibrium particles were all stationary). All global averages (such as those of the squared momenta or the vanishing global average of the gas molecule velocities) of the closed system are stationary and depend only on the gas volume, total energy, and the total number of molecules. In other words, the precise, time evolving, quantum many body state of an equilibrated system or its corresponding detailed classical microstate description is completely immaterial for determining {\it the stationary globally averaged properties of equilibrium systems}. As a ``canonical'' textbook example, the conventional Gibbs-Boltzmann distribution yields the expectation values in the macrostate of an open system defined by only the temperature, volume, and particle number. In integrable systems, additional global (macro) state variables associated with the integrals of motion may appear (as captured by the Generalized-Gibbs ensemble \cite{GGE1,eth3,GGE3}). The existence of unimportant particulars of a very different and, in a sense, much broader sort is further underscored by the universal nature of critical points and applications of the Renormalization Group in iteratively removing microscopic details in order to obtain the fixed point infrared description of the system that depends only on universality class features such as the spatial dimensionality, number of internal order parameter components, and underlying symmetries of the system. The redundancy of microscopic information for the global average of local observables in many body systems was, far more recently, brought to the fore by the Eigenstate Thermalization Hypothesis \cite{eth1,eth2,eth3,eth4,rigol,pol,polkovnikov1,polkovnikov2,srednicki-95}. According to this hypothesis, the expectation values of local observables in trivially stationary many body quantum eigenstates of a given Hamiltonian can be replaced by a far more compact description associated with an effective thermal macrostate in the system defined by the same Hamiltonian. Generally, a many body thermal probability density matrix (that we will write henceforth as $\rho_{\Lambda}$) defines the detailed quantum state of the system $\Lambda$ (whether open or closed). Similar to the classical microstates, this density matrix is not unique. There are multiple quantum many body states (or classical microstates) that all evolve under the same global time independent Hamiltonian $H_{\Lambda}$; the commutator (as well as the corresponding classical Poisson brackets) does not vanish, $\left[H_{\Lambda}, \rho_{\Lambda}\right] \neq 0$. However, although they are different, all of these systems share the same stationary equilibrium expectation values of all local observables. The canonical density matrix, 
\begin{equation}
\label{rcanonical}
\rho_{\Lambda}^{\sf{canonical }}=\frac{e^{-\beta H_{\Lambda}}}{Z_{\Lambda}},
\end{equation}
with the partition function
\begin{equation}
\label{eq:ZL}
Z_{\Lambda}={\sf{Tr}}\left(e^{-\beta H_{\Lambda}}\right),
\end{equation}
where $\beta = \frac{1}{k_{B} T}$ denotes the inverse temperature, yields these stationary global (i.e., macrostate) averages in open thermal systems. That is, unlike $\rho_{\Lambda}$, trivially, of course, the commutator $\left[H_{\Lambda}, \rho_{\Lambda}^{\sf {canonical }}\right]=0$ and all expectation values are stationary for a time independent Hamiltonian. The microstate-to-microstate variability of local observables in equilibrium is also encoded in the canonical ensemble itself in the form of the fluctuations about the average. In ergodic systems, the long time averages of local observables in all equilibrium systems sharing the same state variables become equal to those computed with the single density matrix $\rho_{\Lambda}^{\sf canonical}$. That is, formally, the replacement of density matrices,
\begin{eqnarray}
\label{long-time-eq}
\rho_{\Lambda}^{\sf {canonical }} \rightarrow \rho_{\Lambda}^{\sf {long }-\sf { time-average }} \nonumber
\\ \equiv \lim _{{\cal{T}} \rightarrow \infty} \frac{1}{{\cal{T}}} \int_{0}^{{\cal{T}}} d t^{\prime} e^{-i H_{\Lambda} t^{\prime} / \hbar} \rho_{\Lambda} e^{i H_{\Lambda} t^{\prime} / \hbar},
\end{eqnarray}
leads to no change when computing an average of a local observable. In the spirit of the above lightning review of microstates, macrostates, and state variables, what we will refer to as ``equilibrium averages'' of various local observables in the current work will be defined as the system wide global averages of these quantities. As we just noted, in ergodic systems, the equilibrium averages of local observables also become equal to the long time averages of these quantities (and correspond to those in an ergodic sector of fixed integrals of motion or quantum numbers if symmetry breaking occurred \cite{Nigel}). The off-diagonal matrix elements of $\rho_{\Lambda}$ in the eigenbasis of $H_{\Lambda}$ are those that endow the system with dynamics. Thermalization occurs when the system observables are, effectively, drawn at random from the probability distribution associated with $ \rho_{\Lambda}^{\sf {long }-\sf { time-average }}$. Analogously, in a classical Markov chain type framework, given a constant transition (or so-called stochastic) matrix $W$ iteratively evolving the system state from one time slice to the next, equilibration onsets when the probabilities of being in and transitioning between one microstate to another satisfy detailed balance and become time independent. The equilibrium state can be derived from the steady fixed point values of $\lim_{t \to \infty} W^{t}$ (when this limit exists). Although, at sufficiently long times $t$ in ergodic systems, the probabilities of being in different microstates become independent of the initial state, unless the initial microstate was special \footnote{If the transition matrix $W$ is special in that it has a microstate (a unit vector in the basis of micro states in which $W$ is defined) as an eigenstate and if the initial state is such a microstate then the system will remain trivially stationary.}, the system will keep varying in time with transition rates set by the time independent $W$. The stationarity of the probability distribution of being in different microstates in equilibrium should not be confused as to imply a stationarity of any particular microstate that may describe the classical system. Only the probabilities of being different microstates no longer change at late times $t$. Dynamics in the form of transitions between the different microstates are always present. Similar conclusions may be directly drawn for particle systems by writing down the classical equations of motion and simulating their dynamics. Similarly, in the quantum setting, unless the initial state $\rho_{\Lambda}$ was special (e.g., a projection to an eigenstate or a more general function of $H_{\Lambda}$), the system will continuously evolve in time. In Section \ref{sec:off}, we comment on the values of the off-diagonal matrix elements (in the energy eigenbasis)
of local observables need to assume in order to exhibit dynamics.

Henceforth, in order to make the classical analogs of our results and their underlying physics (including the equations of motion) more transparent, we will typically work in the Heisenberg picture. We will consider arbitrary local operators that we will denote by $Q^H_{i}$ in the large (possibly divergent in size in the thermodynamic limit) system $ \Lambda$ (of a number of particles or spins, etc., $N_{\Lambda}$) and write simple general equations that apply to all such local operators. For the stationary Hamiltonian $H_{\Lambda}$, the Heisenberg picture evolved local operators become 
\begin{equation}
\label{evolveeq}
Q_{i}^{H}(t)=e^{i H_{\Lambda} t / \hbar} Q_{i} e^{-i H_{\Lambda} t / \hbar},
\end{equation}
with $\{Q_i\}$ the Schrodinger picture operators. We will assume that the system is disorder free and thus that $H_{\Lambda}$ is translationally invariant. We will denote the global spatial averages by
\begin{eqnarray}
\label{O:avg}
\overline{Q} \equiv \frac{1}{N_{\Lambda}} \sum_{i=1}^{N_{\Lambda}}\left\langle Q_{i}^{H}(t)\right\rangle \equiv \frac{1}{N_{\Lambda}} \sum_{i=1}^{N_{\Lambda}} {\sf{Tr}}\left(\rho_{\Lambda} Q_{i}^{H}(t)\right).
\end{eqnarray} 
In an open thermal system, following the discussion at the start of this Section, we may trivially replace the latter average with $\frac{1}{N_{\Lambda}} \sum_{i=1}^{N_{\Lambda}} {\sf{Tr}}\left(\rho_{\Lambda}^{\sf{canonical }} Q_{i}^{H}\right)$. Formally, if the global average of the reduced one (or few) body 
density matrix is equal to the canonical density matrix, $\frac{1}{N_{\Lambda}} \sum_{i} \rho_{i}^{\sf{r e d.}}=\rho_{i}^{\sf{red.~ canonical }}$ then we may, trivially, replace the spatial average of local expectation values in an equilibrium system by those computed with the canonical ensemble density matrix. Here, the reduced density matrices $\rho_{i}^{\sf{red.}} \equiv {\sf{Tr}}_{\Lambda-i}\left(\rho_{\Lambda}\right)$ and $\rho_{i}^{\sf {red.~ canonical }} \equiv {\sf{Tr}}_{\Lambda-i}\left(\rho_{\Lambda}^{\sf {canonical }}\right)$ where ${\sf{Tr}}_{\Lambda-i}$ denotes a trace over all of the Hilbert space spanned by $\Lambda$ apart the subspace associated with the support of $Q_i$. This global expectation value in the thermal system is trivially stationary. In Section \ref{sec:stress}, we will further make use of the fact that not only does the global average of Eq. (\ref{O:avg}) approach, in the thermodynamic limit, the canonical ensemble average but that, more generally, any average over a macroscopic number of particles $i$ (with that number diverging in thermodynamic limit) will become equal to the thermal equilibrium ensemble expectation value.  

Notwithstanding the classical intuition that we just reviewed (expected for quantum systems having a classical description), some physicists might define equilibrated thermal systems as having the (trivially stationary) canonical density matrix. Pragmatically, from a technical standpoint, equilibrium dynamics computations (including those employing the Keldysh formalism, thermofield double representation of the canonical density matrix, general related Kubo-Martin-Schwinger conditions, and much else) \cite{das, KMS1,KMS2,KMS3} are typically centered around the canonical density matrix. On a conceptual level, the Eigenstate Thermalization Hypothesis  \cite{eth1,eth2,eth3,eth4,rigol,pol,polkovnikov1,polkovnikov2,srednicki-95} posits that in many thermal systems, the (typically highly entangled) many body eigenstates of the Hamiltonian defining the system readily reproduces the full global equilibrium average. Viewed through such lens, talking about time dependent expectation values of observables may sound a little bit strange- the time derivatives of all such expectation values computed with $\rho_{\Lambda}^{\sf {canonical }}$ vanish and any inequalities involving these (such as those of Eq. (\ref{central1})) become trivial. To further highlight the generality of our results and their physical content also in such idealized situations, we will illustrate (Section \ref{high-moment-derivative}) that not only are the expectation values of ${\sf Tr} \Big(\rho_{\Lambda} \frac{dQ_{i}^{H}}{dt} \Big)$ bounded but that also all higher moments of any such derivatives ${\sf Tr} \Big(\rho_{\Lambda} (\frac{dQ_{i}^{H}}{dt})^{n} \Big)$ satisfy similar inequalities for any power $n>1$. For even $n$, the latter expectation values (similar to the squared velocities in the earlier discussed example of the gas) will become finite also when considering the idealized situation in which the density matrix of the single thermal system is set, by fiat, equal to that of the canonical ensemble (i.e., we will derive inequalities that become nontrivial for $~{\sf Tr} \Big(\rho_{\Lambda}^{\sf {canonical }} (\frac{dQ_{i}^{H}}{dt})^{n} \Big)$ when $n$ is even). Our discussion in Section \ref{high-moment-derivative} will largely be semiclassical. To further illustrate that {\it semiclassical approximations are not necessary} we will, in Sections \ref{t-auto},\ref{sec:diff} and \ref{GsKsec}, first derive and employ bounds on the time derivatives of the equilibrium autocorrelation functions. The time derivatives of the canonical ensemble autocorrelation functions also do not trivially vanish. In general thermal systems harboring spatially decaying connected correlations, ensemble averages can be seen to reproduce the global expectation values \footnote{If particle (or site) $i \equiv i_0$ is decoupled from a distant similar particle $i_{1}$ which in turn is
  	decoupled from $i_{2}, \cdots i_{n'}$, then we may view $i, i_{1}, i_{2}, \ldots, i_{n'}$ as belonging to
  	different realizations of a system $\Lambda$.
  	The ensemble average over these realizations of $\Lambda$ will coincide with the global average of the expectation value
  	of $Q^H$ at sites $i, i_{1}i_{2}, \ldots$ .
  	In such situations in which the correlation length is finite
  	and particles (or sites) $i, i_{1}, i_{2}, \ldots, i_{n'}$ are decoupled,
  	we can see how relations such as
  	\begin{equation}
  		{\sf{Tr}}\left(\rho_{\Lambda}^{\sf {canonical }} Q^H_{i}\right)=\lim _{n' \rightarrow \infty} \frac{1}{n'} \sum_{i=0}^{n'} {\sf{Tr}}\left(\rho_{\Lambda} Q^H_{i_{n'}}\right)
  \end{equation}
  are obtained.}.

We reiterate that in typical semiclassical thermal systems, local averages are indeed not stationary,
\begin{equation}
\left\langle Q_{i}^{H}(t)\right\rangle \equiv {\sf{Tr}}\left(\rho_{\Lambda} Q_{i}^{H}(t)\right) \neq {\sf{Tr}}\left(\rho_{\Lambda}^{\sf{canonical}} Q_{i}^{H}(t)\right),
\end{equation}
while, by contrast, global spatial (and/or long time and/or the ensemble) averages are given by stationary canonical ensemble probability density.  As noted above, in translationally invariant ergodic systems, the global averages of Eq. (\ref{O:avg}) of local observables are also equal to their long time averages,
\begin{equation}
\label{cal-long}
{\overline{Q}} = {\sf{Tr}}\left(\rho_{\Lambda}^{\sf{canonical }} Q 
\right) = {\sf{Tr}}\left(\rho_{\Lambda}^{\sf {long }-\sf { time }-\sf { average }} Q \right),
\end{equation}
computed with the density matrix $\rho_{\Lambda}^{\sf {long }-\sf { time }-\sf { average }}$ 
of Eq. (\ref{long-time-eq}). For this reason we will, with some abuse of notation, also denote this long time average system by the same shorthand ${\overline{Q}}$ that we use to indicate the global average (Eq. (\ref{O:avg})). The equality of Eq. (\ref{cal-long}) between the long time average and the equilibrium average may also hold for non translationally invariant ergodic systems.

We next define $\tilde{H}_{i}^{H} \subset H_{\Lambda}$ to be the set of all terms in 
$H_{\Lambda}$ that do not commute with $Q_{i}^{H}=e^{i H_{\Lambda} t / \hbar} Q_{i} e^{-i H_{\Lambda} t / \hbar}$ and thus (by Heisenberg's equation of motion) contribute to its time derivative,
\begin{equation}
\label{eq:HM}
\frac{d Q_{i}^{H}}{d t}=\frac{i}{\hbar}\left[\tilde{H}_{i}^{H}, Q_{i}^{H}\right],
\end{equation}
and endow $Q^H_{i}$ with dynamics. In systems with local interactions, $\tilde{H}_{i}^{H}$ may have its support on a bounded spatial region and include a finite number of particles/spins/... . With the usual replacement of the commutator of Eq. (\ref{eq:HM}) by Poisson brackets that yields Hamilton's equations of motion, the above discussion trivially applies  to classical systems. An identification of minimal set of terms (those in $\tilde{H}_{i}^{H}$) not commuting with a local observable and thus driving its dynamics is also necessary for proving the Lieb-Robinson bounds \cite{Lieb_Robinson, Bruno, Sergey,alioscia,kaden,Else20}. Conceptually related is also the notion, in Statistics and Machine Learning, of only a small subset of variables (a ``Markov blanket''  \cite{Markovblanket}) that is relevant to a local variable. The full system Hamiltonian is identically the same in the Heisenberg representation, $H_{\Lambda}^{H}=H_{\Lambda}$. 

In several system types that we will study (in Sections \ref{decoupled}, \ref{sec:lowT} and Appendices \ref{sec:RP}), we will be able to express the global Hamiltonian as a sum of related local terms,
\begin{eqnarray}
\label{eq:decoupled+sum}
H_{\Lambda}=\sum_{i^{\prime}=1}^{N_{\mathrm{\Lambda}}^{\prime}} \tilde{H}_{i^{\prime}}^{H}.
\end{eqnarray}
In Eq. (\ref{eq:decoupled+sum}), we added the prime superscript to indicate that the number of terms $N_{\Lambda}^{\prime}$ in this decomposition of the global system Hamiltonian $H_{\Lambda}$ into local Hamiltonians may generally differ from the total number $N_{\Lambda}$ of particles in the system $\Lambda$; the local Hamiltonians $\{\tilde{H}_{i^{\prime}}^{H}\}_{i'=1}^{N_{\Lambda^{\prime}}}$ are not necessarily associated with the $N_{\Lambda}$ individual particles $i$. We stress, however, that in bulk of the current work, we will largely focus on general systems that {\it are {\underline{not}} limited to Hamiltonians of the particular form of Eq. (\ref{eq:decoupled+sum})}. 

Most of our results will rely on a trivial extension of the variance type uncertainty inequalities to mixed states having a density matrix,
\begin{eqnarray}
\label{eq:AB}
\! \! \! \! \! \! \! \! \! \! {\sf{Tr}}\left(\rho_{\Lambda}(\Delta A)^{2}\right) {\sf{Tr}}\left(\rho_{\Lambda}(\Delta B)^{2}\right) \geq \frac{1}{4}\left|{\sf{Tr}}\left(\rho_{\Lambda}[A, B]\right)\right|^{2}. 
\end{eqnarray}
In Eq. (\ref{eq:AB}), $\rho_{\Lambda}$ is a general density matrix, and $A$ and $B$ are arbitrary operators with $\Delta A \equiv A - {\sf Tr} (\rho_{\Lambda} A)$ and with a similar definition for $\Delta B$. 
We may prove this inequality by, e.g., extending the Araki-Lieb purification (or thermofield double) construct 
\cite{lieb-pure,das} to replace density matrix averages by those in pure states. In Appendix \ref{sec:thermofield}, we expand on this procedure in more detail (see also \cite{bound}). As we emphasized above, our focus in the current work will be on thermal density matrices $\rho_{\Lambda}$ associated with open systems. For the derivation of the temporal bounds, we will choose $A$ to be $\tilde{H}_i^{H}$ and $B$ to be $Q_{i}^{H}$ so as to obtain a simple extension of the standard energy-time uncertainty relations to general local observables in such open many body systems, i.e.,
\begin{eqnarray}
\label{eq:uncertain:Q}
{\sf{Tr}}\left(\rho_{\Lambda}(\Delta \tilde{H}_{i}^{H}(t))^{2}\right) {\sf{Tr}}\left(\rho_{\Lambda}\left(\Delta Q_{i}^{H}(t)\right)^{2}\right) \nonumber
\\  \geq \frac{1}{4}\left|{\sf{Tr}}\left(\rho_{\Lambda}\left[\tilde{H}_{i}^{H}(t), Q_{i}^{H}(t)\right]\right)\right|^{2}.
\end{eqnarray}
For specificity, we will, in this work, examine systems with a fixed particle number. By working with other ensembles, our results can extended to other open systems in which not only is energy exchanged with the environment but also the particle number, the volume, etc., may fluctuate as well. 

Lastly, in what follows, we will often employ $\langle\cdot\rangle$ to denote averages computed with $\rho_{\Lambda}$.
These averages correspond to the physically measured expectation values. For similar reasons, all standard deviations $\sigma_Q$ of general operators $Q$ denote those computed with $\rho_{\Lambda}$.

\section{Derivation of time derivative bounds by averaging the uncertainty relations}
\label{derive:time}

In this Section, we will develop the general machinery that we will employ in order to arrive at universal bounds on the rates of change of general quantities. While our derivations are very simple, they might nonetheless seem abstract. To make our results more lucid, complementing the universal inequalities that we derive for various observables in the following Sections, in Appendix \ref{sec:toy}, we analyze the corresponding bounds for two simple toy models: a harmonic solid (Appendix \ref{sec:harmonic}) and an XY spin model (Appendix \ref{sec:XY}).

\subsection{Exact bounds on the expectation values of single operator temporal derivatives}
\label{bound-1t}

We will now explicitly arrive at the local thermal uncertainty relation of Eq. (\ref{central1}) by a sequence of three steps.
We will first (i) write the equations of motion (Eq. (\ref{eq:HM})), then (ii) invoke the variance uncertainty relations of Eq. (\ref{eq:uncertain:Q}) for general local observables in thermal many body systems and, lastly, employ (iii) the defining property (Eq. (\ref{O:avg})) of the thermal equilibrium macrostate. Elements of this calculation appear in \cite{bound}. For simplicity, in what follows, we consider a system of distinguishable particles. As we explain in Appendix \ref{sec:identical}, similar conclusions follow for identical particles.  

We follow the algorithm outlined above. Writing (step (i) of the above mentioned recipe) the equations of motion of Eq. (\ref{eq:HM}) for the rate of change of a general local operator and then bounding the resultant expectation value 
(step (ii)) using Eq. (\ref{eq:uncertain:Q}) yields 
\begin{eqnarray}
\label{first-eq}
&& \Big|\left\langle\frac{d Q_{i}^{H}}{d t}\right\rangle \Big|^{2} \equiv\Big|{\sf{Tr}}\left(\rho_{\Lambda} \frac{d Q_{i}^{H}}{d t}\right)\Big|^{2} \nonumber
\\ && =\Big|{\sf{Tr}}\left(\rho_{\Lambda}\left(\frac{i}{\hbar}\left[H_{\Lambda}, Q_{i}^{H}(t)\right]\right)\right)\Big|^{2} \nonumber
\\ &&=\left|{\sf{Tr}}\left(\rho_{\Lambda}\left(\frac{i}{\hbar}\left[\tilde{H}_{i}^{H}(t), Q_{i}^{H}(t)\right]\right)\right)\right|^{2} \nonumber
\\ && \leq \frac{4}{\hbar^{2}} \sigma_{\tilde{H}_{i}^{H}(t)}^{2} \sigma_{Q_{i}^{H}(t)}^{2}.
\end{eqnarray}
Trivially rearranging Eq. (\ref{first-eq}),
\begin{equation}
\label{timel}
\sigma_{\tilde{H}_{i}^{H}(t)}^{2} \geq \frac{\hbar^{2}}{4} \frac{\left({\sf{Tr}}\left(\rho_{\Lambda} \frac{d Q_{i}^{H}}{d t}\right)\right)^{2}}{\sigma_{Q_{i}^{H}(t)}^{2}}.
\end{equation}
Averaging the last inequality over all $N_{\Lambda}$ particles (or sites in a spin system), 
\begin{eqnarray}
\label{eq:intermediaten}
\frac{1}{N_{\Lambda}} \sum_{i=1}^{N_{\Lambda}} \sigma_{\tilde{H}_{i}^{H}(t)}^{2} 
\geq \frac{\hbar^{2}}{4 N_{\Lambda}} \sum_{i=1}^{N_{\Lambda}} 
\frac{\left({\sf{Tr}}\left(\rho_{\Lambda} \frac{d Q_{i}^{H}}{d t}\right)\right)^{2}}{\sigma_{Q_{i}^{H}(t)}^{2}}.
\end{eqnarray}
We now convert (step (iii)) the time-energy uncertainty relation of Eq. (\ref{eq:intermediaten}) to a thermal equilibrium macrostate inequality. Recalling that the average over all $N_{\Lambda}$ particles in Eq. (\ref{eq:intermediaten}) reproduces the canonical ensemble average in disorder free thermal systems (Eqs. (\ref{O:avg}, \ref{cal-long})), we may write the lefthand side of Eq. (\ref{eq:intermediaten}) as 
\begin{eqnarray}
\frac{1}{N_{\Lambda}} \sum_{i=1}^{N_{\Lambda}} {\sf{Tr}}\left(\rho_{\Lambda}^{\sf {canonical }}\left(\Delta \tilde{H}_{i}^{H}(t)\right)^{2}\right) \nonumber
\\ =\frac{1}{N_{\Lambda}} \sum_{i=1}^{N_{\Lambda}} {\sf{Tr}}\left(\rho_{\Lambda}\left(\Delta \tilde{H}_{i}^{H}(t)\right)^{2}\right).
\end{eqnarray}
Thus, in a thermal system, the local time-energy uncertainty relation of Eq. (\ref{eq:intermediaten}) can be recast as
\begin{eqnarray}
\label{a-central-1}
\frac{1}{N_{\Lambda}} \sum_{i=1}^{N_{\Lambda}} {\sf{Tr}}\left(\rho_{\Lambda}^{\sf {canonical }}\left(\Delta \tilde{H}_{i}^{H}(t)\right)^{2}\right) \nonumber
\\  \geq \frac{\hbar^{2}}{4 N_{\Lambda}} \sum_{i=1}^{N_{\Lambda}} \frac{\left({\sf{Tr}}\left(\rho_{\Lambda} \frac{d Q_{i}^{H}}{d t}\right)\right)^{2}}{{\sf{Tr}}\left(\rho_{\Lambda}\left(\Delta Q_{i}^{H}(t)\right)^{2}\right)}.
\end{eqnarray}
In a translationally invariant system with local Hamiltonians $\{\tilde{H}_{i}^{H}(t)\}_{i=1}^{N_{\Lambda}}$ that are of an identical form for all $i$, the traces ${\sf{Tr}}\left(\rho_{\Lambda}^{\sf {canonical }}\left(\Delta \tilde{H}_{i}^{H}(t)\right)^{2}\right)$ are the same for all $i$. Using the shorthand symbol defined in Eq. (\ref{O:avg}),
\begin{eqnarray}
\label{overdef}
\overline{\mathcal{O}} \equiv \frac{1}{N_{\Lambda}} \sum_{i=1}^{N_{\Lambda}}\left\langle\mathcal{O}_{i}^{H}(t)\right\rangle \equiv \frac{1}{N_{\Lambda}} \sum_{i=1}^{N_{\Lambda}} {\sf{Tr}}\left(\rho_{\Lambda} \mathcal{O}_{i}^{H}(t)\right) \nonumber
\\ =\frac{1}{N_{\Lambda}} \sum_{i=1}^{N_{\Lambda}} {\sf{Tr}}\left(\rho_{\Lambda}^{\sf {canonical }} \mathcal{O}_{i}^{H}\right).
\end{eqnarray}
Eq. (\ref{a-central-1}) then becomes
\begin{equation}
\label{cv1}
k_{B} T^{2} C_{v,i} \geq \frac{\hbar^{2}}{4} \overline{\mathcal{O}}.
\end{equation}
 Here, the effective local heat capacity is given by
  earlier noted translationally invariant (particle index independent) canonical ensemble average
\begin{equation}
\label{cv2}
k_{B} T^{2} C_{v,i} \equiv {\sf{Tr}}\left(\rho_{\Lambda}^{\sf {canonical }}\left(\Delta \tilde{H}_{i}^{H}(t)\right)^{2}\right).
\end{equation}
In Eqs. (\ref{overdef},\ref{cv1}), $\overline{\mathcal{O}}$ denotes the global average of the square local relaxation rate given by
\begin{equation}
\label{cv3}
\mathcal{O}_{i} \equiv \frac{\left\langle\frac{d Q_{i}^{H}}{d t}\right\rangle^{2}}{\left(\sigma_{Q_{i}}^{H}(t)\right)^{2}}.
\end{equation}
Putting all of the pieces together, we arrive at Eq. (\ref{central1}). 
Following the discussion in Section \ref{setup}, the above set of steps may be repeated verbatim in order to further derive a variant of Eq. (\ref{central1}) in which the global average of Eq. (\ref{O:avg}) is replaced by a long time average computed with the probability density matrix $\rho_{\Lambda}^{\sf {long }-\sf { time-average }}$. We thus observe that there is a fundamental quantum limit on the rate
at which any observable $Q_{i}$
may change in an equilibrated
system at a temperature $T$. 
For a single (or more generally few particle) operator $Q^H_{i}$ having a bounded local Hamiltonian 
$\tilde{H}^H_{i} \subset H_{\Lambda}$ that fully generates its dynamics, the lower bound of Eq. (\ref{cv1}) is generally finite and becomes dominated, at vanishing temperatures, by zero point quantum fluctuations. At finite low temperatures, this variance is typically quantum suppressed relative to its semiclassical value. The precise value of Eq. (\ref{cv1}) may then, in such instance, lead to far tighter bounds on the low temperature dynamics than its semiclassical limit (where the effective heat capacity is evaluated classically). In what follows, in order to avoid cumbersome notation we will dispense with the explicit time argument of the Heisenberg picture operators unless necessary.  

In Sections \ref{decoupled}, \ref{sec:viscosity}, and \ref{sec:lowT}, as well as Appendix \ref{sec:RP}, we will discuss situations in which the thermodynamic constant volume heat capacity $C_{v}^{(\Lambda)}$ of the entire system $\Lambda$ appears instead of, or alongside, the effective local heat capacity of Eq. (\ref{cv2}). In particular, in Section \ref{decoupled}, and Appendix \ref{sec:RP}, we will explain how for this occurs for systems with separable Hamiltonians and in the broad class of Reflection Positive systems and, more broadly, theories with positive semi-definite connected correlation functions. The latter new findings augment earlier results \cite{bound} for {\it global observables} $Q$ in which the system thermodynamic heat capacity $C_{v}^{(\Lambda)}$ appears in bounds. \footnote{Specifically, in \cite{bound} it was demonstrated that the relative rate of change at which general global observables $Q$ may vary is bounded is set by the thermodynamic heat capacity, 
\begin{eqnarray}
 \label{Qsee}
 \tau^{-1}_{Q} \equiv \frac{|\langle \frac{dQ}{dt} \rangle|}{\sigma_{Q}} \le \frac{2 T \sqrt{k_{B} C_{v}^{(\Lambda)}}} {\hbar}.
 \end{eqnarray}
 This global relaxation rate is the counterpart of the local relaxation rate of Eq. (\ref{cv3}). Since the global constant volume heat capacity $C_{v}^{(\Lambda)} = T (\frac{\partial S^{(\Lambda)}}{\partial T})_{v}$, this bound is set by the temperature derivative of the system entropy $S^{(\Lambda)}$ (or information content of $\Lambda$ as measured by the 
 Shannon entropy that differs from $S^{(\Lambda)}$ by a trivial constant multiplicative factor (of $1/(k_{B} \ln 2)$)). 
 Similarly, in equilibrated systems, the uniform susceptibility of the global quantity $Q$ is bounded by a value that is further set by the global heat capacity, 
\begin{eqnarray}
\label{fluct-eqq}
\chi_{Q} \ge \frac{\hbar^{2}}{4 k_{B}^{2} T^{3} C_{v}^{(\Lambda)}} \Big| \frac{dQ}{dt}  \Big|^{2}.
\end{eqnarray}}

\subsection{Semiclassical Bounds on Expectation Values of General Moments of Temporal Derivatives}
\label{high-moment-derivative}

We next extend the exact results of Section \ref{bound-1t} to general moments of the time derivatives, ${\sf Tr} \Big(\rho_{\Lambda} (\frac{dQ_{i}^{H}}{dt})^{n} \Big)$ for arbitrary powers $n$. We first consider the case $n=2$. The generalization for general $n$ will then be apparent. Unlike the exact analysis of Section \ref{bound-1t}, the final results that we derive will involve a semiclassical approximation.

We start by trivially invoking the Heisenberg equation of motion for the expectation value of the square of the time derivative of a general local observable, 
\begin{eqnarray}
\label{longABABA}
\Big\langle \Big ( \frac{dQ_{i}^{H}}{dt} \Big)^2  \Big\rangle = \frac{1}{\hbar^2} \Big| \Big\langle  
[\tilde{H}^{H}_{i},Q^H_i][\tilde{H}^{H}_{i},Q^H_i]
\Big\rangle \Big| \nonumber
\\ =  \frac{1}{\hbar^2} \Big| \Big\langle  
[ \Delta \tilde{H}^{H}_{i}, \Delta Q^H_i][ \Delta \tilde{H}^{H}_{i}, \Delta Q^H_i] 
\Big\rangle \Big| \nonumber
\\ \le \frac{4}{\hbar^2} \max \Big \{  \Big| \Big\langle  \Delta \tilde{H}^{H}_{i}  \Delta Q^H_i   \Delta \tilde{H}^{H}_{i}  \Delta Q^H_i  \Big\rangle \Big|, \nonumber
\\ \Big| \Big\langle    \Delta Q^H_i   \Delta \tilde{H}^{H}_{i}  \Delta Q^H_i  \Delta \tilde{H}^{H}_{i} \Big\rangle \Big|, 
\nonumber
\\ \Big| \Big\langle    \Delta Q^H_i   (\Delta \tilde{H}^{H}_{i})^2  \Delta Q^H_i  \Big\rangle \Big|,
\nonumber
\\ \Big| \Big\langle   \Delta \tilde{H}^{H}_{i} ( \Delta Q^H_i)^2  \Delta \tilde{H}^{H}_{i} \Big\rangle \Big| \Big \},
\end{eqnarray}
where, as throughout, $\Delta \tilde{H}^{H}_i \equiv \tilde{H}^{H}_{i} - \langle  \tilde{H}^{H}_i \rangle$ and $\Delta  Q_i^H \equiv
Q_i^H - \langle Q_i^H \rangle$. 
A short calculation (see Appendix \ref{QUB}), already {\it before the semiclassical limit is taken}, illustrates that a scaled sum of the last two expectation values in Eq. (\ref{longABABA}) provides a general bound. In the semiclassical limit, $  \Big\langle  \Delta \tilde{H}^{H}_{i}  \Delta Q^H_i   \Delta \tilde{H}^{H}_{i}  \Delta Q^H_i  \Big\rangle =  \Big \langle \Delta Q^H_i   \Delta \tilde{H}^{H}_{i}  \Delta Q^H_i  \Delta \tilde{H}^{H}_{i} \Big\rangle  =  \Big\langle    \Delta Q^H_i   (\Delta \tilde{H}^{H}_{i})^2  \Delta Q^H_i  \Big\rangle =
 \Big\langle   \Delta \tilde{H}^{H}_{i} ( \Delta Q^H_i)^2  \Delta \tilde{H}^{H}_{i} \Big\rangle$. 

As a matter of principle, Eq. (\ref{longABABA}) and its semiclassical limit hold for arbitrary local operators in general systems $\Lambda$ whether or not these systems happen to be in equilibrium. In equilibrated thermal systems, we may, as in Section \ref{bound-1t}, average these inequalities over the entire system so as to replace expectation values computed with $\rho_{\Lambda}$ by those evaluated with the canonical density matrix $\rho^{\sf canonical}_{\Lambda}$. The resulting canonical averages may be expressed, in the semiclassical limit, as phase space averages with the probability density $\rho^{\sf canonical}_{\Lambda}$ replaced by its corresponding classical form ($\rho_{\Lambda}^{\sf classical ~canonical}$). The classical canonical probability density $\rho_{\Lambda}^{\sf~classical ~canonical}$ factorizes, in phase space, to a part depending on the momenta alone and that depending on positions (i.e., classically, $e^{-\beta H_{\Lambda}} = e^{-\beta \sum_{i=1}^{N_{\Lambda}} \frac{{\bf p}_{i}^{2}}{2m}} e^{-\beta V_{\Lambda}(\{{\bf r}_{i}\})}$). This simple factorization property will prove to be useful for deriving various semiclassical bounds. In (i) Sections \ref{v-bounds} and \ref{sec:stress} and (ii) Section \ref{a-bounds}, in order to obtain (ia) velocity or (ib) the time derivative of the acceleration and (ii) acceleration bounds for a particle of mass $m$ in the absence of a magnetic field and for two-body position dependent potential energies ${\cal V}\left({\bf r}_{i}^{H}, {\bf r}_{j}^{H}\right)$, we will, respectively, (i) set (ia) $Q^H_{i}= r_{i \ell}^{H}$ (a position vector component of a single particle $i$) or (ib) its acceleration in Section \ref{sec:stress} and $\tilde{H}_{i}^{H} = \frac{({\bf p}^{H}_{i})^2}{2m}$ (or a sum of similar kinetic terms in Section \ref{sec:stress}) and (ii) set $Q_{i}^{H} = v_{i \ell}^{H}$ (a velocity component) and $\tilde{H}_{i}^{H}=\sum_{j} {\cal{V}} \left({\bf r}_{i}^{H}, {\bf r}_{j}^{H}\right)$. In the semiclassical limit, in any of these cases, 
\begin{eqnarray}
\label{longABABA'}
&& \frac{1}{N_{\Lambda}} \sum_{i=1}^{N_{\Lambda}} \Big| \Big\langle  \Delta \tilde{H}^{H}_{i}  \Delta Q^H_i   \Delta \tilde{H}^{H}_{i}  \Delta Q^H_i  \Big\rangle \Big| \nonumber
 \\ && = {\sf{Tr}} (\rho_{\Lambda}^{\sf classical ~canonical} (\Delta \tilde{H}^{H}_{i})^2 (\Delta Q_{i}^{H})^{2}) \nonumber
 \\ &&= {\sf{Tr}} (\rho_{\Lambda}^{\sf classical~ canonical} (\Delta \tilde{H}^{H}_{i})^2) \nonumber
 \\  && \times {\sf{Tr}} (\rho_{\Lambda}^{\sf classical~ canonical} (\Delta Q^{H}_{i})^2), \nonumber
 \\ && \equiv k_{B} T^{2} C_{v,i}  {\sf{Tr}} (\rho_{\Lambda}^{\sf classical~ canonical} (\Delta Q^{H}_{i})^2).
 \end{eqnarray}
In the second equality of the above equation, we made explicit the factorization (that arises for the above noted cases (ia), (ib), and (ii)) into two parts: one factor depending on momenta alone and another factor depending solely on particle positions. A factorization similar to that of Eq. (\ref{longABABA'}) also trivially (and more generally) arises for arbitrary $\Delta \tilde{H}_{i}^{H}$ and $Q^{H}_{i}$ that are semiclassical yet do not necessarily depend only on the momenta or only on the spatial coordinates. However the resulting bounds will, normally, be slightly less restrictive (see Appendix \ref{triv-triv-long}).

Substituting Eq. (\ref{longABABA'}) into Eq. (\ref{longABABA}) yields
 \begin{eqnarray}
\label{squared-ineq}
&& \frac{1}{N_{\Lambda}} \sum_{i=1}^{N_{\Lambda}} \Big\langle \Big ( \frac{dQ_{i}^{H}}{dt} \Big)^2  \Big\rangle \nonumber
\\ && \le \frac{4 k_{B} T^{2} C_{v,i} }{\hbar^2}  {\sf{Tr}} (\rho_{\Lambda}^{\sf classical~ canonical} (\Delta Q^{H}_{i})^2).
\end{eqnarray} 
Here, the effective local ``heat capacity'' $ C_{v,i}$ is 
defined by the energy fluctuation variance of Eq. (\ref{cv2}). As detailed in Appendix \ref{QUB}, an analogous relation holds for {\it the exact (i.e., non semiclassical)} averages 
with deformed density matrices (see Eqs. (\ref{ddensity}, \ref{variantquad}) therein).

 The generalization of Eqs. (\ref{longABABA'}, \ref{squared-ineq}) to general higher moments $n>2$ is rather straightforward. Explicitly,
 \begin{eqnarray}
 \label{semi-class-bound-0}
&& \frac{1}{N_{\Lambda}} \sum_{i=1}^{N_{\Lambda}} \Big| \Big\langle  \Big(\frac{dQ_{i}^{H}}{dt} \Big)^{n} \Big\rangle  \Big| = \frac{1}{\hbar^n N_{\Lambda}} \sum_{i=1}^{N_{\Lambda}}\Big| \Big\langle  
[\tilde{H}^{H}_{i},Q^H_i]^n \Big\rangle \Big| \nonumber
\\ && =  \frac{1}{\hbar^n N_{\Lambda}} \sum_{i=1}^{N_{\Lambda}} \Big| \Big\langle  
\Big([ \Delta \tilde{H}^{H}_{i}, \Delta Q^H_i]\Big)^n
\Big\rangle \Big| \nonumber
\\&&  \le \frac{2^n}{\hbar^n N_{\Lambda}} \sum_{i=1}^{N_{\Lambda}}\max \Big \{  \Big| \Big\langle \cdots  \Delta \tilde{H}^{H}_{i}   \Delta \tilde{H}^{H}_{i} \Delta Q^H_i \Delta Q^H_i  \cdots \Big\rangle \Big|, \cdots, \nonumber
\\ && \Big| \Big\langle  \cdots \Delta Q^H_i   \Delta Q^H_i   \Delta \tilde{H}^{H}_{i}  \Delta \tilde{H}^{H}_{i}  \cdots \Big\rangle \Big| \Big \} \nonumber
\\ && =  \Big(\frac{2}{\hbar}\Big)^n  {\sf Tr}\Big(\rho_{\Lambda}^{\sf classical~canonical} |( \Delta Q^H_i)^n ( \Delta \tilde{H}^H_i)^n| \Big) \nonumber
\\ && = \Big(\frac{2}{\hbar}\Big)^n {\sf Tr}\Big(\rho_{\Lambda}^{\sf classical~canonical} |( \Delta Q^H_i)|^n \Big) \nonumber
\\ && \times {\sf Tr}\Big(\rho_{\Lambda}^{\sf classical~canonical} |(\Delta \tilde{H}^H_i)|^n \Big).
\end{eqnarray}
In the last line, we invoked, once again, the factorization of the classical canonical ensemble phase space average. The maximization on the third line refers to that amongst all $2^n$ expectation values having $n$ factors of $\tilde{H}^{H}_i$ and $n$ factors of $Q_{i}^{H}$ in those product strings of total length $(2n)$ that are formed by writing $([\Delta \tilde{H}^{H}_{i},\Delta Q^H_i])^n$ as a sum of monomials. With the replacement 
 $ \frac{1}{N_{\Lambda}} \sum_{i=1}^{N_{\Lambda}}  \Big\langle  \Big(\frac{dQ_{i}^{H}}{dt} \Big)^{2 } \Big\rangle \to \frac{1}{N_{\Lambda}} \sum_{i=1}^{N_{\Lambda}}  \Big\langle  \Big(\frac{dQ_{i}^{H}}{dt} \Big) \Big\rangle^2$,   
the bound of Eq. (\ref{squared-ineq}) is, essentially, identical to the average of Eq. (\ref{first-eq}) over all of the $N_{\Lambda}$ particles in the system (which leads to Eq. (\ref{a-central-1}) and thus, as detailed in Section \ref{bound-1t}, to Eq. (\ref{central1})). Since for any probability density $\rho_{\Lambda}$, we have that
 $ \sum_{i=1}^{N_{\Lambda}}  \Big\langle  \Big(\frac{dQ_{i}^{H}}{dt} \Big) \Big\rangle^2  \le  \sum_{i=1}^{N_{\Lambda}}  \Big\langle  \Big(\frac{dQ_{i}^{H}}{dt} \Big)^{2 } \Big\rangle$,  
 it might seem that the bound of Eq. (\ref{squared-ineq}) automatically implies Eqs. (\ref{central1}, \ref{a-central-1}) and that thus the inequalities of Section \ref{bound-1t} are superfluous. We caution, however, that Eq. (\ref{squared-ineq}) is only valid in the semiclassical limit whereas Eqs. (\ref{central1}, \ref{a-central-1}) are always correct. 

When the local operators $\{Q_{i}^{H}\}$ are of bounded norm $||Q_{i}^{H}|| \le ||Q||$ (with the operator norm $\|Q_{i}^{H}\|$ denoting the largest absolute value amongst those of the eigenvalues of $Q_{i}^{H}$), the top line of Eq. (\ref{longABABA}) implies the general inequality
\begin{eqnarray}
\frac{1}{N_{\Lambda}} \sum_{i=1}^{N_{\Lambda}} \Big\langle \Big ( \frac{dQ_{i}^{H}}{dt} \Big)^2  \Big\rangle && \le \frac{4}{\hbar^2 N_{\Lambda}} \sum_{i=1}^{N_{\Lambda}} \Big \langle \Big( \Delta \tilde{H}^{H}_{i} \Big)^2 \Big \rangle  \Big| \Big|Q \Big| \Big|^2 \nonumber
\\ && \equiv  \frac{4 k_{B} T^{2} C_{v, i}}{\hbar^2} \Big| \Big|Q \Big| \Big|^2.
\label{bounded=Q=2} 
\end{eqnarray}
Similarly, for such bounded local operators,
\begin{eqnarray}
&& \frac{1}{N_{\Lambda}} \sum_{i=1}^{N_{\Lambda}} \Big\langle \Big ( \frac{dQ_{i}^{H}}{dt} \Big)^n  \Big\rangle \nonumber
\\&&  \le \frac{1}{N_{\Lambda}} \Big( \frac{2}{\hbar} \Big)^n \sum_{i=1}^{N_{\Lambda}} \Big \langle \Big( \Delta \tilde{H}^{H}_{i} \Big)^n \Big \rangle  \Big| \Big|Q \Big| \Big|^n \nonumber
\\ && = \Big( \frac{2}{\hbar}   \Big| \Big|Q \Big| \Big| \Big)^n  {\sf{Tr}} \Big(\rho_{\Lambda}^{\sf canonical} \Big( \Delta \tilde{H}^{H}_{i} \Big)^n \Big).
\label{bounded=Q=n} 
\end{eqnarray}

\subsection{Time derivatives of equilibrium autocorrelation functions}
\label{t-auto}

Now we come to basic bounds which we will repeatedly exploit for bounding transport coefficients. Unlike the inequalities that we derived in the earlier Sections, these inequalities are also non-trivial when evaluated with {\it explicitly stationary thermal density matrices} (including the canonical ensemble density matrix). Similar to the inequalities of Section \ref{bound-1t}, these bounds will {\it {\underline{not}} appeal to semiclassical considerations} and hold at all (including arbitrarily low) temperatures. Towards this end, we consider the autocorrelation function $\langle Q_i^{H}(t) Q^H_i(0) \rangle$ with a general Hermitian $Q_i$ and examine its dependence on the time separation $t$. From Heisenberg's equations of motion and the triangle inequality,
\begin{eqnarray}
\label{so-trivial}
&&~~~ \Big|\frac{d}{dt} \Big\langle Q_{i}^{H}(t) Q_{i}^{H}(0) \Big\rangle \Big| \nonumber
\\  && =  \Big|\frac{i}{\hbar} \Big\langle [\Delta \tilde{H}_{i}^{H}(t), Q^{H}_{i}(t)] Q_{i}^{H}(0) \Big\rangle \Big| \nonumber
\\ && \le \frac{1}{\hbar} \Big(  \Big|\Big\langle \Delta \tilde{H}_{i}^{H}(t) Q^{H}_{i}(t) Q_{i}^{H}(0) \Big\rangle\Big|+ \nonumber
\\  &&~~~~~~  \Big|\Big\langle Q_{i}^{H}(t)  \Delta \tilde{H}_{i}^{H}(t) Q^{H}_{i}(0) \Big\rangle \Big| \Big). 
\end{eqnarray}

We next apply an extension of the Cauchy-Schwarz inequality to general operators $A$ and $B$, and density matrices $\rho_{\Lambda}$ (see also Appendix \ref{sec:thermofield}), 
 \begin{eqnarray}
\label{CSg'}
|{\sf{Tr}}(\rho_{\Lambda} (A B))|^2 \le {\sf{Tr}}(\rho_{\Lambda} AA^{\dagger}) {\sf{Tr}}(\rho_{\Lambda} B^{\dagger}B).
\end{eqnarray}
This implies that
\begin{eqnarray}
\label{so-trivial*}
&&\Big|\frac{d}{dt} \Big\langle Q_{i}^{H}(t) Q_{i}^{H}(0) \Big\rangle \Big| \nonumber
  \\
  && ~~ \le \frac{1}{\hbar} \left(  \sqrt{\big\langle \big(\Delta
     \tilde{H}^{H}_{i}(t)\big)^2 \big\rangle}
     \sqrt{\big\langle Q_{i}^{H}(0) \big(Q_{i}^{H}(t)\big)^2
     Q_{i}^{H}(0)\big\rangle}~+\right.   \nonumber
  \\
  && ~~~~\left. \sqrt{\big\langle
         Q_{i}^{H}(t) \big(\Delta\tilde{H}^{H}_{i}(t)\big)^2
         Q_{i}^{H}(t)
         \big\rangle}
         \sqrt{\big\langle \big(Q_{i}^{H}(0)\big)^2 \big\rangle}~\right).
\end{eqnarray}
By the uniform time translational invariance of the averages in the equilibrium thermal
state,
\begin{eqnarray}
  \left\langle
  Q_{i}^{H}(t) \big(\Delta\tilde{H}^{H}_{i}(t)\big)^2
  Q_{i}^{H}(t)\right\rangle&&\nonumber\\
  &&\hspace{-2cm}=
  \left\langle
  Q_{i}^{H}(0) \big(\Delta\tilde{H}^{H}_{i}(0)\big)^2
  Q_{i}^{H}(0)\right\rangle.
\end{eqnarray}
Equilibrium autocorrelation functions invariably peak at $t=0$ due to loss of memory at finite times ($t>0$),
\begin{eqnarray}
\left\langle Q^{H}_{i}(0)
  \left(Q_{i}^{H}(t)\right)^2 Q^{H}_{i}(0) \right\rangle \le
  \left\langle  \left(Q_{i}^{H}(0)\right)^4\right\rangle.
\end{eqnarray}
Using the last two relations,
 Eq. (\ref{so-trivial*}) becomes
 \begin{eqnarray}
 \label{so-so-simple}
   &&\Big|\frac{d}{dt} \Big\langle Q_{i}^{H}(t) Q_{i}^{H}(0) \Big\rangle \Big| \nonumber
   \\
  && ~ \le \frac{1}{\hbar} \left(  \sqrt{\big\langle \big(\Delta
     \tilde{H}^{H}_{i}(t)\big)^2 \big\rangle}
     \sqrt{\big\langle \big(Q_{i}^{H}(0)\big)^4
     \big\rangle}~+\right.   \nonumber
  \\
  && ~~\left. \sqrt{\big\langle
         Q_{i}^{H}(0) \big(\Delta\tilde{H}^{H}_{i}(0)\big)^2
         Q_{i}^{H}(0)
         \big\rangle}
         \sqrt{\big\langle \big(Q_{i}^{H}(0)\big)^2
     \big\rangle}~\right),
     \nonumber\\
   && ~= \frac{1}{\hbar} \left(  \sqrt{k_{B} T^{2} C_{v,i}}
     \sqrt{\big\langle \big(Q_{i}^{H}(0)\big)^4
     \big\rangle}~+\right.   \nonumber
  \\
  && ~~\left. \sqrt{\big\langle
         Q_{i}^{H}(0) \big(\Delta\tilde{H}^{H}_{i}(0)\big)^2
         Q_{i}^{H}(0)
         \big\rangle}
         \sqrt{\big\langle \big(Q_{i}^{H}(0)\big)^2
     \big\rangle}~\right).
 \end{eqnarray}

As simple as it is, Eq. (\ref{so-so-simple}) will indeed enable us to bound transport coefficients (Sections \ref{sec:diff} and \ref{GsKsec}) and to suggest power law bounds on relaxation processes at asymptotically low temperatures (Section \ref{sec:lowT}). We stress that with a substitution of the thermal density matrix by its canonical ensemble form, $\rho_{\Lambda} \to \rho^{\sf canonical}_{\Lambda}$, the time derivative ${\sf{Tr}}(\rho_{\Lambda}^{\sf canonical}
\frac{d Q_{i}^{H}}{dt} Q_{i}^{H}(0))$ in Eq. (\ref{so-trivial}) does
{\it not} trivially vanish (as it does for the simple single operator
derivatives ${\sf{Tr}}(\rho_{\Lambda}^{\sf canonical} \frac{d
  Q_{i}^{H}}{dt})$). The bound appearing in the bottom line of
Eq. (\ref{so-trivial})
simplifies to the time independent expression in Eq. (\ref{so-so-simple}). Identical results apply for the derivatives of general time dependent correlation functions $\langle Q^{H}_{i}(t)Q^{H}_{j}(0) \rangle$ that are not necessarily autocorrelation functions (i.e., $i \neq j$). 
We will return to explicit extensions and applications in Sections
\ref{sec:diff} and \ref{GsKsec}.  
 
\section{Thermalization bounds in theories with no connected correlators}
\label{decoupled}

Nearly all of the results in this paper apply to systems harboring rather general interactions and correlations. Before returning to study these, we briefly discuss the simplest possible case in which the Hamiltonian is a sum of uncorrelated few body decoupled terms  $\{ \tilde{H}_{i^{\prime}}^{H}\}_{i=1}^{N'_\Lambda}$ (Eq. (\ref{eq:decoupled+sum})). The number of these non-interacting terms $N_{\Lambda}^{\prime}$ {\it may generally differ} from the total number of particles or sites $N_{\Lambda}$. This situation is not merely academic. In many interacting theories (e.g., Fermi liquids that we will briefly discuss in Section \ref{sec:lowT}), the system can be cast in terms of non-interacting particles with renormalized masses, charges, and other parameters. For such decoupled systems, we may work with a decomposition of the global Hamiltonian in the form of Eq. (\ref{eq:decoupled+sum}) for which ${\sf{Tr}}\left(\rho_{\Lambda}^{\sf {canonical }}\left(\tilde{H}^{H}_{i^{\prime}} \tilde{H}^{H}_{j^{\prime}}\right)\right)={\sf{Tr}}\left(\rho_{\Lambda}^{\sf {canonical }}\left(\tilde{H}^{H}_{i^{\prime}}\right)\right) \times {\sf{Tr}}\left(\rho_{A}^{\sf {canonical }}\left(\tilde{H}^{H}_{j^{\prime}}\right) \right)$. Since the temperature $T$ and the constant volume heat capacity $C_v^{(\Lambda)}$ determine the variance of the global Hamiltonian $H_{\Lambda}$ of the entire system $\Lambda$, when the Hamiltonians $\{ \tilde{H}_{i^{\prime}}^{H}\}_{i=1}^{N'_\Lambda}$ in the sum of Eq. (\ref{eq:decoupled+sum}) are uncorrelated,
\begin{eqnarray}
k_{B} T^{2} C_{v}^{(\Lambda)}={\sf{Tr}} \left(\rho_{\Lambda}^{\sf {canonical }}\left(\Delta H_{\Lambda}\right)^{2}\right) \nonumber
\\ =\sum_{i^{\prime}=1}^{N^{\prime}_{\Lambda}} {\sf{Tr}}\left(\rho_{\Lambda}^{\sf {canonical }}\left(\Delta \tilde{H}_{i^{\prime}}^{H}\right)^{2}\right).
\end{eqnarray}
Thus, from Eq. (\ref{a-central-1}),
\begin{equation}
\label{f-central-1}
\frac{4}{\hbar^{2}} k_{B} T^{2} C_{v}^{(\Lambda)} \geq \sum_{i^{\prime}=1}^{N_{\Lambda}^{\prime}} \frac{\left\langle\frac{d Q^{H}_{i}}{d t}\right\rangle^{2}}{\sigma_{Q_{i '}^{H}}^{2}}.
\end{equation}
If $\{Q^{H}_{i^{\prime}}\}$ are bounded operators, $||Q^{H}_{i'}|| \leq ||Q||$ then 
$\sigma_{Q^{H}_{i^{\prime}}}^{2} \leq\|Q\|^{2}$, and the global rate of change, averaged over all local observables, 
\begin{equation}
\label{ref:decoupled-final}
\frac{4||Q \|^{2} k_{B} T^{2} c_v}{\hbar^{2}}   \geq  \frac{1}{N_{\Lambda}^{\prime}}  \sum_{i'=1}^{N_{\Lambda}^{\prime}}\left|\left\langle\frac{d Q^{H}_{i'}}{d t}\right\rangle\right|^{2}.
\end{equation}
Here, $c_{v} \equiv \frac{C_{v}^{(\Lambda)}}{N_{\Lambda}^{\prime}}$ is the bona fide {\it thermodynamic specific heat} (not the effective local heat capacity $C_{v,i}$ of the earlier Sections).

As we will discuss in Appendix \ref{sec:RP}, apart from decoupled systems that we focused on in this Section, the bound of Eq. (\ref{ref:decoupled-final}) will also rigorously hold for interacting systems that are Reflection Positive. More generally, Eq. (\ref{ref:decoupled-final}) applies to all systems in which all connected pair correlation functions between the local Hamiltonians $\{ \tilde{H}_{i^{\prime}}^{H}\}_{i=1}^{N'_\Lambda}$ (whose sum forms the total Hamiltonian (Eq. (\ref{eq:decoupled+sum}))) are positive semi-definite. In Section \ref{sec:lowT}), we will derive universal low temperature bounds on the dynamics that follow from Eq. (\ref{ref:decoupled-final}).

\section{Non-relativistic velocity bounds}
\label{v-bounds}
 
Some of the discussion in the preceding Sections might seem a bit artificial inasmuch as the actual physics may be concerned. We next explicitly put our formalism to real use in order to arrive at concrete bounds on various physical quantities. In this Section, we will improve and expand on a derivation provided in \cite{bound} on how quickly particles may move in thermal equilibrium of a non-relativistic system. Our result of Eq. (\ref{vbound12}) implies a bound suggested and extensively empirically tested by \cite{melting-speed} for the particular case of a solid in which atomic displacements are bounded (satisfying the Lindemann bound \cite{Lindemann}). 

We consider a system of non-relativistic particles of mass $m$ in $d$ spatial dimensions interacting via general ({\it not necessarily two-particle nor local}) position dependent forces. In order to derive the velocity bound, we will choose $Q_{i}=r_{i \ell}$ to be the position coordinate for any fixed arbitrary Cartesian label $\ell=1,2, \ldots, d$ and particle index $i=1,2, \ldots, N_{\Lambda}$.
In the absence of magnetic fields, the Hamiltonian is a sum of kinetic and potential terms, $H_{\Lambda}=K+V$ with the potential energy $V=V\left(\left\{{\bf r}_{i}\right\}\right)$ being a general function depending on spatial coordinates (and thus commuting with $Q_i$). There is only {\it a single} term in the Hamiltonian $H_{\Lambda}$ that does not commute with $Q^H_i$, namely,
\begin{eqnarray}
\label{HILI}
\tilde{H}_{i}^{H}(t)=\frac{\left(p_{i \ell}^{H}(t)\right)^{2}}{2 m}. 
\end{eqnarray}

\subsection{Expectation values of the velocities}
\label{Expect-v}

At low temperatures, the effective heat capacities $C_{v,i}$ (as well as the thermodynamic heat capacity $C^{(\Lambda)}_{v}$) may be strongly suppressed relative to its classical value due to quantum effects. This is explicitly illustrated in the solvable harmonic solid example of Appendix \ref{sec:harmonic}, see, e.g., Eq. (\ref{harmonic-final} in particular). In what follows, we obtain exact generous upper velocity bound in general (not necessarily solvable) non-relativistic systems by considering their semiclassical limit.

With the convention of Eq. (\ref{O:avg}), inserting the Hamiltonian of Eq. (\ref{HILI}) into Eq. (\ref{cv1}), we arrive at
\begin{equation}
\label{speeeed} 
\frac{\hbar^{2}}{4} \overline{\left(\sigma_{r_{i \ell}^{H}}\right)^{-2}\left|\left\langle d r_{i \ell}^{H} / d t\right\rangle\right|^{2}} \leq k_{B} T^{2} C_{v, i}.
\end{equation}
At sufficiently high temperatures where the classical equipartition theorem applies, the 
effective heat capacity (Eq. (\ref{cv2}) that is associated with the kinetic term of 
Eq. (\ref{HILI}) is $C_{v,i} = k_B/2$.
Thus, at these temperatures, 
Eqs. (\ref{cv1},\ref{cv2},\ref{cv3}) imply that
\begin{equation}
\label{vbound12}
\overline{\left(\sigma_{r_{i \ell}^{H}}\right)^{-2}\left|\left\langle d r_{i \ell}^{H} / d t\right\rangle\right|^{2}}^{1 / 2} \leq \frac{k_{B} T \sqrt{2}}{\hbar}.
\end{equation}
For an isotropic medium, the ratio on the lefthand side of Eq. (\ref{vbound12}) is the same for all Cartesian directions $\ell$ and, consequently, we may drop this subscript.
As discussed in Sections \ref{setup} and \ref{bound-1t}, in ergodic systems, instead of performing the global average of Eq. (\ref{O:avg}), we may alternatively use the expectation value as computed with the long time average density matrix $\rho_{\Lambda}^{\sf {long }-\sf { time-average }}$ that appeared on the righthand side of Eqs. (\ref{long-time-eq}, \ref{cal-long}). If we do so then, on the lefthand side of Eq. (\ref{vbound12}), we may further drop the Heisenberg picture superscript $H$ and the angular brackets denoting the average with $\rho_{\Lambda}$. 
In Table \ref{real-values}, we contrast the bound of Eq. (\ref{vbound12}) with empirical values of Aluminum assuming a constant value of the individual particle ($i$) fluctuations $\sigma_{r_{i \ell}^{H}}$. At melting, the magnitude of the fluctuations $\sigma_{r_{i \ell}^{H}}$ in the crystal is given by the product of the Lindemann constant and the lattice constant. In a semiclassical gas or fluid, one may similarly expect that, at any moment in time, the fluctuation $\sigma_{r_{i \ell}^{H}}$ of the single particle location evaluated with $\rho_{\Lambda}$ to be smaller than the inter-particle spacing. For an isotropic medium, assuming constant $\sigma_{r_{i \ell}^{H}}$, the inequality of Eq. (\ref{vbound12}) becomes a bound on the particle root mean square (rms) speed. Since, in a classical system, the rms speed is set by $\sqrt{k_{B} T/m}$, Eq. (\ref{vbound12}) implies bounds on the fluctuations $\sigma_{r_{i \ell}^{H}}$ of the single particle locations. We will turn to these shortly.
 
We pause momentarily to make general remarks on the physical meaning of our respective exact and approximate bounds of Eqs. (\ref{speeeed},\ref{vbound12}) and to contrast these with other known bounds. The fundamental Lieb-Robinson bounds \cite{Lieb_Robinson,Bruno,Sergey,alioscia,kaden,Else20} establish ``speed limits'' on the propagation of information (or, equivalently, on correlations), and thus on (quasi)particle speed in lattice systems having short range interactions of bounded norm that are defined in finite dimensional Hilbert spaces. By contrast, the bounds of Eqs. (\ref{speeeed},\ref{vbound12}) and their extensions (elaborated on in substantial detail in the next subsections and in Appendix \ref{velocity-bound-App}) {\it do not, generally, require} these constraints nor, as we will demonstrate in this work, are they solely confined to to the realm of speed limits. Qualitatively, the finite energy fluctuations at any physical (i.e., non-divergent) temperature provide a cutoff on the possible velocities that plays the role of the norm of operators for which the standard Lieb-Robinson bounds apply. In the limit of infinite temperatures for bounded operators, we can replace the standard deviation of the effective local Hamiltonians $\tilde{H}_{i^{\prime}}^{H}$ by their operator norm (i.e., we may replace $k_{B} T^{2} C_{v, i}$ in Eq. (\ref{speeeed}) by the bounded operator norm 
$|| \tilde{H}_{i^{\prime}}^{H}||^{2}$). This leads to a bound which is slightly more similar to the standard, temperature independent, Lieb-Robinson type inequality on the speed. In a solid, the standard deviation of atomic displacements $\sigma_{r_{i \ell}^{H}}$ is bounded from above by the lattice constant times a factor of order unity (the Lindemann ratio). This, in turn, yields a bound on the velocities $\left\langle v_{i \ell}\right\rangle$. In Appendix \ref{velocity-bound-App}  we will illustrate that Eq. (\ref{vbound12}) further suggests that high temperature thermal equilibrium is possible only if the global average ($\bar{\sigma}_{r_{i \ell}^{H}}$) of the single particle position standard deviations satisfies the inequality
\begin{equation}
\label{srih}
   \bar{\sigma}_{r_{i \ell}^{H}}  \ge  \frac{\hbar}{\sqrt{2 m k_{B} T}} = \frac{\lambda_T}{2 \sqrt{\pi}}.
\end{equation}
Here, $\lambda_T$ is the thermal de-Broglie wavelength of Eq. (\ref{dBeq}). 

At general temperatures, the standard deviation $\sigma_{r_{i \ell}^{H}}$ is not a consequence of zero-point quantum fluctuations. Rather, $\sigma_{r_{i \ell}^{H}}$ represents the standard deviation of the $\ell$-th Cartesian position coordinate of particle $i$ when it is computed with the full (finite temperature) density matrix $\rho_{\Lambda}$. As such, this standard deviation of the position is typically dominated, at high temperatures, by classical thermal fluctuations. 

As we will emphasize in Appendix \ref{sec:comments} and illustrate in a trivially solvable example in Appendix \ref{sec:harmonic}, the choice of the local Hamiltonian of Eq. (\ref{HILI}) is not unique. Other local Hamiltonians may be chosen as those that drive the dynamics of the spatial coordinates. For certain systems (e.g., the harmonic oscillator of Appendix \ref{sec:harmonic}), some of these other local Hamiltonians may yield bounds that are even stronger than those derived in the current subsection.

\subsection{Expectation values of higher moments of the velocity}
\label{sec:v-high}
In this subsection, we will employ the inequalities of Section \ref{high-moment-derivative} to bound higher moments of the velocities. As the reader may anticipate, no essential changes will appear in the final results as compared to the above bounds for the expectation value of the velocity itself. The virtue of even moments of the velocity is that they do not vanish when computed with the canonical ensemble probability density. 

\subsubsection{Compact coordinates}
\label{sec:compact}
We briefly discuss the velocity bounds associated with compact coordinates. We consider a compact self-adjoint operator ${\sf q}_{i}$ of bounded norm $ ||{\sf q}_{i}|| \le || {\sf q}_{}||$ that is a periodic function of an angular coordinate $\varphi$ associated with a principal moment of inertia tensor component. For deriving bounds on its rate of change, we will follow our earlier inequalities setting $Q^H_{i}={\sf q}_{i}^{H}$ therein. Now, the dynamics of ${\sf q}^H_{i}$ are generated by the kinetic $\tilde{H}_{i}^{H} = (p^{H}_{\varphi})^{2}/(2 I_{\varphi})$ with $p^{H}_{\varphi} \equiv J^{H}_{z}$ being an angular momentum operator (where the $z$ direction is the one about which the rotation angle $\varphi$ is defined and $I_{\varphi}$ is the corresponding moment of inertia component). We then have
\begin{eqnarray}
\Big\langle \Big ( \frac{d{\sf q}_{i}^{H}}{dt} \Big)^2  \Big\rangle \le \frac{4}{\hbar^2} \Big \langle \Big( \Delta \tilde{H}^{H}_{i} \Big)^2 \Big \rangle  \Big| \Big| {\sf q}  \Big| \Big|^2 \nonumber
\\ \equiv  \frac{4 k_{B} T^{2} C_{v, i}}{\hbar^2} \Big| \Big|{\sf q} \Big| \Big|^2.
\label{bounded=Q=2'} 
\end{eqnarray}
Similar to our earlier velocity bound of Section \ref{Expect-v}, the effective local heat capacity $C_{v, i}$ is that associated with the kinetic energy yet now involving a single angular degree of freedom. Here, too, in the classical high temperature limit $C_{v,i} = k_{B}/2$. Thus, at sufficiently high temperatures where the equipartition theorem applies, 
\begin{eqnarray}
\label{compact-v}
\Big\langle \Big ( \frac{d{\sf q}_{i}^{H}}{dt} \Big)^2  \Big\rangle \le 2 \Big(\frac{k_{B} T}{\hbar}\Big)^2 ||{\sf q}||^2.
\end{eqnarray}
Eq. (\ref{compact-v}) affords a higher moment inequality relative to that presented in \cite{bound} (in which the average $\Big|\Big\langle  \frac{d{\sf{q}}_{i}^{H}}{dt}   \Big\rangle\Big|$ was bounded).

\subsubsection{Cartesian coordinates}
\label{sec:Descartes}

Next, we turn to bounds on linear velocities. As discussed in Section \ref{high-moment-derivative}, in the semiclassical limit, Eq. (\ref{squared-ineq}) implies Eqs. (\ref{central1}, \ref{a-central-1}). Thus, in this regime, we will reproduce the bounds of Section \ref{Expect-v}. Different from Section \ref{Expect-v}, in the high temperature semiclassical regime, our results will be rigorous; we will not need to invoke any assumptions concerning relations between higher local square momentum and more pronounced localization. To elucidate the derivation of the general bound, we set, as in Section \ref{Expect-v}, $Q_{i}^{H} = r^{H}_{il}$ with the local Hamiltonian of Eq. (\ref{HILI}) generating its dynamics. In the semiclassical limit, we may replace 
averages of the type of Eq. (\ref{longABABA'}) when these are evaluated by the canonical density matrix of Eq. (\ref{rcanonical}) by trivial phase space averages,
\begin{eqnarray}
\label{factorxHx}
{\sf{Tr}}(\rho_{\Lambda}^{\sf canonical}   \tilde{H}^{H}_{i}  \Delta Q^H_i   \tilde{H}^{H}_{i} \Delta Q^H_i  ) \nonumber
\\ = \frac{\int d^{dN_{\Lambda}}r~~ \Delta r_{i l}^{2}~~ e^{-\beta V({{\bf r}_{1}, \cdots {\bf r}_{N})}}}{\int d^{dN_{\Lambda}}r~~~ e^{-\beta V({{\bf r}_{1}, \cdots {\bf r}_{N})}}}  \nonumber
 \\ \times
 \frac{\int dp_{il}~~ \frac{p_{i l}^{2}}{2m}~~ e^{-\beta  \frac{p_{il}^{2}}{2m}}}  {\int dp_{il}~~~ e^{-\beta  \frac{p_{il}^{2}}{2m}}} \nonumber
 \\ =  \frac{k_{B} T}{2} {\sf Tr}( \rho_{\Lambda}^{\sf classical~ canonical} \Delta r^2_{il} ).
  \end{eqnarray}
In order to avoid complicated notation, in Eq. (\ref{factorxHx}) and the remainder of this Section, we drop the ``$H$'' superscript from the position coordinate fluctuations $\Delta r_{il}^{H}$ whenever classical averages are performed. Similar to the discussion following Eq. (\ref{srih}), $ {\sf Tr}( \rho_{\Lambda}^{\sf classical~canonical} \Delta r^2_{il} )$ is now the variance of the position coordinate when evaluated with the classical canonical probability density matrix $\rho_{\Lambda}^{\sf classical~ canonical}$ when appropriate at high temperatures.  Analogous to the general relation of Eq. (\ref{longABABA'}), taken together Eq. (\ref{longABABA}) and Eq. (\ref{factorxHx}) imply that
  \begin{eqnarray}
\label{longABABA''}
\frac{1}{N_{\Lambda} } \sum_{i=1}^{N_{\Lambda}} \Big\langle \Big ( \frac{dr_{il}^{H}}{dt} \Big)^2  \Big\rangle \nonumber 
\\ \le \frac{2 (k_B T)^2}{\hbar^2} {\sf Tr}( \rho_{\Lambda}^{\sf classical~ canonical} \Delta r^2_{il} ).
\end{eqnarray}
Since, classically, 
\begin{eqnarray}
\label{kTm}
\! \! \! \! \! \! \! \! \! \! \! \! \! \! \! \! \! \! \! \! \! \! \!
\frac{1}{N_{\Lambda} } \sum_{i=1}^{N_{\Lambda}} \Big\langle \Big ( \frac{dr_{il}^{H}}{dt} \Big)^2  \Big\rangle 
&& = {\sf{Tr}} \Big(\rho_{\Lambda}^{\sf classical~canonical}  \Big ( \frac{dr_{il}}{dt} \Big)^2 \Big)  \nonumber
\\  &&= \frac{k_{B} T}{m},
\end{eqnarray}
we arrive at an analog of Eq. (\ref{srih}) in the semiclassical limit. That is, in thermal equilibrium, combining Eqs. (\ref{longABABA''}, \ref{kTm}) at sufficiently high temperatures where the system becomes semiclassical, the average square particle displacements are strictly bounded, 
\begin{equation}
\label{srih'}
\frac{1}{N_{\Lambda}}  \sum_{\ell=1}^{d} \sum_{i=1}^{N_{\Lambda}} \langle (\Delta r^H_{il})^2 \rangle  \ge  d \frac{ \hbar^2}{2 m k_{B} T} = d \frac{\lambda_T^2}{4 \pi}.
\end{equation}
As the factor of $\hbar$ on the righthand side of Eq. (\ref{srih'}) indicates, quantum fluctuations lead to a lower bound on the particle displacements. 

Following the steps outlined in Section \ref{high-moment-derivative}, bounds on $\frac{1}{N_{\Lambda}} \sum_{i=1}^{N_{\Lambda}}  \langle \Big( \frac{dr_{il}^{H}}{dt} \Big)^n \rangle$ with $n>2$ may be reproduced analogously to those derived for $n=2$ above with identical conclusions. Multiplying all of our velocity inequalities, whether linear or angular, by the mass or moment of inertia, trivially leads to linear or angular momentum bounds.

\section{Acceleration (and Force) Bounds}
\label{a-bounds}

\subsection{Expectation values of the acceleration}
\label{aa-bounds}

\subsubsection{General schematics}
\label{sec:diff'}
In this Section, we will set the local quantity $Q_{i}^{H}$ to be a Cartesian velocity component, $Q_{i}^{H}=v^H_{i \ell}=\frac{p^H_{i \ell}}{m}$. Once this is done, Eq. (\ref{a-central-1}) leads to bounds on the acceleration components $a_{i \ell}^{H}=\frac{d v_{i \ell}^{H}}{dt}$. In what follows, the Cartesian component index $\ell$ will be fixed.
We consider generic systems with no external fields that are governed by global Hamiltonians of the form 
\begin{eqnarray}
\label{eq:HLAMBDA}
H_{\Lambda}=\sum_{i=1}^{N_{\Lambda}} \frac{{\bf p}_{i}^{2}}{2 m}+\frac{1}{2} \sum_{i, j} {\cal{V}} \left({\bf r}_{i}, {\bf r}_{j}\right) \equiv K + V.
\end{eqnarray}
As in Section \ref{v-bounds}, $K$ and $V$ denote, respectively, the kinetic and potential energy contributions. For such Hamiltonians $H_{\Lambda}$, the {\it minimal} $\tilde{H}_{i}^{H} \subset H_{\Lambda}$ that endows $Q^H_{i}$ with dynamics is now 
\begin{eqnarray}
\label{tileV}
\tilde{H}_{i}^{H}=\sum_{j} {\cal{V}} \left({\bf r}_{i}^{H}(t), {\bf r}_{j}^{H}(t)\right) \equiv V_i^{H}(t).
\end{eqnarray}
As in Eq. (\ref{eq:HLAMBDA}) (and Section \ref{high-moment-derivative}), we will employ ${\cal{V}}$ to denote the pair interactions and use $V$ to indicate the global potential energy. With all of the above in tow, Eq. (\ref{a-central-1}) now implies a trivial bound on the acceleration,
\begin{eqnarray}
\label{aa-central-1}
\frac{1}{N_{\Lambda}} \sum_{i=1}^{N_{\Lambda}} {\sf{Tr}}\left(\rho_{\Lambda}^{\sf {canonical }}\left( V_i^H\right)^{2}\right) \nonumber
\\  \geq \frac{\hbar^{2}}{4 N_{\Lambda}} \sum_{i=1}^{N_{\Lambda}} \frac{\left({\sf{Tr}}\left(\rho_{\Lambda} \frac{d v_{i \ell}^{H}}{d t}\right)\right)^{2}}{{\sf{Tr}}\left(\rho_{\Lambda}\left(v_{i \ell}^{H}(t)\right)^{2}\right)}.
\end{eqnarray}
Physically, Eq. (\ref{aa-central-1}) asserts that the average of the square of the relaxation rate of the velocities is bounded from above by the average of $\frac{4}{\hbar^2} {\sf{Tr}}\left(\rho_{\Lambda}^{\sf {canonical }}\left( V_i^H\right)^{2}\right)$. The single particle velocities appear in the denominator of Eq. (\ref{aa-central-1}). 

We remark, once again, that (as we will further elaborate on in Appendix \ref{sec:comments}) multiple local Hamiltonians can, generally, be taken in order to obtain dynamical bounds. In the current case, using other Hamiltonians instead of Eq. (\ref{tileV}) may, potentially, lead to inequalities that are more stringent than Eq. (\ref{aa-central-1}). We next turn to the consequences of  Eq. (\ref{aa-central-1}).

\onecolumngrid
\subsubsection{Diffusion Constant Bounds}
\label{sec:diff}
\bigskip
\noindent {\underline{\it Quantum Bounds.}}

Before proceeding further, we examine, in some detail, a qualitative (and quantitative) corollary of the result that we just derived. Eq. (\ref{aa-central-1}) suggests that, since the velocity cannot relax arbitrarily quickly, there may be an inherent fundamental lower limit on the rate of change of the velocity autocorrelation function and on the integral of the velocity autocorrelation function- the diffusion constant. Indeed, a trivial application of Eq. (\ref{a-central-1}) leads to bounds associated with the diffusion constant in an isotropic medium as computed from the Green-Kubo relation \cite{Green1954,KMS2} which expresses the diffusion constant as the integral of the velocity autocorrelation function,
\begin{eqnarray}
\label{eq:diffusion}
D =  \int_{0}^{\infty} dt ~ {\sf{Tr}} (\rho_{\Lambda}^{\sf canonical}~ v_{i \ell}^{H}(t)~ v_{i \ell}^{H}(0)) \nonumber
\\ \equiv \int_{0}^{\infty} dt~ G_{v}(t).
\end{eqnarray}
The autocorrelation function $G_v$ has been very extensively studied in numerous ``classical'' works, e.g., \cite{liqGv1,liqGv2,liqGv3,liqGv4,liqGv5,liqGv6,liqGv7,liqGv8}. The velocity autocorrelation function is positive at short times (decaying from its maximum at time $t=0$). Similar to simple diffusing Brownian particles with Langevin dynamics, in many systems, such as dilute gases, an exponential decay in $t$ persists at all times and thus $G_v$ is positive semi-definite (vanishing only at asymptotically long times). However, there are numerous other systems in which $G_v(t)$ may assume negative values. These typically lead to observed oscillations about zero at long times in solids, liquids, and dense gases in which the oscillatory velocity autocorrelation {\textcolor{red} may be triggered by restoring forces and caging effects. In liquids and gases,} these oscillatory tails usually do not contribute substantially to the integral of Eq. (\ref{eq:diffusion}). In solids, oscillations in the velocity autocorrelation functions may become more dominant. Trivial sign changes of the autocorrelation function also appear for gases in an external magnetic field due to cyclotron motion. In what follows, we will derive bounds on a related quantity 
\begin{eqnarray}
\label{eqnD+}
D_{+} \equiv  \int_{0}^{t^v} dt~ G_{v}(t).
\end{eqnarray}
In Eq. (\ref{eqnD+}), we set $t^v$ is the smallest positive time at which $G_v$ vanishes; if $G_v >0$ for all finite times
then $t^v = \infty$ and $D_{+}$ becomes the exact diffusion constant, $D_{+} = D$ \footnote{In certain circles, the time $t^v$ in Eq. (\ref{eqnD+}) is referred to as to onset time of the so-called ``correlation hole'' \cite{Lev86} in which the autocorrelation function drops below its asymptotic long time value (which, in the current case, amounts to the first zero of the velocity autocorrelation function).}. In the remainder of this subsection, we formalize and apply the results of Section \ref{t-auto}. Specifically, in the context of our diffusion bounds, we assume similar to our discussion in Section \ref{t-auto} that, for $t \ge 0$, the expectation value 
\begin{eqnarray}
\label{H4}
&& {\sf{Tr}} (\rho_{\Lambda}^{\sf canonical} (v_{i \ell}^{H}(0))^4)  \nonumber
\\&&  \ge {\sf{Tr}} (\rho_{\Lambda}^{\sf canonical}  v_{i \ell}^{H}(0) (v_{i \ell}^{H}(t))^2 v_{i \ell}^{H}(0))).
\end{eqnarray}
Analogously, as mentioned above, the expectation value $G_v(0) \ge G_v(t)$. In equilibrium systems, the maxima of these expectation values are indeed always achieved at $t=0$. We now will derive our exact lower bound on $D_{+}$ following the same recipe provided in Section \ref{t-auto} with all of the details highlighted. Applying Eq. (\ref{so-so-simple}) to the canonical density matrix $\rho_{\Lambda}^{\sf canonical}$,

\begin{eqnarray}
\label{aa-central-111}
  \Big|\frac{d  G_v}{dt} \Big|
  && = \frac{1}{\hbar}
     \left|
     {\sf Tr}\left(\rho_{\Lambda}^{\sf canonical} [\tilde{H}_i^{H}(t),
     v_{i \ell}^{H}(t)] v_{i \ell}^{H}(0)\right)\right|
     \nonumber\\
     && = \frac{1}{\hbar}
     \left|
     {\sf Tr}\left(\rho_{\Lambda}^{\sf canonical} \Delta\tilde{H}_i^{H}(t)
     v_{i \ell}^{H}(t) v_{i \ell}^{H}(0)\right) - {\sf Tr}\left(\rho_{\Lambda}^{\sf canonical}  v_{i \ell}^{H}(t)  \Delta\tilde{H}_i^{H}(t)
    v_{i \ell}^{H}(0)\right)
     \right| \nonumber
     \\ && \le \frac{1}{\hbar}
     \Big(\left|
     {\sf Tr}\left(\rho_{\Lambda}^{\sf canonical} \Delta\tilde{H}_i^{H}(t)
     v_{i \ell}^{H}(t) v_{i \ell}^{H}(0)\right) \right| + \left| {\sf Tr}\left(\rho_{\Lambda}^{\sf canonical}  v_{i \ell}^{H}(t)  \Delta\tilde{H}_i^{H}(t)
    v_{i \ell}^{H}(0)\right)  
     \right|  \Big)       \nonumber    
 \\ &&   \le \frac{1}{\hbar} \left(
     \sqrt{{\sf Tr}\Big(\rho_{\Lambda}^{\sf canonical} \big(\Delta V_{i}^{H}(0)\big)^2\Big)}
     \sqrt{{\sf Tr}\Big(\rho_{\Lambda}^{\sf canonical}
     v_{il}^{H}(0) \big(v_{il}^{H}(t)\big)^2 v_{il}^{H}(0)  \Big)}
     +
     \right.\nonumber\\&& ~~\left.
     \sqrt{{\sf Tr}\Big(\rho_{\Lambda}^{\sf canonical}
     v_{il}^{H}(t)
     \big(\Delta V_{i}^{H}(t)\big)^2
     v_{il}^{H}(t)
     \Big)}
     \sqrt{{\sf Tr}\Big(\rho_{\Lambda}^{\sf canonical}
                          \big(v_{il}^{H}(0)\big)^2\Big)}\right)
                          \nonumber       
 \\ &&   \le \frac{1}{\hbar} \left(
     \sqrt{{\sf Tr}\Big(\rho_{\Lambda}^{\sf canonical} \big(\Delta V_{i}^{H}(0)\big)^2\Big)}
     \sqrt{{\sf Tr}\Big(\rho_{\Lambda}^{\sf canonical}
     \big(v_{il}^{H}(0)\big)^4\Big)}
     +
     \right.\nonumber\\&& ~~\left.
     \sqrt{{\sf Tr}\Big(\rho_{\Lambda}^{\sf canonical}
     v_{il}^{H}(0)
     \big(\Delta V_{i}^{H}(0)\big)^2
     v_{il}^{H}(0)
     \Big)}
     \sqrt{{\sf Tr}\Big(\rho_{\Lambda}^{\sf canonical}
                          \big(v_{il}^{H}(0)\big)^2\Big)}\right),
                          \nonumber\\
  \Rightarrow
  \Big|\frac{d  G_v}{dt} \Big|
  && \le
     \frac{1}{\hbar}
     \left(
     \sqrt{{\sf Tr}\Big(\rho_{\Lambda}^{\sf canonical} \big(\Delta V_{i}^{H}\big)^2\Big)
     {\sf Tr}\Big(\rho_{\Lambda}^{\sf canonical}
     \big(v_{il}^{H}\big)^4\Big)}
     +
     \sqrt{{\sf Tr}\Big(\rho_{\Lambda}^{\sf canonical}
     v_{il}^{H}
     \big(\Delta V_{i}^{H}\big)^2
     v_{il}^{H}
     \Big){\sf Tr}\Big(\rho_{\Lambda}^{\sf canonical}
     \big(v_{il}^{H}\big)^2\Big)}
     \right)
     ,
     \nonumber\\
\end{eqnarray}
 In the second equality, we implicitly shifted $ \tilde{H}_{i}^{H}(t) \to \Delta \tilde{H}^{H}_{i}(t)$ leaving the commutator unchanged. The subsequent steps above are identical to those employed in Eqs. (\ref{so-trivial}) and (\ref{so-so-simple}). In going from the third to the fourth lines we employed the Cauchy-Schwarz inequality of Eq. (\ref{CSg'}) and noted, once again, that what drives the dynamics of $v_{i \ell}^{H}(t)$ is, as in our earlier
calculations, $\tilde{H}_{i}^{H}(t) = V^H_{i}$. Finally, we inserted Eq. (\ref{H4}) and the time independence of the equilibrium average ${\sf Tr}\Big(\rho_{\Lambda}^{\sf canonical}
     v_{il}^{H}(t) \big(\Delta V_{i}^{H}(t)\big)^2 v_{il}^{H}(t)
     \Big)$ which allowed us to replace this equilibrium average by its value at $t=0$. In the last line, we made the time independence of our bound explicit.  

Since $G_v$ is positive semi-definite in the integration domain of
Eq. (\ref{eqnD+}), we have, for times $0\le t \le t^v$, that  
\begin{eqnarray}
\label{Gvt}
  G_v(t)
  \ge \max 
  \left\{
  G_{v}(0)-t\left|
  \frac{dG_v}{dt}
  \right|_{\max},0\right\}.
\end{eqnarray}
Here, $\left|
  \frac{dG_v}{dt}
  \right|_{\max}$ denotes the maximal derivative over all times $t \in [0,t^v]$. 
Integrating the righthand side of Eq. (\ref{Gvt}) yields a precise lower bound 
\begin{eqnarray}
\label{D+min}
D_{+} \ge \frac{ \big(G_{v}(0)\big)^2}{2\Big|\frac{dG_v}{dt}\Big|_{\max}}.
\end{eqnarray}
Recalling Eq. (\ref{aa-central-111}) this, in turn, implies that 
\begin{eqnarray}
\label{minDiff}
D_{+} 
\ge 
\hbar\frac{\big(G_{v}(0)\big)^2}{2}
     \left(
     \sqrt{{\sf Tr}\Big(\rho_{\Lambda}^{\sf canonical}
  \big(\Delta V_{i}^{H}\big)^2\Big){\sf Tr}\Big(\rho_{\Lambda}^{\sf
  canonical} 
     \big(v_{il}^{H}\big)^4\Big)}
     +
     \sqrt{{\sf Tr}\Big(\rho_{\Lambda}^{\sf canonical}
     v_{il}^{H}
     \big(\Delta V_{i}^{H}\big)^2
     v_{il}^{H}
     \Big){\sf Tr}\Big(\rho_{\Lambda}^{\sf canonical}
     \big(v_{il}^{H}\big)^2\Big)}
  \right)^{-1}\hspace{-0.4cm}.
  \nonumber\\
\end{eqnarray}
The inequality of Eq. (\ref{minDiff}) is exact for general Hamiltonians
of the form of Eq. (\ref{eq:HLAMBDA}). In its derivation, we made no appeal to any approximation. We will shortly examine its
semiclassical limit. Before doing so, we make several
qualitative comments. As the factor of $\hbar$ in the numerator
Eq. (\ref{minDiff}) makes plain, this is an inherently quantum limit
on the diffusion capturing quantum fluctuations. Quantum dynamics (in
the form of Heisenberg's equations of motion in the current derived bound on $D_+$
manifestly contain a factor of $\hbar$) may trigger fluctuations
that prohibit the diffusion constant from vanishing. In effect, in
deriving the bound, we have examined the time integral of a velocity
autocorrelation function that drops with a maximal allowed slope from its value at time
$t=0$ to zero within a minimal time $t_{v} = t_{\min}$ given by our
uncertainty inequalities.

\bigskip
\bigskip
\bigskip
\bigskip

\noindent {\underline{\it Semiclassical Bounds.}}

Semiclassically, the above bound 
 simplifies further. Returning to Eq. (\ref{aa-central-111}), we explicitly write in this limit,
\begin{eqnarray}
\label{aa-central-111+!}
  \Big|\frac{d  G_v}{dt} \Big|
  && = \frac{1}{\hbar}
     \left|
     {\sf Tr}\left(\rho_{\Lambda}^{\sf canonical} [\Delta\tilde{H}_i^{H}(t),
     v_{i \ell}^{H}(t)] v_{i \ell}^{H}(0)\right)\right|
     \nonumber\\
     && = \frac{1}{\hbar}
     \left|
     {\sf Tr}\left(\rho_{\Lambda}^{\sf canonical} \Delta\tilde{H}_i^{H}(t)
     v_{i \ell}^{H}(t) v_{i \ell}^{H}(0)\right) - {\sf Tr}\left(\rho_{\Lambda}^{\sf canonical}  v_{i \ell}^{H}(t)  \Delta\tilde{H}_i^{H}(t)
     v_{i \ell}^{H}(0)\right)
     \right|
     \nonumber
     \\   && \le \frac{2}{\hbar}
   \max\Big\{ \left|
     {\sf Tr}\left(\rho_{\Lambda}^{\sf canonical} \Delta\tilde{H}_i^{H}(t)
     v_{i \ell}^{H}(t) v_{i \ell}^{H}(0)\right)\right|,  \left|
     {\sf Tr}\left(\rho_{\Lambda}^{\sf canonical}  v_{i \ell}^{H}(t) \Delta\tilde{H}_i^{H}(t)
      v_{i \ell}^{H}(0)\right)\right| \Big\} \nonumber
      \\ && \sim \frac{2}{\hbar}
  \left|
     {\sf Tr}\left(\rho_{\Lambda}^{\sf classical~canonical} \Delta\tilde{H}_i^{H}(t)
     v_{i \ell}^{H}(t) v_{i \ell}^{H}(0)\right)\right| = \frac{2}{\hbar}
  \left|
     {\sf Tr}\left(\rho_{\Lambda}^{\sf classical~canonical}
     v_{i \ell}^{H}(t)  \Delta\tilde{H}_i^{H}(t) v_{i \ell}^{H}(0)\right)\right|.
     \end{eqnarray}
     Here, $\sim$ highlights where the semiclassical limit first appears. Each of the two (identical) classical averages appearing in the last line line may be bounded via (sequential applications of) the Cauchy-Schwarz inequality. Specifically, for any three Hermitian operators $A,B$ and $C$ that, in all expectation values, mutually commute with one another (including, trivially, physical observables in the classical averages that we are considering),
\begin{eqnarray}
\label{trivvval}
  \hspace{-0.4cm}
  \left|
   {\sf Tr} \left(\rho_{\Lambda}^{\sf classical~canonical}
  ABC
  \right)  \right|
  \le
  \sqrt{{\sf Tr} \left(\rho_{\Lambda}^{\sf classical~canonical} A^2\right)}
  \sqrt{{ \sf Tr} \left(\rho_{\Lambda}^{\sf classical~canonical} B^2C^2\right)} \nonumber
  \\ 
  \le
   \sqrt{{\sf Tr} \left(\rho_{\Lambda}^{\sf classical~canonical} A^2\right)}
  \sqrt[4]{ {\sf Tr} \left(\rho_{\Lambda}^{\sf classical~canonical} B^4 \right)}
 \sqrt[4]{ {\sf Tr} \left(\rho_{\Lambda}^{\sf classical~canonical} C^4 \right)} .
\end{eqnarray}
Since the classical average is invariant under all permutations of $A, B$ and $C$ (e.g., those appearing in the last line of Eq. (\ref{aa-central-111+!})), the tightest upper bound on the product is formed by the smallest of the resulting Cauchy-Schwarz bounds of Eq. (\ref{trivvval}). That is, 
\onecolumngrid\noindent
\begin{eqnarray}
  \label{abcClassical}
 && \left|
   {\sf Tr} \left(\rho_{\Lambda}^{\sf classical~canonical}
  ABC
  \right)  \right|
  \nonumber
  \\  && \le  \sqrt[4]{ {\sf Tr} \left(\rho_{\Lambda}^{\sf classical~canonical} A^4 \right)}
   \sqrt[4]{ {\sf Tr} \left(\rho_{\Lambda}^{\sf classical~canonical} B^4  \right)}
     \sqrt[4]{ {\sf Tr} \left(\rho_{\Lambda}^{\sf classical~canonical} C^4  \right)}
 \nonumber\\ 
   \times&& \min\left\{
  \frac{\sqrt{ {\sf Tr} \left(\rho_{\Lambda}^{\sf classical~canonical} A^2 \right)}}{\sqrt[4]{  {\sf Tr} \left(\rho_{\Lambda}^{\sf classical~canonical} A^4 \right)}}, 
    \frac{\sqrt{ {\sf Tr} \left(\rho_{\Lambda}^{\sf classical~canonical} B^2 \right)}}{\sqrt[4]{  {\sf Tr} \left(\rho_{\Lambda}^{\sf classical~canonical} B^4 \right)}},
      \frac{\sqrt{ {\sf Tr} \left(\rho_{\Lambda}^{\sf classical~canonical} C^2\right)}}{\sqrt[4]{  {\sf Tr} \left(\rho_{\Lambda}^{\sf classical~canonical} C^4 \right)}}
                          \right\}.
\end{eqnarray}
We recapitulate that in this simple general inequality we take the $\min$
(or $\max$ when the commutator appears in the denominator) in order
to obtain the tightest bounds. Any one of the factors inside the
argument of the $\min$ would, on its own, yield a valid inequality. 

In the high temperature classical limit, given the Hamiltonian of
Eq. (\ref{eq:HLAMBDA}), the probability distribution
$\rho_{\Lambda}^{\sf classical~ canonical}$ is Gaussian in the
velocities. Thus, the (time independent) thermal
averages of $(v_{i \ell}^H)^2$ and $(v_{i \ell}^{H})^4$ are,
respectively, given by $G_{v}(0) = \frac{k_{B} T}{m}$ and
${\sf{Tr}}
(\rho_{\Lambda}^{\sf classical~canonical} (v_{i \ell}^{H})^4) = 3
\Big(\frac{k_{B} T}{m} \Big)^2$. Inserting (i) $A =  \Delta\tilde{H}_i^{H}(t) = \Delta V_i^{H}(t),~ B = v_{i \ell}^{H}(t)$, and $C=   v_{i \ell}^{H}(0)$ (or any permutation thereof) into Eq. (\ref{abcClassical}), (ii) Invoking Eq. (\ref{D+min}), (iii) recognizing the time independence of equilibrium averages of the terms in the resulting bound, and (iv)  the above noted temperature dependent equilibrium averages of $(v_{i \ell}^2)$ and $(v_{i \ell}^{H})^4$ leads to
\begin{eqnarray}
\label{minDiff+}
D_{+} \ge \frac{ \hbar k_{B} T}{4\sqrt{3}~m~\sqrt[4]{{\sf
  Tr}\Big(\rho_{\Lambda}^{\sf classical~ canonical} \big(\Delta
  V_{i}^{H}\big)^4\Big)}}
     \max\left\{
  \frac
  {\sqrt[4]{{\sf Tr}\left(
  \rho_{\Lambda}^{\sf classical~canonical}
  \big(\Delta V_i^{H}\big)^4
  \right)}}
  {\sqrt{{\sf Tr}\left(
  \rho_{\Lambda}^{\sf classical~canonical}
  \big(\Delta V_i^{H}\big)^2
  \right)}}
  ,\sqrt[4]{3}
     \right\}.
\end{eqnarray}
This inequality hints at a possible bound on the viscosity in such equilibrated systems. Indeed, whenever the Stokes-Einstein relation holds for high temperature thermal liquids of spherical particles of radius $R$, Eq. (\ref{minDiff+}) (strictly speaking, a bound on $D_{+}$ not on the diffusion constant $D$) further suggests an {\it upper bound on the viscosity} set by the reciprocal of Planck's constant, 
\begin{eqnarray}
\label{etaup}
\eta = \frac{k_{B} T}{6 \pi R D} \lesssim \frac{2m
  ~\sqrt[4]{{\sf Tr}(\rho_{\Lambda}^{\sf classical~ canonical}
  (\Delta V_{i}^{H})^4)}}{\pi\sqrt{3}~R~\hbar}
     \min\left\{
  \frac{\sqrt{{\sf Tr}\left(
  \rho_{\Lambda}^{\sf classical~canonical}
  \big(\Delta V_i^{H}\big)^2
  \right)}}{\sqrt[4]{{\sf Tr}\left(
  \rho_{\Lambda}^{\sf classical~canonical}
  \big(\Delta V_i^{H}\big)^4
  \right)}},\frac{1}{\sqrt[4]{3}}
     \right\}.
  \nonumber\\
\end{eqnarray}
\twocolumngrid\noindent
 A sufficiently high value of the viscosity might thus point to a necessary violation of the Stokes-Einstein relation.     
     
In systems with bounded interactions,
${\sf{Tr}}\left(\rho_{\Lambda}^{\sf {canonical }}\left(
    \Delta V_i^H\right)^{p}\right) \le || V_{i}^{H}||^p$, with $p$ being a
natural number, while, in the classical limit,  for power law interactions, ${\sf{Tr}}\left(\rho_{\Lambda}^{\sf {classical~canonical }}\left( \Delta V_i^H\right)^{p}\right)$ scales as $(k_{B} T)^p$. Eq. (\ref{minDiff+}) is a rigorous semiclassical bound on $D_{+}$ and thus a bound on the diffusion constant $D$ when all substantial velocity autocorrelation functions $G_{v}(t)$ contributing to Eq. (\ref{eq:diffusion}) are positive or, far more generally, when the integral of Eq. (\ref{eq:diffusion}) may be bounded by a finite dimensionless constant $A_D= {\cal{O}}(1)$ multiplying Eq. (\ref{eqnD+}) (i.e., $D \ge A_D ~D_{+})$. As reviewed at the start of this subsection, in most systems \cite{liqGv1,liqGv2,liqGv3,liqGv4,liqGv5,liqGv6,liqGv7}, the oscillations of $G_v$ at long times are not the dominant contributions to the integral of Eq. (\ref{eq:diffusion}) and thus the above bound with a constant $A_D$ (close to $1$) applies. We caution that the negative velocity autocorrelation function contributions to the integral of Eq. (\ref{eq:diffusion}) may render the real diffusion $D$ constant smaller than $D_{+}$ (Eq. (\ref{eqnD+})) and thus our bound to it. Indeed, in the solid phase, oscillatory velocity autocorrelation functions of the ions are anticipated. Accordingly, the diffusion constants in crystalline systems become far smaller than $D_{+}$. However, in all fluids and gases that we examined, our rigorous semiclassical bound on $D_{+}$
was also a bound on the diffusion constant $D$ (see, e.g., Table \ref{real-values} for water). In Section \ref{GKsec}, we will extend the above derivation to other general transport coefficients. 

Section \ref{OTOCBound}, premised on a conjectured link between chaos bounds and autocorrelation functions, suggests a general qualitative bound for the diffusion constant, viscosities, and other transport coefficients.  

\subsubsection{Acceleration bounds in semiclassical systems}

We now return to derive general bounds on particle accelerations. In what follows, we write, following the definition in Eq. (\ref{overdef}),  $\quad {\sf{Tr}}\left(\rho_{\Lambda}\left(v_{i \ell}^{H}(t)\right)^{2}\right) \equiv \overline{v^{2}}+\delta\left(\left(v_{i \ell}^{H}\right)^{2}\right)$ where $\overline{v^{2}}$ marks the system wide average of a squared linear velocity component ($\ell$). By the classical equipartition theorem, at high temperatures, regardless of the specifics of the interactions, the global average of the square particle velocity, $\overline{v^{2}} \equiv {\sf{Tr}}\left(\rho_{\Lambda}^{\sf {classical~canonical }} v_{i \ell}^{2}\right)=\frac{k_{B} T}{m}$.
  For operators 
  $V^{H}_{i}$ that are of bounded norm,
  we have from Eq. (\ref{aa-central-1}), 
  $\left\langle a_{i \ell}^{H}\right\rangle^{2} \leq \frac{4|| V^{H}_i||^{2}}{\hbar^{2}}\left(\frac{k_{B} T}{m}+\delta\left(\left(v_{i \ell}^{H}\right)^{2}\right)\right)$.
  Since (identically) $\sum_{i=1}^{N_{\Lambda}} \delta\left(\left(v_{i \ell}^{H}\right)^{2}\right)=0$, in the high temperature when the classical equipartition is valid, averaging this local bound over all sites (or particles) $i$, \begin{equation}
\label{local-VA}
\overline{\left\langle a_{i \ell}^{H}\right\rangle^{2}} \leq \frac{4 k_{B} T \|V^{H}_{i} \|^{2}}{m \hbar^{2}}.
\end{equation} 
Eq. (\ref{local-VA}) constitutes {\it a universal bound on the average squared acceleration}. Similar to the discussion following Eq. (\ref{vbound12}), we may replace the global average of Eq. (\ref{O:avg}) by the long time average density matrix $\rho_{\Lambda}^{\sf {long }-\sf { time-average }}$. In we do so then we will remove the Heisenberg picture superscript $H$ and the angular brackets marking the average with $\rho_{\Lambda}$ on the lefthand side of Eq. (\ref{local-VA}).

Further yet, sans any global averages, we have the exact inequality
\begin{eqnarray}
\label{potential-eq}
\langle a_{i \ell}^{H}\rangle^{2} \le \frac{4}{\hbar^{2}} \sigma_{{V}_{i}^{H}(t)}^{2} \sigma_{v_{i \ell}^{H}(t)}^{2},
\end{eqnarray}
where, as throughout this work, the simply denoted variances on the righthand side of Eq. (\ref{potential-eq}) are computed with the density matrix $\rho_{\Lambda}$. In the high temperature (i.e., classical) limit, the variance of the velocity for any systems with a Hamiltonian of the form of Eq. (\ref{eq:HLAMBDA}),  
\begin{eqnarray}
\label{potential-eq1}
\frac{1}{N_{\Lambda}} \sum_{i=1}^{N_{\Lambda}} \sigma_{v_{i \ell}^{H}(t)}^{2}  = 
\frac{k_{B} T}{m}.
\end{eqnarray} 

If, for energy densities set by the temperature $T$ in the region of interest, the potential energy varies as a power law of the spatial distances then, in the semiclassical limit where computations with the classical canonical ensemble are valid, both the average kinetic and average potential energies will scale linearly in the temperature $T$. In such instances, if the potential and kinetic energies are comparable, we may, for the said temperatures $T$, bound the variance of $\tilde{H}_i^H$ by that of the local kinetic energy term $K^H_{i}=\frac{\left(p_{i \ell}^{H}(t)\right)^{2}}{2 m}$ times (in the regime where classical canonical averages may be performed) a temperature independent constant ${\sf C}$ that is of order unity, i.e., 
\begin{eqnarray}
\label{potential-eq2}
\sum_{i=1}^{N_{\Lambda}} \sigma^2_{\tilde{H}_{i}^{H}} \le {\sf C} \sum_{i=1}^{N_{\Lambda}} \sigma_{K_i^H}^{2}.
\end{eqnarray} 
This yields
\begin{eqnarray}
\label{upper-a}
&& \overline{\langle a_{i \ell}^{H}\rangle^{2}} \le \frac{4}{\hbar^{2} N_{\Lambda}} \sum_{i=1}^{N_{\Lambda}} \sigma_{\tilde{H}_{i}^{H}(t)}^{2} \sigma_{v_{i \ell}^{H}(t)}^{2} 
\le \frac{2{\sf C}(k_{B} T)^3}{m\hbar^{2}}.
\end{eqnarray}
The largest upper bound in Eq. (\ref{upper-a}) applies when the global average of Eq. (\ref{potential-eq1}) may be invoked in the inequalities of Eqs. (\ref{potential-eq}, \ref{potential-eq2}).

\subsection{Expectation values of higher moments of the acceleration and force}
\label{sec:a-higher}
As discussed in Section \ref{high-moment-derivative}, in the semiclassical limit, Eq. (\ref{squared-ineq}) is equivalent to Eqs. (\ref{central1}, \ref{a-central-1}). Thus, in semiclassical regime, we will reproduce the results of Section \ref{aa-bounds}. Indeed, in the high temperature limit, with $Q_{i}^{H}$ set equal to the particle velocity components, Eq. (\ref{squared-ineq}) becomes
 \begin{eqnarray}
\label{squared-ineq'}
\frac{1}{N_{\Lambda}} \sum_{i=1}^{N_{\Lambda}} \Big\langle \Big ( \frac{d v_{il}^{H}}{dt} \Big)^2  \Big\rangle \nonumber
\\ \le \frac{4 k_{B} T^{2} C_{v,i} }{\hbar^2}  {\sf{Tr}} (\rho_{\Lambda}^{\sf classical~ canonical} (\Delta v^{H}_{il})^2) \nonumber
\\ =  \frac{4 k_{B}^2 T^{3} C_{v,i} }{m \hbar^2}. 
\end{eqnarray}
In the last line of Eq. (\ref{squared-ineq'}), we invoked, as we have earlier, the high temperature average, $ {\sf{Tr}} (\rho_{\Lambda}^{\sf classical~ canonical} (\Delta v^{H}_{il})^2) = \frac{k_{B} T}{m}$. Here, $k_{B} T^{2} C_{v,i} = {\sf{Tr}} (\rho_{\Lambda}^{\sf classical~canonical} (\tilde{H}_{i}^{H})^2)$ with, as in Section \ref{aa-bounds}, $\tilde{H}_{i}^{H}=\sum_{j} V\left({\bf r}_{i}^{H}, {\bf r}_{j}^{H}\right) \equiv V^{H}_{i}$. For a pair potential $V$ that is of a general power law form in the coordinates, the classical variance ${\sf{Tr}} (\rho_{\Lambda}^{\sf classical~canonical} (V_{i}^{H})^2)$ is quadratic in $(k_B T)$. For interactions $||V_{i}^{H}||$ of bounded norm, Eq. (\ref{squared-ineq'}) trivially implies a bound on the average squared acceleration, 
$\frac{1}{N_{\Lambda}} \sum_{i=1}^{N_{\Lambda}} \Big\langle \Big ( \frac{d v_{il}^{H}}{dt} \Big)^2  \Big\rangle 
\le \frac{4 k_{B} T ||V^H_i||^2 }{m \hbar^2}$.  
Similar to the closing discussion of Section \ref{sec:v-high}, we remark that bounds on the expectation values of $\Big ( \frac{dr_{il}^{H}}{dt} \Big)^n$ with $n>2$ may be reproduced analogously to those derived for $n=2$ above (Section \ref{high-moment-derivative}).

Since the force on the $i$-th particle, ${{\bf \sf{f}}}^H_i = m {{\bf a}}^H_i$, all of the above bounds on the acceleration translate into those on the forces in systems that are in thermal equilibrium. Thus, e.g., in such systems with bounded potentials, we trivially have 
\begin{eqnarray}
\label{squared-ineq''f}
\frac{1}{N_{\Lambda}} \sum_{i=1}^{N_{\Lambda}} \Big\langle \Big ({\sf f}_{il}^{H} \Big)^2  \Big\rangle  \le \frac{4 m k_{B} T ||V^H_i||^2 }{ \hbar^2}.  
\end{eqnarray}
In general spatial dimension $d$, the upper bound on the average squared force per particle is, trivially, $d$ times larger than that of Eq. (\ref{squared-ineq''f}) for a particular Cartesian component of the force. Similar to Eq. (\ref{srih'}), we can write the ratio of the upper bound on the average squared force of 
Eq. (\ref{squared-ineq''f}) to the square norm of the local effective potential energy $||V^H_i||^2$ in terms of the thermal de-Broglie wavelength $\lambda_{T}$ of Eq. (\ref{dBeq}). That is, $\frac{1}{N_{\Lambda}} \sum_{i=1}^{N_{\Lambda}} \Big\langle \Big ({\sf f}_{il}^{H} \Big)^2  \Big\rangle/||V^H_i||^2 \le 8 \pi/\lambda_{T}^2$. In Section \ref{sec:space} and Appendix \ref{higher_grad}, we will very generally demonstrate that, in semiclassical thermal systems, gradients of arbitrary functions divided by the functions themselves (i.e., the derivatives of the logarithm of general functions) cannot have a norm exceeding ${\cal{O}}(1/\lambda_{T})$. This constraint is consistent with the result of Eq. (\ref{squared-ineq''f}) when this general function is taken to be $V_{i}^{H}$ (for which the partial derivatives are the force components ${{\bf \sf{f}}}^H_{i \ell}$). We emphasize that, given our derivation employing the uncertainty inequalities, in {\it all of the bounds} derived in this Section, the norm $||V_i^{H}(t)||$ as well as $\sqrt{ {\sf{Tr}} (\rho_{\Lambda}^{\sf classical~ canonical} (V_{i}^{H})^2)}$ may be replaced by the standard deviation of $\tilde{H}_{i}^{H}= V_{i}^{H}$ evaluated with the classical canonical probability density.  

As in Section \ref{sec:compact}, all of the bounds derived for the linear acceleration may be replicated for the angular acceleration (and, following a multiplication by the moment of inertia, on the torque).

\section{Bounds on acceleration (and force or stress) rates}
\label{sec:stress}
We will now derive bounds on the rates of change of the acceleration (and hence on the force or stress) rates. Towards this end, we first note that given the general many body Hamiltonian of Eq. (\ref{eq:HLAMBDA}). With, as in Section \ref{a-bounds}, $a_{i \ell}^{H}$, the $\ell$th Cartesian component of the acceleration of the $i$-th particle, the Heisenberg (essentially Newtonian) equations of motion read 
\begin{eqnarray}
\label{ai-eq}
a_{i \ell}^{H} = 
- \frac{1}{m} \frac{ \partial V_{i}^{H} }{\partial r_{i \ell}^{H}} =  - \frac{1}{m} \frac{ \partial V_{i}(\{{\bf r}^{H}_{j}\})}{\partial r_{i \ell}^{H}}.
\end{eqnarray}
Eq. (\ref{ai-eq}) is written so as to elucidate that the {\it minimal} local Hamiltonian $V_{i}^{H}$ determining the acceleration of particle $i$ is none other than the static function $V_i$ of the time dependent Heisenberg picture position operators $\{{\bf r}^{H}_{j}\}$. Thus, the time derivatives of the acceleration components are
\begin{eqnarray}
\label{ai1-eq}
 \frac{d a_{i \ell}^{H}}{dt} &=&
-  \frac{1}{m}  \sum_{j \ell'} \frac{d r^H_{j \ell'}}{dt}  \frac{\partial^{2} V_{i}^{H}}{\partial r_{i \ell}^{H} \partial r^H_{j \ell'}} =  -\frac{1}{m} \sum_{j \ell'} \frac{p_{j \ell'}^{H}}{m} \frac{\partial^{2} V_{i}^{H}}{\partial r_{i \ell}^{H} \partial r^H_{j \ell'}} \nonumber
\\ &=& \frac{i}{\hbar} [  \sum_{j \ell'} \frac{(p_{j \ell'}^{H})^{2}}{2m}, a_{i \ell}^{H} ].
\end{eqnarray}
As Eq. (\ref{ai1-eq}) makes clear, the relevant minimal local Hamiltonian that evolves the acceleration of particle $i$ in time is $\tilde{H}_{i}^{H} = \sum_{j \ell'} \frac{(p_{j \ell'}^{H})^{2}}{2m}$. The pertinent particles $j$ (and Cartesian components $\ell'$) appearing in the latter sum are of those particles $j$ that interact with particle $i$ with non-negligible gradients (relative to the Cartesian components $\ell'$ of ${\bf r}^{H}_{j}$)
so that they have substantial contributions to the sum in Eq. (\ref{ai1-eq}).  Eq. (\ref{ai1-eq}) thus implies  Eq. (\ref{first-eq}) with the said identification of the appropriate Hamiltonian of $\tilde{H}_{i}^{H}= \sum_{j} \frac{(p_{j \ell}^{H})^{2}}{2m}$ for our local quantity of interest currently being the acceleration, $Q_{i}^{H} = a_{i \ell}^{H}$. From Eq. (\ref{a-central-1}), 
\begin{eqnarray}
\label{a-central-11}
&& {\sf{Tr}} \Big (\rho_{\Lambda}^{\sf {canonical }} \big(\sum_{j \ell'} \frac{(p_{j \ell'}^{H})^{2}}{2m} \big)^{2} \Big) -
\Big({\sf{Tr}} \Big (\rho_{\Lambda}^{\sf {canonical }} \sum_{j \ell'} \frac{(p_{j \ell'}^{H})^{2}}{2m} \Big) \Big)^2\nonumber
\\ &&  \geq \frac{\hbar^{2}}{4 N_{\Lambda}} \sum_{i=1}^{N_{\Lambda}} \frac{\left({\sf{Tr}}\left(\rho_{\Lambda} \frac{d  a_{i \ell}^{H}}{d t}\right)\right)^{2}}{{\sf{Tr}}\left(\rho_{\Lambda}\left( \Delta a_{i \ell}^{H}\right)^{2}\right)}.
\end{eqnarray}
We now consider the case when there is a finite number of terms ${\sf k}$ that appear in the $(j \ell')$ sums of Eqs. (\ref{ai-eq}, \ref{ai1-eq}) (and thus of the kinetic energy sum $ \tilde{H}^H_i = \sum_{j \ell'} \frac{(p_{j \ell'}^{H})^{2}}{2m}$). For a system in $d$ spatial dimensions, in which each particle interacts with ${\sf z}$ other particles, ${\sf k} = {\sf z} d$. (We refer to ${\sf z}$ as the ``effectvive coordiation number''. The latter may diverge in systems with long range interactions.) For the full many body Hamiltonian $H_{\Lambda}$ of Eq. (\ref{eq:HLAMBDA}), in the classical limit, 
 \begin{eqnarray}
 \label{fluct-eqa}
&&  {\sf{Tr}} \Big (\rho_{\Lambda}^{\sf {classical~canonical }} \big(\sum_{j \ell'} \frac{(p_{j \ell}^{H})^{2}}{2m} \big)^{2} \Big) \nonumber
\\  &&- \Big({\sf{Tr}} \Big (\rho_{\Lambda}^{\sf {classical~canonical }} \sum_{j \ell'} \frac{(p_{j \ell}^{H})^{2}}{2m} \Big) \Big)^2 \nonumber
\\ && = \frac{{\sf k}(k_{B} T)^2}{2}. 
 \end{eqnarray}
Substituting Eq. (\ref{fluct-eqa}) into the upper bound of Eq. (\ref{a-central-11}),
 \begin{eqnarray}
 \label{a-rate-eq}
 2 {\sf k} \Big(\frac{k_{B} T}{\hbar} \Big)^2 \geq \frac{1}{N_{\Lambda}} \sum_{i=1}^{N_{\Lambda}} \frac{\left({\sf{Tr}}\left(\rho_{\Lambda} \frac{d  a_{i \ell'}^{H}}{d t}\right)\right)^{2}}{{\sf{Tr}}\left(\rho_{\Lambda}\left( \Delta a_{i \ell'}^{H}\right)^{2}\right)}.
 \end{eqnarray}
For systems with finite range interactions, ${\sf k}$ is finite. When the pair interactions are long ranged, we may truncate ${\sf k}$ at finite values at the price of increasing error in the evaluation of both the numerator and denominator of Eq. (\ref{fluct-eqa}) from Eq. (\ref{ai1-eq}) and Eq. (\ref{ai-eq}) respectively. 
 Heisenberg's equations of motion for the system of uniform masses reproduce the Newtonian ones and we may replace $a^H_{i \ell'}$ by the $\ell'$-th Cartesian component of the force ${{\bf {\bf \sf{f}}}}^H_{i}$ acting on the $i$-th particle (i.e., $a^H_{i \ell'} \to {\sf f}^H_{i \ell'} = m a^H_{i \ell'}$). In other words,
 \newline
 
 $\bullet$ Eq. (\ref{a-rate-eq}) implies that, in semiclassical thermal equilibrium, force components cannot, on average, fluctuate at a rate exceeding $\frac{k_{B} T}{\hbar} \sqrt{2{\sf k}}$. 
 \newline
 
In an analogous manner, force contributions to the temporal rate of change of general continuum stress tensor components may be similarly bounded. 

We conclude this Section by reiterating yet again that, as stressed in Appendix \ref{sec:comments}, numerous Hamiltonians other than the below minimal $\tilde{H}_{i}^{H}$  may be considered. In various instances, these might to bounds stronger than those derived above.

\onecolumngrid\noindent

\section{Bounds on General Transport Coefficients}
\label{GsKsec}
In Section \ref{t-auto}, we illustrated how bounds on the time derivatives of general autocorrelation functions may be derived. Building on this, a particular Green-Kubo relation \cite{Green1954,KMS2} was invoked in Section \ref{sec:diff} to examine the diffusion constant. In this Section, we sketch what occurs for general transport coefficients $\gamma$ in local theories and illustrate how this may be applied to viscosity and electrical conductivity bounds.

\subsection{General formalism}
\label{GKsec}
Generally, the Green-Kubo relations connect transport coefficients to integrals of the autocorrelation functions of the time derivatives of operators $Y^{H}$ (see Table \ref{tab:GY}), 
\begin{eqnarray}
\label{eq:GY}
\gamma=  \int_{0}^{\infty} dt ~ {\sf{Tr}} (\rho_{\Lambda}^{\sf canonical}~ \dot{Y}^{H}(t)~ \dot{Y}^{H}(0)) \nonumber
\\ \equiv \int_{0}^{\infty} dt~ G_{\dot{Y}}(t).
\end{eqnarray} 
Similar to Section \ref{sec:diff}, we will bound 
\begin{eqnarray}
\label{gamma+eqeq}
\gamma_{+} \equiv \int_{0}^{t^{\dot{Y}}} dt~ G_{\dot{Y}}(t),
\end{eqnarray}
with $t_{\dot{Y}}$ denoting the smallest positive time at which $G_{\dot{Y}}(t)$ vanishes. 
In what follows, we extend the analysis of Section \ref{sec:diff} in order to bound general transport coefficients. Specifically, we will consider observables that are global averages of local quantities (see Table \ref{tab:GY}), 
\begin{eqnarray}
\label{eqYI}
Y^{H} =  \frac{ {\sf{n}} }{N_{\Lambda}} \sum_{i=1}^{N_{\Lambda}} Y_{i}^{H},
\end{eqnarray}
 with ${\sf n} \equiv N_{\Lambda}/{\sf V}$ being the number density of Eq. (\ref{etanhh}). Here, as in Section \ref{sec:diff}, ${\sf V}$ marks the volume. In Eq. (\ref{eqYI}), $\{Y_{i}^{H}\}$ are local operators associated with individual particles $i$. For instance (see Table \ref{tab:GY}), for the shear viscosity $(\eta)$, the relevant local single particle operator is $Y^{i}_{H} = r^H_{i\ell}p^H_{i\ell'}$ (and $\dot{Y}_{i}^{H}$ is the sum of a local kinetic Hamiltonian and a virial). 
We associate, as in the earlier Sections, with each such local $Y^{i}_{H}$ a local Hamiltonian $\tilde{H}^{H}_{i}$.  
\bigskip

\noindent {\underline{\it{Quantum Bounds.}}}
\newline
We now invoke the uncertainty relation of
Eq. (\ref{eq:uncertain:Q}) with $Q_{i}^{H}(t) = (\dot{Y}^{H}(t)
\dot{Y}^{H}(0))$.
Doing so, we have
\begin{eqnarray}
\label{gammaeq1}
\gamma_{+} \equiv \int_{0}^{t^{\dot{Y}}} G_{\dot{Y}}(t') dt' \nonumber
\\ = \int_{0}^{t^{\dot{Y}}} {\sf{Tr}}(\rho_{\Lambda}^{\sf{canonical}} ~\dot{Y}^{H}(0) \dot{Y}^{H}(t')) dt'  \nonumber
\\ \ge \frac{1}{2} G_{\dot{Y}}(0) t_{\min}.
\end{eqnarray}
Here, $t_{\min}$ denotes the time for $G_{\dot{Y}}$ to drop from its maximal value at time $t=0$ to zero if $ G_{\dot{Y}}(0)$ were to vary with time at the highest rate allowed by the uncertainty relations. Similar to Section \ref{sec:diff}, we will find  $t_{\min}$ by bounding, from above, the time derivative of $G_{\dot{Y}}$ and then examining the minimal time required for $G_{\dot{Y}}$ to drop from its value at $t=0$ to its vanishing value $t=t^{\dot{Y}}$ assuming this maximal allowed drop of $G_{v}(t)$. By particle symmetry, all of the pair correlators \newline
\begin{eqnarray}
\label{samei} 
{\sf{Tr}}(\rho_{\Lambda}^{\sf canonical}   (\dot{Y}_{i}^{H}(t)) (\dot{Y}^{H}(0))) 
\\ \mbox{ are particle $i$ independent.} \nonumber
\end{eqnarray}
\bigskip

Putting all of the pieces together, we arrive at a bound on the minimal drop-off time $t_{\min}$ appearing in Eq. (\ref{gammaeq1}),  
\begin{eqnarray}
\label{tmin}
\frac{1}{t_{\min}} 
  = \max_{t} \Big|\frac{{\sf{Tr}}(\rho_{\Lambda}^{\sf canonical}
  \sum_{i=1}^{N_{\Lambda}} \frac{d}{dt} (\dot{Y}_{i}^{H}(t))
  \dot{Y}^{H}(0))}
  {{{\sf{Tr}}(\rho_{\Lambda}^{\sf canonical}  \sum_{i=1}^{N_{\Lambda}}
  (\dot{Y}_{i}^{H}(0) \dot{Y}^{H}(0)))}} \Big|
  =  \max_{t}
  \Big|\frac{{\sf{Tr}}(\rho_{\Lambda}^{\sf canonical}  \frac{d}{dt}(
  \dot{Y}_{i}^{H}(t)) \dot{Y}^{H}(0))} 
  {{\sf{Tr}}(\rho_{\Lambda}^{\sf canonical}  (\dot{Y}_{i}^{H}(0) \dot{Y}^{H}(0)))} \Big|,
\end{eqnarray} 
where we used the $i$ independence of the ratio in
Eq. (\ref{samei}). For each of the operators
$\{Y^{H}_{i}\}_{i=1}^{N_{\Lambda}}$,
\begin{eqnarray}
\label{YYYYY}
\Big|{\sf{Tr}}(\rho_{\Lambda}^{\sf canonical}
  \frac{d}{dt}(\dot{Y}_{i}^{H}(t)) \dot{Y}^{H}(0)) \Big| 
  = \frac{1}{\hbar} \Big|{\sf{Tr}}(\rho_{\Lambda}^{\sf canonical}
  [\Delta \tilde{H}^{H}_{i}(t), \dot{Y}_{i}^{H}(t)
  ]\dot{Y}^{H}(0))\Big| &&
  \nonumber\\ && \hspace{-9cm}
  \le  \frac{1}{\hbar} \left(
  \sqrt{{\sf{Tr}}\Big(\rho_{\Lambda}^{\sf canonical}  \big(\Delta
  \tilde{H}^{H}_{i}(t)\big)^2\Big) {\sf{Tr}}\Big(\rho_{\Lambda}^{\sf canonical}
  \dot{Y}_{i}^{H}(t) \big(\dot{Y}^{H}(0)\big)^2  \dot{Y}_{i}^{H}(t)}\Big)
  \right. \nonumber\\ && \hspace{-7cm}\left.
                 +\sqrt{{\sf{Tr}}\Big(\rho_{\Lambda}^{\sf canonical}  \big(\dot{Y}^{H}(0)\big)^2\Big) {\sf{Tr}}\Big(\rho_{\Lambda}^{\sf canonical}
  \dot{Y}_{i}^{H}(t) \big(\Delta
  \tilde{H}^{H}_{i}(t)\big)^2  \dot{Y}_{i}^{H}(t)}\Big)
  \right),
                 \end{eqnarray}
where we invoked Eq. (\ref{CSg'}) for the two terms $({\sf Tr}(\rho_{\Lambda}^{\sf canonical}
\Delta \tilde{H}^{H}_{i}(t) \dot{Y}_{i}^{H}(t)
  \dot{Y}^{H}(0))$) and  $({\sf Tr}(\rho_{\Lambda}^{\sf canonical}
 \dot{Y}_{i}^{H}(t) \Delta \tilde{H}^{H}_{i}(t)
  \dot{Y}^{H}(0))$ whose difference constitutes the expectation value ${\sf{Tr}}(\rho_{\Lambda}^{\sf canonical}
  [\Delta \tilde{H}^{H}_{i}(t), \dot{Y}_{i}^{H}(t)
  ]\dot{Y}^{H}(0))$.  From Eqs. (\ref{gammaeq1},\ref{tmin},\ref{YYYYY}), explicitly underscoring the time independence of the canonical equilibrium average of ${\sf{Tr}}\Big(\rho_{\Lambda}^{\sf canonical}  \big(\Delta
  \tilde{H}^{H}_{i}(t)\big)^2\Big)$ and ${{\sf{Tr}}\Big(\rho_{\Lambda}^{\sf canonical}
  \dot{Y}_{i}^{H}(t) \big(\Delta
  \tilde{H}^{H}_{i}(t)\big)^2  \dot{Y}_{i}^{H}(t)}\Big)$, we have 
\begin{eqnarray}
  \label{gammagamma}
  \gamma_{+} 
  \ge 
  \frac{\hbar}{2}
  G_{\dot{Y}}(0)  
  \left|{\sf{Tr}}\Big(\rho_{\Lambda}^{\sf canonical} \dot{Y}_{i}^{H}(0)
  \dot{Y}^{H}(0)\Big)\right|
  \times&&\nonumber\\
  \nonumber\\ && \hspace{-3cm}
  \left(
  \sqrt{{\sf{Tr}}\Big(\rho_{\Lambda}^{\sf canonical}  \big(\Delta
  \tilde{H}^{H}_{i}(0)\big)^2\Big) {\sf{Tr}}\Big(\rho_{\Lambda}^{\sf canonical}
  \dot{Y}_{i}^{H}(t) \big(\dot{Y}^{H}(0)\big)^2  \dot{Y}_{i}^{H}(t)}\Big)
                 \right.
\nonumber\\ && \hspace{-2cm}\left.
                                  +\sqrt{{\sf{Tr}}\Big(\rho_{\Lambda}^{\sf
                                  canonical}
                                  \big(\dot{Y}^{H}(0)\big)^2\Big)
                                  {\sf{Tr}}\Big(\rho_{\Lambda}^{\sf
                                  canonical} 
                                  \dot{Y}_{i}^{H}(0) \big(\Delta
                                  \tilde{H}^{H}_{i}(0)\big)^2  \dot{Y}_{i}^{H}(0)}\Big)
                                  \right)^{-1}.
\end{eqnarray}

\twocolumngrid\noindent
This general bound constitutes an analog of Eq. (\ref{minDiff}) for transport coefficients associated with {\it global}~ $Y_{H}$ that are of the form of Eq. (\ref{eqYI}). Similar to our discussion in the earlier Sections, we may invoke Eq. (\ref{cv2}) to explicitly express the bound in terms of an effective local heat capacity and the temperature. We underscore that in systems with local interactions, the requisite $\{\tilde{H}_{i}^{H}\}$ are local for the typical operators $\{Y_{i}^{H}\}$ appearing in Table \ref{tab:GY}. In Secions \ref{sec:viscosity} and \ref{sec:conduct}, we discuss applications of the above derived inequalities for obtaining lower bounds on the bulk and shear viscosities and electrical conductivity. As in Section \ref{sec:diff}, we further stress that all of the bounds derived thus far in this Section apply at all temperatures and {\it are free of semiclassical approximations}. 
The factor of $\hbar$ in the numerator of Eq. (\ref{gammagamma}) illustrates that, similar to our particular results for the diffusion constant, other transport coefficients may be bounded from below by {\it finite values having their origins in quantum fluctuations}. 

\bigskip

\begin{table*}
  \begin{tabular}{c|c|c|c|m{0.22\textwidth}}
    {\large Transport coefficient} & {\large $\gamma$} & {\large $Y$}
    & {\large $\dot{Y}$} & \parbox{0.22\textwidth}{\large Comments} \\
    \hline
    Diffusion constant, $D$ & $D$ & $r_{i\ell}$ & $v_{i\ell}$ & \parbox[c][1.1in]{0.22\textwidth}{The canonical ensemble velocity autocorrelation is homogeneous and isotropic, being the same for all particles $i\in\{1,2,\dots,N_{\Lambda}\}$ and Cartesian components $\ell \in\{1,2, \ldots, d\}$.}\\
    \hline
    Shear viscosity, $\eta$ & $\frac{k_BT}{{\sf V}} \eta$ &
                                                      $\frac{1}{{\sf V}}\sum\limits_ir_{i\ell}p_{i\ell'}$
    &
      $\frac{1}{{\sf V}}\sum\limits_i\left(\frac{p_{i\ell}p_{i\ell'}}{m_i}+r_{i\ell}{\sf f}_{i\ell'}\right)$
                         & \parbox[c][0.52in]{0.22\textwidth}{$\dot{Y}$ is an
                           off-diagonal ($\ell \neq\ell'$) component of the stress
                           tensor.}\\
    \hline
    Bulk viscosity, $\zeta$ & $\frac{k_BT}{{\sf V}} \zeta$ & ${\sf{Q}}-\langle
                                                       {\sf{Q}}\rangle_{\sf canonical}$ &
                                                                    $P-\langle
                                                                   P\rangle_{\sf canonical}$ & 
                                                      \parbox[c][0.6in]{0.22\textwidth}{${\sf{Q}}=\frac{1}{d{\sf V}}\sum\limits_i{\bf r}_i\cdot{\bf p}_i$,\\$P=\frac{1}{d{\sf V}}\sum\limits_i\left(\frac{{\bf p}_i^2}{m_i}+{\bf r}_i\cdot {{\bf \sf{f}}}_i\right)$}
    \\
    \hline
    Thermal conductivity, $\kappa$ & $\frac{k_BT^2}{{\sf V}} \kappa$ & $\frac{1}{{\sf V}}\sum\limits_ir_{i\ell}
                                                                 \left(\epsilon_i-\left\langle\epsilon_i\right\rangle_{\sf canonical}\right)$
    & $\frac{1}{{\sf V}}\sum\limits_i\left[
      v_{i\ell}\left(\epsilon_i-\left\langle\epsilon_i\right\rangle_{\sf canonical}\right)
      +r_{i\ell}\dot{\epsilon}_i\right]$
                         & \parbox[c][0.62in]{0.22\textwidth}{The (canonical probability density) autocorrelation function is the same for all $d$ Cartesian components.}\\
    \hline
    Electrical conductivity, $\sigma$ & $\frac{k_BT}{{\sf V}} \sigma$ 
    & $\frac{1}{{\sf V}}\sum\limits_{i=1}^{N_{\Lambda}} e_{i} r_{i \ell}$
    & $\frac{1}{{\sf V}}\sum\limits_{i=1}^{N_{\Lambda}} e_{i} v_{i \ell}$
                         & \parbox[c][0.5in]{0.22\textwidth}{$\dot{Y}$ is a component of the average electric
                           current density.}\\
  \end{tabular}
\caption{
  The pertinent quantity $Y(t)$ appearing in Eq. (\ref{eq:GY}) for some transport coefficients.
  Here, ${\bf r}_i$, ${\bf v}_i$, and, ${\bf p}_i$ are the
  position, velocity, and, momentum of the $i^{\mbox{\tiny th}}$ paricle,
  respectively. The subscript, $\alpha$, represents the Cartesian
  components of the corresponding quantities. The force on the
  $i^{\mbox{\tiny th}}$ paricle, and the energy associated with it, are
  denoted by ${{\bf \sf{f}}}_i$ and $\epsilon_i$, respectively, $e.g$, for a
  system with pairwise interactions,
  $\epsilon_i=\frac{{\bf p}_i^2}{2m}+\sum\limits_{j}{\cal V}_{ij}$.
  In this table,
  $\langle X\rangle_{\sf canonical}$ is ${\sf{Tr}} \left(\rho_{\Lambda}^{\sf
      canonical}~ X \right)$. All the operators in this table are Heisenberg picture
  operators (the superscript $H$ is omitted for brevity).
  }
\label{tab:GY}
\end{table*}

\noindent {\underline{\it{Semiclassical Bounds.}}}
\newline
We now consider the semiclassical limit. We follow the same sequence of steps as in Section \ref{sec:diff}. Replicating the steps of Eq. (\ref{aa-central-111+!}) to Eq. (\ref{YYYYY}), 
\onecolumngrid
\begin{eqnarray}
\label{ZYZYZ}
&& \Big|{\sf{Tr}}(\rho_{\Lambda}^{\sf canonical}
  \frac{d}{dt}(\dot{Y}_{i}^{H}(t)) \dot{Y}^{H}(0)) \Big| 
   \le \frac{1}{\hbar} \Big(\Big|{\sf{Tr}}(\rho_{\Lambda}^{\sf canonical}
  \Delta \tilde{H}^{H}_{i}(t) \dot{Y}_{i}^{H}(t)
  \dot{Y}^{H}(0))\Big| + \Big|{\sf{Tr}}(\rho_{\Lambda}^{\sf canonical}
   \dot{Y}_{i}^{H}(t) \Delta \tilde{H}^{H}_{i}(t)
  \dot{Y}^{H}(0))\Big|  \Big)
  \nonumber\\ && 
  \le  \frac{2}{\hbar} \max  \Big\{ \Big(\Big|{\sf{Tr}}(\rho_{\Lambda}^{\sf canonical}
  \Delta \tilde{H}^{H}_{i}(t) \dot{Y}_{i}^{H}(t)
  \dot{Y}^{H}(0))\Big| , \Big|{\sf{Tr}}(\rho_{\Lambda}^{\sf canonical}
   \dot{Y}_{i}^{H}(t) \Delta \tilde{H}^{H}_{i}(t)
  \dot{Y}^{H}(0))\Big|  \Big) \Big \}
   \nonumber\\ 
   && \sim \frac{2}{\hbar}  \Big|{\sf{Tr}}(\rho_{\Lambda}^{\sf classical~canonical}
  \Delta \tilde{H}^{H}_{i}(t) \dot{Y}_{i}^{H}(t)
  \dot{Y}^{H}(0))\Big| = \frac{2}{\hbar}  \Big|{\sf{Tr}}(\rho_{\Lambda}^{\sf classical~canonical}
   \dot{Y}_{i}^{H}(t) \Delta \tilde{H}^{H}_{i}(t)
  \dot{Y}^{H}(0))\Big|.
     \end{eqnarray}
      Here, as in Eq. (\ref{aa-central-111+!}), $\sim$ denotes the semiclassical limit. 
     Thus, from Eq. (\ref{tmin}),
     \begin{eqnarray}
     \label{tminclass}
     t_{\min} \ge \frac{\hbar}{2} \Big| \frac{{\sf{Tr}}(\rho_{\Lambda}^{\sf classical~canonical}~  \dot{Y}_{i}^{H}(0) \dot{Y}^{H}(0))}{{\sf{Tr}}(\rho_{\Lambda}^{\sf classical~canonical}~
  \Delta \tilde{H}^{H}_{i}(t) \dot{Y}_{i}^{H}(t)
  \dot{Y}^{H}(0))} \Big|.
     \end{eqnarray}

Next, we apply Eq. (\ref{abcClassical}) for the classical canonical average appearing in the denominator of Eq. (\ref{tminclass}) and plug the resultant bound into Eq. (\ref{gammaeq1}). This yields
\onecolumngrid\noindent
\begin{eqnarray}
  \label{gammagammaClassical}
  \gamma_{+} 
  \ge 
  \frac{\hbar}{4}
  \frac{{\sf Tr}\Big(\rho_{\Lambda}^{\sf
  classical~canonical}\big(\dot{Y}^{H}\big)^2\Big)
  \left|{\sf{Tr}}\Big(\rho_{\Lambda}^{\sf classical~canonical} \dot{Y}_{i}^{H}
  \dot{Y}^{H}\Big)\right|}{\sqrt[4]{
  {\sf Tr}\Big(\rho_{\Lambda}^{\sf
  classical~canonical}\big(\dot{Y}^{H}\big)^4\Big)
  {\sf Tr}\Big(\rho_{\Lambda}^{\sf
  classical~canonical}\big(\dot{Y}_{i}^{H}\big)^4\Big)
  {\sf Tr}\Big(\rho_{\Lambda}^{\sf
  classical~canonical}\big(\Delta \tilde{H}^{H}_{i}\big)^4\Big)
  }}
  \times
  &&
     \nonumber\\
  \nonumber\\
   &&\hspace{-13cm}
                 \max\left\{
                 \frac{\sqrt[4]{{\sf Tr}\Big(\rho_{\Lambda}^{\sf
                 classical~canonical}\big(\dot{Y}^{H}\big)^4\Big)}}{\sqrt{{\sf
                 Tr}\Big(\rho_{\Lambda}^{\sf
                 classical~canonical}\big(\dot{Y}^{H}\big)^2\Big)}}, 
                 \frac{\sqrt[4]{{\sf Tr}\Big(\rho_{\Lambda}^{\sf
                 classical~canonical}\big(\dot{Y}_i^{H}\big)^4\Big)}}{\sqrt{{\sf
                 Tr}\Big(\rho_{\Lambda}^{\sf
                 classical~canonical}\big(\dot{Y}_i^{H}\big)^2\Big)}}, 
                 \frac{\sqrt[4]{{\sf Tr}\Big(\rho_{\Lambda}^{\sf
                 classical~canonical}\big(\Delta \tilde{H}^{H}_{i}\big)^4\Big)}}{\sqrt{{\sf
                 Tr}\Big(\rho_{\Lambda}^{\sf
                 classical~canonical}\big(\Delta \tilde{H}^{H}_{i}\big)^2\Big)}}
      \right\}.
\end{eqnarray}
\twocolumngrid\noindent
Here, we highlighted the time independence of the resulting equilibrium averages by omitting the time arguments. 
\bigskip

\noindent {\underline{\it{Order of Magnitude of General Semiclassical Bounds.}}}
\newline
In what follows, we will only discuss the order of magnitude of Eq. (\ref{gammagammaClassical}) to see what it may physically imply in generic semiclassical situations. The arguments of the $\max$ function in Eq. (\ref{gammagammaClassical}) are of order unity. This yields a factor of the
dimensionless correlation coefficient between $\dot{Y}^H$ and
$\dot{Y}^H_i$, i.e., 
\begin{eqnarray}
\frac{
  \left|{\sf{Tr}}\Big(\rho_{\Lambda}^{\sf classical~canonical} \dot{Y}_{i}^{H}
  \dot{Y}^{H}\Big)\right|}{\sqrt[4]{
  {\sf Tr}\Big(\rho_{\Lambda}^{\sf
  classical~canonical}\big(\dot{Y}^{H}\big)^4\Big)
  {\sf Tr}\Big(\rho_{\Lambda}^{\sf
  classical~canonical}\big(\dot{Y}_{i}^{H}\big)^4\Big)
   }} \nonumber
\end{eqnarray}
which, again, is typically of order unity. We are then left with the following order of magnitude bound,
\begin{eqnarray}
  \gamma_+
  &\ge&
  {\cal O}\left(
  \frac{\hbar~{\sf Tr}\Big(\rho_{\Lambda}^{\sf classical~canonical}\big(\dot{Y}^{H}\big)^2\Big)}{\sqrt[4]{{\sf Tr}\Big(\rho_{\Lambda}^{\sf
        classical~canonical}\big(\Delta \tilde{H}^{H}_{i}\big)^4\Big)}}
  \right).
  \label{gammagammaClassicalNonrigor}
  \end{eqnarray}
 Assuming (a) the fourth root of the fourth moment of the local Hamiltonian fluctuations $\Delta \tilde{H}^{H}_{i}$ to be of the same order of magnitude as the standard deviation and that (b) the potential energy contributions to $\Delta \tilde{H}^{H}_{i}$ are of the same order of magnitude or smaller than those of the kinetic energy fluctuations (which, by classical equipartition, are of the order of $k_{B}T$), we obtain that
  \begin{eqnarray}
  \gamma_+
  &\ge&
  {\cal O}\left(
  \frac{\hbar}{k_B T}
  \sigma^2_{\dot{Y}}
  \right).
  \label{gammagammaClassicalNonrigor1}
\end{eqnarray}
In this limit, the variance $\sigma^2_{\dot{Y}}$ is classical, i.e., ${\sf Tr}\Big(\rho_{\Lambda}^{\sf classical~canonical}\big(\dot{Y}^{H}\big)^2\Big)$ \footnote{We emphasize that $\sigma_{\dot{Y}}$ so defined is the standard deviation of $\dot{Y}^{H}$ since the corresponding (classical) canonical ensemble average of any time derivative trivially vanishes, 
${\sf Tr}\Big(\rho_{\Lambda}^{\sf classical~canonical} ~\dot{Y}^{H} \Big) =0$.}.
  
In what follows, we apply this formalism to the analysis of the viscosity and electrical conductivity. In Section \ref{OTOCBound}, we illustrate how, in the semiclassical limit, a possible qualitative link between Lyapunov exponents and autocorrelation functions may lead to general bounds on the transport coefficients.

\subsection{Viscosity Bounds}
\label{sec:viscosity}

In this subsection, we will derive rigorous viscosity bounds within our general setting. These exact inequalities augment insightful pioneering suggestions that there might be fundamental bounds involving the viscosity  \cite{KSS1,KSS2,Tom,nnbk,jan-book,Matteo2019,jan-planck,KT}.

As in the general discussion of subsection \ref{GKsec}, the Green-Kubo relations \cite{Green1954,KMS2} express the {\it bulk viscosity} as the integral of the pressure fluctuation autocorrelation function \cite{Hoover1980}, 
\begin{eqnarray}
\label{etaVH}
\zeta && = \frac{ {\sf V}}{k_B T} \int_{0}^{\infty} dt ~{\sf Tr} \Big(\rho_{\Lambda}^{\sf canonical} ~\Delta P^H(0) \Delta P^H(t) \Big) \nonumber
\\ && \equiv \frac{ {\sf V}}{k_B T} \int_{0}^{\infty} dt ~G_{P}(t) . 
\end{eqnarray}
Here, $ \Delta P^H \equiv P^H- {\sf Tr}(\rho^{\sf canonical} P^H)$ are the Heisenberg picture pressure fluctuations and ${\sf V}$ is the system volume. 

The results of subsection \ref{GKsec} may now be applied. With $t^{P}_{0}$ denoting the shortest time for which $G_{P}(t)$ vanishes, repeating the derivation of Eqs. (\ref{eqnD+},\ref{aa-central-111},\ref{Gvt},\ref{minDiff},\ref{minDiff+}) leads to a rendition of Eq. (\ref{gammagamma}). Written longhand,

\onecolumngrid
\begin{eqnarray}
\label{eq:eta+long}
\zeta_+  &\equiv&  \frac{{\sf{V}}}{k_B T} \int_{0}^{t^{P}_{0}} dt
                  ~G_{P}(t) \nonumber\\
         &\ge& 
               \frac{\hbar {\sf{V}} G_{P}(0)~ |{\sf{Tr}}(\rho_{\Lambda}^{\sf
               canonical} (\Delta P_{i}^{H}(0) \Delta P^{H}(0))) |}{4 k_B T}
               \times\nonumber\\
         && \hspace{1cm}
            \left(
            \sqrt{{\sf{Tr}}\Big(\rho_{\Lambda}^{\sf canonical}  \big(\Delta
            \tilde{H}^{H}_{i}(0)\big)^2\Big) {\sf{Tr}}\Big(\rho_{\Lambda}^{\sf canonical}
            \Delta P_{i}^{H}(t) \big(\Delta P^{H}(0)\big)^2\Delta P_{i}^{H}(t)}\Big)
            \right. \nonumber\\
         && \hspace{2cm}\left.
            +\sqrt{{\sf{Tr}}\Big(\rho_{\Lambda}^{\sf
            canonical}
            \big(\Delta P^{H}(0)\big)^2\Big)
            {\sf{Tr}}\Big(\rho_{\Lambda}^{\sf
            canonical} 
            \Delta P_{i}^{H}(0) \big(\Delta
            \tilde{H}^{H}_{i}(0)\big)^2  \Delta P_{i}^{H}(0)}\Big)
            \right)^{-1} \nonumber
            \\            &\ge& 
            \frac{\hbar {\sf{V}} G_{P}(0)~ |{\sf{Tr}}(\rho_{\Lambda}^{\sf
               canonical} (\Delta P_{i}^{H}(0) \Delta P^{H}(0))) |}{4 k_B T 
            \sqrt{{\sf{Tr}}\Big(\rho_{\Lambda}^{\sf canonical}  \big(\Delta
            \tilde{H}^{H}_{i}(0)\big)^2\Big) {\sf{Tr}}\Big(\rho_{\Lambda}^{\sf canonical}
            \Delta P_{i}^{H}(t) \big(\Delta P^{H}(0)\big)^2\Delta P_{i}^{H}(t)}\Big)
            }.
\end{eqnarray}
\twocolumngrid\noindent
For equilibrium pressure fluctuations \cite{Hiura}, 
\begin{equation}
\label{GPeqq}
G_{P}(0) \equiv \sigma_P^2 = \frac{k_{B} T^{2}}{C_{v}^{(\Lambda)}} \Big(\frac{\partial P}{\partial T}\Big)_v^2.
\end{equation}
The pressure $P= \frac{1}{d {\sf V }} \sum_{i=1}^{N_{\Lambda}}
\Big(\frac{{\bf p}^2_{i}}{m} + {\bf r}_{i} \cdot {\sf f}_{i}\Big) = \frac{{\sf n}}{d}  \frac{1}{N_{\Lambda}}  \sum_{i=1}^{N_{\Lambda}} \Big(\frac{{\bf p}^2_{i}}{m} + {\bf r}_{i} \cdot {\sf f}_{i}\Big)$ with ${\sf n} \equiv \frac{N_{\Lambda}}{\sf V}$ the number density. Given a general many body Hamiltonian of the form of Eq. (\ref{eq:HLAMBDA}), the minimal local Hamiltonian $\tilde{H}^{H}_{i} \subset H_{\Lambda}$ contributing, in the notation of Section \ref{GKsec} to the rate of change of $P_{i} \equiv \frac{1}{d} \Big(\frac{{\bf p}^2_{i}}{m} + {\bf r}_{i} \cdot {\sf f}_{i}\Big) $ is now
\begin{eqnarray}
\label{hhvis}
\tilde{H}^{H}_{i}  = \sum_{j} \frac{{\bf p}^2_{j}}{2m}  + V_{i}^{H},
\end{eqnarray}
with the operator $V_{i}^{H}$ of Eq. (\ref{tileV}). We briefly elaborate on the physics behind Eq. (\ref{hhvis}). As emphasized in Section \ref{v-bounds}, the single particle kinetic energy $\frac{{\bf p}_{i}^{2}}{2m}$ is the only term in $H_{\Lambda}$ that does not commute with the particle position ${\bf r}_{i}$ and thus contributes to its time derivative. In Section \ref{sec:stress}, we described how the kinetic terms of all particles that interact with particle $i$ contribute to the time derivative of the total force ${\sf f}_{i}$ that this particle $i$ experiences. Thus, in the sum of Eq. (\ref{hhvis}), the set of $j$ values to be summed over now contains the particle index $i$ itself as well as those of all other particles that interact with particle $i$. Lastly, as underscored in Section \ref{aa-bounds}, the potential energy $V_{i}^{H}$ of Eq. (\ref{tileV}) is the only term contributing to the time derivative of $(\frac{{\bf p}^2_{i}}{m})$. For a Hamiltonian of the form of Eq. (\ref{eq:HLAMBDA}), the probability density 
$\rho_{\Lambda}^{\sf classical~canonical}$ is a product of a Gaussian
in the momenta and a probability distribution depending only on the
spatial coordinates. The spatial coordinate and momenta contributions
to the variance of $\tilde{H}^{H}_{i}$ then add in quadrature.
If, as in Section \ref{sec:stress}, the number of particles that interact appreciably with particle $i$ is ${\sf z}$ then in the classical limit,
\begin{eqnarray}
\label{Hforpressure}
\! \! \! \! \! \! \! \! \! \! {\sf{Tr}} (\rho_{\Lambda}^{\sf classical ~canonical} (\Delta \tilde{H}_{i}^{H})^{2}) =  (k_{B} T)^2 d ({\sf z} +1)/2 \nonumber
\\  +{\sf {Tr}}(\rho_{\Lambda}^{\sf classical~canonical} (V_{i}^{H})^2).
\end{eqnarray} 
Inserting Eqs. (\ref{GPeqq}, \ref{Hforpressure}) into Eq. (\ref{eq:eta+long}) yields a general rigorous viscosity bound.

To obtain order of magnitude estimates, we assume, in Eq. (\ref{Hforpressure}), that the potential ($V_{i}^{H}$) contributions to be comparable or smaller than the kinetic ones, and take the ratio 
\begin{eqnarray}
 \frac{|{\sf{Tr}}(\rho_{\Lambda}^{\sf canonical} (P_{i}^{H}(0) P^{H}(0))) |}{ \sqrt{{\sf{Tr}}(\rho_{\Lambda}^{\sf canonical} (P_{i}^{H}(0) P^{H}(0))^2)}} 
 \end{eqnarray}
in Eq. (\ref{eq:eta+long}) to be of order unity (as it is for local Gaussian pressure fluctuations). We further set, as it is for an ideal gas, the pressure $P = {\cal{O}}({\sf{n}} k_{B} T)$ and the heat capacity $C^{(\Lambda)}_v = {\cal{O}}({\sf n} d k_B {\sf V})$ (for the ideal gas, the heat capacity is half of this value). Subsequently inserting Eqs. (\ref{GPeqq}, \ref{Hforpressure}) into Eq. (\ref{eq:eta+long}) produces, up to the above noted numerical factors, 
\begin{eqnarray}
\label{upeta:eq}
  \frac{\zeta k_BT}{V} &\ge& {\cal
                             O}\left(\hbar\frac{\sigma_P^2}{
                             \sqrt{{\sf Tr}
                             \Big(\rho_{\Lambda}^{\sf classical
                             ~canonical} \big(\Delta
                             \tilde{H}_{i}^{H}\big)^{2}\Big)}}\right)
                             \nonumber\\
  \Rightarrow
  \zeta &\ge& {\cal O}\left(
              \frac{{\sf V}}{k_BT}
              \frac{\hbar}{k_BT\sqrt{d ({\sf z} +1)}}
              \frac{{\sf n}\big(k_BT\big)^2}{{\sf V}d}\right)
  \nonumber\\
  \Rightarrow
  \zeta &\ge& {\sf n}\hbar\times{\cal O}\left(d^{-3/2}({\sf z}+1)^{-1/2}\right).
\end{eqnarray}
In Table \ref{real-values}, we contrast this bound on the bulk viscosity with the experimentally measured bulk viscosity of water.

In a vein similar to the above calculations for the bulk viscosity, we may replace the pressure fluctuation autocorrelation function in Eq. (\ref{etaVH}) by one of the stress autocorrelation function in order to obtain lower bounds on the shear viscosity $\eta$; the relevant local Hamiltonian for the calculation of the shear viscosity bound, once again, given by Eq. (\ref{hhvis}). This results in a lower bound on $\eta$ which is of the form of the righthand side of Eq. (\ref{upeta:eq}). In Section \ref{semithermbound},
we will illustrate how tighter yet less rigorous bounds on the shear viscosity may be obtained. 

As will be further highlighted in Appendix \ref{sec:comments}, numerous local Hamiltonians other than that of Eq. (\ref{hhvis}) may be further considered. These Hamiltonians will lead to additional bounds. Lastly, we note that in Section \ref{OTOCBound} we sketch how, in the semiclassical limit, a conjectured link between chaos bounds and the autocorrelation functions may lead to general bounds on the viscosity.

\subsection{Electrical Conductivity Bounds} 
\label{sec:conduct}
In the absence of magnetic fields, the (longitudinal) electrical conductivity of electronic (and other charged) fluids may be similarly computed via the general recipe of subsection \ref{GKsec} when applied to the currents along Cartesian directions $\ell$, 
\begin{eqnarray}
\label{econducivity}
\dot{Y}_{\ell} = J_{\ell} = \sum_{i=1}^{N_{\Lambda}} e_{i} v_{i \ell},
\end{eqnarray}
with $e_{i}$ the charge of particle $i$. As emphasized in the main
text, the minimal relevant part of the full many body Hamiltonian
$H_{\Lambda}$ that endows the velocities $\dot{Y}_{i \ell} = e_{i}
v_{i \ell}$ with dynamics is the interaction term $\tilde{H}^{H}_{i} =
V_i^{H}(t)$ of Eq. (\ref{tileV}).
Again, using the arguments that led to
Eq. (\ref{gammagammaClassicalNonrigor}), and
omitting numerical factors (including fundamental constants), the semiclassical high temperature scaling of the resistivity is bounded from above by 
\begin{eqnarray}
\label{rho-estimate}
\! \! \! \! \! \! \! \! \! \! \! \rho \leq {\cal{O}} \Big(\sqrt{{\sf{Tr}}(\rho_{\Lambda}^{\sf canonical}  (V_i^{H})^2) - ({\sf{Tr}}(\rho_{\Lambda}^{\sf canonical}  V_i^{H})})^2 \Big).
\end{eqnarray}
An explicit calculation of the variance in Eq. (\ref{rho-estimate}) is beyond the scope of the current work.
We will qualitatively discuss asymptotic high temperature behavior and bad metal behavior in Section \ref{sec:highT}.

\section{Spatial Gradient Bounds}
\label{sec:space}
 In this Section, we derive a dual of the Section \ref{derive:time} when time derivatives are replaced by spatial derivations with the Hamiltonians replaced by the momenta. Although all of our results apply to these verbatim, we will not explicitly focus on the Heisenberg picture operators. Towards this end, we return, once again, to the trivial extension (Eq. (\ref{eq:AB})) of the uncertainty inequalities to mixed states defined by a density matrix
and choose $A=p_{i \ell}$ and $B=f\left(\left\{r_{j, \ell^{\prime}}, p_{j, \ell^{\prime}}\right\}\right)$ with
 $\left(1 \leq j \leq N\right.$ and $\left.1 \leq \ell^{\prime} \leq d\right)$.
Here, $f$ is an arbitrary function of the phase space variables (i.e., of the position and momentum operators). This yields
\begin{equation}
\label{eq:pxh}
\sigma_{p_{i \ell}}^{2} \sigma_{f}^{2} \geq \frac{\hbar^{2}}{4}\left|\left\langle\frac{\partial f}{\partial r_{i \ell}}\right\rangle\right|^{2}.
\end{equation}
Eq. (\ref{eq:pxh}) is a trivial extension of (textbook type) bounds on local gradients of any function (especially well known for pure states) to general mixed states. We next further rewrite Eq. (\ref{eq:pxh}),
\begin{equation}
\left\langle p_{i \ell}^{2}\right\rangle \geq \sigma_{p_{i \ell}}^{2} \geq \frac{\hbar^{2}}{4 \sigma_{f}^{2}}\left|\left\langle\frac{\partial f}{\partial r_{i \ell}}\right\rangle\right|^{2} \geq \frac{\hbar^{2}}{4\left\langle f^{2}\right\rangle}\left|\left\langle\frac{\partial f}{\partial r_{i \ell}}\right\rangle\right|^{2},
\end{equation}
and subsequently average over the entire system to obtain
\begin{equation}
\label{eq:globalspace}
\frac{1}{N_{\Lambda}} \sum_{i=1}^{N_{\Lambda}}\left\langle p_{i \ell}^{2}\right\rangle \geq \frac{1}{N_{\Lambda}} \sum_{i=1}^{N_{\Lambda}} \frac{\hbar^{2}}{4\left\langle f^{2}\right\rangle}\left|\left\langle\frac{\partial f}{\partial r_{i \ell}}\right\rangle\right|^{2}.
\end{equation}
 Replacing the global average of the equilibrium system of Eq. (\ref{eq:globalspace}) by an expectation value computed with the canonical density matrix $\rho_{\Lambda}^{\sf canonical}$ yields
 \begin{equation}
 \label{alwaysgrad}
{\sf{Tr}}\left(\rho_{\Lambda}^{\sf {canonical }} p_{i \ell}^{2}\right) \geq \frac{1}{N_{\Lambda}} \sum_{i=1}^{N_{\Lambda}} \frac{\hbar^{2}}{4\left\langle f^{2}\right\rangle}\left|\left\langle\frac{\partial f}{\partial 
r_{i \ell}}\right\rangle\right|^{2}.
\end{equation}
The lefthand side of this last inequality is the canonical ensemble average of the squared single particle (and Cartesian component) momentum. The above derivation of Eq. (\ref{alwaysgrad}) invoked no assumptions and is always valid.

We now turn to higher moments. Reproducing the derivation of Section \ref{high-moment-derivative} with the replacement of $\tilde{H}^{i}_{H}$ by $p_{i \ell}$, we arrive at an analogue of Eq. (\ref{squared-ineq}) that is, similarly, generally valid only in the semiclassical limit. 
That is,
 \begin{eqnarray}
\label{squared-ineq-grad}
 \frac{1}{N_{\Lambda}} \sum_{i=1}^{N_{\Lambda}} \Big\langle \Big ( \frac{\partial f}{\partial r_{i \ell}} \Big)^2  \Big\rangle \le \frac{4}{\hbar^2}  {\sf{Tr}}(\rho_{\Lambda}^{\sf classical~ canonical} p_{i \ell}^2)  \nonumber
\\ \times {\sf{Tr}} (\rho_{\Lambda}^{\sf classical~ canonical} f^2).
\end{eqnarray} 
At high temperatures where equipartition holds, ${\sf{Tr}}\left(\rho_{\Lambda}^{\sf {canonical }} p_{i \ell}^{2}\right)=m k_{B} T$. Putting all of the pieces together, we then have
\begin{equation}
\label{eq:grad-bound=}
\frac{1}{N_{\Lambda}} \sum_{i=1}^{N_{\Lambda}} \frac{\left\langle(\frac{\partial f}{\partial r_{i \ell}})^2\right\rangle}{\left\langle f^{2}\right\rangle} \leq \frac{4 m k_{B} T}{\hbar^{2}}=\frac{8 \pi}{\lambda_{T}^{2}},
\end{equation}
where, as throughout, $\lambda_{T}$ is the thermal de-Broglie wavelength of Eq. (\ref{dBeq}) \footnote{For a function $f$ of general Heisenberg picture position operators, we may, in all of the equations of this Section, substitute the Schrodinger picture operators by their corresponding form in the Heisenberg picture, $p_{i\ell} \to p^{H}_{i \ell}$ and $r_{i \ell} \to   r_{i \ell}^{H}$, to obtain identical inequalities.}. In Appendices \ref{higher_grad},\ref{sec:2-point}, and \ref{sec:generalcorrelatorgrad}, we discuss extensions of this construct in deriving higher order gradient, general correlator, and general spatio-temporal derivative bounds.

\section{Bounds on electromagnetic and other gauge field strengths}
\label{sec:em}
We now consider one of the many results that the inequalities of Section \ref{sec:space} imply. Towards this end,  we take the function $f$ to be a time independent electrostatic potential $f=\phi$ (with, in our time-independent Hamiltonian system $H_{\Lambda}$, a stationary vector potential ${\bf A}$). In this case, the gradients $\frac{\partial \phi}{\partial r_{i \ell}}=-E_{i \ell}$ are the components of the electric field acting on particle $i$ along the $\ell$-th Cartesian direction. Eq. (\ref{eq:grad-bound=}) then becomes
\begin{equation}
\label{eq:field}
\frac{1}{N_{\Lambda}} \sum_{\ell=1}^{d} \sum_{i=1}^{N_{\Lambda}} \frac{\left\langle(\frac{\partial \phi}{\partial r_{i \ell}})^2\right\rangle}{\left\langle\phi^{2}\right\rangle}=\frac{1}{N_{\Lambda}} \sum_{i=1}^{N_{\Lambda}} \frac{\left\langle{\bf E}^2_{i}\right\rangle}{\left\langle\phi^{2}\right\rangle} \leq \frac{8 \pi d}{\lambda_{T}^{2}}.
\end{equation}
In a system of uniform particle density, the lefthand side of Eq. (\ref{eq:field}) carries a clear physical meaning since the total stored in the electric field $\frac{1}{2} \epsilon_{0} \int d^{3} r ~{\bf E}^{2}$. If the inequality of Eq. (\ref{eq:field}) is violated then the system cannot be in equilibrium.
We caution that this inequality only holds at sufficiently high temperatures
where the classical equipartition theorem holds. We may similarly derive bounds for magnetic field strengths (setting $f$ to be Euclidean components of the vector potential ${\bf A}$).

Along related lines, for general gauge theories, the uncertainty relation of Eq. (\ref{eq:AB}) implies (with the operators $A$ and $B$ set equal to the covariant derivatives $D_{\mu}$ and $D_{\nu}$ respectively) that
\begin{eqnarray}
\label{field-kinetic}
\! \! \! \! \! \! \! \! \! \! \! \! \! \! \langle D_{\mu}^{\dagger} D_{\mu} \rangle \langle D^{\dagger}_{\nu} D_{\nu} \rangle \ge  \frac{1}{4} |\langle [D_{\mu}, D_{\nu}] \rangle |^2 \equiv \frac{ g^2\langle F_{\mu \nu}\rangle^2}{4},
\end{eqnarray}
where the covariant derivative $D_{\mu} \equiv \partial_\mu - i g A_{\mu}$ with $A_\mu$ gauge field components, $g$ the associated coupling constant, and $\langle F_{\mu \nu} \rangle$ the expectation value of the associated field strength
($F_{\mu \nu} =  - \frac{i}{g} [A_{\mu}, A_{\nu}]$). The Greek indices $\mu$ and $\nu$ span both spatial ($\mu =1,2, \ldots, d$) and temporal ($\mu =0$) components and $\partial_{\mu}$ marks the derivatives relative to these. The lefthand side of Eq. (\ref{field-kinetic}) is set by products of the local kinetic energy expectation values while the righthand side is the energy density associated with the field ($F_{\mu \nu}$). Eq. (\ref{field-kinetic}) has a transparent geometric meaning and can thus be related to gravity; the covariant derivatives $D_{\mu}$ are associated with parallel transport (with the gauge connections $A_\mu$ replaced by Christoffel symbols) and $F_{\mu \nu}$ with the Ricci curvature tensor $R_{\mu \nu}$. In this context, Eq. (\ref{field-kinetic}) implies an upper bound on the possible curvature that scales with the product of effective energy densities given by $ \langle D_{\mu}^{\dagger} D_{\mu} \rangle$. Replicating the derivation of Section \ref{high-moment-derivative}, we may similarly bound $\langle F_{\mu \nu}^2 \rangle$ and higher moments of the gauge field strength.  

Eq. (\ref{field-kinetic}) holds universally (both in equilibrium and in out of equilibrium settings) and illustrates that, similar to Eq. (\ref{eq:field}), gauge fields strengths cannot be arbitrarily large. In thermal systems in which the global kinetic energy averages (those of $\langle D_{\mu}^{\dagger} D_\mu \rangle$) are set by temperature, Eq. (\ref{field-kinetic}) can be rearranged (similar to our derivation of Eq. (\ref{eq:globalspace}) and its corollary of Eq. (\ref{eq:field})) so as to illustrate that the global average of the ratio of the field strength energy density to the kinetic energy is bounded from above by a function of the temperature. 

\section{Chaos bounds}
\label{sec:chaos}

In recent years, largely inspired by \cite{juan-martin}, finite temperature bounds on quantum analogs of classical chaotic dynamics have been proposed. Penetrating earlier work examined chaos as captured by the classical Poisson bracket identity \cite{Larkin},
\begin{equation}
\label{pqp}
\{ q(t),  p(0)\}=\frac{\partial q(t)}{\partial q(0)},
\end{equation}
with, $q(t')$ and $p(t')$ here denoting the conjugate classical position and momentum at time $t'$. As seen from the right hand side of Eq. (\ref{pqp}), this Poisson bracket directly monitors the sensitivity of the system evolution to its initial conditions. Eq. (\ref{pqp}) motivated \cite{juan-martin,RSL,Larkin} investigations of ``two time commutators'' (or ``Out of Time Order Correlation functions'' (OTOC)) of the type
\begin{equation}
\label{CTM}
C(t)=-\left\langle[W^H(t), V^H(0)]^{2}\right\rangle.
\end{equation}
Much work has centered on examining a deformed  version of one of the four terms contributing to $C(t)$, the amended ``regularized OTOC'', given by ${\sf{Tr}}(y V^H(0) y W^H(t) y V^H(0) y  W^H(t))$ with $y^{4} = e^{-\beta H_{\Lambda}}/{\sf{Tr}}(e^{-\beta H_{\Lambda}})$ on which rigorous results have been obtained \cite{juan-martin,subleading} using rather modest physical assumptions (which imply analyticity, positivity, and Schwarz reflection properties). At long times, the expectation value of the commutator of Eq. (\ref{CTM}) saturates to a constant. The OTOCs were first introduced \cite{Larkin} to asses the reliability of of quasi-classical analysis for studying superconductivity (notably, investigating the vertex corrections to the current). The OTOCs (and, notably, their intermediate time evolution) were studied in expansive detail by numerous investigators. Illuminating extensions of the OTOCs have been investigated, e.g., \cite{saso2}. The time evolution of the OTOCs enables a definition of general Lyapunov type exponents in quantum systems. Various theories (including those of black holes \cite{juan-martin,op-scram1,Shenker}, the Sachdev-Ye-Kitaev model \cite{SY-SYK,K-SYK,Juan-SYK,0-SYK,1-SYK,2-SYK,3-SYK,4-SYK,5-SYK} and, possibly, systems without quasiparticles \cite{PatelSachdev}) may quickly thermalize and saturate conjectured bounds \cite{juan-martin} on Lyapunov type exponents. There are intriguing connections between the chaotic dynamics probed by these correlation functions and the scrambling of initial local information into non-local degrees of freedom \cite{op-scram1,op-scram2,Zanardi}, the Eigenstate Thermalization Hypothesis \cite{Foini,MM19}, and to free particle propagation in various geometries \cite{Kurchan}.  
In Section \ref{sec:chaos1}, we will derive direct Lyapunov exponent bounds and in
Section \ref{appOTOC}, we will bound (within the semiclassical limit) the evolution of two time commutators of Eq. (\ref{CTM}). In Section \ref{OTOCBound}, we sketch a general qualitative connection between the Lyapunov exponent inequalities and the transport coefficient bounds. 

\subsection{Direct semiclassical Lyapunov exponent bounds without using the OTOC}
\label{sec:chaos1}
In this subsection, we will not rely on the correlators of Eq. (\ref{CTM}) and attendant regularized OTOCs as proxies for Eq. (\ref{pqp}). Rather, we will provide direct inequalities on the rate of change of any phase space operator (or semiclassical coordinate). Towards this end, following the results that we derived in the earlier Sections, we consider two thermal translationally invariant thermal systems (labelled, henceforth,``1'' and ``2'') which differ minimally in their initial density matrices $\rho_{\Lambda}^{(1)}$ and $\rho_{\Lambda}^{(2)}$ and examine the dynamics at general times $t>0$. 
The strategy that we will follow will be to consider two copies of the system (each associated with the different initial boundary conditions or associated density matrices) that occupy the same volume $\Lambda$ but do not interact with one another. These copies (which we unimaginatively alluded to above as ``1'' and ``2'') can, in a classical cartoon, be pictorially thought of as of composed of ``blue'' and ``red'' particles. Initially, the system of blue particles and the system of red particles will be close to each other with only minor variations in their initial state. The two systems will evolve simultaneously in the same volume. After some time $t$ we will measure the evolved distances in configuration or phase space between the corresponding particles in the blue and red systems. In this picture, the Hamiltonian defining the system is simply a decoupled sum of a Hamiltonian $H_{\Lambda}$ for the blue particles and an identical looking Hamiltonian $H_{\Lambda}$ that acts only on the red particles. We can then look at the distance between a blue and a red particle location as an observable $Q_i$ and, following the tools developed in the previous sections, bound this distance as a function of time. 

To transform the above cartoon into a rigorous bound, we proceed with a few definitions. We first construct the hybrid density matrix 
\begin{eqnarray}
\label{rho1|2}
\rho_{\Lambda}^{1 \cup 2} \equiv \rho_{\Lambda}^{(1)} \otimes \rho_{\Lambda}^{(2)}
\end{eqnarray}
defined on two copies of the system $\Lambda$. For $a=1,2$, the density matrix $\rho_{\Lambda}^{(a)}$ describes the probability density of quantities $\{Q_{i}^{a}\}$. That is, there are two copies ($a=1,2$) of the system that are independent of each other. Eq. (\ref{rho1|2}) is {\it a definition} for the hybrid density matrix that may be applied for general systems of different underlying particle statistics. We will consider this hybrid system (given by the initial $t=0$ Schrodinger picture density matrix $\rho_{\Lambda}^{1 \cup 2}$) to evolve according to a Hamiltonian $H^{\Lambda}_{12}$ defined by the direct sum
\begin{eqnarray}
\label{H1|2}
H_{\Lambda}^{(1\cup 2)} \equiv H_{\Lambda}^{(1)} + H_{\Lambda}^{(2)}.
\end{eqnarray}
In Eq. (\ref{H1|2}), the Hamiltonians $\{H_{\Lambda}^{(a)}\}$ (with $a=1,2$) are identical replicas of the global Hamiltonian $H_{\Lambda}$ of the earlier Sections. These two Hamiltonians govern the dynamics of the Heisenberg picture operators $\{Q_{i}^{H(a)}\}$ and trivially commute with $\{Q_{i}^{H(a')}\}$ for $a' \neq a$. Thus, the operator $Q_{i}^{H(a)}$ evolves only according to $H_{\Lambda}^{(a)}$, i.e., $Q_{i}^{H(a)} (t) = e^{i H^{(a)}_{\Lambda} t/\hbar} Q_{i} e^{-i H^{(a)}_{\Lambda} t/\hbar}$. Paralleling the earlier Sections, we define local Hamiltonians $\tilde{H}^{H(1)}_{i}$ and $\tilde{H}^{H(2)}_{i}$ for which the commutator $[Q_{i}^{H(1)}, \tilde{H}^{H(1)}_{i}] =  [Q_{i}^{H(1)}, H_{\Lambda}^{(1)}]$) and $[Q_{i}^{H(2)}, \tilde{H}^{H(2)}_{i}] =  [Q_{i}^{H(2)}, H_{\Lambda}^{(2)}]$. Lastly, we introduce the shorthand for the difference 
\begin{eqnarray}
\label{Q1|2}
Q_{i}^{(1-2)} \equiv Q_{i}^{(1)} - Q_i^{(2)},
\end{eqnarray}
that monitors the variability between the two copies of the same observable $Q_{i}$ when these two copies have different initial conditions and governed by identical looking Hamiltonians $\{H_{\Lambda}^{(a)}\}$. 

With all of the above definitions in tow, we now derive simple Lyapunov exponent inequalities. At general positive times $t>0$, the difference between the expectation value of $Q_{i}^{H}$ when two different initial ($t=0$) conditions (specified by $\rho_{\Lambda}^{(1)}$ and $\rho_{\Lambda}^{(2)}$) are imposed becomes 
${\sf Tr}\Big((\rho_{\Lambda}^{(1)} - \rho_{\Lambda}^{(2)})  Q_{i}^{H}(t)\Big)$ with the time evolved $Q_{i}^{H}(t) = e^{iH_{\Lambda} t/\hbar} Q_{i} e^{-iH_{\Lambda}t/\hbar}$. Given Eqs. (\ref{rho1|2}, \ref{H1|2}, \ref{Q1|2}), this deviation probes the disparate dynamics given two different initial conditions. This quantity may, alternatively, be expressed as
\begin{eqnarray}
\label{diffchaos}
\! \! \! \! \! \! \! {\sf Tr} \Big((\rho_{\Lambda}^{(1)} - \rho_{\Lambda}^{(2)}) Q_{i}^{H}(t) \Big) = {\sf Tr} \Big(\rho^{1 \cup 2}  Q_{i}^{H (1-2)}(t) \Big), 
\end{eqnarray}
with 
\begin{eqnarray}
\label{Q(1-2)t}
Q_{i}^{H (1-2)}(t) \equiv e^{i H_{\Lambda}^{(1\cup 2)}  t/\hbar} Q_{i}^{(1-2)}  e^{-i H_{\Lambda}^{(1\cup 2)}  t/\hbar}.
\end{eqnarray}
With the definition of Eq. (\ref{Q(1-2)t}),
\begin{eqnarray}
\label{tdiffchaos}
{\sf Tr} \Big((\rho_{\Lambda}^{(1)} - \rho_{\Lambda}^{(2)}) \frac{dQ_{i}^{H}}{dt} \Big) = {\sf Tr} \Big(\rho_{\Lambda}^{1 \cup 2}  \frac{d Q_{i}^{H (1-2)}}{dt}\Big).
\end{eqnarray}
Thus, we arrive at
\begin{eqnarray}
\label{longobvi}
\! \! \! \! \! \! \! && \frac{{\sf Tr} \Big((\rho_{\Lambda}^{(1)} - \rho_{\Lambda}^{(2)}) (\frac{dQ_{i}^{H}}{dt})^2 \Big)}{{\sf Tr} \Big((\rho_{\Lambda}^{(1)} - \rho_{\Lambda}^{(2)}) (Q_{i}^{H}(t))^2 \Big)}  \nonumber
\\ 
\! \! \! \! \! \! \!  && =  \frac{{\sf Tr} \Big(\rho_{\Lambda}^{1 \cup 2}  
(\frac{dQ_{i}^{H(1-2)}}{dt})^2 \Big)}
{{\sf Tr} \Big(\rho_{\Lambda}^{1 \cup 2} (Q_{i}^{H(1-2)}(t))^2 \Big)}  \nonumber
\\ \! \! \! \! \! \! \! && \le  \frac{4}{\hbar^2} {\sf Tr} \Big(\rho_{\Lambda}^{1 \cup 2}  (( \Delta \tilde{H}^{H(1)}_{i}(t))^2+ ( \Delta \tilde{H}^{H(2)}_{i}(t))^2) \Big),
\end{eqnarray}
where, for both replicas $a=1,2$, we define in the above, $ \Delta \tilde{H}^{H(a)}_{i}(t) \equiv (\tilde{H}^{H(a)}_{i}(t)  - {\sf Tr}(\rho_{\Lambda}^{(1\cup 2)}  \tilde{H}^{H(a)}_{i}(t) ))
= (\tilde{H}^{H(a)}_{i}(t)  - {\sf Tr}(\rho_{\Lambda}^{(a)}  \tilde{H}^{H(a)}_{i}(t) ))$. The last inline equality follows from the definitions of Eq. (\ref{rho1|2}) and of the Hamiltonians $\{H_{\Lambda}^{(a)}\}_{a=1,2}$. The equality in Eq. (\ref{longobvi}) is an outcome of Eqs. (\ref{diffchaos}, \ref{tdiffchaos}). The final inequality in Eq. (\ref{longobvi}) is a consequence of the local time-energy uncertainty relation of Section \ref{derive:time} applied to the hybrid system formed by the two replicas; this final inequality is a rendition of Eq. (\ref{squared-ineq}) applicable to semiclassical systems. In a similar manner, given the definitions of Eqs. (\ref{rho1|2}, \ref{H1|2}), 
\begin{eqnarray}
\label{obvious!!}
 \! \! \! \! \! \! \! \! \! \! \! \! \! \! {\sf Tr} \Big(\rho_{\Lambda}^{1 \cup 2}  ( \Delta \tilde{H}^{H(a)}_{i}(t))^2 \Big) && = {\sf Tr} \Big(\rho_{\Lambda}^{a}  ( \Delta \tilde{H}^{H(a)}_{i}(t))^2 \Big). 
 \end{eqnarray}
 Inserting Eq. (\ref{obvious!!}) into Eq. (\ref{longobvi}) and averaging over all particles $i$ yields
  \begin{eqnarray}
\label{longobvi11}
&& \frac{1}{N_{\Lambda}} \sum_{i=1}^{N_{\Lambda}} \frac{{\sf Tr} \Big((\rho_{\Lambda}^{(1)} - \rho_{\Lambda}^{(2)}) (\frac{dQ_{i}^{H}}{dt})^2 \Big)}{{\sf Tr} \Big((\rho_{\Lambda}^{(1)} - \rho_{\Lambda}^{(2)}) (Q_{i}^{H}(t))^2 \Big)}  \nonumber
\\ && \le  \frac{4}{\hbar^2 N_{\Lambda}} \sum_{i=1}^{N_{\Lambda}} \sum_{a=1,2} {\sf Tr} \Big(\rho_{\Lambda}^{a}  ( \Delta \tilde{H}^{H(a)}_{i}(t))^2 \Big).
\end{eqnarray}
In thermal equilibrium, the two averages $ \frac{1}{N_{\Lambda}} \sum_{i=1}^{N_{\Lambda}} {\sf Tr} \Big(\rho_{\Lambda}^{a}  ( \Delta \tilde{H}^{H(a)}_{i}(t))^2 \Big)$ are the same in both replicas $a=1$ and $2$ (with both given by Eq. (\ref{cv2})). Thus, with the definition of Eq. (\ref{cv2}) for the local heat capacity, we can then rewrite Eq. (\ref{longobvi11}) as
  \begin{eqnarray}
\label{longobvi111}
&& \frac{1}{N_{\Lambda}} \sum_{i=1}^{N_{\Lambda}} \frac{{\sf Tr} \Big((\rho_{\Lambda}^{(1)} - \rho_{\Lambda}^{(2)}) (\frac{dQ_{i}^{H}}{dt})^2 \Big)}{{\sf Tr} \Big((\rho_{\Lambda}^{(1)} - \rho_{\Lambda}^{(2)}) (Q_{i}^{H}(t))^2 \Big)}  \nonumber
\\ && \le  \frac{8 k_B T^2 C_{v,i}}{\hbar^2} \equiv \lambda_{L}^{2}. 
\end{eqnarray}
We next follow the standard recipe for monitoring chaotic dynamics. We ``launch'' the system when it is given two very similar initial conditions and compute the difference in the expectation values of various observes at positive times in order to assess whether the system is chaotic. At general intermediate times $t$, the standard deviation associated with the difference of the same measured observable in these two different replicas is much smaller than the expectation value of that difference (for, e.g., particle displacements, the latter increases as the particles separate from each other on evolution while the magnitude of the thermal fluctuations does not increase), $\Big[{\sf Tr} \Big((\rho_{\Lambda}^{(1)} - \rho_{\Lambda}^{(2)}) (Q_{i}^{H}(t))^2 \Big) - 
\Big( {\sf Tr}(\rho_{\Lambda}^{(1)} - \rho_{\Lambda}^{(2)}) (Q_{i}^{H}(t)) \Big)^2\Big] \equiv \sigma^{2}_{Q_{i}^{H(1-2)}} \ll 
\Big( {\sf Tr}(\rho_{\Lambda}^{(1)} - \rho_{\Lambda}^{(2)}) (Q_{i}^{H}(t)) \Big)^2$. Noting that
${\sf Tr} \Big((\rho_{\Lambda}^{(1)} - \rho_{\Lambda}^{(2)}) (\frac{dQ_{i}^{H}}{dt})^2 \Big) \ge  \Big({\sf Tr} \Big((\rho_{\Lambda}^{(1)} - \rho_{\Lambda}^{(2)}) (\frac{dQ_{i}^{H}}{dt}) \Big) \Big)^2$, we observe that 
 Eq. (\ref{longobvi111}) implies that
\begin{eqnarray}
\label{longobvi1111}
 && \frac{1}{N_{\Lambda}} \sum_{i=1}^{N_{\Lambda}}\Big[ \frac{d}{dt} \Big( \ln {\sf Tr} \Big((\rho_{\Lambda}^{(1)} - \rho_{\Lambda}^{(2)}) Q_{i}^{H} \Big) \Big) \Big]^2 \nonumber
\\ &&  = \frac{1}{N_{\Lambda}} \sum_{i=1}^{N_{\Lambda}}  \Big[ \frac{{\sf Tr} \Big((\rho_{\Lambda}^{(1)} - \rho_{\Lambda}^{(2)})  \frac{dQ_{i}^{H}}{dt} \Big) }{{\sf Tr} \Big((\rho_{\Lambda}^{(1)} - \rho_{\Lambda}^{(2)}) Q_{i}^{H}(t) \Big)}\Big]^2  \nonumber
\\ && \le  \lambda_{L}^{2}. 
\end{eqnarray}
The above bound applies to any local observable associated with the $N_{\Lambda}$ single particles $i$. We may trivially compound these single particle expectation values into a vector spanning all of the particles in the system. The bound of Eq. (\ref{longobvi1111}) restricts, component by component, how rapidly these configuration space vectors can diverge from one another \footnote{Our results are actually stronger than those suggested by the semiclassical analysis alone that led to (\ref{longobvi1111}). This equation and all that followed from Eq. (\ref{longobvi}) are applicable to semiclassical systems (since Eq. (\ref{longobvi}) is a consequence of the semiclassical Eq. (\ref{squared-ineq})). We work within the semiclassical regime since the notions of chaos (including the definition of the Lyapunov exponent) are better described classically. Fundamentally, however, in both the real basic quantum as well as the semiclassical setting, the very existence of different initial conditions allowing for the common textbook definition of Lyapunov exponents does allow one to simply set both initial conditions (or associated thermal probability densities $\rho_{\Lambda}^{(1)}$ and $\rho_{\Lambda}^{(2)}$ (with these defined as in Section \ref{setup})) to be the same (i.e., with both of these probability densities being equal to $\rho_{\Lambda}^{\sf canonical}$). Thus, in order to define the problem so as to address what is typically measured classically when making small changes to the initial conditions, one cannot readily set the initial probability densities to be the canonical probability density since this will yield a vanishing result for the measured ${\sf Tr}\Big((\rho_{\Lambda}^{(1)} - \rho_{\Lambda}^{(2)} ) Q_{i}^{H}(t) \Big)$. Our construct of $\rho_{\Lambda}^{1 \cup 2}$ and proof in Section \ref{sec:chaos} circumvents these problems. With all of the above cautionary remarks now made, we note that along lines identical to those pursued in the current Section \ref{sec:chaos}, we may alternatively derive Eq. (\ref{longobvi1111}) as an exact inequality at all times by directly invoking Eqs. (\ref{central1}, \ref{a-central-1}) within a more general (i.e., not necessarily only semiclassical) framework.}. 
\bigskip

This leads us to a simple general corollary:
\newline
\bigskip

$\bullet$ At arbitrary intermediate times, the globally averaged deviation in the expectation value of a local observable caused by a change in initial conditions
$  {\sf Tr} \Big((\rho_{\Lambda}^{(1)} - \rho_{\Lambda}^{(2)}) ~Q_{i}^{H}(t) \Big)$ cannot, on average, increase more rapidly than $e^{\lambda_{L} t}$ up to a time independent multiplicative prefactor. We thus now identify the ratio $\lambda_{L}$ of Eq. (\ref{longobvi111}) as the maximal Lyapunov exponent. 
\newline
\bigskip

Applying Eq. (\ref{longobvi1111}) to the position coordinates of particles along a fixed Cartesian axis (Section \ref{Expect-v}), we find that in thermal systems comprised of $N_{\Lambda}$ particles in $d$ spatial dimensions, at intermediate times, when averaged over the entire system, the divergence in the distance between particle trajectories that were perturbed at time $t=0$ cannot exceed a time independent constant multiplying $e^{\lambda_{L} t}$ with, in the semiclassical regime, $\lambda_{L} \le \frac{2  k_{B} T \sqrt{d}}{\hbar}$. Similar bounds may be derived for the evolution of the momenta (following Section \ref{a-bounds}) and general observables. Our exact upper bound on the maximal Lyapunov exponent may be contrasted with the slightly higher conjectured  bound of  \cite{juan-martin} according to which 
$\lambda_{L}$ is smaller than $ \frac{2 \pi k_{B} T}{\hbar}$. In three spatial dimensions ($d=3$), our non-relativistic upper bound of $\frac{2  k_{B} T \sqrt{d}}{\hbar}$ for configuration space trajectories is smaller by a factor of $\sqrt{3}/\pi  \sim 0.551$ from the conjecture of \cite{juan-martin}. In ultra-relativistic systems, the local Hamiltonian is linear (instead of quadratic) in the momentum ($\tilde{H}^{H}_{i} = c |\vec{p}_{i}|$) and the effective local heat capacity $C_{v,i}$ assumes a value double that in the non-relativistic limit that we largely focus on in the current work; this leads to an additional increase in our above upper bound by a factor of $\sqrt{2}$. By using Eqs. (\ref{long-time-eq}, \ref{cal-long}), the system averages over the particle index $i$ may be replaced by time averages provided that the associated time windows are sufficiently wide so that the long time averages coincide with those over finite times. As we will discuss in Section \ref{quantum-thermalization}, the lower bound on these averaging times can become exceedingly short. In what follows, we examine inequalities for finite averaging times.

\onecolumngrid
\subsection{Semiclassical Chaos Bounds on the OTOC}
\label{appOTOC}

In this subsection, we extend the results of Section \ref{sec:chaos} to illustrate how bounds may be obtained on the two-time commutator of Eq. (\ref{CTM}) 
\begin{eqnarray}
\label{CTM1}
C(t)=&&-\left\langle[W^H(t), V^H(0)]^{2}\right\rangle \nonumber
\\ = &&-
  \left\langle
    V^H(0)W^H(t)V^H(0) W^H(t)
  \right\rangle 
-
  \left\langle
    W^H(t)V^H(0)W^H(t)V^H(0)
  \right\rangle  \nonumber
  \\ &&+ \left\langle  W^H(t)(V^H(0))^2 W^H(t)  \right\rangle  + \left\langle  V^H(0) (W^H(t) )^2  V^H(0) \right\rangle.
\end{eqnarray}
When the density matrix associated with the expectation value $\left\langle V^H(0)W^H(t)V^H(0) W^H(t) \right\rangle$ is  factorized and symmetrically placed between the operators, the expectation value is altered to become the regularized OTOC \cite{juan-martin} defined by ${\sf{Tr}}(y V^H(0) y W^H(t) y V^H(0) y  W^H(t))$ with $y^{4} = e^{-\beta H_{\Lambda}}/{\sf{Tr}}(e^{-\beta H_{\Lambda}})$ for which elegant rigorous results have been obtained \cite{juan-martin,subleading}. In what follows, we directly analyze the commutator $C(t)$ to derive a semiclassical bound that is applicable to the logarithmic derivative of each of the four terms appearing in the second and third line of Eq. (\ref{CTM1}). To illustrate the general idea, we explicitly bound the logarithmic derivative of the first of these terms. By repeating the below steps, identical results apply to all four terms. We start by explicitly writing down the derivative,  
\begin{eqnarray}
&& \frac{d}{dt}  \left\langle
    V^H(0)W^H(t)V^H(0) W^H(t)
  \right\rangle \nonumber
  \\ && = \frac{i}{\hbar} \Big(\left\langle
    V^H(0)
    [\tilde{H}_W^H(t),W^H(t)]
    V^H(0) W^H(t)
  \right\rangle + \left\langle
    V^H(0)W^H(t)V^H(0) 
     [\tilde{H}_W^H(t),W^H(t)]
  \right\rangle \Big).
\end{eqnarray}
Similar to our earlier calculations, $\tilde{H}_W^H(t) \subset H_{\Lambda}$ is the part of the Hamiltonian that endows
$W^H(t)$ with its full dynamics, i.e., $[\tilde{H}_W^H,W^H] = [H_{\Lambda}, W^H]$. To obtain the strongest bounds, we will replace $\tilde{H}_W^H \to  \Delta \tilde{H}_W^H \equiv \tilde{H}_W^H - \langle \tilde{H}_W^H \rangle$. Analogous to Eq. (\ref{longABABA}),
\begin{eqnarray}
\label{longACAC}
&& \Big|\frac{\frac{d}{dt}  \left\langle
    V^H(0)W^H(t)V^H(0) W^H(t)
  \right\rangle}{  \left\langle
    V^H(0)W^H(t)V^H(0) W^H(t)
  \right\rangle} \Big| \nonumber
  \\ && = \frac{1}{\hbar}  \frac{ \Big|\left\langle
    V^H(0)
    [\Delta \tilde{H}_W^H(t),W^H(t)]
    V^H(0) W^H(t)
  \right\rangle \Big|+  \Big|\left\langle
    V^H(0)W^H(t)V^H(0) 
     [\Delta \tilde{H}_W^H(t),W^H(t)]
  \right\rangle \Big|}{\Big|\left\langle
    V^H(0)W^H(t)V^H(0)   W^H(t)
  \right\rangle \Big|} \nonumber 
  \\ && \le \frac{4}{\hbar  \Big|\left\langle
    V^H(0)W^H(t)V^H(0)   W^H(t)
  \right\rangle \Big|}  \max  \Big \{ \Big|\left\langle
    V^H(0)
   \Delta  \tilde{H}_W^H(t) W^H(t)
    V^H(0) W^H(t)
  \right\rangle \Big| , \Big| \left\langle
    V^H(0)
    W^H(t) \Delta \tilde{H}_W^H(t)
    V^H(0) W^H(t)
  \right\rangle \Big| , \nonumber
  \\ &&  ~~~~~~~~~~~~~~~~~~~~~~~~~\Big|\left\langle
    V^H(0)W^H(t)V^H(0) 
    \Delta \tilde{H}_W^H(t) W^H(t)
  \right\rangle \Big|,  \Big|\left\langle
    V^H(0)W^H(t)V^H(0) 
     W^H(t) \Delta \tilde{H}_W^H(t)
  \right\rangle \Big|  \Big \}.
  \end{eqnarray}
  In the semiclassical limit, all four expectation values in the argument of the maximum of Eq. (\ref{longACAC}) are identically the same. Thus,
  \begin{eqnarray}
\Big|\frac{\frac{d}{dt}  \left\langle
    V^H(0)W^H(t)V^H(0) W^H(t)
  \right\rangle}{  \left\langle
    V^H(0)W^H(t)V^H(0) W^H(t)
  \right\rangle} \Big|  \le \frac{4 \Big|
  \left\langle
  (V^H(0) W^H(t))^2  \Delta \tilde{H}_W^H(t)  \right\rangle  \Big|}{\hbar \Big|   \left\langle    (V^H(0) W^H(t))^2
   \right\rangle \Big |},
  \end{eqnarray}
    where $\langle \cdot  \rangle$ now denotes an average with the classical probability density $\rho_{\Lambda}^{\sf classical~canonical}$. Applying, in this limit, Eq. (\ref{CSg'}) with $A=  (V^H(0) W^H(t))^2 $ and $B=\Delta  \tilde{H}_W^H(t) $,
    \begin{equation}
    \label{classicalOTOC}
 \Big|\frac{\frac{d}{dt}  \left\langle
    V^H(0)W^H(t)V^H(0) W^H(t)
  \right\rangle}{  \left\langle
    V^H(0)W^H(t)V^H(0) W^H(t)
  \right\rangle} \Big|  \le  \frac{4}{\hbar}\sqrt{
    \left\langle
    \Big(\Delta \tilde{H}_W^H(t)\Big)^2
  \right\rangle}
  \frac{\sqrt{\left\langle \Big(W^H(t)V^H(0)\Big)^4
    \right\rangle}}{\left\langle \Big(W^H(t)V^H(0)\Big)^2
    \right\rangle}.
\end{equation}
Replicating the above sequence of steps for the other three terms in the second and third line of Eq. (\ref{CTM1}), it is seen that each of the four contributions to two-time commutator $C(t)$ cannot increase in time with an exponent that is larger than the righthand side of Eq. (\ref{classicalOTOC}). We next consider systems with local interactions and local operators $V^H$ and $W^H$. If $W^H$, and therefore, $\tilde{H}_W^H$, involves a small finite number
of order unity of degrees of freedom,
then the variance of $\tilde{H}_W^H$ as computed with the classical thermal probability density $\rho_{\Lambda}^{\sf classical~canonical}$, i.e., 
$\left\langle
  \Big(\Delta \tilde{H}_W^H(t)\Big)^2
\right\rangle$, will be of the order of $\Big(k_BT\Big)^2$.
Furthermore, if the ratio
$\langle (W^H(t)V^H(0))^4\rangle/
\langle(W^H(t)V^H(0))^2\rangle^2$ is of order unity then from Eq. (\ref{classicalOTOC}) and its analogs, each of the contributions to $C(t)$ in 
Eq. (\ref{CTM1}) (and thus $C(t)$ as a whole) will not be able to increase in time with an exponent larger than $\lambda_{L} = {\cal{O}}(k_{B} T/\hbar)$. This bound is consistent with the precise results obtained in subsection \ref{sec:chaos1} (and Eq. (\ref{longobvi111}) therein in particular).  \\

\twocolumngrid

\subsection{Bounds on Transport Coefficients from the Chaos Bounds}
\label{OTOCBound} 
We next sketch how the final transport coefficient bounds (Sections \ref{sec:diff} and \ref{GsKsec}) may be reinterpreted in terms of the chaos bounds. This will further flesh out earlier conjectured connections between chaos and transport that were raised when studying electronic systems \cite{kapitulnik}. Specifically, as we will briefly illustrate, if the Lyapunov exponent bound of \cite{juan-martin} can be qualitatively extended then, semiclassically, these will lead to lower bounds on general Green-Kubo type integrals. Towards this end, returning to the general formalism of Section \ref{GsKsec} for the computation of general transport coefficients, we first note that in the semiclassical limit, the autocorrelation functions $\langle \dot{Y}^H(0) \dot{Y}^H(t) \rangle$ decay, approximately, exponentially in time ($\propto e^{-t/t_{d}}$) with a ``dissipation time'' $t_{d} = 1/\lambda_{L}$ \cite{juan-martin}. This then implies that
\begin{eqnarray}
\label{|GK|}
&& \gamma \equiv \int_{0}^{\infty} dt ~   
\langle \dot{Y}^{H}(0) \dot{Y}^{H}(t)  \rangle
 \nonumber
\\ && \gtrsim \int_{0}^{\infty} dt  ~ e^{-  \lambda_L t}~   \Big\langle \Big( (\dot{Y}^{H}(0))^2  \Big \rangle \nonumber
\\ = && \frac{1}{\lambda_{L}}  ~   \Big\langle (\dot{Y}^{H}(0))^2  \Big \rangle 
\end{eqnarray}
In this bound, the factor of $(t_{\min}/2)$ that appeared in Eq. (\ref{gammaeq1}) has, effectively, been replaced by $t_{d} = \lambda_{L}^{-1}$. In the second line of Eq. (\ref{|GK|}), the semiclassical limit has been taken. The integral of Eq. (\ref{|GK|}) carries the same units as the Green-Kubo integral of Eq. (\ref{eq:GY}). If we take $\dot{Y}$ to be a Cartesian velocity component of a single particle $v_{i \ell}$ (as in the integral of the velocity autocorrelator leading to the diffusion constant discussed in Sections \ref{sec:diff} and \ref{GKsec}) then, semiclassicaly, for Hamiltonians of the form of Eq. (\ref{eq:HLAMBDA}), the canonical thermal average 
$ \Big\langle (\dot{Y})^2 \Big\rangle =  \Big\langle v_{i \ell}^2 \Big\rangle
=  (k_{B} T/m)$. If we insert the bound of \cite{juan-martin}, $\lambda_L \le \frac{2 \pi k_B T}{\hbar}$ then we find that Eq. (\ref{|GK|}) implies that, in thermal semiclassical systems, the diffusion constant obeys a simple universal inequality 
\begin{eqnarray}
\label{DOTOC}
D  \gtrsim \frac{\hbar}{ 2 \pi m}.
\end{eqnarray} 
In the derivation of this bound, the velocity autocorrelation function is assumed to decay in time with no negative contributions to the integral of Eq. (\ref{|GK|}). In defining $D_{+}$ in Eq. (\ref{eqnD+}) as the integral of the velocity autocorrelation function up to its first zero ($t_{v}$), such negative autocorrelation contributions were similarly not present. The inequality of Eq. (\ref{DOTOC}) assuming exponential decay of the velocity autocorrelation function having a value of $t_{v}= \infty$ is not as rigorous as the exact inequalities of Eqs. (\ref{minDiff}, \ref{minDiff+}) where no such assumption was made. In semiclassical systems in which the moments of $(\Delta V_{i}^{H})$ in Eq. (\ref{minDiff+}) are set by the respective powers of $(k_{B} T)$ times numbers of order unity, the bound of Eq. (\ref{minDiff+})
will become similar to that of Eq. (\ref{DOTOC}). When long range interactions and correlations between many particles are present (e.g., those between distant ions in a crystal), the moments of $(\Delta V_{i}^{H})$ may become significantly larger than $(k_{B} T)$ and the resulting bound of
Eq. (\ref{minDiff+}) can become appreciably smaller than that of Eq. (\ref{DOTOC}). Similar to Section \ref{sec:diff} (in particular, Eq. (\ref{etaup}) therein), Eq. (\ref{DOTOC}) suggests an upper bound on the viscosity whenever the Stokes-Einstein relation holds. Specifically, for particles of radius $R$,
\begin{eqnarray}
\label{etaup'}
\eta = \frac{k_{B} T}{6 \pi R D} \lesssim \frac{m k_{B} T}{3 \hbar R}
\end{eqnarray}
In Table \ref{real-values}, we compare our bounds of Eqs. (\ref{DOTOC},\ref{etaup'}) with empirical values for water \footnote{Violations of Eq. (\ref{DOTOC}) are found in crystalline systems. As we remarked earlier, during the long time oscillatory motion of ions in a crystal, the velocity autocorrelation function does not exhibit a simple exponential decay in time and notable negative contributions to the Green-Kubo integral can arise. The behavior of the long time velocity autocorrelation function is not universal. Depending on system details, oscillatory long time tails of the autocorrelation function may also lead to positive contributions to the Green-Kubo integral.}. 

Analogously, in thermal relativistic systems, since $\Big\langle v_{i \ell}^2 \Big\rangle \le c^2$, the diffusion constant $D \gtrsim \frac{\hbar c^2}{2 \pi k_{B} T}$. Proceeding, in a similar manner, to bound the shear and bulk viscosities using the identification of $\dot{Y}$ in 
Table \ref{tab:GY}, we find that, semiclassically, 
\begin{eqnarray}
\label{etazetachaos}
\eta \ge {\cal{O}}( \mathrm{n} \hbar), \nonumber
\\ \zeta  \ge {\cal{O}}( \mathrm{n} \hbar).
\end{eqnarray} 
Inequalities congrous to Eq. (\ref{etazetachaos}) are produced by inserting the more specific (quantity $Q$ dependent) Lyapunov exponent bounds of Section \ref{sec:chaos} instead of the general bound of \cite{juan-martin}. That is, we arrive anew at our order of magnitude bound
of Eq. (\ref{etanhh}) and its likes. It is worth emphasizing the limitation of our derivation in this subsection. An assumption of a semiclassical single exponential decaying autocorrelation function is, in many instances, incorrect (it, e.g., does not allow for situations in which the autocorrelation function exhibits a ``correlation hole''  \cite{Lev86} and becomes negative nor for more general oscillations of the autocorrelation functions about their long time vanishing value). In Sections \ref{sec:diff} and \ref{GsKsec}, we bounded the positive contributions $\gamma_{+}$ (Section \ref{GKsec}) to the autocorrelation function yet did not assume the autocorrelation function to be a simple exponential (nor to be bounded from below by an exponential). Instead, we employed a rigorous bound on the time derivative of the autocorrelation function that, by the uncertainty relation, is universally valid.

\bigskip
\onecolumngrid

\section{Time averaged bounds and how they lead to Mean Free Path Inequalities and the Ioffe-Regel limits}
\label{sec:finitetime}

In what follows, we examine systems having well defined quasiparticles. We will illustrate how the the Ioffe Regel criterion arises, within our general finite temperature framework, from the position- momentum uncertainty relations. Towards this end, we now write anew the general variance uncertainty relation,
\begin{equation} 
\label{general}
{\sf{Tr}}\left(\hat{\rho}(\Delta A)^{2}\right)  {\sf{Tr}}\left(\hat{\rho}(\Delta B)^{2}\right) \geq \frac{1}{4}|{\sf{Tr}}(\hat{\rho}[A, B])|^{2},
\end{equation}
with $\hat{\rho}$ a general density matrix (that includes that of a thermal system that we largely focused on thus far in the current work). As emphasized earlier, when $\hat{\rho} = \rho_{\Lambda}$, this inequality becomes that of Eq. (\ref{eq:AB}). We will now consider another density matrix $\hat{\rho} \equiv \rho_{\tau} \equiv \frac{1}{\tau} \int_{0}^{\tau} d t^{\prime} \mathcal{U}\left(t^{\prime}\right) \rho_{\Lambda} \mathcal{U}^{\dagger}\left(t^{\prime}\right)$ with the evolution operator $\mathcal{U}\left(t^{\prime}\right)=e^{-i H_{\Lambda} t^{\prime} / \hbar}$. For any operator $W$, the trace
\begin{eqnarray}
\label{long-mpf}
&& {\sf{Tr}}\left(\rho_{\tau} W\right) \equiv \frac{1}{\tau} \int_{0}^{\tau} d t^{\prime} {\sf{Tr}}\left(\mathcal{U}\left(t^{\prime}\right) \rho_{\Lambda} \mathcal{U}^{\dagger}\left(t^{\prime}\right) W\right) \nonumber
 \\ \! \! \! \! \! \! \! \! \! \! \! \!  && =\frac{1}{\tau} \int_{0}^{\tau} {\sf{Tr}}\left(\rho_{\Lambda} \mathcal{U}^{\dagger}\left(t^{\prime}\right) W \mathcal{U}\left(t^{\prime}\right)\right) \nonumber
\\ && \equiv \frac{1}{\tau} \int_{0}^{\tau} {\sf{Tr}}\left(\rho_{\Lambda} 
W^{H}\left(t^{\prime}\right)\right) d t^{\prime}
\end{eqnarray}
yields the average of the expectation value (as computed with $\rho_{\Lambda}$) of the Heisenberg picture operator$W^{H}$ over the time interval $[0, \tau]$. Plugging everything back into Eq. (\ref{general}),
\begin{eqnarray}
\label{general-t-average}
&& \left(\frac{1}{\tau} \int_{0}^{\tau} d t^{\prime}~ {\sf{Tr}}\left(\rho_{\Lambda}\left(A^{H}\left(t^{\prime}\right)\right)^{2}\right) -\left(\frac{1}{\tau} \int_{0}^{\tau} d t^{\prime}~{\sf{Tr}}\left(\rho_{\Lambda} A^{H}\left(t^{\prime}\right)  \right)\right)^{2}\right) \nonumber
\\ && \times
\left(\frac{1}{\tau} \int_{0}^{\tau} d t^{\prime}~ {\sf{Tr}}\left(\rho_{\Lambda}\left(B^{H}\left(t^{\prime}\right)\right)^{2}\right) -\left(\frac{1}{\tau} \int_{0}^{\tau} d t^{\prime}~{\sf{Tr}}\left(\rho_{\Lambda} B^{H}\left(t^{\prime}\right)  \right)\right)^{2}\right) \nonumber
\\ && \geq \frac{1}{4 \tau^{2}}\left|{\sf{Tr}}\left(\rho_{\Lambda} \int_{0}^{\tau} d t^{\prime}\left[A^{H}\left(t^{\prime}\right), B^{H}\left(t^{\prime}\right)\right]\right)\right|^{2}.
\end{eqnarray}
Eq. (\ref{general-t-average}) holds for all averaging times $\tau$ and is free of any assumptions. Previously, we discussed the $\tau \rightarrow \infty$ limit associated with thermal averages in ergodic systems. 

We next consider what occurs when {\it well-defined particles (or quasiparticles) exist} and that these scatter
after a mean-free path $\ell_{m . f . p .}$ and concomitant mean-free time $\tau$. We will set $A^{H}$ and $B^{H}$ to be the conjugate and dual position and momentum of a particle undergoing collisions with such mean-free path parameters. Doing so, we find that Eq. (\ref{general-t-average}) implies that for rectilinear motion between mean-free collisions,
\begin{eqnarray}
\label{mpf1}
&& \left(\frac{1}{\tau} \int_{0}^{\tau} d t^{\prime}~ {\sf{Tr}}\left(\rho_{\Lambda}\left(x^{H}\left(t^{\prime}\right)\right)^{2}\right) -\left(\frac{1}{\tau} \int_{0}^{\tau} d t^{\prime}~  {\sf{Tr}}\left(\rho_{\Lambda} x^{H}\left(t^{\prime}) \right)
\right)\right)^{2}\right) \nonumber 
\\ && \times \left(\frac{1}{\tau} \int_{0}^{\tau} d t^{\prime}~ {\sf{Tr}}\left(\rho_{\Lambda}\left(p^{H}\left(t^{\prime}\right)\right)^{2}\right) -\left(\frac{1}{\tau} \int_{0}^{\tau} d t^{\prime}~  {\sf{Tr}}\left(\rho_{\Lambda} p^{H}\left(t^{\prime}) \right)
\right)\right)^{2}\right) \nonumber
\\ && \geq \frac{1}{4 \tau^{2}}\left|{\sf{Tr}}\left(\rho_{\Lambda} \int_{0}^{\tau} d t^{\prime}\left[x^{H}\left(t^{\prime}\right), p^{H}\left(t^{\prime}\right)\right]\right)\right|^{2}=\frac{\hbar^{2}}{4}.
\end{eqnarray}
For semiclassical ballistic motion at constant velocity, 
\begin{eqnarray} 
\label{mpf2}
\frac{1}{\tau} \int_{0}^{\tau} {\sf{Tr}}\left(\rho_{\Lambda}\left(x^{H}\left(t^{\prime}\right)\right)^{2}\right) d t^{\prime}-\left(\frac{1}{\tau} \int_{0}^{\tau} {\sf{Tr}}\left(\rho_{\Lambda}\left(x^{H}\left(t^{\prime}\right)\right) \right) d t^{\prime}\right)^{2} =\frac{\ell_{m . f . p .}^{2}}{12}.
\end{eqnarray}
The factor of $\frac{1}{12}$ has its origins in the same simple integral as that for the moment of inertia of a thin rod. We then have the bound
\begin{equation}
\label{mpf3}
\frac{\ell_{m . f . p .}^{2}}{12 \tau} \int_{0}^{\tau} {\sf{Tr}}\left(\rho_{\Lambda}\left(p^{H}\left(t^{\prime}\right)\right)^{2}\right) d t^{\prime} \geq \frac{\hbar^{2}}{4}.
\end{equation}
If the average value of the squared momentum during this time is $(\hbar k)^{2},$ then 
for such ballistic like particle motion to be possible,
\begin{equation}
\label{mpf4}
k~ \ell_{m . f . p .} \geq \sqrt{3}.
\end{equation}
This condition for a uniform speed semiclassical description that follows from the rigorous inequality of 
Eq. (\ref{mpf1}) is reminiscent of the Ioffe-Regel criterion for metallic behavior
($k_{F} \ell_{m . f . p .} \gtrsim 1$ with $k_{F}$ the Fermi wave-vector). Quantum mechanically, an inequality
\begin{eqnarray}
\label{mpf5}
 \left(\frac{1}{\tau} \int_{0}^{\tau}   d t^{\prime} ~{\sf{Tr}}\left(\rho_{\Lambda}\left(x^{H}\left(t^{\prime}\right)\right)^{2}\right)-\left(\frac{1}{\tau} \int_{0}^{\tau} d t^{\prime} ~ {\sf{Tr}}\left(\rho_{\Lambda}\left(x^{H}\left(t^{\prime}\right)\right) \right)^{2}\right) \right) >\frac{\ell_{m . f . p .}^{2}}{12}
\end{eqnarray}
might, generally, be possible. That is, in deviating from the semiclassical limit, the inequality is not as restrictive.
In the cuprates, there is no resistivity saturation \cite{saturate1,NFL,planck1,planck4,bad,badmetal} and the 
Ioffe-Regel bound is indeed violated. Whenever long-lived quasiparticles no longer exist, inequalities (\ref{mpf1},\ref{mpf2},\ref{mpf3},\ref{mpf4}) become void. 
 \newline
\twocolumngrid

\section{Low temperature bounds in quasiparticle systems}
\label{sec:lowT}

We next focus on low temperature bounds in systems with uncorrelated separable local Hamiltonians of the form of Eq. (\ref{eq:decoupled+sum}). A companion pedagogical lightning review of quasiparticle Fermi systems (with which our low temperature bounds can be contrasted) is given in Appendix \ref{Fermigas}. When the decomposition of Eq. (\ref{eq:decoupled+sum}), $H_{\Lambda}=\sum_{i^{\prime}} \tilde{H}_{i^{\prime}}^{H}$, applies with the few body operators $\{\tilde{H}_{i^{\prime}}^{H}\}$ displaying no finite connected correlations amongst themselves (as evaluated with $\rho^{\sf canonical}_{\Lambda}$), we can revert to the bound discussed in Section \ref{decoupled}. The bona fide thermodynamic heat capacity $C_{v}^{(\Lambda)}$ of $\Lambda$ provides an upper bound on the rates of change of general observables. In what follows, we will use simple thermodynamic relations for the heat capacity to examine the asymptotic low temperature bounds in such systems. We will focus on Eq. (\ref{central1}) and the inequality of Eq. (\ref{f-central-1}) that it leads to for theories with uncorrelated separable local Hamiltonians. The ``locality'' need not be in real space- it may also be, e.g., in Fourier space as we will shortly discuss. 

When the system temperature $T \to 0^{+}$, its heat capacity must (as dictated by the third law of thermodynamics) drop to $C_{v}^{(\Lambda)} =0$ no less slowly than linear in the temperature. This is so since, at low temperatures, the constant volume heat capacity of the system $C_{v}^{(\Lambda)}=T (\frac{\partial S^{(\Lambda)}}{\partial T})_{v}$ (with $S^{(\Lambda)}$ the entropy of the system) may, in the absence of a zero temperature non analyticity, be Taylor expanded around the above noted vanishing zero temperature value of the heat capacity. This implies that $\sqrt{k_{B} T^{2} C^{(\Lambda)}_{v}}$ must vanish, at least, as fast as $T^{3 / 2}$ when the temperature $T$ tends to zero. Eq. (\ref{ref:decoupled-final}) and the derivations in Sections \ref{sec:diff} and \ref{GKsec} then assert that as the temperature $T \to 0$, relaxation rates must, similarly, vanish as $T^{3 / 2}$, if not more rapidly,
\begin{equation}
\label{eq:lowTbound}
\lim_{T \to 0^+} T^{3/2} \tau > 0.
\end{equation}
Eq. (\ref{eq:lowTbound}) constitutes {\it  a universal low $T$ bound} on relaxation rates in systems with separable {\it{uncorrelated}} local Hamiltonians-  i.e., {\it simple quasiparticle systems}. We caution that this inequality does not hold for strongly correlated systems. As we will explain in Appendix \ref{sec:RP}, the bounds of Section \ref{decoupled} with the thermodynamic heat capacity $C_{v}^{(\Lambda)}$ rear their head also for certain Reflection Positive systems discuss in the Appendix (Eq. (\ref{RP:L1})) and broader theories in which the connected correlation functions between the local Hamiltonians $\{ \tilde{H}_{i^{\prime}}^{H}\}$ are positive semi-definite. From this, it follows that Eq. (\ref{eq:lowTbound}) applies not only to systems with separable uncorrelated  local Hamiltonians but rather also to {\it all interacting} Reflection Positive systems (and more general theories with positive semi-definite connected correlators). 

\section{Asymptotic high temperature bounds}
\label{sec:highT}

In typical metals, primarily due to the electron-phonon scattering \cite{sarma}, the resistivity increases linearly with temperature and then saturates at high temperatures consistent with the Ioffe-Regel limit discussed in Section \ref{sec:finitetime}. This resistivity saturation \cite{saturate1} is absent in so-called ``bad metals'' that include the normal state of the cuprate superconductors and a host of other systems \cite{NFL,planck1,planck4,bad,badmetal,Legros18A} (including insightful models \cite{AssaBose}). Elegant work \cite{VadimT} illustrated how, as a general rule, linear in $T$ electrical resistivity is mandated, by the Kubo formula, at asymptotically high temperatures. When fit to a Drude model  
the scattering time in bad metals was, for some systems, found to be exceedingly close to the Planckian time $\tau_{Planck} = (\hbar/(k_{B} T))$, e.g., \cite{Legros18A}. 

Inspired by these and related findings, in this Section, we will, only very qualitatively, discuss bounds on the rates of relaxation rates of general local observables at high temperatures. Our formal asymptotic results need not, of course, necessarily carry any implications for real metals. To arrive at these bounds, we return to Eq. (\ref{central1}) and observe that if the relevant effective high temperature heat capacity saturates to a constant (Dulong-Petit type) value then for {\it any} local quantity $Q_i^{H}$, the corresponding relaxation time (as given by Eq. (\ref{cv3})),
\begin{equation}
\label{linearTeq}
\tau^{-1} \leq \mathcal{O}(T).
\end{equation}
We stress that the bounds of Eq. (\ref{linearTeq}) are only restrictive on asymptotic high temperature scaling. 
For an effective general high temperature $C_{v,i}(T)$ that depends on the temperature in a non-trivial manner, Eqs. (\ref{cv1},\ref{cv2},\ref{cv3},\ref{squared-ineq}) imply that the relaxation rate of all associated observables must, {\it at asymptotically high temperatures} $T$, scale in such a way as to satisfy $\tau^{-1} \leq \mathcal{O}(T \sqrt{C_{v,i}(T)})$. 

We next further discuss the asymptotic scaling of the effective heat capacity for a general Hamiltonian $\tilde{H}^{H}_{i}$ for a general local observable that is a function of the local position ${\bf r}_{i}$, momentum ${\bf p}_{i}$ and (similar to our analysis in Sections \ref{sec:stress} and \ref{sec:viscosity}) possibly also the forces ${{\bf {{\bf \sf{f}}}}}_{i}$. As in our earlier discussions, in the high temperature limit, the canonical ensemble variance of the local Hamiltonian $\left(\Delta \tilde{H}_{i}^{H}\right)^{2}$ driving the dynamics of the observable $Q_{i}^{H}$ can be evaluated classically. The associated classical probability density $\rho_{\Lambda}^{\sf {classical~canonical}}$ may be factorized into a part that depends on the momenta and a part depending on the spatial coordinates. If both the local Hamiltonian $\tilde{H}_{i}^{H}$ and the global system Hamiltonian $H_{\Lambda}$ may, as in the Hamiltonian of Eq. (\ref{eq:HLAMBDA}), be expressed as a sum of a term that depends on the momentum alone (i.e., a kinetic energy in the absence of the magnetic field) and a second term that depends solely on the spatial coordinates (potential energy contributions in both Hamiltonians) then, in a classical high temperature limit defined by $\rho_{\Lambda}^{\sf {classical~canonical}}$, the variance of the local Hamiltonian $\tilde{H}_{i}^{H}$ will become a sum of the two respective contributions,
\begin{eqnarray} 
\label{eq:highTcan}
k_{B} T^{2} C_{v, i} \equiv {\sf{Tr}}
\big(\rho_{\Lambda}^{\sf {classical~canonical }
}\left(\Delta \tilde{H}_{i}^{H}\right)^{2}\big)
\nonumber
\\  ={\sf{Tr}}\left(\rho_{\Lambda}^{\sf {classical~canonical }}\left(\Delta\left( \sum_{j} \frac{({\bf p}_{j}^{H})^2}{2 m}\right) \right)^{2} \right) \nonumber
\\+{\sf{Tr}}\left(\rho_{\Lambda}^{\sf {classical~canonical}}\left(\Delta V_{i}^{H}\right)^{2} \right).
\end{eqnarray}
The first term in Eq. (\ref{eq:highTcan}) may be evaluated with the kinetic energy contribution to $\rho_{\Lambda}^{\sf {classical~canonical }}$ that is Gaussian in the momentum while the second term will be computed with the Boltzmann weight $e^{-\beta V}$ (where $V$ is the potential energy contribution of Eq. (\ref{eq:HLAMBDA})) integrated over all spatial coordinates. As in Sections \ref{v-bounds}, \ref{sec:stress} and \ref{sec:viscosity}, the sum over $j$ in the kinetic term includes all (of the ${\sf z}$) particles that interact with $i$ as well as the particle $i$ itself.  The variance of the kinetic term in $\tilde{H}_{i}^{H}$, 

\begin{eqnarray}
\label{kinhighT}
{\sf{Tr}}\left(\rho_{\Lambda}^{\sf {classical ~canonical }}\left(\Delta\left(\sum_{j} \frac{({\bf p}_{j}^{H})^{2}}{2 m}\right)\right)^{2}\right) \nonumber
\\ =\frac{d({\sf z}+1)}{2}\left(k_{B} T\right)^{2}.
\end{eqnarray}
 If the potential $V^H_{i}$ is bounded then in Eq. (\ref{eq:highTcan}), ${\sf{Tr}}\left(\rho_{\Lambda}^{\sf {canonical}}\left(\Delta V_{i}^{H} \right)^{2} \right) \le \left\|V^H_{i}\right\|$. 
Conversely, if the interactions (and thus $V_{i}^{H}$) are of a power law form in the spatial coordinates then
$k_{B} T^{2} C_{v, i} \equiv {\sf{Tr}}\left(\rho_{\Lambda}^{\sf {classical~canonical }}\left(\Delta \tilde{H}_{i}^{H}\right)^{2}\right)=\mathcal{O}\left(\left(k_{B} T\right)^{2}\right)$. Eqs. (\ref{cv1}, \ref{cv2}, \ref{cv3}) imply that at high temperatures, where Eq. (\ref{kinhighT}) applies, Eq. (\ref{linearTeq}) must be satisfied. 

As just remarked above, in the cuprates and other ``bad metals'', the resistivity scales linearly in the temperature $T$  \cite{NFL,planck1,planck4,badmetal,Legros18A} consistent with this bound. We make no pretense, however, to claim that our  analysis is related to intriguing ``bad metal'' behaviors. 
 The electrical response may, instead, be bounded following the schematics outlined in Section \ref{sec:conduct}. As discussed earlier, Eq. (\ref{rho-estimate}) provides a qualitative estimate for the upper bound on the resistivity. If the variance of local interaction potential $V_{i}^{H}$ tends to a constant value at high temperatures then standard resistivity saturation may be anticipated. If the variance of $V_{i}^{H}$ is not bounded as the temperature increases then neither is our estimate of Eq. (\ref{rho-estimate}).

\bigskip
\bigskip

\section{Thermalization and Measurements}
 \label{quantum-thermalization}
 
 \subsection{The quantum case}
 
 \subsubsection{Thermalization times, time averaged measurements and equilibration}
 \label{sec:therm_time_int}

We now expand on the finite time averages considered in Section \ref{sec:finitetime} and turn to thermalization time bounds. We will further comment on possible links to quantum measurements \cite{NMott,johnvN,measure_dyn}. The inequalities derived in this Section relate to numerous earlier works on thermalization times, e.g., \cite{nnbk,juan-martin, bound, op-scram1,op-scram2,ent-growth,typical0,typical}. In what follows, we will briefly make stronger connections that were suggested in \cite{bound}. Our results point to a possible relation between short time measurement averages and thermal expectation values (that, as we explain below, may be expressed to those in single eigenstates) in ergodic systems. 

As was noted in our discussion of chaos in Section \ref{sec:chaos} as well as in our bounds on the diffusion constant and other transport coefficients, the inequality of Eq. (\ref{central1})
not only implies a bound on how rapidly various local observables $Q^H_i$ may grow but also on how fast they can decay to their equilibrium expectation values. This aspect of time reversal is trivially captured by the bounds on the absolute value of the derivative $\left|\left\langle\frac{d Q^H_{i}}{d t}\right\rangle\right|$. In general, the tightest bounds on the relaxation rates will be different for each individual $Q^H_i$. These bounds may be obtained by finding the local Hamiltonians $\tilde{H}_{i}^{H}$ that fully generate the dynamics of  $Q^H_i$ while having the smallest variance ${\sf{Tr}}\left(\rho_{\Lambda}(\Delta \tilde{H}_{i}^{H}(t))^{2}\right)$. As in Section \ref{sec:finitetime}, we will now turn again to {\it short finite time} averages given by
\begin{eqnarray}
\label{finiteTime}
 \frac{1}{{\cal{T}}} \int_{0}^{{\cal{T}}} {\sf{Tr}}\left(\rho_{\Lambda} Q_{i}^{H}(t)\right) d t,
\end{eqnarray}
with an arbitrary (short) averaging time ${\cal{T}}$. It is important to pause and emphasize here that although the full system Hamiltonian $H_{\Lambda}$ is time independent, the local Heisenberg picture Hamiltonian $\tilde{H}^{H}_{i}(t)$ is time dependent. Since $[H_{\Lambda},  Q_{i}^{H}] = [\tilde{H}_{i}^{H}, Q_{i}^{H}]$ (where, as just noted, $\tilde{H}_{i}^{H}$ is, in general, time dependent), the average of Eq. (\ref{finiteTime}) may be expressed as 
\begin{eqnarray}
\label{finite_time_average}
 \frac{1}{{\cal{T}}} \int_{0}^{{\cal{T}}} {\sf{Tr}}\left(\rho_{\Lambda}  {\mathbbm{U}}^{\dagger}_{i}(t)   Q_{i}^{H}(0) 
  {\mathbbm{U}}_{i}(t)   \right) d t.
\end{eqnarray}
Different from the full system evolution operator $\mathcal{U}\left(t^{\prime}\right)$ of Eq. (\ref{long-mpf}), the relevant local evolution operator ${\mathbbm{U}}_{i}(t)$ appearing in Eq. (\ref{finite_time_average}) that solves Eq. (\ref{eq:HM}) is given by the time ordered exponential of the local $(-i \tilde{H}^{H}_{i} t/\hbar)$, i.e., by the Dyson type series
\begin{eqnarray}
\label{uieqn}
\! \! \! \! \! \! \! \! {\mathbbm{U}}_{i}(t) &&\equiv {\mathbbm{1}} - \frac{i}{\hbar} \int_{0}^{t} dt_1~ \tilde{H}^{H}_{i}(t_1) \nonumber
\\ &&- \frac{1}{\hbar^{2}} \int_{0}^{t} dt_1 \int_{0}^{t_1} dt_2~ \tilde{H}^{H}_{i}(t_1)~ \tilde{H}^{H}_i(t_2) +  \ldots ~ .
\end{eqnarray}
In Eqs. (\ref{finite_time_average}, \ref{uieqn}), we put into effect the central theme of the current work: although the dynamics are generated by the full Hamiltonian $H_{\Lambda}$ of the entire system, the only relevant terms of the Hamiltonian for the dynamics of the local operator $Q_{i}^{H}$ (and for the above time average) are those of the local Hamiltonian $\tilde{H}^{i}_{H}$. Now, here is a simple yet important point. If a measurement is averaged (performed) over a time interval ${\cal{T}}$ with ${\cal{T}} \gtrsim \hbar/\sigma_{\tilde{H}_{i}^{H}}$ then in the integral of Eq. (\ref{finite_time_average}), the oscillatory phases, borne by a finite variance $\sigma_{\tilde{H}_{i}^{H}}^2 = {\sf Tr} ( \rho_{\Lambda} (\Delta \tilde{H}_{i}^{H})^2)$ in the state $\rho_{\Lambda}$, will, approximately, average to zero. We stress that here $\sigma_{\tilde{H}_{i}^{H}}$ denotes the {\it standard deviation of the {\underline{local Hamiltonian}}} $\tilde{H}_{i}^{H}$ governing the dynamics of particle $i$ \footnote{Strictly speaking, we should look at the frequency variance of the integrand appearing in Eq. (\ref{finite_time_average}) and recall that the Hamiltonian $\tilde{H}^{H}_{i}(t)$ defining the time ordered exponential ${\mathbbm{U}}_{i}(t)$ is time dependent. We approximate the standard deviation by the time independent average of $(k_{B}T^{2} C _{v,i})$ as defined by Eq. (\ref{cv2}) over all $i$.}. In such a case, the short finite time average of Eq. (\ref{finite_time_average}) will remain unchanged also in the ${\cal{T}} \to \infty$ limit. 

As emphasized in Section \ref{setup}, the long time limit of Eq. (\ref{finiteTime}) is none other than the equilibrium average, 
\begin{equation}
\label{canon:}
{\sf{Tr}}\left(\rho_{\Lambda}^{\sf{canonical }} Q_{i}^{H}\right)=\lim _{{\cal{T}} \rightarrow \infty} \frac{1}{{\cal{T}}} \int_{0}^{{\cal{T}}} {\sf{Tr}}\left(\rho_{\Lambda} Q_{i}^{H}(t)\right) d t.
\end{equation}
Thus, for times 
\begin{eqnarray}
\label{suff-T}
{\cal{T}} \gtrsim \frac{\hbar}{\sigma_{\tilde{H}_{i}^{H}}} \equiv \frac{\hbar}{\sqrt{k_{B} T^2 C_{v,i}}}, 
\end{eqnarray}
with the effective local heat capacity of Eq. (\ref{cv2}), the corresponding time averaged measurement of $Q_{i}$ will become equal to its equilibrium thermal expectation value. That is,  
\begin{eqnarray}
\label{lllong}
 \lim_{\cal{T} \to \infty}
\frac{1}{\cal{T}}
 \int_{0}^{\cal{T}} dt~  {\sf{Tr}} (e^{-iH_{\Lambda}t/\hbar} \rho_{\Lambda} e^{iH_{\Lambda}t/\hbar} Q_{i}) 
 \nonumber
 \\ = \sum_{n,m} \rho_{nm} Q_{mn} 
\lim_{\cal{T} \to \infty}
\frac{1}{\cal{T}} \int_{0}^{\cal{T}}  dt~ e^{-i(E_{n}-E_{m})t/\hbar} \nonumber
\\ =  \sum_{n} \rho_{nn} Q_{nn} .
\end{eqnarray}
In Eq. (\ref{lllong}), $\rho_{nn}$ and $Q_{nn}$ mark, respectively, the diagonal matrix elements of $\rho_{\Lambda}$ and $Q_{i}$ in the eigenbasis of the many body Hamiltonian $H_{\Lambda}$. 
The long time average of the integral vanishes unless the corresponding energy eigenvalues of $H_{\Lambda}$ are the same, $E_{n} = E_{m}$. When degeneracies appear, we can work in the eigenbasis formed by diagonalizing $Q$ in the eigenbasis of fixed (degenerate) energy. Eq. (\ref{lllong}) may be alternatively also rigorously arrived at even without performing any long time integrations of the phase factors by realizing that the long time average of the Heisenberg picture operator $\Big(\lim_{\cal{T} \to \infty} \frac{1}{\cal{T}} \int_{0}^{\cal{T}} dt~Q^H_{i}(t) \Big)$ is, by construction, identically a conserved time independent operator (that must, by Heisenberg's equations of motion, trivially commute with $H_{\Lambda}$) and thus diagonal in the eigenbasis of $H_{\Lambda}$ \footnote{ Whenever $H_{\Lambda}$ has degeneracies, we may construct a common diagonal eigenbasis of this long time average operator and $H_{\Lambda}$.}. Computing the expectation value of this integral by multiplying with the density matrix $\rho_{\Lambda}$ and taking the trace then leads to the final diagonal form in Eq. (\ref{lllong}) \cite{bound}. Eq. (\ref{lllong}) trivially enables an analog \cite{bound,glass} of a well known result of long time averages and the Eigenstate Thermalization Hypothesis \cite{eth1,eth2,eth3,eth4,rigol,pol,polkovnikov1,polkovnikov2,book-far} to long time averages for general density matrices $\rho_{\Lambda}$ in terms of weighted thermal equilibrium expectation values.

Thus, whenever the bound of Eq. (\ref{suff-T}) is satisfied,
the equivalence of Eqs. (\ref{finiteTime}, \ref{finite_time_average}, \ref{canon:}, \ref{lllong}),
then leads to the simple conclusion that
\bigskip

{\bf{(I)}} A measurement over a time window of width 
\begin{eqnarray}
\label{mintime}
{\cal T} \gtrsim \frac{\hbar}{\sqrt{k_{B} T^2 C_{v,i}}}
\end{eqnarray}
or longer may produce the equilibrium expectation value of a general local observable. The expectation value is, in turn, equal to a weighted average (Eq. (\ref{lllong})) of the expectation values of this observable in individual eigenstates of the Hamiltonian $H_{\Lambda}$ that produces its dynamics (Eq.(\ref{evolveeq})). Since, in equilibrium, the density matrix is a delta function in the energy density, the latter weighted single eigenstate expectation values appearing in Eq. (\ref{lllong}) are associated with eigenstates of the same exact energy density. 
\bigskip 

Furthermore if, as posited by Eigenstate Thermalization Hypothesis \cite{eth1,eth2,eth3,eth4,rigol,pol,polkovnikov1,polkovnikov2,book-far}, the diagonal matrix elements $Q_{nn}$ are given by a smooth function of the energy density then since the thermal density matrix $\rho$ is a delta function in the energy density,
 Eqs. (\ref{finiteTime}, \ref{finite_time_average}, \ref{canon:}, \ref{lllong}) will imply that
\bigskip

{\bf{(II)}} A time averaged measurement of a local observable $Q_i$ over a time interval of length ${\cal T}$ or longer may yield the expectation value of $Q_i$ in a {\it single eigenstate} of $H_{\Lambda}$.  
\bigskip

Amongst these two conclusions, {\bf{(I)}} is the more careful statement. As we just explained, conclusion {\bf{(II)}} is a natural corollary of the Eigenstate Thermalization Hypothesis for the diagonal elements of $Q_i$ in the eigenbasis of $H_{\Lambda}$. Deviations from the Eigenstate Thermalization Hypothesis \cite{eth1,eth2,eth3,eth4,rigol,pol,polkovnikov1,polkovnikov2,book-far} may arise in various cases. Since, from Eq. (\ref{mintime}), measurements over time intervals that are of width that is greater than or equal to the minimal ${\cal{T}}$ produce thermal equilibrium expectation values, we identify $\frac{\hbar}{\sqrt{k_{B} T^2 C_{v,i}}}$ as a lower bound on the thermalization time as discerned by the local observable $Q_{i}$. We emphasize that the thermalization time that we derived above is that of the minimal requisite averaging time for equilibration of local observables. As such, it may be a strict lower bound on the global thermalization time. Truly thermal systems exhibit thermal behavior and correlations on all scales. In experiments, however, one typically probes thermalization as measured by local observables such as the operators $Q_i$ that we focused on above.

\subsection{Qualitative consequences of the equivalence of short time average measurements and equilibrium (or eigenstate) expectation values}
We now qualitatively expand on possible corollaries of conclusions {\bf{(I)}} and {\bf{(II)}} of the last subsection. Towards that end, we consider what may transpire when the global Hamiltonian $H_{\Lambda}$ may include a dominant coupling $(\lambda_{i} Q_{i})$ between a local observable $Q^i$ and a macroscopic probe in the case of such measurements. As we discussed in some detail in the previous Sections, typically, for a local $Q^H_{i}$ with associated $C_{v,i}  = {\cal{O}}(k_{B})$, the standard deviation of the corresponding $\sigma_{\tilde{H}_{i}^{H}} = {\cal{O}}(k_{B} T)$. Thus, if conclusion {\bf{(II)}} holds, the requisite time beyond which we may thermalize and effectively ``collapse'' to an 
eigenstate of $H_{\Lambda}$ will be  ${\cal{O}}(\hbar/(k_{B} T))$. More cautiously, following conclusion {\bf{(I)}}, an effective ``collapse'' of the expectation of local observables to an average over eigenstates of the same energy density may arise on this time scale. That is, inasmuch local observables are concerned, the time averaged density matrix will be effectively that of the equilibrium system (emulating a delta function as a function of the energy density). Since, as noted in Section \ref{central}, at room temperature, $\hbar/(k_{B} T)= 2.5 \times 10^{-14}$ seconds, the thermalization time may be short and appear (yet certainly does not need) to be nearly instantaneous. Colloquially, the state of the system may exhibit ``rapid precessions'' about the its long time average (in which the expectation values are equal to Hamiltonian eigenstate expectation values) with these precessions canceling out at times larger than the thermalization time. A key ingredient in the above derivation is that, inasmuch as a local observable $Q^H_{i}$ is concerned, the evolution with (as we now reiterate) the {\it local} Heisenberg picture Hamiltonian $\tilde{H}_{i}^{H}$ (specifically, with the evolution operator ${\mathbbm{U}}_{i}(t) $ of Eq. (\ref{uieqn})) used in our inequality is the same as evolution with the global Hamiltonian $H_{\Lambda}$. Thus, effectively, we are looking at equilibration times (and equilibrium expectation values) of $Q^H_{i}$ as computed within the entire macroscopic system $\Lambda$. Our derived bound is that of minimal time scale for a ``collapse'' to equilibrium expectation values in the sense described above. The use of the (optimal) local $\tilde{H}_{i}^{H} \subset H_{\Lambda}$ capturing the dynamics of $Q_{i}^{H}$ leads, as remarked at the start of this Section, to (the tightest) general bounds on the equilibration times (see Appendix \ref{sec:comments}). 

Unlike macroscopic systems in which the thermalization time may diverge (with putative large effective free energy barriers), {\it small finite size systems} have analytic free energy densities and are (for ordinary interactions and random initial states) typically ergodic. Indeed, these shortest thermalization times are, for general local quantities $Q_i^H$, bounded by the insertion of the typically system size independent $\tilde{H}^{H}_i$ in Eq. (\ref{timel}). These small systems may include few particle, spin, etc., collections that are coupled to an experimental probe. Beyond bounds alone, the exact dynamics of general local quantities are governed by their corresponding local Hamiltonians $\tilde{H}_i^H$. 

Equilibration is typically driven and characterized by entropy maximization \cite{Jaynes1,Jaynes2} subject to the global constraint of fixed energy (and other constraints whenever present). Heat exchange between objects that come into contact with one another typically ceases when equilibrium is reached and entropy maximization is achieved. Entropic effects may further lift degeneracies and stabilize those states, either classical or quantum \cite{OBD,OBD1,NBCB,BCN,QOBDK}, that allow for more numerous softer low energy fluctuations over other states. A naive implementation of the entropy maximization maxim suggests that eigenstates may be similarly favored. Eigenstates have a larger number of states in their vicinity that share the same energy than states that are not eigenstates. Towards this end, we explicitly write a general pure state $| \psi \rangle = \sum_{n} c_{n} | \phi_{n} \rangle$ in the eigenbasis of $H_{\Lambda}$ (the eigenstates $\{|\phi_{n} \rangle\}$ have energies $\{E_{n}\}$). The energy $E= \langle \psi| H_{\Lambda} | \psi \rangle = \sum_{n} |c_{n}|^{2} E_{n} $ will disperse linearly in the in the deviations of the amplitudes $\{\delta c_{n} \}$ about a general state $| \psi^{(0)} \rangle$ of amplitudes $\{c_{n}^{(0)}\}$ in which the amplitudes for at least two non-degenerate states are non-vanishing. The situation is different for any of the eigenstates of $H_{\Lambda}$ for which the energy changes are quadratic in the amplitude fluctuations \footnote{This a trivial extension of the variational principle. For fluctuations about the $m$-th eigenstate of the Hamiltonian, the energy of the state $| \psi \rangle = \sum_{n \neq m} (c_{n}^{(0)} + \delta c_{n} )| \phi_{n} \rangle + \sqrt{1- \sum_{n \neq m} |(c_{n}^{(0)} + \delta c_{n})|^{2}} |\phi_{m} \rangle$ will vary only quadratically in $\{\delta c_{n}\}$ about that of $|\psi^{0} \rangle$. That is, for such states, the energy $E = \sum_{n} |\delta c_{n}|^{2} E_{n} +  (1- \sum_{n \neq m} |\delta c_{n}|^{2}) E_{m}$. By contrast, fluctuations about states that are not eigenstates of the Hamiltonian will generally yield contributions to the energy that are linear in $\{\delta c_{n}\}$. }. Thus, there may be more states of fixed energy that are formed by (these soft quadratic) small fluctuations about an eigenstate of $H_{\Lambda}$ than those formed by fluctuations about states that are not eigenstates of the Hamiltonian (where the energy changes will, generally, be linear in amplitude deviations).

\subsection{Semiclassical thermalization times in non-relativistic and ultra-relativistic systems}
\label{semithermbound}
We next return to pedestrian semiclassical considerations to illustrate why, in two quintessential weakly interacting systems, the thermalization times (the time ${\cal{T}}$ beyond which the long time average are, essentially, the equilibrium expectation values) indeed cannot be smaller than ${\cal{O}}(\hbar/k_{B} T)$. Common lore asserts that, different from electronic quasiparticle systems (see also Appendix \ref{Fermigas}) where, from the quantum Boltzmann equation, the scattering rate scales as $T^2$, the Planckian rate ${\cal{O}}(k_{B} T/\hbar)$ might constitute a universal bound that is also applicable in strongly interacting systems that are devoid of well defined quasiparticles. The simple calculations in this subsection will illustrate how the, linear in temperature, inverse Planckian time scale emerges naturally as a lower bound on thermalization times in weakly interacting systems. Towards this end, we first consider a dilute three-dimensional classical gas (a gas in which the mean free path is larger than the de-Broglie wavelength) that was, initially, out of equilibrium. Clearly, in order to thermalize, particles must, on average, collide (at least once) with one another. Thus, for such gases, the de-Broglie wavelength $h/|\vec{p}|$ of the particles divided by their speed $|\vec{p}|/m$ constitutes a physical lower bound on the thermalization time. The associated single particle momentum space average of the latter ratio is trivially of the order of ${\cal{O}}(\hbar/k_{B} T)$ (i.e., up to a factor of $(2 \pi)$) since \footnote{Here we consider a non-relativistic system given by Eq. (\ref{eq:HLAMBDA}). As usual for such Hamiltonians with a potential energy that depends only on particle positions, the classical canonical probability density $\rho_{\Lambda}^{\sf classical~canonical}$ factorizes into real and momentum space distributions. The expectation value of any single particle ratio
$(hm/\vec{p}^2)$ involves only the single particle momentum space average displayed in Eq. (\ref{classicalh}).}
\begin{equation}
\label{classicalh}
\frac{\int d^{3}\vec{p}~~ e^{- \beta \vec{p}^{2}/(2m)}~~ (hm/\vec{p}^2)}{\int d^{3}\vec{p}~~  e^{- \beta \vec{p}^{2}/(2m)}} = \frac{h}{k_{B} T}.
\end{equation}
The lower bound on the collision time $\tau$ constitutes,
for the reason described above (particles must experience at least one collision in order to achieve thermalization), a lower bound on the thermalization time ${\cal T}$.
This hints, as is indeed the case, that Eq. (\ref{classicalh}) further suggests other bounds (including those associated with transport coefficients). For instance, the shear viscosity $\eta$ of a classical gas \cite{nnbk,Reif} is simply related to the collision relaxation time $\tau$ (that is, in turn, bounded from below by Eq. (\ref{classicalh})),
\begin{equation}
\label{nhT}
    \eta = {\sf n} k_{B} T \tau.
\end{equation}
Typically, as a function of temperature, the shear viscosity minimum occurs in the lower temperature regime of the gaseous phase as the system transitions from a viscosity that is monotonically decreasing in temperature (the fluid) to a viscosity that is monotonically increasing in temperature (the higher temperature gaseous phase). Substituting the lower bound on the gas collision time of Eq. (\ref{classicalh}) into Eq. (\ref{nhT}) therefore implies that the shear minimal viscosity of the non-degenerate classical gas (and therefore of general semiclassical systems over all temperatures) satisfies the inequality
\begin{eqnarray}
\label{nhb}
\eta \gtrsim {\sf n} h.
\end{eqnarray}
This simplified lower bound on the shear viscosity conforms to the more cautious order of magnitude bound of Eq. (\ref{etanhh}). Indeed, up to factors involving the spatial dimensionality $d$ and an effective coordination number ${\sf z}$, Eq. (\ref{nhb}) and its analogs for other quantities are similar to the (numerically weaker yet) more rigorous bounds derived via the formalism of Section \ref{GsKsec} and the less rigorous inequalities the we derived by an application of the chaos bounds of Section \ref{appOTOC}. In Table \ref{real-values}, we compare Eq. (\ref{nhb}) with the empirical shear viscosity of water. The bulk viscosity \cite{Tisza1941,Muller2018} and other transport coefficients are similarly related to the relaxation time. 

Nearly identical results appear for ultra-relativistic gases where the particle speeds $\simeq c$ and the minimal equilibration time is set by the average of $h/(c|\vec{p}|)$ which similarly becomes \footnote{The mean of Eq (\ref{relativistich}) is a classical canonical expectation value associated with a general position dependent potential energy augmenting the sum of single particle ultra-relativistic kinetic energies $c \sum_{i=1}^{N_{\Lambda}} |\vec{p}_{i}|$. For such classical canonical averages (as in the non-relativistic system of Eq. (\ref{classicalh})), the probability density $\rho_{\Lambda}^{\sf classical~canonical}$ factorizes into real and momentum space parts with the average of the single particle $h/(c|\vec{p}|)$ set by the momentum space integral ratio of Eq. (\ref{relativistich}).} in thermal equilibrium,  
\begin{equation}
\label{relativistich}
\frac{\int d^{3}\vec{p}~~ e^{- \beta c |\vec{p}|}~~ h/(c|\vec{p}|)}{\int d^{3}\vec{p}~~  e^{- \beta c |\vec{p}|}} = \frac{h}{2k_{B} T}.
\end{equation}
Qualitatively consistent with the above trend of longer thermalization times of the ultra-relativistic classical ideal gas as compared to its non-relativistic counterpart and Eq. (\ref{mintime}), the heat capacity of the ultra-relativistic gas is larger than that of the non-relativistic gas (being twice as large by the equipartition theorem). 
 We emphasize that the intuitive argument that we presented in this subsection is not, at all, rigorous \footnote{Among other approximations, apart from performing classical phase (momentum) space averages (Eqs. (\ref{classicalh}, \ref{relativistich})), we wish to highlight that we assumed that the kinetic energy per particle following the collisions (interactions) with the other particles remains of the same order of magnitude as the average value set by the thermal average at temperature $T$ after equilibration} nor general \footnote{ Our focus in deriving Eqs. (\ref{classicalh}, \ref{relativistich}) has been on simplified (semi-)classical systems. These considerations do not apply to numerous quantum extensions of classical systems, including, e.g, the dilute classical gases that we discussed. For instance, as is well known, in dilute degenerate gases at ultra cold temperatures, cross-sections can become large (as in the divergent scattering length at the Feshbach resonance marking the BCS-BEC crossover e.g., \cite{BEC-BCS-R},\cite{BEC-BCS}) and the mean-free path may then be very small.}.

\subsection{Absence of rapid thermalization in glass forming liquids and other systems}
\label{sec:absent}
The lower bounds on the equilibration times that we derived in the earlier subsections need not be saturated. For instance, in chemical reactions \cite{eyring} and nucleation processes \cite{Ken-book}, the dynamics are governed by free energy barriers that may become sizable and lead to slow dynamics. In Eyring's reaction rate theory \cite{eyring}, a Planckian type time scale multiplies an exponential in the Gibbs free energy activation barrier $\Delta G$ (i.e., the inverse reaction rate is $\frac{h}{k_{B} T} e^{\beta \Delta G}$); the minimal time scale here is Planckian yet the latter exponential factor can become very notable. Such forms can be motivated for more general phenomena, e.g., \cite{nnbk,eyringViscosity}. For some systems, ergodicity and equilibration to the bona fide canonical ensemble expectation value associated with the microscopic Hamiltonian $H_{\Lambda}$ may indeed take an exceptionally long time to achieve. This occurs in, e.g., supercooled glass forming fluids \cite{bound,glass,Nick2,Ludo,Glass2}. As we briefly discuss in Appendix \ref{minprethermal}, the considerations that led us to Eq. (\ref{mintime}) may be further extended to provide bounds on minimal time scales in systems with nearly stationary non-equilibrium probability densities.

\section{Off-diagonal matrix elements of local
operators and dynamics}
\label{sec:off}
This Section conceptually builds on general notions reviewed in Section \ref{setup}. Our focus will now be on the part of the Eigenstate Thermalization Hypothesis concerning {\it off-diagonal} matrix elements of local operators in the energy eigenbasis (not on the diagonal elements of such operators that we discussed in Section \ref{quantum-thermalization}) and how these may constrain the dynamics of local observables. Our computations relate to those in \cite{srednicki-95} yet their conclusions differ. Similar to known lore and some earlier descriptions, e.g., \cite{Tolya1,Tolya2,Foini}, we find that, albeit erratically varying, the off-diagonal matrix elements of local operators cannot be completely random in the energy eigenbasis. Different from earlier works, however, {\it our considerations are not limited to thermal systems nor to the analysis of the dynamics of systems that will thermalize starting from an initial athermal state}. We illustrate that in general states, even those of {\it non-equilibrium systems}, no dynamics may be locally observable if the off-diagonal matrix elements of local operators are completely random. The dynamics in such systems lead to bounds on the deviation from randomness.

To simplify the notation (especially since we will deal with matrix element subscripts) in this Section, we will not use the local subscripts $i$ (nor global super/superscripts $\Lambda$) that we employed in the earlier Sections. 

The final conclusion that we arrive at in this Section applies to general random (and thus typically non-equilibrium) density matrices $\rho$. We now specialize to (general Hermitian) local operators $Q$. According to the Eigenstate Thermalization Hypothesis \cite{eth1,eth2,eth3,eth4,rigol,pol,polkovnikov1,polkovnikov2}, for such operators, in the eigenbasis of the Hamiltonian $H$, the off-diagonal ($m \neq n$) matrix elements 
\begin{eqnarray}
\label{off-ansatz}
Q_{m n} \sim f(E_{mn}, \omega_{mn}) \frac{R_{m n}}{e^{S  (E_{mn}) /\left(2 k_{B}\right)}}.
\end{eqnarray} 
Here, $R_{m n}$ is a random complex number that is drawn from a distribution ${\cal{P}}$ of a standard deviation of unit size and mean zero and $S$ is the entropy of the equilibrium system defined by the Hamiltonian $H$ at an average energy $E_{mn}=\left(E_{m} + E_{n}\right) / 2$. It is important to emphasize that Eq. (\ref{off-ansatz}) is a general hypothesis for the matrix elements of local observables in the eigenbasis of the $N-$particle Hamiltonian and thus can be applied {\it for general states} of the system (whether in equilibrium or not). By hermiticity of the local observable, $R_{mn} = R_{nm}^{*}$. In Eq. (\ref{off-ansatz}), the frequency $\omega_{mn} \equiv \frac{E_{m}-E_{n}}{\hbar}$ denotes the energy difference between the eigenstates labeling the off-diagonal elements. The magnitude of the continuous function $f(E_{mn}, \omega_{mn})$ is set by the thermal standard deviation $\sqrt{{\sf{Tr}}\left(\rho^{\sf {canonical }} Q^{2}\right) - ({\sf{Tr}}\left(\rho^{\sf {canonical}} Q \right)^2)}$ of the local observable. Since the latter standard deviation is of order unity (i.e., system size independent) for a local observable, so is the function $f(E_{mn}, \omega_{mn})$. The Eigenstate Thermalization Hypothesis further assumes (as is to be expected) the factor $f(E_{mn}, \omega_{mn})$ to be negligible if $\omega_{mn}$ is larger than the typical energy width. 


We next insert the Eigenstate Thermalization Hypothesis ansatz of Eq. (\ref{off-ansatz}) into the Heisenberg equations of motion. Doing so, we see that for an {\it arbitrary}  local observable $Q$,
\begin{eqnarray}
\! \! \! \! \! \! \! \! \! \! \! && \left\langle\frac{d Q}{d t}\right\rangle =\frac{i}{\hbar} {\sf{Tr}}\left(\rho[H, Q]\right)
\nonumber
\\  \! \! \! \! \! \! \! \! \! \! \! && 
 \sim \frac{i}{\hbar} \sum_{n, m} \left(E_{m}-E_{n}\right)  \rho_{n m} \nonumber
 \\ &&~~~~~~~~ \times f\left(E_{mn}, \omega_{mn} \right) \frac{R_{m n}}{e^{S /\left(2 k_{B}\right)}} \nonumber
 \\ && ~~\equiv \sum_{n,m} \rho_{nm} \dot{Q}^{ETH}_{mn} \equiv {\sf{Tr}}(\rho ~\dot{Q}^{ETH}).
 \label{pashut}
\end{eqnarray}
Here, $\dot{Q}^{ETH}_{mn}$ are the matrix elements of the commutator $\frac{i}{\hbar}[H,Q]$ with the insertion of 
Eigenstate Thermalization Ansatz of Eq. (\ref{off-ansatz}). 
Since $\{R_{mn}\}$ are assumed to be uncorrelated random numbers of vanishing mean and unit variance, the sum of Eq. (\ref{pashut}) (which can be viewed as that over a ``random walk'' (of individual steps $(nm)$ of size $(\rho_{nm} \dot{Q}^{ETH}_{mn})$ in the complex plane)) must, by the additivity of the variance for the decoupled variables in the sum, have a typical modulus that scales as the square root of the number of terms times the absolute value of a typical term. A simple counting of relevant terms (or, equivalently, the number of steps in the above ``random walk'') of pair states $(nm)$ while invoking the probability density normalization constraint over these states (leading to individual probability density factors that scale as the reciprocal of the number of relevant states in the sum (i.e., the possible values of $n$ or of $m$)) then suggests that the typical value of $\langle\frac{d Q}{d t}\rangle$ is exponentially small in the system size (scaling as $e^{-S/(2 k_{B})}$). In what follows, we make such arguments precise. As the number of individual terms $(nm)$ becomes large (as it does for a macroscopic system having numerous state pairs for which the matrix elements of $Q$ are finite), the squared norm of Eq. (\ref{pashut}) (similar to random walk type sums) may be replaced by its average over different draws of the set of random variables $\{R_{mn}\}$ from
the distribution ${\cal{P}}$. The error incurred in replacing the sum by its average vanishes in the limit of a large number of individual terms that are drawn from the distribution ${\cal{P}}$ in the sum of Eq. (\ref{pashut}). Thus, for large systems $\Lambda$, 
\begin{eqnarray}
\left\langle\frac{d Q}{d t}\right\rangle^2  \sim  \Big[({\sf{Tr}}(\rho ~\dot{Q}^{ETH}))^2 \Big]_{{\cal{P}}},
 \label{pashut+}
\end{eqnarray}
where $[ Y ]_{{\cal{P}}}$ denotes the average of $Y$ computed with the distribution ${\cal{P}}$ (and $\sim$ highlights, once again, the use of the Eigenstate Thermalization Hypothesis of Eq. (\ref{off-ansatz})). As we explain below, Eq. (\ref{off-ansatz}) implies that in the thermodynamic limit, a typical term in the sum of Eq. (\ref{pashut}) is exponentially small in the system size (as the entropy is extensive, $S= {\cal{O}}(N)$),
\begin{eqnarray}
\label{pashut1}
 \dot{Q}^{ETH}_{mn} &&\equiv \frac{i}{\hbar}  (E_{m}-E_{n})~ f(E_{mn}, \omega_{mn}) \nonumber
 \\ &&~~~~~~~~~~~~~~~~~~~~~~~~~~~ \times\frac{R_{m n}}{e^{S /\left(2 k_{B}\right)}} \nonumber
  \\ &&= {\cal{O}}(e^{-S/(2k_{B})}).
 \end{eqnarray} 
In writing the second equality in Eq. (\ref{pashut1}), we noted the following three scales:\\
 
 (i) In any physical theory, the relevant system energies and thus the energy differences $(E_{m}-E_{n})$ (or frequencies $\omega_{mn}$) may, at most, scale as a power of the system size. \\
 
 (ii) The function $f(E_{mn}, \omega_{mn}) $ is of order unity.  \\ 
  
 (iii) The random complex numbers $\{R_{mn}\}$ drawn from the distribution ${\cal{P}}$ are of typical unit norm and vanishing average. \\ 

Given Eq. (\ref{pashut}), this then implies that 
\begin{eqnarray}
&& \left\langle\frac{d Q}{d t}\right\rangle^2  \sim  \Big[({\sf{Tr}}(\rho ~\dot{Q}^{ETH}))^2 \Big]_{{\cal{P}}} \nonumber
\\ && = \sum_{nmn'm'} \rho_{nm} \rho_{n'm'} [\dot{Q}^{ETH}_{mn} \dot{Q}^{ETH}_{m'n'}]_{\cal{P}} \nonumber
\\ && = \sum_{nmn'm'} \rho_{nm} \rho_{n'm'} \delta_{mn'} \delta_{nm'} [|\dot{Q}^{ETH}_{mn}|^2]_{\cal{P}} \nonumber
\\ && = \sum_{nn'} \rho_{nn'} \rho_{n'n} \times {\cal{O}}(e^{-S/k_{B}}) \nonumber
\\&& \le  {\cal{O}}(e^{-S/k_{B}}).
 \label{pashut''}
\end{eqnarray}
The third line in Eq. (\ref{pashut''}) follows from the absence of correlations between the random variables $\{R_{mn}\}$,  the matrix elements $\dot{Q}^{ETH}_{mn}$ are also uncorrelated with each other. Thus, unless the state pairs $(nm)$ and $(m'n')$ are the same, the average $[R_{mn} R_{m'n'}]_{\cal{P}} =0$. Furthermore, if $m=m'$ and $n=n'$, the expectation value $[R_{mn}^{2}]_{\cal{P}}$ vanishes since the phase of $R_{mn}$ is random. The only bilinear in $R$ that has a non-vanishing expatiation value is that given by the ``contraction'' $[R_{nn'} R_{n'n}]_{\cal{P}} = 1$ set by the assumed unit variance
of these random variables. In the fourth line of Eq. (\ref{pashut}), we inserted Eq. (\ref{pashut1}), which given that $[[R_{mn}|^2]_{\cal{P}} =1$ (all too explicitly) implies that
\begin{eqnarray}
 &&[|\dot{Q}^{ETH}_{mn}|^2]_{\cal{P}} \nonumber
 \\  &&= \frac{1}{\hbar^2} 
  (E_{m}-E_{n})^2 (f(E_{mn}, \omega_{mn}))^2 \frac{[|R_{m n}|^2]_{\cal{P}}}{e^{S /\left(k_{B}\right)}} \nonumber
  \\ && = \frac{1}{\hbar^2} 
  (E_{m}-E_{n})^2 (f(E_{mn}, \omega_{mn}))^2 e^{-S /k_{B}} \nonumber
  \\ && = {\cal{O}}(e^{-S /k_{B}}).
\end{eqnarray}
 Finally, in deriving the (literal) bottom line scaling of the bound of Eq. (\ref{pashut''}), we invoked the following inequality for general density matrices $\rho$,
\begin{eqnarray}
  && \sum_{nn'} \rho_{nn'} \rho_{n'n} = {\sf{Tr}}(\rho^2)  = \sum_{a} p_{a}^{2} 
  \nonumber
  \\ &&\le \sum_{a} p_{a} =1.
  \label{rhomn}
\end{eqnarray}
Here, $a$ denotes the eigenstates of the probability density matrix $\rho$ (with $p_{a}$ being the corresponding probability eigenvalues). From Eq. (\ref{pashut''}), we indeed see that since
\begin{eqnarray}
\lim_{N \to \infty}
e^{-S /k_{B}} = 0,
\end{eqnarray}
in order to have a finite time derivatives of observables, one must have correlations between the energy differences and the matrix elements. We now return to the complex plane ``random walk'' analogy briefly discussed after Eq. (\ref{pashut}) and close our circle of ideas. The Kronecker delta factors in the third line of Eq. (\ref{pashut''}) and resultant variance ($ [|\dot{Q}^{ETH}_{mn}|^2]_{\cal{P}}$) \footnote{The average $ [\dot{Q}^{ETH}_{mn}]_{\cal{P}} =0$ since $[R_{mn}]_{\cal{P}} =0$. Thus, the variance is indeed $[|\dot{Q}^{ETH}_{mn}|^2]_{\cal{P}} - ([\dot{Q}^{ETH}_{mn}]_{\cal{P}})^2  = [|\dot{Q}^{ETH}_{mn}|^2]_{\cal{P}}$.} sum capture the uncorrelated individual steps in this ``random walk''. The normalization of the probability density factors that were alluded to in that analogy is fleshed out in Eq. (\ref{rhomn}).  Similar conclusions will be drawn if Eq. (\ref{off-ansatz}) is assumed to hold only for sufficiently small finite energy differences $|\omega_{mn}| \le \omega^{*}$ and is violated for larger $|\omega_{mn}|$. In such a case, we may repeat all of the above steps {\it mutatis mutandis} and note that the argument of the sum in Eq. (\ref{pashut''}) is positive semi-definite. Therefore, if, in Eq. (\ref{pashut''}), we sum only over the subset of frequencies associated with $|\omega_{mn}|, |\omega_{m'n'}| \le \omega^*$, then we will similarly obtain that the contribution of these frequencies to $\langle dQ/dt \rangle^{2}$ vanishes in the thermodynamic limit. 

Setting $Q(t) = e^{iHt/\hbar} Q e^{-iHt/\hbar}$ and $Q(0)=Q$ and repeating the considerations that led to Eq. (\ref{pashut''}), one sees that the assumption of Eq. (\ref{off-ansatz}) analogously suggests that
\begin{eqnarray}
\label{autot}
  \Big\langle \frac{dQ(t)}{dt} Q(0) \Big\rangle^2  \le {\cal{O}}(e^{-2S/k_{B}}) \to_{N \to \infty} 0.
\end{eqnarray}
That is, similar to Eq. (\ref{pashut''}), in a general macroscopic system, the autocorelation function of {\it any} local observable $Q$ cannot vary in time unless the assumption of Eq. (\ref{off-ansatz}) is amended by non-random contributions $\{F_{mn}\}$ to its off-diagonal matrix elements, i.e., 
\begin{eqnarray}
\label{Fcorrect}
Q_{m n} \sim f(E_{mn}, \omega_{mn}) \frac{R_{m n}}{e^{S  (E_{mn}) /\left(2 k_{B}\right)}} + F_{mn}.
\end{eqnarray}
In order to obtain physically meaningful results, the nature of these additional contributions $\{F_{mn}\}$ {\it cannot} be such that resulting expectation value of the time derivative $\langle dQ/dt \rangle$ of each observable $Q$ and of all other time dependent correlation functions will become trivially independent of the system state $\rho$.

We reiterate that the above conclusion {\it does not hinge on the general random state $\rho$ being an equilibrium state nor one approaching equilibrium}. Thus, if Eq. (\ref{off-ansatz}) holds with truly random numbers $\{R_{mn}\}$ then, in all physical states $\rho$ uncorrelated with these random numbers, no dynamics of any local quantity $Q$ may be observed.

Employing Eq. (\ref{Fcorrect}) to calculate $\langle \frac{dQ(t)}{dt} Q(0) \rangle^2$ in states known to exhibit dynamics of $Q$ will generally lead to constraints on the matrix elements $F_{mn}$. We remark that it is not surprising that, on average for typical cases, Eq. (\ref{off-ansatz}) may appear to hold with a seemingly random $R_{mn}$. This is so since, up to readily calculable diagonal element contributions, the averages over off-diagonal matrix elements of $Q$ and general functions thereof can be computed with a density matrix $\rho = | \psi \rangle \langle \psi |$ associated with a pure state 
\begin{eqnarray}
\label{pureshell}
| \psi \rangle = e^{-S/(2k_{B})} \sum_{n} | \phi_{n} \rangle \langle \phi_{n} |
\end{eqnarray}
formed by an equal amplitude superposition of all eigenstates  $| \phi_{n} \rangle$ of the Hamiltonian that lie within a ``shell'' of fixed energies. The width of this energy shell over which the sum in Eq. (\ref{pureshell}) extends is set by that value of $\omega_{mn}$ beyond which the function $f(E_{mn}, \omega_{mn})$ in Eq. (\ref{off-ansatz}) is assumed to be negligible. Thus, average properties of the matrix elements $Q_{mn}$ are emulated by computing expectation values within such equal amplitude pure states $| \psi \rangle$. That is, for a general function ${\sf g}$ of the observables $\{Q_i\}$ and their derivatives, the expectation value
\begin{eqnarray}
\label{gg}
\langle \psi| {\sf g} | \psi \rangle = e^{-S/k_{B}} \sum_{n} \langle \phi_{n} | {\sf g} | \phi_{n} \rangle.
\end{eqnarray}
However, the quantity on the righthand side
of Eq. (\ref{gg}) is the microcanonical ensemble average of 
${\sf g}$ over an energy shell that yields the respective equilibrium expectation values. Since observables may appear stochastic in equilibrium, it is natural to anticipate that, on average over shells of fixed energies, the statistics of the matrix elements $Q_{mn}$ conforms to Eq. (\ref{off-ansatz}). Picking the function ${\sf g}$ to be $Q^2$ for a local operator $Q$, Eq. (\ref{gg}) will then suggest that the off-diagonal matrix elements may indeed be of the form of Eq. (\ref{off-ansatz}) in order to obtain a variance that is of order unity. The requirement of the variance to be of order unity  was initial logic that underlied the Eigenstate Thermalization Hypothesis ansatz  \cite{eth1,eth2,eth3,eth4,rigol,pol,polkovnikov1,polkovnikov2} of Eq. (\ref{off-ansatz}). 

In a related vein, we may consider general time dependent correlation functions between different operators (not solely the (auto)correlation function of a single operator with itself at different times). We may further consider spatial instead of temporal evolution of local observables and their correlations. In order to have distance dependent spatial correlations between local observables, if the off-diagonal matrix elements of local observables are random in the energy eigenbasis then this randomness must be correlated at different spatial sites in order to generate a meaningful, distance dependent, correlation function. 

Apart from determining the dynamics, the contributions of the off-diagonal elements of local observables may be crucial in other regards (whether or not they obey 
Eq. (\ref{off-ansatz})). Indeed, without these contributions,
as we now explain, one will paradoxically obtain that if the equilibrium expectation values of general local operators $Q$ are continuous smooth function of the energy density then the expectation values of all such operators will be identically the same when computed in any thermal ensemble having a local Hamiltonian. To see this, consider two local Hamiltonians $H_1 = \sum_{i}  h^{i'}_1$ and $H_2=  \sum_{i} h^{i'}_2$ where $\{h^{i'}_{1}\}$ and $\{h^{i'}_2\}$ are local operators associated with sites or local volumes $i'$ (whose number is of the order of the system size $N$). We will refer to systems associated with these Hamiltonians as ``system 1'' and ``system 2''. When evaluated in the thermal equilibrium of system 2, the connected correlation function ${\sf G}_1^{i'j'} \equiv \langle h_{1}^{i'} h_{1}^{j'} \rangle - \langle h^{i'}_{1} \rangle \langle h^{j'}_1 \rangle$
will tend to zero as the distance $|i'-j'| \to \infty$ (typically doing so with either a finite correlation length exponential decay or an algebraic decay). The variance of the energy density $(H_1/N)$ may be expressed as the sum of connected correlation function over all pairs, $\sigma_{H_{1}/N}^{2} = \frac{1}{N^{2}} \sum_{i'j'} {\sf G}_1^{i'j'}$. The decay of connected correlation function implies, in turn, that in any thermal state of system 2 the energy density associated with $(H_1/N)$ is sharp (i.e., the variance of $(H_1/N)$ tends to zero in the thermodynamic limit) \footnote{For the particular case of an exponential decay of correlations with distance in ${\sf G}_1^{i'j'}$, the variance of $(H_1/N)$ scales as $1/N$ while for an algebraic decay of ${\sf G}_1^{i'j'} \propto |i'-j'|^{p}$ for a system in $d$ spatial dimensions, the variance of $(H_1/N)$ scales as $N^{-d/p}$. In both cases, the variance tends to zero as $N \to \infty$. More generally, this variance will vanish whenever the correlations decay to zero in the limit of large distances.}. Thus, in any equilibrium state of system 2, both energy densities $(H_1/N)$ and $(H_2/N)$ are sharp. The latter sharpness of $(H_2/N)$ is just the statement that the energy density of system 2 in any of its thermal states is a well defined intensive state variable \footnote{For completeness, we remark that there are numerous situations involving open systems in which intensive state variables are not sharp and divergent long range connected correlations ${\sf G}^{i'j'}$ exist. These appear, e.g., in systems that are coupled to an external drive \cite{bound} as well as the condensate number fluctuations in a Bose condensate within the grand canonical ensemble (where the chemical potential term plays the role of an external bath that couples to all particles in the system). For such an open system with a chemical potential, below the Bose-Einstein condensate temperature, the standard deviation $\sigma_{N_{0}}$ of the total number of particles in the condensate and the number of particles in the condensate are the of the scale of the number of particles in the system, $\sigma_{N_{0}} = {\cal{O}}(N)$. Thus, the condensate fraction $N_{0}/N$ is not sharp and the associated connected condensate number correlation functions do not decay to zero at long distances. Similar physics also appears in the classical mean spherical model below its critical temperature in which the mean spherical constraint is implemented by a Lagrange multiplier (serving the role of an effective chemical potential).}. By flipping $1 \leftrightarrow 2$ and repeating the above steps, it is seen that for any, similarly provable, sharp value of $(H_2/N)$ in system 1 there is a corresponding sharp energy density (the expectation value of $(H_1/N)$) of system 1 itself. A stronger yet statement would follow whenever the Eigenstate Thermalization Hypothesis
holds for diagonal elements discussed in Sections \ref{setup} and \ref{sec:therm_time_int} \cite{eth1,eth2,eth3,eth4,rigol,pol,polkovnikov1,polkovnikov2,book-far} (i.e., whenever the expectation value of $Q$ in any eigenstate of a Hamiltonian $H$ will be equal to the thermal average of $Q$ in the system defined by
this Hamiltonian). Whenever the Eigenstate Thermalization Hypothesis applies, $H_{1}/N$ will have a vanishing standard deviation in each eigenstate of $H_2$
and vice versa ($H_2/N$ will have a vanishing standard deviation in each eigenstate of $H_1$). In other words, the average energy density $(H_1/N)$ when computed in system 2 defined by $H_2$ would be a function of the energy density $(H_2/N)$ in the same system (the values of the latter energy density define the equilibrium states of
system 2 at different temperatures). This further implies that if the off-diagonal contributions of $Q$ in the eigenbasis of $H_2$ are omitted then the thermal average of {\it any local operator} $Q$ computed in system 1 at one temperature $T_1$ would be identically the same as the thermal average of $Q$ when computed with $H_2$ (system 2)
at, generally, a different temperature $T_2$ whenever the expectation value of $(H_1/N)$ evaluated in system 2 at temperature $T_1$ matches the energy density of system 2 at temperature $T_2$. Clearly, such an equality cannot hold for {\it all} local observables $Q$ simultaneously. This illustrates once again the possible importance of the off-diagonal contributions of $Q$ (in either eigenbasis)
and that of having dependencies of equilibrium expectation values on quantities other than only the energy density. In the classical arena (where all averages performed over microstates are ``diagonal''), this paradox underscores that the individual microstate expectation values cannot depend on the energy density alone. The above considerations can be extended to situations in which expectation values are assumed to depend on only a finite number of additional intensive variables other than the energy density alone. 

To recap, the results of this Section suggest that the off-diagonal matrix elements of local observables cannot be completely random for dynamics and nontrivial spatio-temporal correlations to appear. This hints that although Eq. (\ref{off-ansatz}) is consistent with the global Gaussian distribution of the norm of the off-diagonal matrix elements when averaged over all eigenstates \cite{offdiag}, more intricate deviations from Eq. (\ref{off-ansatz}) might need to appear if local dynamics are to be observed in general random non-equilibrium states. We explained, however, why on average 
the statistics of the off-diagonal matrix elements may appear to be random. We further commented on the importance of the off-diagonal matrix elements and/or dependencies of equilibrium expectation values on quantities other than the energy density and a finite number of other state variables alone.

\section{Conclusions} 
\label{sec:conclusions}
In summary, we illustrated how the application of the uncertainty relations to quantum thermal systems leads to  
numerous bounds on {\it local} observables, their dynamics, Lyapunov exponent, transport coefficients, spatial gradients, and general correlation functions. Some of our conclusions 
are summarized in Table \ref{real-values}. These illustrate that our bounds are often close to empirical values. Complementing the quantities highlighted in this Table, we described how other transport coefficients may be bounded (e.g., Eq. (\ref{rho-estimate}) for the order of magnitude of the upper bound on the electrical resistivity of metals) and other attributes of the dynamics (e.g., the acceleration of Eq. (\ref{potential-eq})) may be bounded.  
The bulk of our derived transport coefficient bounds {\it do not} assume the existence of quasiparticles. In Section \ref{sec:lowT}, we discussed universal low temperature bounds in quasiparticle systems. 

 {{\bf Acknowledgments}} \newline
 We are grateful to discussions with and encouragement by Boris Fain, Itamar Kimchi, Kater Murch, Laimei Nie, Shmuel Nussinov, Philip Phillips, Mark Srednicki, and Jan Zaanen. We are, in particular, we are indebted to numerous conversations with Flavio Nogueira and Alexander Seidel and are grateful for critical reading of the manuscript and to a written comment by Alexander Seidel that is captured in Appendix \ref{QUB}. The work is an outgrowth of research supported by NSF DMR-1411229 (ZN) which has since been terminated. We further wish to acknowledge the Aspen Center for Physics, which is supported by NSF grant PHY-1607611, where this work was completed.
 
\appendix

\section{Review of the Araki-Lieb construct and its implications for finite temperature uncertainty relations}
\label{sec:thermofield}

In this Appendix, we briefly review Araki and Lieb's original purification (also known as the thermofield double) construct \cite{lieb-pure,das}. We then explain \cite{bound} how this construct immediately implies that uncertainty inequalities proven for pure states apply to general mixed states. The final result of the below derivation- that the uncertainty relations carry over to mixed states- is well known. 

Let the eigenvectors and eigenvalues of $\rho_{\Lambda}$ be $\{|c\rangle\}$ and $\left\{p_{c}\right\}$ respectively. Thus, 
\begin{equation}
\rho_{\Lambda}=\sum_{c} p_{c}|c\rangle\langle c|.
\end{equation}
Define vectors $\left\{\left|w_{c}\right\rangle\right\}$ in a space orthogonal to that spanned by $\{|c\rangle\}$ such that
\begin{equation}
\left\langle w_{c} \mid w_{c^{\prime}}\right\rangle=\delta_{c, c^{\prime}}
\end{equation}
The volume $\bar{\Lambda}$ on which $\left\{\left|w_{c}\right\rangle\right\}$ have their spatial support is different from $\Lambda$ (i.e., $\bar{\Lambda} \cap \Lambda=\emptyset$). Now define the pure state
\begin{equation}
|\psi\rangle \equiv \sum_{c} \sqrt{p_{c}}\left(|c\rangle \otimes\left|w_{c}\right\rangle\right).
\end{equation}
If the states $\left\{\left|w_{c}\right\rangle\right\}$ have their spatial support on $\bar{\Lambda}$ then the partial trace of the pure state density matrix 
\begin{equation}
{\sf{Tr}}_{\bar{\Lambda}}(|\psi\rangle\langle\psi|)=\rho_{\Lambda}.
\end{equation}
Putting all of the pieces together, this implies that any uncertainty inequality proven for pure states (over $\Lambda \cup \bar{\Lambda}$) mandates a corresponding inequality for operators defined on $\Lambda$ where averages are computed with a general density matrix $\rho_{\Lambda}$. This establishes Eq. (\ref{eq:AB}).

Stated alternatively, the uncertainty inequality that we employ throughout the current work for mixed states is intimately related to the Cauchy-Schwarz inequality associated with a Hilbert-Schmidt (trace) inner product. By setting $|u \rangle \equiv A^\dagger | \psi \rangle$ and $|v \rangle \equiv B| \psi \rangle$ with two general operators $A$ and $B$, the results of this Section illustrate that Cauchy-Schwarz inequality  $|\langle u | v \rangle|^2 \le \langle u | u \rangle \langle v | v \rangle$ leads to the trace inequality of Eq. (\ref{CSg'}).

\section{Non semiclassical quadratic moment bounds}
\label{QUB}

Returning to the bound of Eq. (\ref{longABABA}), we observe that the expectation value of the anticommutator
\begin{eqnarray}
\label{2term}
 && \Big\langle  \{ \Delta \tilde{H}^{H}_{i}  \Delta Q^H_i, \Delta Q^H_i \Delta \tilde{H}^{H}_{i} \} \Big\rangle \nonumber
 \\ && =  \Big\langle   \Delta \tilde{H}^{H}_{i} ( \Delta Q^H_i)^2  \Delta \tilde{H}^{H}_{i} \Big\rangle +   \Big\langle    \Delta Q^H_i   (\Delta \tilde{H}^{H}_{i})^2  \Delta Q^H_i  \Big\rangle \nonumber
 \\&&  = \frac{1}{4} \Big \langle \Big( \{ \Delta \tilde{H}^{H}_{i},  \Delta Q^H_i \} + [ \Delta \tilde{H}^{H}_{i},  \Delta Q^H_i ]\Big) \nonumber
 \\ && \times  \Big( \{ \Delta \tilde{H}^{H}_{i},  \Delta Q^H_i \} - [ \Delta \tilde{H}^{H}_{i},  \Delta Q^H_i ]\Big) \nonumber
 \\ && + \Big( \{ \Delta \tilde{H}^{H}_{i},  \Delta Q^H_i \} - [ \Delta \tilde{H}^{H}_{i},  \Delta Q^H_i ]\Big) \nonumber
\\  && \times \Big( \{ \Delta \tilde{H}^{H}_{i},  \Delta Q^H_i \} + [ \Delta \tilde{H}^{H}_{i},  \Delta Q^H_i ]\Big)  \Big \rangle \nonumber
\\ && = \frac{1}{2} \Big \langle  \{ \Delta \tilde{H}^{H}_{i},  \Delta Q^H_i \}^2  \Big \rangle - \frac{1}{2}  \Big \langle  [ \Delta \tilde{H}^{H}_{i},  \Delta Q^H_i ]^2  \Big \rangle \nonumber
\\ &&  \ge - \frac{1}{2}  \Big \langle  [ \Delta \tilde{H}^{H}_{i},  \Delta Q^H_i ]^2  \Big \rangle.
\end{eqnarray}
Thus,
\begin{eqnarray}
\label{2term+}
-  \Big \langle  [ \Delta \tilde{H}^{H}_{i},  \Delta Q^H_i ]^2  \Big \rangle  \le 2  \Big\langle  \{ \Delta \tilde{H}^{H}_{i}  \Delta Q^H_i, \Delta Q^H_i \Delta \tilde{H}^{H}_{i} \} \Big\rangle.  \nonumber 
\end{eqnarray}
On the righthand side of Eq. (\ref{2term}), both expectation values 
\begin{eqnarray}
\label{Q1}
&& \Big\langle   \Delta \tilde{H}^{H}_{i} ( \Delta Q^H_i)^2  \Delta \tilde{H}^{H}_{i} \Big\rangle \nonumber
\\ = 
&& \Big\langle   (\Delta Q^H_i  \Delta \tilde{H}^{H}_{i})^{\dagger}  (\Delta Q^H_i  \Delta \tilde{H}^{H}_{i})  \Big\rangle  \ge 0
\end{eqnarray}
and 
\begin{eqnarray}
\label{Q2}
&& \Big\langle   \Delta Q^H_i  (\Delta \tilde{H}^{H}_{i})^2 \Delta Q^H_i \Big\rangle \nonumber
\\  = 
&& \Big\langle   ( \Delta \tilde{H}^{H}_{i}    \Delta Q^H_i )^{\dagger}  (\Delta \tilde{H}^{H}_{i} \Delta Q^H_i)  \Big\rangle \ge 0
\end{eqnarray}
are manifestly positive semi-definite. 

Given a general thermal density matrix $\rho_{\Lambda}$, we may define the two density matrices 
\begin{eqnarray}
\label{ddensity}
\rho_{\Delta \tilde{H}^{H}_{i}} && \equiv \frac{1}{{{\sf{Tr}} (\rho_{\Lambda} \Delta \tilde{H}^{H}_{i})^2}}  \Delta \tilde{H}_{i}^{H} \rho_{\Lambda} \Delta \tilde{H}^{H}_{i}, \nonumber
\\ \rho_{\Delta Q_i^H} &&\equiv \frac{1}{{{\sf{Tr}} (\rho_{\Lambda} \Delta Q_i^H)^2}}  \Delta Q_i^H \rho_{\Lambda} \Delta Q_i^H.
\end{eqnarray}
One may readily verify that both $\rho_{\Delta \tilde{H}^{H}_{i}}$ and $\rho_{\Delta Q} $ have a trace of unity and are positive semi-definite and thus indeed constitute probability density matrices. Averages computed with the density matrices of Eq. (\ref{ddensity}), in particular those in Eqs. (\ref{Q1},\ref{Q2}), correspond to expectation values calculated in an initial thermal state $\rho_{\Lambda}$ that has, subsequently, been ``locally heated'' (or ``cooled'') by a perturbation $\tilde{H}^{H}_{i}$ or $Q_i^H$ (with the variance associated with these perturbations set by their equilibrium standard deviation at the temperature $T$ associated with the thermal state $\rho_{\Lambda}$). Given the density matrices of Eq. (\ref{ddensity}) and using the cyclic invariance of the trace, we may recast Eq. (\ref{2term}) as
\begin{eqnarray}
\label{variantquad}
&& \frac{\hbar^2}{2} {\sf{Tr}} (\rho_{\Lambda}  (\frac{d Q^H_i}{dt})^2) \nonumber
\\ && =  -  \frac{1}{2} {\sf{Tr}} (\rho_{\Lambda}  [ \Delta \tilde{H}^{H}_{i},  \Delta Q^H_i ]^2  ) \nonumber
\\  && \le 
{\sf{Tr}}(\rho_{\Delta \tilde{H}^{H}_{i}} (\Delta Q^H_i )^2)~  {\sf{Tr}}(\rho_{\Lambda} (\Delta \tilde{H}^H_i)^2)
\nonumber 
\\ && + {\sf{Tr}}(\rho_{\Lambda} (\Delta Q^H_i)^2) ~ {\sf{Tr}}(\rho_{\Delta Q^H_i} (\Delta \tilde{H}^{H}_{i})^2).
\end{eqnarray}
We underscore that ${\sf{Tr}}(\rho_{\Lambda} (\Delta \tilde{H}^H_i)^2)$ is set by the effective heat capacity employed in the main text (Eq. (\ref{cv2})) and that ${\sf{Tr}}(\rho_{\Delta Q^H_i} (\Delta \tilde{H}^{H}_{i})^2)$ is the variance of the local Hamiltonian when computed with the modified density matrix $\rho_{\Delta Q^H_i}$. Eq. (\ref{variantquad}) and simple extensions thereof constitute analogs of the uncertainty relations of Section \ref{high-moment-derivative} for higher order moments of the time derivative of a general observable that are free of a semiclassical or other approximations. By deforming the thermal density matrix to be the density matrices of Eq. (\ref{ddensity}), the semiclassical inequality of Eq. (\ref{longABABA'}) trivially becomes the exact bound of Eq. (\ref{variantquad}). The inequality of Eq. (\ref{longABABA'}) is further generalized in Appendix \ref{triv-triv-long}.

\section{semiclassical quadratic and higher moment bounds for general operators}
\label{triv-triv-long}
In this Appendix, we explicitly note how Eq. (\ref{longABABA'}) generalizes in the classical limit for commuting $\tilde{H}_{i}^{H}$ and observables $Q_{i}^{H}$ that do {\it not} depend only the spatial coordinates or only on the momenta. Here, an average of the form of Eq. (\ref{longABABA'}) becomes
\begin{eqnarray}
\label{longABABA'!}
 \frac{1}{N_{\Lambda}} \sum_{i=1}^{N_{\Lambda}}  \Big\langle  \Delta \tilde{H}^{H}_{i}  \Delta Q^H_i   \Delta \tilde{H}^{H}_{i}  \Delta Q^H_i  \Big\rangle \nonumber
\\  =  {\sf{Tr}}(\rho_{\Lambda}^{\sf classical ~canonical} ( \Delta \tilde{H}^{H}_{i}  \Delta Q^H_i  \Delta \tilde{H}^{H}_{i}  \Delta Q^H_i ) ) \nonumber
 \\  = {\sf{Tr}} (\rho_{\Lambda}^{\sf classical ~canonical} (\Delta \tilde{H}^{H}_{i})^2 (\Delta Q^H_i)^2) \nonumber
\\  \le \sqrt{ {\sf{Tr}} (\rho_{\Lambda}^{\sf classical ~canonical} (\Delta \tilde{H}^{H}_{i})^4)} \nonumber
\\  \times \sqrt{ {\sf{Tr}} (\rho_{\Lambda}^{\sf classical~ canonical} (\Delta Q^H_i)^4).}
 \end{eqnarray}
In the first equality in Eq. (\ref{longABABA'!}), we invoked Eq. (\ref{O:avg}). In the second equality, we used the commutativity in the classical limit. In the inequality in the fourth line of Eq. (\ref{longABABA'!}), we invoked the Cauchy-Schwarz relation of Eq. (\ref{CSg'}), with $A=( \Delta \tilde{H}^{H}_{i})^2$ and $B=( \Delta Q^H_i)^2$, trivially applied to the classical canonical ensemble probability density averages. Eq. (\ref{longABABA'!}) is similar to our semiclassical result of Eq. (\ref{longABABA'}) for separable probability densities and local observables $Q_{i}^{H}$ and associated local Hamiltonians $\tilde{H}_{i}^{H}$ that each depend only on the spatial coordinates or only on the momenta. However, since 
\begin{eqnarray}
\! \! \! \! \! \! \! && {\sf{Tr}}( \rho_{\Lambda}^{\sf classical~ canonical} (\Delta \tilde{H}^{H}_{i})^4) \nonumber
\\    \ge  && ({\sf{Tr}}(\rho_{\Lambda}^{\sf classical canonical} (\Delta \tilde{H}^{H}_{i})^2))^2,
 \end{eqnarray}
 and 
 \begin{eqnarray}
\! \! \! \! \! \! \!  \! \! \! \! \! \! \! &&  {\sf{Tr}} (\rho_{\Lambda}^{\sf classical~ canonical} (\Delta Q^H_i)^4)
\nonumber
\\ \ge &&  ({\sf{Tr}}  (\rho_{\Lambda}^{\sf classical~ canonical} (\Delta Q^H_i)^2))^2, \nonumber
 \end{eqnarray}
 the upper bound of Eq. (\ref{longABABA'!}) is, normally, less restrictive than Eq. (\ref{longABABA'}) (as anticipated for the more general $\{Q_{i}^{H}\}$ and $\{\tilde{H}^{H}_{i}\}$). We now briefly discuss the non-relativistic kinetic term that is quadratic in the momentum (Eq. (\ref{eq:HLAMBDA})). By Wick's theorem, for an associated $\rho_{\Lambda}^{\sf classical~ canonical}$ that is Gaussian in the momentum, for such a kinetic $ \tilde{H}^{H}_{i})$, the average $ {\sf{Tr}} (\rho_{\Lambda}^{\sf classical~canonical} (\Delta \tilde{H}^{H}_{i})^4) = 3( {\sf{Tr}} (\rho_{\Lambda}^{\sf classical~canonical} (\Delta \tilde{H}^{H}_{i})^2)^2)$. Thus, the term involving the squared variance of the local Hamiltonian of Eq. (\ref{longABABA'!}) is larger (by a factor of $\sqrt{3}$) as compared to that in Eq. (\ref{longABABA'}). Eq. (\ref{longABABA'!}) can be extended to higher ($n>2$) moments of $\Delta Q_{i}^{H}$.

\section{Localized Identical Particles}
\label{sec:identical}

Our central inequalities in the main text relied on exact operator uncertainty relations. These relations held for arbitrary states $\rho_{\Lambda}$ whether these describe identical or distinguishable particles or whether these particles are localized or itinerant. As such, the inequalities that we derived were universal. The existence of an index $i$ labeling a local Hamiltonian $\tilde{H}^{H}_{i}$ associated with a local observable $Q^{H}_{i}$ does not, of course, imply that the particles in the system are distinguishable. Prior to taking semiclassical limit, all of our uncertainty based inequalities were exact. These also include our bounds on the autocorrelation function and their implications for transport coefficients. For the particular situation in which particles are localized in some space, the index $i$ may naturally provide additional information regarding the localization (as it does in a semiclassical limit in which particle locations may be specified). Indeed, in various experimental setups for measuring single particle velocities or other properties, individual particles may need to be resolved. We briefly elaborate on this situation in the current Appendix. The localization that we discuss now may be that in real, momentum, or any other space. For definitiveness, in what follows, we allude to localization in real space. With a trivial change of spatial labels by others, the results apply verbatim in momentum or any other space in which localized particles exist. We examine the single particle density matrix,
\begin{equation}
\rho_{1} \equiv {\sf{Tr}}_{2,3, \ldots, N_{\Lambda}}\left(\rho_{\Lambda}\right).
\end{equation}
If particles may be spatially resolved then, in the position space representation, the diagonal elements $\rho_{1}\left({\bf x}_{1}\right)$ form 
a function of ${\bf x}_{1}$ that is equal to the sum of $N_{\Lambda}$ spatially localized functions that are centered about positions $\left\{{\bf X}_{i}\right\}_{i=1}^{N_{\Lambda}}$. Given, for each $i$, the normalized distribution $\mathcal{G}_{i t}\left({\bf x}_{1}-\bar{X}_{i}\right)$, the single body probability density
\begin{equation}
\rho_{1}\left({\bf x}_{1}, t\right) \equiv \operatorname{Tr}_{2,3, \ldots, N_{\Lambda}}\left(\rho_{\Lambda}(t)\right)=\frac{1}{N_{\Lambda}} \sum_{i=1}^{N_{\Lambda}} \mathcal{G}_{i t}\left({\bf x}_{1}-{\bf X}_{i}(t)\right)
\end{equation}
peaks at ${\bf x}_{1}=X_{i}$. The contribution of each particle $i$ to the single body probability density $\rho_{1}$ must be of the above general form. The single particle distribution $\mathcal{G}_{i t}\left({\bf x}_{1}-{\bf X}_{i}(t)\right)$ need not be of identical functional form for different $i$. For example, at a particular spatial location, a particle can be in a higher excited stated than at another location. The particles may, at times $t$, be spatially well defined (apart from possibly a set of time intervals of measure zero) if at all such $t$, the maximal standard deviation
$\Delta_{\mathcal{G}}^{\max }$ of $\mathcal{G}_{i t}$ is far smaller than 
a fixed shortest inter-particle distance ${\sf a}$.

A key role in what follows will be played a Gaussian $\mathcal{F}$ that is centered about the origin (with $\mathcal{F}(0)=1)$ that is of standard deviation $ {\sf a} \gg \sigma_{\mathcal{F}} \gg \Delta_{\mathcal{G}}^{\max }$. The derivatives of Gaussians of such specific standard deviation will be finite only where the probability density $\rho_{1}$ (and $\rho_{\Lambda}$) is vanishingly small. Hence, for a general spatial function $f$, the expectation value of the commutator ${\sf{Tr}}\left(\rho_{\Lambda}\left[f, \mathcal{F}\left({\bf x}_{j}-{\bf X}_{j} (t)\right)\right]\right)$ is negligible \footnote{We stress that here $f$ is a general function of the coordinates and their derivatives. The derivative (as well as finite higher order derivatives) of the Gaussian $\mathcal{F}$ is non-vanishing at a distance $\sigma_{F}$ where the density matrix vanishes. It follows that the expectation value of the commutator $[f, \mathcal{F}$] will vanish. This is so since (i) where the derivative of the Gaussian is finite the density matrix  vanishes and (ii) where the density matrix is finite the derivative of the Gaussian is vanishingly small.}. Since all time dependence of $\mathcal{F}$ arises from that of ${\bf X}_{j}(t)$, the expectation value of the time derivative ${\sf{Tr}}\left(\rho_{\Lambda} \frac{\partial \mathcal{F}}{\partial t}\right)$ vanishes.

The Gaussian $\mathcal{F}$ assumes the role of a broadened delta function selecting positions ${\bf x}_j$ close to ${\bf X}_j$. For single body operators $Q_{i}^{H}$, the results derived in the main text for distinguishable particles
are unaltered by the replacement 
$\tilde{Q}_{i}^{H}(t) \rightarrow\left(\tilde{Q}_{i}^{H}(t) \times \mathcal{F}\left({\bf x}_{i}^{H}(t)-{\bf X}_{i}(t)\right)\right)$. For, e.g., our velocity
bounds of Section \ref{v-bounds}, the only term in $H_{\Lambda}$ that endows
$\left({\bf x}_{i}^{H} \times \mathcal{F}\left({\bf x}_{i}^{H}(t)-{\bf X}_{i}(t)\right)\right)$
with dynamics is, as before, $\tilde{H}_{i}^{H}=\frac{\left({\bf p}_{i}^{H}\right)^{2}}{2 m}.$
Given our definitions and construct, this and our other earlier Hamiltonians $\tilde{H}_{i}^{H}(t) \subset H_{\Lambda}$ similarly generate the dynamics of other general
$\left(\mathcal{Q}_{i}^{H}(t) \times \mathcal{F}\left({\bf x}_{i}^{H}(t)-{\bf X}_{i}(t)\right)\right)$.
Rather explicitly, from Heisenberg's equations of motion,
\onecolumngrid
\begin{eqnarray}
\label{longob}
&&{\sf{Tr}}\left(\rho_{\Lambda} \frac{d}{d t}\left(\tilde{Q}_{i}^{H} \times \mathcal{F}\left({\bf x}_{i}^{H}-{\bf X}_{i}(t)\right)\right)\right) \nonumber
\\= && \frac{i}{\hbar} {\sf{Tr}}\left(\rho_{\Lambda}\left[H_{\Lambda}^{H},\left(\tilde{Q}_{i}^{H} \times \mathcal{F}\left({\bf x}_{i}^{H}-{\bf X}_{i}(t)\right)\right)\right]\right) \nonumber
\\ && +{\sf{Tr}}\left(\rho_{\Lambda} \tilde{Q}_{i}^{H} \frac{\partial \mathcal{F}\left({\bf x}_{i}^{H}-{\bf X}_{i}(t)\right)}{\partial t}\right) \nonumber
\\ && = \frac{i}{\hbar} {\sf{Tr}}\left(\rho_{\Lambda}\left[H_{\Lambda}^{H},\left(\tilde{Q}_{i}^{H} \times \mathcal{F}\left({\bf x}_{i}^{H}-{\bf X}_{i}(t)\right)\right)\right]\right)  \nonumber
\\ && = \frac{i}{\hbar} {\sf{Tr}}\left(\rho_{\Lambda}\left[H_{\Lambda}^{H}, \tilde{Q}_{i}^{H}\right] \times \mathcal{F}\left({\bf x}_{i}^{H}-{\bf X}_{i}(t)\right)\right).
\end{eqnarray}
In the second equality of Eq. (\ref{longob}), we invoked the vanishing expectation value of the derivatives of $\mathcal{F}$ when these are computed with $\rho_{\Lambda}$. We now return to the general uncertainty inequality
\begin{eqnarray}
 && \left({\sf{Tr}}\left(\rho_{\Lambda}\left(\Delta \tilde{H}_{i}^{H}(t)\right)^{2}\right)\right) \times \left({\sf{Tr}}\left(\rho_{\Lambda}\left(\Delta\left(\mathcal{Q}_{i}^{H}(t) \times \mathcal{F}\left({\bf x}_{i}^{H}(t)-{\bf X}_{i}(t)\right)\right)\right)^{2}\right)\right) \nonumber
\\  
&& \geq \frac{1}{4} 
{\sf{Tr}}\left(\rho_{\Lambda}\left[\tilde{H}_{i}^{H}(t), \mathcal{Q}_{i}^{H}(t) \times \mathcal{F}\left({\bf x}_{i}^{H}(t)-{\bf X}_{i}(t)\right)\right]\right)
^{2} \nonumber
\\ 
&& =\frac{1}{4}\left|{\sf{Tr}}\left(\rho_{\Lambda}\left(\left[\tilde{H}_{i}^{H}(t), \mathcal{Q}_{i}^{H}(t)\right] \times \mathcal{F}\left({\bf x}_{i}^{H}(t)-{\bf X}_{i}(t)\right)\right)\right)\right|^{2} \nonumber
\\  
&& =\frac{\hbar^{2}}{4}\left({\sf{Tr}}\left(\mathcal{F}\left({\bf x}_{i}^{H}(t)-{\bf X}_{i}(t)\right) \times \rho_{\Lambda} \times \frac{d Q_{i}^{H}}{d t}\right)\right)^{2}.
\end{eqnarray}
This implies that
\begin{eqnarray}
k_{B} T^{2} C_{v,i} \equiv {\sf{Tr}}\left(\rho_{\Lambda}^{\sf {canonical }}\left(\Delta \tilde{H}_{i}^{H}(t)\right)^{2}\right) =\frac{1}{N_{\Lambda}} \sum_{i=1}^{N_{\Lambda}}\left({\sf{Tr}}\left(\rho_{\Lambda}\left(\Delta \tilde{H}_{i}^{H}(t)\right)^{2}\right)\right)  \nonumber
\\ \geq \frac{\hbar^{2}}{4 N_{\Lambda}} \sum_{i=1}^{N_{\Lambda}} \frac{\left({\sf{Tr}}\left(\rho_{\Lambda} \times \frac{d Q_{i}^{H}}{d t} \times 
\mathcal{F}\left({\bf x}_{i}^{H}(t)-{\bf X}_{i}(t)\right)\right)^{2}\right.}{{\sf{Tr}}\left(\rho_{\Lambda}\left(\mathcal{Q}_{i}^{H}(t) \times \mathcal{F}\left({\bf x}_{i}^{H}(t)-{\bf X}_{i}(t)\right)\right)^{2}\right)}.
\end{eqnarray}
\twocolumngrid
The multiplication by $\mathcal{F}$ acts as a projection operator that leads to the identification of $x_{i}^{H}$ with the particular coordinate ${\bf X}_{i}$. The above sketch may be further refined.

In the simple bounds that we wrote in this Appendix, we alluded, for concreteness, to the velocity bounds of Section \ref{v-bounds}. Similar constructs may be replicated, nearly verbatim, in momentum (or other) space(s) whenever well-defined localized particles (in the associated space) may be defined.

\section{Different bounds associated with the non-unique choice the local Hamiltonians}
\label{sec:comments}

We now highlight the non-uniqueness of the choice of the operator $\tilde{H}_{i}^{H}$ giving rise to the dynamics of $Q^{H}_{i}$. This non-uniqueness may be leveraged in order to find optimal sets of local Hamiltonians $\{\tilde{H}_i^H\}$ that give rise to the strongest bounds on the expectation values of the time derivatives of $Q^H_i$ and their moments. 

As noted in the main text of the paper, we may choose $\tilde{H}_{i}^{H}$ to be the {\it minimal} subset of terms in $H_{\Lambda}$ that endow $Q_{i}^{H}$ with dynamics. That is, $\frac{d Q_{i}^{H}}{d t}=\frac{i}{\hbar}\left[H_{\Lambda}, Q_{i}^{H}\right]=\frac{i}{\hbar}\left[\tilde{H}_{i}^{H}, Q_{i}^{H}\right]$ with $\tilde{H}_{i}^{H} \subset H_{\Lambda}$. 
However, any such $\tilde{H}_{i}^{H} \subset H_{\Lambda}$ may be augmented by additional terms that do not alter the equations of motion.  
  In general, in order to obtain the tightest bounds, we may seek an addition $W^H_{i}$ for which the variance ${\sf{Tr}}\left(\rho^{\sf{canonical}}_{\Lambda}\left(\tilde{H}^H_{i}+W^H_{i}\right)^{2}\right)-\left({\sf{Tr}}\left(\rho^{\sf{canonical}}_{\Lambda}\left(\tilde{H}^H_{i}+W^H_{i}\right)\right)^{2}\right)$ will be the smallest. Apart from commuting with $Q_{i}^{H}$ (i.e., $[W^{H}_{i}, Q^{H}_i]=0$) so as to leave the equations of motion invariant, the operators $W^H_i$ over which we may minimize the variance are completely arbitrary \footnote{In those special cases in which we will seek a specific decomposition of the form of Eq. (\ref{eq:decoupled+sum}), we will attempt to find a transformation $\tilde{H}_{i'}^{H} \to \tilde{H}_{i'}^{H} + W^H_{i'}$ minimizing ${\sf{Tr}}\left(\rho^{\sf{canonical}}_{\Lambda}\left(\tilde{H}^H_{i'}+W^H_{i'}\right)^{2}\right)-\left({\sf{Tr}}\left(\rho^{\sf{canonical}}_{\Lambda}\left(\tilde{H}^H_{i'}+W^H_{i'}\right)\right)^{2}\right)$ and also further 
 satisfying the constraint $\sum_{i'} W^H_{i'} =0$.}. In a harmonic solid model example that we will examine in Appendix \ref{sec:harmonic}, we will explicitly demonstrate that keeping more terms than the minimal ones may lead to far stronger bounds. In an example that we will examine in Appendix \ref{sec:harmonic}, we will illustrate that keeping the full Hamiltonian associated with a given mode instead of only the associated kinetic term for that mode, will give rise to a more stringent upper bound on the velocity that tends to zero in the low temperature limit. The latter vanishing behavior is universally expected since at low temperatures when the system veer towards its ground state(s) for which the energy fluctuations trivially vanish. In such a situation, we will simply set $\tilde{H}_{i}^{H}$ to be larger subset of $H_{\Lambda}$ (i.e., containing more of the terms that appeared in $H_{\Lambda}$). The variance ${\sf{Tr}}\left(\rho_{\Lambda}^{\sf {canonical }}  \left( \Delta \tilde{H}_{i}^{H}\right)^{2}\right)$ can, typically, be readily bounded. In the semiclassical limit (similar to associated exact quantum calculations), similar to \cite{boundary-bounds1} (and \cite{boundary-bounds2}), we may choose all external fields $\{\phi_{j}\}$ (or phase space degrees of freedom) not associated with the support of $\tilde{H}_{i}^{H}$ to assume fixed values $\phi_{j} = \overline{\phi}_{j}$ that minimize the variance of $\tilde{H}_{i}^{H}$ when the latter is computed with a probability density $ \propto e^{-\beta H_{\Lambda}}|_{\overline{\phi}}$ with those fixed external values. This calculation yields the classical variance of $\tilde{H}_{i}^{H}$ computed within a local whose Hamiltonian is system given by $H_{\Lambda}$ with all the said fields $\phi_{j}$ not in the domain of support of the local Hamiltonian $\tilde{H}_{i}^{H}$ set equal to the constant values $\overline{\phi}_{j}$. 

We now briefly comment on the diametrically opposite limit- that of vanishing temperature. General systems (including toy textbook type models, e.g., particles in a box, the Hydrogen atom, harmonic oscillators)
typically display zero point quantum fluctuations. These fluctuations does not imply, however, that the local velocities and other general time derivatives of observables 
must be finite at zero temperature. When the system obtains its ground state energy, the density matrix becomes the canonical one associated with zero temperature (in this case, a projection to the ground state manifold) and all expectation values are stationary (and their time derivatives vanish).

\section{Quasiparticle Fermi Systems}
\label{Fermigas}
In this Appendix, we regress to a simple realization of these inequalities discussed in Section \ref{sec:lowT} -- that of a limiting case of quasiparticle systems of decoupled fermions,
\begin{equation}
\label{e-free}
H_{\Lambda}=\sum_{i} \frac{\left({\bf p}_{i}^{H}\right)^{2}}{2 m^*} \equiv \sum_{i} \tilde{H}_{i}^{H}.
\end{equation}
In Fermi gases (and liquids), a hybrid wave-vector (${\bf k}$) and spin $(\sigma$) index will assume the role of the generic index $i$ in labeling the uncorrelated Hamiltonians in Eq. (\ref{e-free}). In what follows, for the benefit of readers from other fields, we briefly review well known rudiments and then contrast our results of Section \ref{sec:lowT} with known low temperature behaviors. Although, in solids, electrons may strongly interact, as is underscored by Fermi liquid theory, in many instances, up to unimportant corrections, the system may be understood as that of non-interacting quasiparticles with renormalized parameters that is adiabatically connected to the Fermi gas of non-interacting electrons. The Sommerfeld expansion of the Fermi gas remains applicable. In these liquids, various quantities such as the mass, electron spectral weight, Land\'e  g factor, and the compressibility may all be renormalized. In particular, the mass $m^*$ in Eq. (\ref{e-free}) is the effective electron mass which, in ``heavy fermion'' compounds, can be several orders of magnitude larger than the bare electronic mass \cite{NFL}. Such Fermi liquid behavior is prevalent \cite{NFL,AGD,LL,coleman}. In solids, the electronic dispersion is periodic and more complicated functions of the momentum appear. However, with the exception of graphene and other ``Dirac materials'' \cite{Dirac_Sasha}, the electronic dispersion typically tends to the above free electron form of Eq. (\ref{e-free}) for small wave-vectors $\vec{k}$ 
 \cite{AMbook,GYbook,stevensimons}. For the Fermi gas Hamiltonian of Eq. (\ref{e-free}), for a fixed spin polarization 
$\sigma$, the energy associated with each such sector of wave-vector ${\bf k}$ 
is $\epsilon_{k} n_{k}$ with $n_{k}$ the occupancy of the state of wave-vector ${\bf k}$ and corresponding single particle energy 
\begin{eqnarray}
\label{ek-free}
\epsilon_{{\bf k}}=\frac{\hbar^{2} {\bf k}^{2}}{2 m^*}.
\end{eqnarray}
The occupancy of a given single particle state of wave-vector ${\bf k}$ follows a binomial distribution with probabilities $f_{\epsilon}$ and $(1-f_{\epsilon})$ and associated variance
\begin{equation}
\sigma_{\tilde{H}_{k}^{H}}^{2}=\epsilon_{k}^{2} f_{\epsilon_{k}} \left(1-f_{\epsilon_{k}} \right),
\end{equation}
with $f_{\epsilon_{k}}=\frac{1}{1+e^{\beta(\epsilon_{k}-\mu)}}$ being the Fermi function. 
In the above briefly noted Fermi liquids, there are additional quadratic terms augmenting the linear (in $n_{k}$) energy contributions $\epsilon_{k} n_{k}$ of the Fermi gas.

Given the character of Fermi gases, the relations of Section \ref{decoupled} and, in particular, those just discussed in Section \ref{sec:lowT}, apply since the local (in momentum space) Hamiltonians $\{\tilde{H}_{i}^{H}\}$ are uncorrelated. Enforcing the total number and energy constraints via the chemical potential and temperature, each sector of fixed wave-vector ${\bf k}$ is decoupled from the others. Since the single state number operator is positive semi-definite and of unit norm and, for fixed spin polarization $\sigma$, the variance 
\begin{equation}
\sigma_{n_{k}}^{2} \leq 1,
\end{equation}
Eq. (\ref{ref:decoupled-final}) implies the inequality
\begin{eqnarray}
\label{eq:derivative-number}
\frac{4}{\hbar^{2}} k_{B} T^{2} \frac{C_{v}^{(\Lambda)}}{N_{\Lambda}} \geq \frac{1}{N_{\Lambda}} \sum_{{\bf k}}\left|\left\langle\frac{d n_{k}}{d t}\right\rangle\right|^{2}. 
\end{eqnarray}
In the Fermi gas, for temperature far smaller than the Fermi temperature, $T \ll T_{F}$, the electronic (or other fermion) 
heat capacity $C_{v}^{e l}=\gamma T$ with a constant $\gamma$  \cite{NFL,AGD,LL,coleman} and thus
\begin{equation}
\label{electron0cv}
 \quad C_{v}=\mathcal{O}(T).
\end{equation}
In solids, at low temperatures, the linear electronic heat capacity of Eq. (\ref{electron0cv}) dominates over other (phonon, spin, etc.) excitations \cite{AMbook}. We emphasize that in Eq. (\ref{eq:derivative-number}), $C_{v}^{(\Lambda)}$ denotes the global thermodynamic heat capacity. Following the above discussion and the results of Section \ref{sec:lowT}, we see that in the $T \ll T_{F}$ limit, the relaxation rates may satisfy Eq. (\ref{eq:lowTbound}). Consistent with Fermi liquid theory, the resistivity of conventional metals scales as
\begin{equation}
\label{Fermires}
\rho=\mathcal{O}\left(T^{2}\right).
\end{equation}
Eq. (\ref{Fermires}) adheres to Eq. (\ref{eq:lowTbound}) in the $T \rightarrow 0$ limit. The above schematic analysis may be replicated for general dispersions $\epsilon_{{\bf k}}$ other than that of Eq. (\ref{ek-free}). Below the Bloch-Gruineisen temperature \cite{B1,G1}, electron-phonon interactions may lead to a more rapid decrease of the resistivity with temperature (that of $\rho=\mathcal{O}\left(T^{5}\right)$) for which the bound of Eq. (\ref{eq:lowTbound}) is, once again, trivially satisfied (even more so than for Fermi liquid theory). Meaningful non-trivial dynamics (and response functions) generally require the introduction of interactions with the general bounds similar to those in the discussion following Eq. (\ref{econducivity}) with the very qualitative estimate of Eq. (\ref{rho-estimate}). 

\section{High temperature velocity bounds in equilibrium systems}
\label{velocity-bound-App}

In this Appendix, we explicitly derive the velocity bounds alluded to in Section \ref{Expect-v} and arrive at Eq. (\ref{srih}). The rather trivial calculation in this Appendix will involve an assumption. Eq. (\ref{vbound12}) implies that
\begin{equation}
\label{eq:eqv}
\frac{1}{N_{\Lambda}} \sum_{i=1}^{N_{\Lambda}}\left\langle v_{i \ell}\right\rangle^{2}\left(\sigma_{r_{i \ell}^{H}}\right)^{-2} \leq 2\left(\frac{k_{B} T}{\hbar}\right)^{2}
\end{equation}
Writing the single particle expectation values as a sum of their system wide global average (Eq. (\ref{O:avg})) and fluctuations about these, $\left\langle v_{i \ell}\right\rangle^{2}=\overline{v^{2}}+\delta\left(\left\langle v_{i \ell}\right\rangle^{2}\right)$ and $\left(\sigma_{r_{i l}^{H}}\right)^{-2}=\left(\overline{\sigma^{-2}}\right)+\delta\left(\sigma_{r_{i l}^{H}}^{-2}\right)$,
we have from Eq. (\ref{eq:eqv}) that
 \begin{eqnarray}
&& \frac{1}{N_{\Lambda}} \sum_{i=1}^{N_{\Lambda}}\left\langle v_{i \ell}\right\rangle^{2}\left(\sigma_{r_{i \ell}^{H}}\right)^{-2} \nonumber
\\ && =\frac{1}{N_{\Lambda}} \sum_{i=1}^{N_{\Lambda}}\left(\overline{v^{2}}+\delta\left(\left\langle v_{i \ell}\right\rangle^{2}\right)\right)\left(\left(\overline{\sigma^{-2}}\right)+\delta\left(\sigma_{r_{i l}^{H}}^{-2}\right)\right)  \nonumber
\\ && \leq 2\left(\frac{k_{B} T}{\hbar}\right)^{2}.
\end{eqnarray}
Since $\overline{\mathcal{O}}$ denotes a global average of $\left\{\left\langle\mathcal{O}_{i}\right\rangle\right\}$
 i.e., $\overline{\mathcal{O}} \equiv \frac{1}{N_{\Lambda}} \sum_{i=1}^{N_{\Lambda}}\left\langle\mathcal{O}_{i}\right\rangle$, 
  the global sum of the local fluctuations from the global average identically vanishes,
  $\sum_{i=1}^{N_{\Lambda}} \delta\left(\left\langle\mathcal{O}_{i}\right\rangle\right)=0$. Thus, trivially, $\sum_{i=1}^{N_{\Lambda}} \delta\left(\left\langle v_{i \ell}\right\rangle^{2}\right)=\sum_{i=1}^{N_{\Lambda}} \delta\left(\sigma_{r_{i l}^{H}}^{-2}\right)=0$, or $\frac{1}{N_{\Lambda}} \sum_{i=1}^{N_{\Lambda}}\left\langle v_{i \ell}\right\rangle^{2}\left(\sigma_{r_{i l}^{H}}^{-2}\right)=\overline{v^{2}} \overline{\left(\sigma^{-2}\right)}+\frac{1}{N_{\Lambda}} \sum_{i=1}^{N_{\Lambda}}\left(\delta\left(\left\langle v_{i \ell}\right\rangle^{2}\right)\right)\left(\delta\left(\sigma_{r_{i l}^{H}}^{-2}\right)\right) \leq 2\left(\frac{k_{B} T}{\hbar}\right)^{2}$. A higher local square momentum (or squared velocity deviation $\left.\delta\left(\left\langle v_{i \ell}^{2}\right\rangle\right)\right)$ relative to the global average,
is associated with a more spatially localized state (and thus larger $\left.\delta\left(\sigma_{r_{i l}^{H}}^{-2}\right)\right)$. Thus, the average local correlation of $\frac{1}{N_{\Lambda}} \sum_{i=1}^{N_{\Lambda}}\left(\delta\left(\left\langle v_{i \ell}\right\rangle^{2}\right)\right)\left(\delta\left(\sigma_{r_{i j}^{H}}^{-2}\right)\right)$
is expected to be positive. If that is the case, $\overline{v^{2}} \leq \frac{2}{\left(\overline{\sigma^{-2}}\right)}\left(\frac{k_{B} T}{\hbar}\right)^{2}$. 
By the Arithmetic Mean - Harmonic Mean Inequality, $\overline{\sigma_{r_{i l}^{H}}^{-2}} \geq(\bar{\sigma}_{r_{i l}^{H}})^{-2}$, and thus
\begin{equation}
\label{eq:speed_limit}
    {\overline{v_{i l}^{2}}^{1/2}} \leq \frac{k_{B} T \bar{\sigma}_{r_{i l}^{H}} \sqrt{2}}{\hbar}.
    \end{equation}
   In classical thermal equilibrium, for any potential energy $V$, the average (Eq. (\ref{kTm})) $ \overline{v_{il}^2}^{1/2}= \sqrt{\frac{k_{B} T}{m}}$. Plugging this average velocity into Eq. (\ref{eq:speed_limit}) yields Eq. (\ref{srih}).

\section{Bounds in Reflection Positive and other systems}
\label{sec:RP}
Reflection Positivity is satisfied by general Euclidean field theories (where it may be viewed as an imaginary time analog of unitarity) and lattice theories \cite{Reflection-Positivity1,Reflection-Positivity2-1,Reflection-Positivity2-2,Reflection-Positivity3}. As we will discuss in this Appendix, whenever Reflection Positivity is present our bounds become more potent. In particular, as noted in Section \ref{decoupled} and established here, in Reflection Positive systems, the variance of the local Hamiltonian $\tilde{H}_{i'}^H$ that drives the dynamics of observables $Q_{i'}^{H}$, can be directly related to the true thermodynamic specific heat associated with the global system whose dynamics is governed by $H_{\Lambda}$. This, in a sense, will generalize bounds of the type that we found in Section \ref{decoupled} also for interacting theories. Systems that are ``Reflection Positive'' satisfy the following inequality
\begin{equation}
\label{eq:RP+}
\left\langle f \theta_{P} f\right\rangle \geq 0
\end{equation}
for general real functions $f$. The reflection operator $\theta_{P}$ is defined as follows. For any $f$, 
\begin{equation}
\label{RP++}
\theta_{P} f(\{\Omega({{\bf x})}\}) \equiv f\left(\left\{\Omega\left(\theta_{P} {{\bf x}}\right)\right\}\right),
\end{equation}
with $\Omega$ general operators labelled by the positions ${{\bf x}}$
and where $\left(\theta_{P} {\bf x}\right)$ marks the mirror image
of point ${{\bf x}}$ when reflected in a plane $P$. The positivity of Eq. (\ref{eq:RP+}) generally holds for Hamiltonians (or actions) of the form
\begin{equation}
\label{HFPF}
H_{\Lambda}=F+\theta_{P} F.
\end{equation}
In such systems, a Boltzmann probability density (or Euclidean path integral measure) becomes a reflection symmetric form product $\rho_{\Lambda}^{\sf {canonical }}=\frac{e^{-\beta H}}{Z}=\frac{e^{-\beta \bar{F}}}{\sqrt{Z}} \theta_{P}\left(\frac{e^{-\beta F}}{\sqrt{Z}}\right)$ that ensures positivity. For any two points $i'$ and $j'$, we may choose the plane $P$ to be the perpendicular bisector of the segment $(i' j')$. For such a perpendicular bisecting plane $P$,  
\begin{eqnarray}
\label{i'j'eq}
i'= \theta_{P} j'.
\end{eqnarray} 
When applied to the general Hamiltonians of Eq. (\ref{eq:decoupled+sum}, \ref{HFPF}), in Reflection Positive systems, the inequality of Eq. (\ref{eq:RP+})) becomes
\begin{equation}
\label{positive-connect}
{\sf{Tr}}\left(\rho_{\Lambda}^{\sf {canonical }}\left(\Delta \tilde{H}_{i^{\prime}}^{H}\right)\left(\Delta \tilde{H}_{j^{\prime}}^{H}\right)\right) \geq 0.
\end{equation}
This inequality is saturated (i.e., becomes an equality) when the Hamiltonian of Eq. (\ref{eq:decoupled+sum}) is a sum of decoupled terms (the situation discussed in Section \ref{decoupled}). Physically, Eq. (\ref{positive-connect}) maintains that the connected correlation function (or covariance) between the local Hamiltonians $\tilde{H}_{i^{\prime}}^{H}$ and $\tilde{H}_{j^{\prime}}^{H}$ associated with the two reflection related points $i'$ and $j'$ (Eq. (\ref{i'j'eq})) is positive semi-definite. To give a flavor of what this broadly means, we  very briefly discuss standard critical phenomena. In typical critical systems away from their transition points, connected correlation functions (such as that of Eq. (\ref{positive-connect})) may exponentially decay to zero, with a finite correlation length, as the spatial separation $|i^{\prime} - j^{\prime}|$ tends to infinity and decay algebraically in the distance $|i^{\prime} - j^{\prime}|$ at the critical transition point. This is indeed what transpires in uniform short range ferromagnets (that may be trivially expressed in the form of Eq. (\ref{HFPF}). We stress that Eq. (\ref{positive-connect}) does not imply finite positive connected correlations for all $|i^{\prime} - j^{\prime}|$. Rather, this inequality merely asserts that the connected correlations cannot be negative. The minimal Hamiltonians $ \tilde{H}_{i^{\prime}}$, in a partition of the form of Eq. (\ref{eq:decoupled+sum}), for which Eq. (\ref{positive-connect}) will be satisfied may depend on the temperature defining $\rho_{\Lambda}^{\sf {canonical }}$.

On a lattice, if the local Hamiltonian $\tilde{H}^{H}_{i'}$ is defined on one (or, often, also several) lattice site(s) then we may continue to trivially choose, as just discussed, the plane $P$ to be the perpendicular bisector between  
$i'$ and $j'$, e.g., the perpendicular bisector between the center of plaquettes for $\tilde{H}^{H}_{i'}$ that are certain plaquette operators. However, generally, we cannot make all local Hamiltonians $\tilde{H}^{H}_{i'}$ related to each other by reflections in such perpendicular bisecting planes $P$. In the continuum limit, such planar reflections may more readily link the local Hamiltonian (or Euclidean space Lagrangian) densities. As in 
Eq. (\ref{eq:decoupled+sum}), we may express the global Hamiltonian $H_{\Lambda}$ in terms of local Hamiltonians $ \tilde{H}^{H}_{i^{\prime}}$ that drive the dynamics of the observables $Q_{i'}^{H}$.
Only a single local Hamiltonian $\tilde{H}^{H}_{i^{\prime}}$ in the sum of Eq. (\ref{eq:decoupled+sum}) does not commute with the local quantity $Q^{H}_{i^{\prime}}$. Similar to our analysis of non-interacting theories (Section \ref{decoupled}), the number of these local Hamiltonians $N_{\Lambda^{\prime}}$ may be smaller than the total number of sites or particles $N_{\Lambda}$ in the system. This is so since $\tilde{H}^H_{i^{\prime}}$ may need to be a multi-site (or multi-particle) operator if $Q^H_{i^{\prime}}$ features dynamics generated by more than one local term in $H_{\Lambda}$. 

With all of the above stated caveats for lattice theories, we will take Eqs. (\ref{i'j'eq},\ref{positive-connect}) to constitute our working definition of the particular Reflection Positive systems under consideration in this Appendix. That is, we consider Reflection Positive theories with Hamiltonians of the form of Eq. (\ref{eq:decoupled+sum}) where any pair of sites $i'$ and $j'$ are related by a suitable reflection (Eq. (\ref{i'j'eq})). Applying Eq. (\ref{a-central-1}) to the decomposition of Eq. (\ref{eq:decoupled+sum}),
\begin{eqnarray}
\label{RPPPPP}
\frac{1}{N_{\Lambda}^{\prime}} \sum_{i^{\prime}=1}^{N_{\Lambda}^{\prime}} {\sf{Tr}}\left(\rho_{\Lambda}^{\sf {canonical }}\left(\Delta \tilde{H}_{i^{\prime}}^{H}(t)\right)^{2}\right) \nonumber
\\ \geq \frac{\hbar^{2}}{4 N_{\Lambda}^{\prime}} \sum_{i^{\prime}=1}^{N_{\Lambda}^{\prime}} \frac{\left({\sf{Tr}}\left(\rho_{\Lambda} \frac{d Q_{i}^{\prime H}}{d t}\right)\right)^{2}}{{\sf{Tr}}\left(\rho_{\Lambda}\left(\Delta Q_{i^{\prime}}^{H}(t)\right)^{2}\right)}.
\end{eqnarray}
From Reflection Positivity, the inequality of Eq. (\ref{positive-connect}) applies to the individual local Hamiltonians in Eq. (\ref{RPPPPP}). Combined with translational invariance, this implies that 
\begin{eqnarray}
\label{RP-final}
k_{B} T^{2} C_{v}^{(\Lambda)} ={\sf{Tr}}\left(\rho_{\Lambda}^{\sf {canonical }}\left(\Delta H_{\Lambda}\right)^{2}\right) \nonumber
\\  ={\sf{Tr}}\left(\rho_{\Lambda}^{\sf {canonical }}\left(\sum_{i^{\prime}=1}^{N'_{\Lambda}}\left(\Delta \tilde{H}_{i^{\prime}}^{H}\right)\right)^{2}\right) \nonumber
\\  \geq N_{\Lambda}^{\prime}\left({\sf{Tr}}\left(\rho_{\Lambda}^{\sf {canonical }}\left(\Delta \tilde{H}_{i^{\prime}}^{H}\right)\right)^{2}\right),
\end{eqnarray}
where $\Delta H_{\Lambda} \equiv H_{\Lambda} - Tr(\rho_{\Lambda}^{\sf {canonical }} H_{\Lambda})$. In Eq. (\ref{RP-final}), $C_{v}^{(\Lambda)}$ is the constant volume heat capacity of the entire system $\Lambda$.
We caution that the Reflection Positivity property of Eq. (\ref{positive-connect}) and thus the inequality of 
Eq. (\ref{RP-final}) cannot hold universally. For instance, at zero temperature the global system $\Lambda$ lies in its ground state- a state that need not be an eigenstate of any of the local Hamiltonians $\tilde{H}^{H}_{i}$ (and thus these Hamiltonians will have a finite variance). When Eq. (\ref{RP-final}) is satisfied, repeating the derivation of Section \ref{bound-1t},
we arrive at an analog of Eq. (\ref{cv1}),
\begin{equation}
\label{RP:L1}
k_{B} T^{2} c_{v} \geq \frac{\hbar^{2}}{4} \overline{\mathcal{O}}.
\end{equation}
In Eq. (\ref{RP:L1}), $c_{v} \equiv \frac{C_{v}^{(\Lambda)}}{N_{\Lambda}^{\prime}}$ {\it is the thermodynamic specific heat} (not the effective local heat capacity of Eq. (\ref{cv1})). As in Section \ref{bound-1t}, in Eq. (\ref{RP:L1}), $\overline{\mathcal{O}}=\frac{1}{N_{\Lambda}^{\prime}} \sum_{i^{\prime}=1}^{N_{\Lambda}^{\prime}} \mathcal{O}_{i^{\prime}}$ with $\mathcal{O}_{i^{\prime}} \equiv \frac{\left\langle\frac{d Q_{i'}^{H}}{d t}\right\rangle^{2}}{\left(\sigma_{Q_{i^{\prime}}}^{H}(t)\right)^{2}}$. Eqs. (\ref{RP-final}, \ref{RP:L1}) are the central results of this Appendix. These inequalities illustrate,  how, similar to the decoupled systems of Section \ref{decoupled} (and Eq. (\ref{ref:decoupled-final}) in particular), in these interacting theories, the rates of change of general local observables $Q_{i}^{H}$ are universally bounded by the specific heat. If the specific heat $c_{v}=\mathcal{O}\left(k_{B}\right)$ then the relaxation rates for any local operator will be bounded by $\mathcal{O}\left(\frac{k_{B} T}{\hbar}\right)$. 

In ferromagnetic systems (of arbitrary interaction range and also in the absence of uniform couplings), it follows from the second Griffiths inequality \cite{Reflection-Positivity1,Grif1,Grif2,Grif3,Grif4}, that Eq. (\ref{positive-connect}) is satisfied. This, in turn, leads to Eq. (\ref{RP-final}) with conclusions identical to those of the Reflection Positive case. Of course, the dynamics of ferromagnets may be rather trivial. Ferromagnets form a particular subset of frustration free systems for which the ground states of $H_{\Lambda}$ are also ground states of all of the local Hamiltonians $\{\tilde{H}_{i^{\prime}}^{H}\}$.

We finally turn to broader nontrivial physical instances in which counterparts of Eq. (\ref{positive-connect}) appear- i.e., systems in which all connected correlation functions of the local Hamiltonians $\{\tilde{H}_{i'}^H\}$ forming the total Hamiltonian $H_{\Lambda}$ (Eq. (\ref{eq:decoupled+sum}))) are positive semi-definite. Here, the connected correlator of Eq. (\ref{positive-connect}) need not be between Hamiltonians that are related to one another by the reflections of Eq. (\ref{i'j'eq}). Indeed, as it may be readily verified, if the global Hamiltonian may be partitioned (Eq. (\ref{eq:decoupled+sum})) into local Hamiltonians such that Eq. (\ref{positive-connect}) is satisfied for all local Hamiltonian pairs $\tilde{H}_{i^{\prime}}^{H}$ and $\tilde{H}_{j^{\prime}}^{H}$ then Eq. (\ref{RP:L1}) will follow. Thus, in all such theories with positive semi-definite connected correlation functions, we obtain a bound involving the thermodynamic specific heat (instead of the effective local heat capacity $C_{v, i}$ of Eq. (\ref{cv2})). For these more general cases, the bound of Eq. (\ref{RP:L1}) involving the thermodynamic specific heat still applies. The consequences of Eq. (\ref{RP:L1}) are manifold. In particular, the low temperature bounds discussed in Section \ref{sec:lowT}) (specifically Eq. (\ref{eq:lowTbound})) hold universally.

\section{High order gradient inequalities}
\label{higher_grad}

\subsection{Spatial gradients}
\label{high:spatgrad}

Returning to the uncertainty inequality of Eq. (\ref{eq:AB}), we next set, for general functions $f$, the two relevant operators to be $A=p_{i \ell}$ and $B=\frac{\partial^{n} f\left(\left\{r_{j, \ell^{\prime}}, p_{j, \ell^{\prime}}\right\}\right)}{\partial r_{i \ell}^{n}}$. For an arbitrary natural number $n$, Eq. (\ref{eq:AB}) then implies that
\begin{equation} 
\label{spathigh}
\sigma_{p_{i \ell}}^{2} \sigma_{\frac{\partial^n f}{\partial r_{i \ell}^{n}}}^{2} \geq \frac{\hbar^{2}}{4}\left|\left\langle\frac{\partial^{n+1} f}{\partial r_{i \ell}^{n+1}}\right\rangle\right|^{2}.
\end{equation}

For the many-body Hamiltonian of Eq. (\ref{eq:HLAMBDA}), at sufficiently high temperatures $T$ where the classical equipartition theorem holds, Eq. (\ref{spathigh}) then implies that
$m k_{B} T={\sf{Tr}}\left(\rho_{\Lambda}^{\sf {canonical }} p_{i \ell}^{2}\right)=\frac{1}{N_{\Lambda}} \sum_{i=1}^{N_{\Lambda}} {\sf{Tr}}\left(\rho_{\Lambda} p_{i \ell}^{2}\right) \geq \frac{1}{N_{\Lambda}} \sum_{i=1}^{N_{\Lambda}} \sigma_{p_{i \ell}}^{2} \geq \frac{\hbar^{2}}{4 N_{\Lambda}} \sum_{i=1}^{N_{\Lambda}}\left|\frac{\left\langle\frac{\partial^{n+1} f}{\partial r_{i \ell}^{n}}\right\rangle}{\sigma_{\frac{\partial^{n} f}{\partial r_{i \ell}^{n}}}}\right|^{2}$. Thus, trivially extending the $n=0$ case of Eq. (\ref{alwaysgrad}), 
\begin{eqnarray}
\label{highertrveq29}
&& \frac{1}{N_{\Lambda}} \sum_{i=1}^{N_{\Lambda}} \frac{\left\langle\frac{\partial^{n+1} {f}}{\partial r_{i \ell}^{n+1}}\right\rangle^{2}}{\left\langle\left(\frac{\partial^{u} f}{\partial r_{i \ell}^{n}}\right)^{2}\right\rangle} \leq \frac{1}{N_{\Lambda}} \sum_{i=1}^{N_{\Lambda}}\left|\frac{\left\langle\frac{\partial^{n+1} f}{\partial r_{i \ell}^{n+1}}\right\rangle}{\sigma_{\frac{\partial^{u} f}{\partial r_{i \ell}^{n}}}}\right|^{2} \nonumber
\\  && \leq \frac{4 m k_{B} T}{\hbar^{2}}=\frac{8 \pi}{\lambda_{T}^{2}}.
\end{eqnarray}

As told, here $f$ is a completely arbitrary differentiable function. It may, e.g., be the total mass or 
charge density of all particles or potentials generated by moving charges that are in equilibrium (Section \ref{sec:em}). In equilibrium, all such functions cannot have average relative gradients that exceed $\mathcal{O}\left(1 / \lambda_{T}\right)$. Inequality of Eq. (\ref{highertrveq29}) only holds at sufficiently high temperatures
where the classical equipartition theorem holds. 

We may similarly set in Eq. (\ref{eq:AB}) the operators $A=p_{i \ell}^{n}$ with integer $n$ and take $B$ to be a general phase space function. The relevant momenta variances in Eq. (\ref{eq:AB}) may be computed. When evaluated with the classical canonical probability density $\rho_{\Lambda}^{\sf{classical~canonical}}$ for the Hamiltonian of Eq. (\ref{eq:HLAMBDA}), the classical high temperature variances of $p_{i \ell}^{n}$ are, for odd $n$,
\begin{eqnarray} 
\label{odd-p-class}
\sigma_{p_{i \ell}^{n}}^{2}
=\left(2 m k_{B} T\right)^{n}(2 n-1) ! !,
\end{eqnarray}
and for even $n$, 
\begin{eqnarray}
\label{even-p-class}
\sigma_{p_{i \ell}^{n}}^{2}=
 \left(2 m k_{B} T\right)^{n}\left((2 n-1) ! !-((n-1) ! !)^{2}\right).
\end{eqnarray}

This leads us to recognize that 

$\bullet$ In thermal systems, spatial gradients cannot, typically, be larger than $\mathcal{O}\left(\frac{1}{\lambda_{T}}\right)$.

Thus, if, e.g., a lattice is in equilibrium then its lattice constant must, typically, be larger than the thermal de Broglie wavelength associated with the ionic mass $\mathcal{O}\left(\lambda_{T}\right)$. This is consistent with our earlier  bound of Eq. (\ref{eq:grad-bound=}). At lower temperatures, the variance of the squared momentum is, generally, different and is not set by the classical results of Eqs. (\ref{odd-p-class}, \ref{even-p-class}).

\subsection{High order time derivatives}
We next consider high order time derivatives. Following the same recipe as in Appendix \ref{high:spatgrad}, we substitute $A=\tilde{H}_{i}^{H}$ and $B=\frac{\partial^{n} Q_{i}^{H}}{\partial t^{n}}$ in Eq. (\ref{eq:AB}). 
with $\tilde{H}_{i}^{H}$ completely generating all of the dynamics of $Q_{i}^{H}$.

\begin{eqnarray}
&&  k_{B} T^{2} C_{v, i} \nonumber
\\  && \equiv {\sf{Tr}}\left(\rho_{\Lambda}^{\sf {canonical }}\left(\tilde{H}_{i}^{H}\right)^{2}\right)-\left({\sf{Tr}}\left(\rho_{\Lambda}^{\sf {canonical }}\left(\tilde{H}_{i}^{H}\right)\right)^{2} \right) \nonumber
\\  && =\frac{1}{N_{\Lambda}} \sum_{i=1}^{N_{\Lambda}}\left({\sf{Tr}}\left(\rho_{\Lambda}\left(\tilde{H}_{i}^{H}\right)^{2}\right)-\left({\sf{Tr}} \rho_{\Lambda} \tilde{H}_{i}^{H}\right)^{2}\right) \nonumber
\\ && \geq \frac{\hbar^{2}}{4 N_{\Lambda}} \sum_{i=1}^{N_{\Lambda}} \frac{\left({\sf{Tr}}\left(\rho_{\Lambda} \frac{\partial^{n+1} Q_{i}^{H}}{\partial t^{n+1}}\right)\right)^{2}}{{\sf{Tr}}\left(\rho_{\Lambda}\left(\frac{\partial^{n} Q_{i}^{H}}{\partial t^{n}}\right)^{2}\right)}.
\end{eqnarray}

For a local $\tilde{H}_{i}^{H},$ the high temperature variance is $\mathcal{O}\left(\left(k_{B} T\right)^{2}\right)$. Thus, similar to our earlier results,

$\bullet$ In thermal systems, at sufficiently high temperatures where classical equipartition holds, temporal rates of change typically cannot be faster than $\mathcal{O}\left(\frac{k_{B} T}{\hbar}\right)$. At lower temperatures, the variance of the squared momentum is, generally, different.

\section{Bounds on two-point correlators}
\label{sec:2-point}

\subsection{Time derivatives}
Our inequalities (for both temporal and spatial derivatives) may be extended to operators that are the average of local operators, i.e.,
\begin{equation}
\mathcal{Q}^H_{i} \rightarrow q^H_{\lambda} \equiv \frac{1}{N_{\lambda}} \sum_{i \in \lambda} \mathcal{Q}^H_{i},
\end{equation}
with $\lambda \subset \Lambda$ an arbitrary subvolume of $N_{\lambda}$ particles or sites. 
Since
\begin{equation} 
\label{spatGQQ}
\sigma_{q_{\lambda_{k}}^{H}}^{2}=\frac{1}{N_{\lambda}^{2}} \sum_{i, j \in \lambda_{k}}\left(\left\langle\mathcal{Q}_{i}^{H} \mathcal{Q}_{j}^{H}\right\rangle-\left\langle\mathcal{Q}_{i}^{H}\right\rangle\left\langle\mathcal{Q}_{j}^{H}\right\rangle\right),
\end{equation}
the uncertainty of $Q_{i}^{H}$ as computed with $\rho_{\Lambda}$ is related to an average of two-point correlators on the righthand side of Eq. (\ref{spatGQQ}) when evaluated with the same density matrix.
Thus, from Eq. (\ref{a-central-1}),
\begin{eqnarray}
\label{t-derivative-corr}
&& \frac{4 k_{B} T^{2} C_{v, \lambda}}{\hbar^{2}} \nonumber
\\ && \! \! \! \! \! \! \! \! \! \! \! \! \! \! \! \! \! \! \!  \geq \lim _{{\cal{T}} \rightarrow \infty} \frac{1}{{\cal{T}}} \int_{0}^{{\cal{T}}} d t \frac{\left|\frac{1}{N_{\lambda}} \sum_{i \in \lambda}\left\langle d \mathcal{Q}^H_{i} / d t\right\rangle\right|^{2}}{\frac{1}{N_{\lambda}^{2}} \sum_{i, j 
\in \lambda}\left(\left\langle\mathcal{Q}^H_{i} \mathcal{Q}^H_{j}\right\rangle-\left\langle\mathcal{Q}^H_{i}\right\rangle\left\langle\mathcal{Q}^H_{j}\right\rangle\right)}.
\end{eqnarray}
In Eq. (\ref{t-derivative-corr}), $\langle\cdot\rangle$ denotes the average computed with $\rho_{\Lambda}$ and, similar to Eq. (\ref{cv2}), $k_{B} T^{2} C_{v, \lambda} \equiv {\sf{Tr}}\left(\rho_{\Lambda}^{\sf {canonical }}\left(\Delta \tilde{H}_{\lambda}^{H}\right)^{2}\right)
$ where $\tilde{H}_{\lambda}^{H}$ now contains all terms in the full Hamiltonian $H_{\Lambda}$ that provide the dynamics of $q_{\lambda}^{H}$, i.e.,
\begin{eqnarray}
\left[\tilde{H}_{\lambda}^{H}, q_{\lambda}^{H}\right]=\left[H_{\Lambda}, q_{\lambda}^{H}\right].
\end{eqnarray}

\subsection{Spatial derivatives}
Let us label the center of mass of each of the volume $\{\lambda_{k}\}$ over which we average the correlation functions by $\{x^H_{k}\}$. From Eq. (\ref{spatGQQ}), for spatial derivatives,
\begin{eqnarray} 
&& \frac{8 \pi N_{\lambda}}{\lambda_{T}^{2}}=\frac{4 m N_{\lambda} k_{B} T}{\hbar^{2}} \nonumber
\\ \!\!\!\!\!\!\!\!\!\!\!\!\!\!  \geq \lim _{{\cal{T}} \rightarrow \infty} && \frac{1}{{\cal{T}}} \int_{0}^{{\cal{T}}}  \frac{dt~\left|\left\langle\partial q_{\lambda_{k}}^{H} / \partial x_{k}\right\rangle\right|^{2}}{\frac{1}{N_{\lambda}^{2}} \sum_{i, j \in \lambda_{k}}\left\langle\mathcal{Q}_{i}^{H} \mathcal{Q}_{j}^{H}\right\rangle-\left\langle\mathcal{Q}_{i}^{H}\right\rangle\left\langle\mathcal{Q}_{j}^{H}\right\rangle}.
\end{eqnarray}
The origin of the factor of $N_{\lambda}$ on the lefthand side is that the mass of $\lambda_{k}$ (whose center of mass is $x^H_{k}$) is $N_{\lambda} 
m$. 

\section{Bounds on correlators of spatio-temporal gradients of general operators}
\label{sec:generalcorrelatorgrad}
In the semiclassical limit, we may bound general correlators of the form 
\begin{eqnarray}
\label{eq:gradb}
\Big\langle \frac{\partial A^H}{\partial r^{H}_{\mu}} ~ \frac{\partial B^H}{\partial r^H_{\nu}}~ \frac{\partial C^H}{\partial r^H_{\kappa}}\cdots {\cal{Y}} \Big\rangle,
\end{eqnarray} 
with $A^H, B^H, C^H, \cdots$ general Heisenberg picture operators (that may also be identical to one another or, e.g., correspond to local operators that are separated relative to one another in space (as well as in time (in a manner similar to our discussion of the autocorrelator $G_{\dot{Y}}$ of Eq. (\ref{eq:GY})))) and ${\cal{Y}}$ is a general operator. In what follows, we will assume that the above correlator is a product of $n$ temporal or spatial gradients. In Eq. (\ref{eq:gradb}), $r^H_{\mu =0}$ labels time while $r^{H}_{\mu=\ell = 1,2, \ldots, d}$ is a Cartesian component of the position coordinate on which the operator $A^{H}$ depends. To bound Eq. (\ref{eq:gradb}), we simply note that we may replace (i) temporal derivatives such as $\frac{\partial A^H}{\partial r^{H}_{\mu}} = \frac{i}{\hbar} [\tilde{H}^{H}_{A}, A^H]$ with $\tilde{H}^H_{A} \subset H_{\Lambda}$ denoting the local Hamiltonian providing the dynamics of $A^{H}$ and (ii) spatial derivatives $\frac{\partial A^H}{\partial r_{\ell}^{H}} = - \frac{i}{\hbar}[p^{H}_{\ell}, A^H]$. Whenever the global system wide average of Eq. (\ref{eq:gradb}) or its long time average corresponds to a canonical ensemble average (as discussed in Section \ref{setup} for a single operator time derivative) then we may readily derive a bound. Replacing, in this fashion, the gradients in Eq. (\ref{eq:gradb}) by commutators and replicating the semiclassical considerations of Section \ref{high-moment-derivative}, an upper bound on absolute value of the correlator Eq. (\ref{eq:gradb}) is given by
\begin{eqnarray}
\label{aver:ABC}
\Big(\frac{2}{\hbar}\Big)^n  {\sf Tr}\Big(\rho_{\Lambda}^{\sf classical~canonical} \Big|( \Delta A^H) (\Delta B^H) (\Delta C^H) \nonumber
\\ \times p^{H}_{\mu} p^{H}_{\nu} p^{H}_{\kappa} \cdots  {\cal{Y}}\Big| \Big). 
\end{eqnarray}
For semiclassical particle systems, the expectation value of Eq. (\ref{aver:ABC}) becomes a phase space average.

\section{Simple examples}
\label{sec:toy}
In this Appendix, we carry out pedagogical calculations that explicitly apply the bound of Eq. (\ref{central1}) to two toy models. 

\subsection{The harmonic solid}
\label{sec:harmonic}
Our first examined model is that of the harmonic solid. As is well known, harmonic oscillator ground states saturates the usual uncertainty relations (i.e., the uncertainty bounds turn to equalities). In this Appendix, we analyze our temporal inequalities for finite temperature realizations of this system and its trivial many body extension where the normal modes decouple. As emphasized throughout this work, the global systems $\Lambda$ that we consider (including the harmonic solid that we study now) are open and in contact with a thermal bath. The canonical density matrix of the system $\Lambda$ is then a product of the canonical density matrices for each of the individual eigenmodes. In any single harmonic mode (phonon), we will label the normal mode coordinate and associated conjugate momentum by $x^H$ and $p^H$ (which will be implicitly assumed to correspond to the mode of fixed frequency $\omega$). If we focus only on an individual normal mode displacement $x^{H}$ (which for consistency of notation with the earlier parts of the paper we label $Q^H_i$ where now $i$ is not the particle index but rather labels the mode of fixed frequency $\omega$) then the relevant part
of the Hamiltonian $H_{\Lambda}$ endowing $x^H$ with dynamics is
\begin{equation}
\tilde{H}_{i}^{H}=\frac{\left(p^{H}\right)^{2}}{2 m}.
\end{equation}
By contrast, for this single decoupled harmonic mode in the open global system $\Lambda$, the full Hamiltonian
($\tilde{H}_i^{H} \subset H_{\omega} \subset H_{\Lambda}$) governing its dynamics is
\begin{equation}
H_{\omega}=\frac{p^{2}}{2 m}+\frac{1}{2} m \omega^{2} x^{2}.
\end{equation}
In the $n$-th eigenstate of the oscillator, the expectation value of $p^4$ is
%

%
%
\begin{equation}
\left\langle n\left|p^{4}\right| n\right\rangle=\left(\frac{m \hbar \omega}{2}\right)^{2}\left[6 n^{2}+6 n+3\right].
\end{equation}
The variance of $\tilde{H}_{i}^{H}$ computed with the canonical density matrix $\rho^{\sf {canonical}}$ of Eqs. (\ref{rcanonical},\ref{eq:ZL}) is thus
\begin{eqnarray}
\label{long-osc1} 
&& \sigma_{\frac{(p^{H})^{2}}{2 m}}^{2}=  {\sf{Tr}}\left(\rho^{\sf canonical}_{\Lambda}\left(\frac{(p^{H})^{2}}{2 m}\right)^{2}\right)-{\sf{Tr}}\left(\rho^{\sf canonical}_{\Lambda} \frac{p_{H}^{2}}{2 m}\right)^{2} \nonumber
\\  \! \! \! \! \! \! \! \! \! \! \! \! \! \! \! \! \! \! \! \! \! && =\frac{1}{4 m^{2}}\left(\frac{m \hbar \omega}{2}\right)^{2}  \Big[6 ~{\sf{Tr}}\left(\rho^{\sf canonical}_{\Lambda} n^{2}\right)+6 ~{\sf{Tr}}\left(\rho^{\sf canonical}_{\Lambda} n \right)\nonumber
\\  && ~~~~~~~~~~~~~~~~~~~~~~~ +3   -\left(2\left({\sf{Tr}}\left(\rho^{\sf canonical}_{\Lambda} n\right)\right)+1\right)^{2} \Big].
\end{eqnarray}
 
Before proceeding, we very briefly review standard results that we will employ. The partition function of the harmonic oscillator,
 \begin{equation}
 \label{Zosc}
Z_{\Lambda}=\sum_{\mathrm{n}=0}^{\infty} e^{-\beta \hbar \omega\left(\mathrm{n}+\frac{1}{2}\right)}=\frac{e^{-\beta \hbar \omega / 2}}{1-e^{-\beta \hbar \omega}},
\end{equation}
and the well known finite temperature average the number operator appearing in Eq. (\ref{long-osc1}) is ${\sf{Tr}}\left(\rho_{\Lambda}^{\sf {canonical }} n \right)=\frac{1}{e^{\beta \hbar \omega}-1}$ (i.e., the Bose function for vanishing chemical potential or the Planck type distribution). The variance of the number operator ${\sf{Tr}}\left(\rho_{\Lambda}^{\sf {canonical }}\left(n^{2}\right)\right)-\left({\sf{Tr}}\left(\rho_{\Lambda}^{\sf {canonical }}(n \right)\right)^{2}) 
=\frac{\partial^{2}}{\partial(\beta \hbar \omega)^{2}} \ln \mathcal{Z}$.

From the above, it follows that ${\sf{Tr}}\left(\rho_{\Lambda}^{\sf {canonical }}\left(n^{2}\right)\right)=\frac{e^{\beta \hbar \omega}}{\left(e^{\beta \hbar \omega}-1\right)^{2}}+\frac{1}{\left(e^{\beta \hbar \omega}-1\right)^{2}}$. An elementary calculation then illustrates that, at an arbitrary temperature $T$, the variance of the kinetic energy in a single mode of $H_{\Lambda}$ given by Eq. (\ref{long-osc1}) is none other than the square of the internal energy $U \equiv {\sf{Tr}}\left(\rho^{\sf canonical}_{\Lambda} H_{\Lambda}\right) (=   \hbar \omega ({\sf{Tr}}\left(\rho^{\sf canonical}_{\Lambda} n \right) + 1/2)$) at that temperature,
\begin{eqnarray}
\label{osc:var:2}
&& \sigma_{\frac{(p^{H})^{2}}{2 m}}^{2}  =\frac{(\hbar \omega)^{2}}{8}
\left[3~ {\sf{Tr}}\left(\rho^{\sf canonical}_{\Lambda}\left(n^{2}\right)\right)-2\left({\sf{Tr}}\left(\rho^{\sf canonical} _{\Lambda}(n)\right)\right)^{2} \right.\nonumber
\\ && \left.+\left({\sf{Tr}}\left(\rho^{\sf canonical}_{\Lambda}(n)\right)\right)+1 \right] \nonumber
\\ 
\! \! \! \! \! \! \! \! \! \! \! \! \! \! \! \! \! \! \! \! \! &&  =\frac{(\hbar \omega)^{2}}{8}\left[3\left(\frac{1+e^{\beta \hbar \omega}}{\left(e^{\beta \hbar \omega}-1\right)^{2}}\right)-2 \frac{1}{\left(e^{\beta \hbar \omega}-1\right)^{2}}+\frac{1}{e^{\beta \hbar \omega}-1}+1\right] 
\nonumber
\\ && =\frac{(\hbar \omega)^{2}}{8} \coth^2\left(\frac{\beta \hbar \omega}{2}\right) \nonumber
  \\ && =\frac{1}{2} U^{2}.
\end{eqnarray}
This relation concerning the variance of the kinetic energy at finite temperature complements the well known result that the average of the kinetic energy is half of the total energy in a harmonic mode. As Eq. (\ref{osc:var:2}) illustrates, the variance of the kinetic term is, at low temperatures, higher than that of the full Hamiltonian. This is a general property that appears whenever we use the thermodynamic heat capacity defined by the Hamiltonian $H_{\Lambda}$ (employed in Section \ref{decoupled} and \ref{sec:lowT}) and Appendix \ref{sec:RP} instead of the effective heat capacity of Eq. (\ref{cv2}). The variance of the full Hamiltonian in the thermal state that it defines (and thus the thermodynamic heat capacity) always vanishes in the low T limit (since the system approaches a (ground state) eigenstate or mixture of degenerate states). 

Taylor expanding Eq. (\ref{osc:var:2}) in the inverse temperature, we have $\sigma_{\frac{(p^H)^{2}}{2 m}}^{2} \sim \frac{1}{2 \beta^{2}}+\cdots=\frac{\left(k_{B} T\right)^{2}}{2}+\cdots $, where the ellipsis denote terms of order $\mathcal{O}\left(\beta^{-1}\right)$ or higher \footnote{In the classical high temperature regime, the probability density $\rho_{\Lambda}^{\sf {classical~canonical }}=\frac{e^{-\beta H_{\Lambda}}}{Z_{\Lambda}}$, with the harmonic oscillator Hamiltonian $H_{\Lambda}$, factorizes into a Gaussian in the momenta multiplying a Gaussian in the spatial coordinates and Eq. (\ref{osc:var:2}) also immediately follows from a trivial application of Wick's theorem.}. We next use the shorthand $\bar{Q}_{i}$ to denote a long time average by invoking Eq. (\ref{cal-long}), ${\sf{Tr}}\left(\rho_{\Lambda}^{\sf {canonical }} Q_{i}^{H}\right)=\lim_{{\cal{T}} \rightarrow \infty} \frac{1}{{\cal{T}}} \int_{0}^{{\cal{T}}} {\sf{Tr}}\left(\rho_{\Lambda} Q_{i}^{H}(t)\right) d t \equiv \bar{Q}_{i}$. When $Q_i=x_H$ is the single mode coordinate, the Hamiltonian $\tilde{H}_i^{H}=\frac{\left(p^{H}\right)^{2}}{2 m} \subset H_{\omega} \subset H_{\Lambda}$ fully generates the dynamics of $Q_i^H$. Recognizing that the earlier global system average inequalities may be replaced by long time averages (since both may be computed with $\rho_{\Lambda}^{\sf {canonical }}$), 
\begin{eqnarray}
\label{eq:eq:harmonic}
&& \overline{\sigma_{\tilde{H}^{H}}^{2}} \geq \nonumber
\\ && \lim _{{\cal{T}} \rightarrow \infty} \frac{\hbar^{2}}{4  {\cal{T}}} \int_{0}^{{\cal{T}}} \frac{dt\left({\sf{Tr}}\left(\rho_{\Lambda} \frac{d Q^{H}}{d t}\right)\right)^{2}}{{\sf{Tr}}\left(\rho_{\Lambda}\left(Q^{H}\right)^{2}\right)-\left({\sf{Tr}}\left(\rho_{\Lambda} Q^{H}\right)\right)^{2}}.
\end{eqnarray}
Substituting Eq. (\ref{osc:var:2}) for the variance of $\tilde{H}^{H}$,
\begin{eqnarray}
\label{eq:harmonic}
\frac{(\hbar \omega)^{2}}{8} \coth^{2}\left(\frac{\beta \hbar \omega}{2}\right) \geq  \nonumber
\\ \! \! \! \! \! \!  \lim _{{\cal{T}} \rightarrow \infty}  \frac{\hbar^{2}}{4 {\cal{T}}} \int_{0}^{{\cal{T}}} \frac{ dt \left({\sf{Tr}}\left(\rho_{\Lambda} \frac{d x^{H}}{d t}\right)\right)^{2}}{{\sf{Tr}}\left(\rho_{\Lambda}\left(x^{H}\right)^{2}\right)-\left({\sf{Tr}}\left(\rho_{\Lambda} x^{H}\right)\right)^{2}}.
\end{eqnarray}
The new bound of Eq. (\ref{eq:harmonic}) holds for any individual harmonic normal mode of frequency $\omega$ 
and coordinate $x^H$. The full set of normal mode coordinates in a harmonic solid is related, by a unitary transformation, to the individual displacements of all atoms. If an atomic displacement in 
the lattice cannot exceed the lattice constant ``${\sf a}$'' times the Lindemann ratio $c_{L}$ \cite{Lindemann} (typically, $c_L \sim 0.1$) then the variance 
\begin{equation}
{\sf{Tr}}\left(\rho_{\Lambda}\left(\Delta x^{H}\right)^{2}\right)<(c_{L} {\sf a})^{2}.
\end{equation}
Here, $\Delta x^{H} \equiv\left(x^{H}-{\sf{Tr}}\left(\rho_{\Lambda} x^{H}\right)\right)$ denotes the deviation of $x^{H}$ from its equilibrium average. Thus, the denominator in the integrand of Eq. (\ref{eq:harmonic}) is bounded by $(c_L {\sf a}) ^{2}$, 
\begin{eqnarray}
\label{final-T-harmonic-0}
\lim _{{\cal{T}} \rightarrow \infty} \frac{1}{{\cal{T}}} \int_{0}^{{\cal{T}}}\left({\sf{Tr}}\left(\rho_{\Lambda} \frac{d x^{H}}{d t}\right)\right)^{2} d t \nonumber
\\  \leq \frac{(\omega {\sf a} c_{L})^{2}}{2} \coth^{2}\left(\frac{\beta \hbar \omega}{2}\right).
\end{eqnarray}
That is, in a harmonic solid, the long time average of the squared normal mode velocity 
is rigorously bounded by the bottom line of Eq. (\ref{final-T-harmonic-0}). 

As we underscored in Appendix \ref{sec:comments}, the choice of the local Hamiltonian $\tilde{H}_{i}^{H}$ providing the dynamics of $Q_i^H$ is not unique. Some choices of $\tilde{H}_{i}^{H}$ may lead to stronger inequalities. Another bound is obtained by setting $\tilde{H}^{H}_i$ to be the full single mode Hamiltonian $H_{\omega}$. The righthand side of Eq. (\ref{eq:eq:harmonic}) is then replaced by $k_{B} T^{2} C_{\sf {Einstein }}$ where $C_{\sf{Einstein }}$ is the heat capacity in the Einstein model \cite{Einstein} (the  thermodynamic heat capacity for this single mode oscillator defined by the full Hamiltonian $H_{\Lambda}$),
\begin{equation}
\label{Albert}
C_{\sf {Einstein }}=\frac{(\beta \hbar \omega)^{2} e^{\beta \hbar \omega}}{\left(e^{\beta \hbar \omega}-1\right)^{2}} k_{B}.
\end{equation}
At low temperatures, this thermodynamic specific heat type bound when using the full single mode Hamiltonian $H_{\omega}$ is tighter than that obtained with a minimal kinetic energy $\tilde{H}_i^{H}$. This is so since the individual kinetic and potential energy fluctuations in the harmonic oscillator are negatively correlated (whereas the variance of $H_{\omega}$ in its own ground state must vanish). Inserting the Einstein heat capacity for a normal mode of angular frequency $\omega$,
\begin{eqnarray}
\label{harmonic-final}
&&  \lim _{{\cal{T}} \rightarrow \infty} \frac{1}{{\cal{T}}} \int_{0}^{{\cal{T}}}\left({\sf{Tr}}\left(\rho_{\Lambda} \frac{d x^{H}}{d t}\right)\right)^{2} d t \nonumber
 \\ && \leq \frac{4 {\sf{a}}^{2} c_{L}^{2}}{\hbar^{2}}\left[k_{B} T^{2} \frac{(\beta \hbar \omega)^{2} e^{\beta \hbar \omega}}{\left(e^{\beta \hbar \omega}-1\right)^{2}} k_{B}\right] \nonumber
\\ =&& \left(2 c_{L} {\sf{a}} \omega\right)^{2} \frac{e^{\beta \hbar \omega}}{\left(e^{\beta \hbar \omega}-1\right)^{2}}.
\end{eqnarray}
The exponential decay of the bound with the inverse temperature (scaling as $e^{-\beta \hbar \omega}$ when $\beta \hbar \omega \gg 1$) captures the quantum depression of the Einstein heat capacity of Eq. (\ref{Albert}). Beyond the specifics of our toy model, in general low temperature systems, there is a quantum suppression of mode excitations (leading to tighter bounds on the dynamics).  In the diametrically opposite high temperature limit, the righthand side forming the upper bound in Eq. (\ref{harmonic-final}) now tends to zero as the inverse temperature $\beta \rightarrow 0$. In this high temperature limit, the bound of Eq. (\ref{harmonic-final}) reads
\begin{equation}
\label{speedharmonic}
\lim _{{\cal{T}} \rightarrow \infty} \frac{1}{{\cal{T}}} \int_{0}^{{\cal{T}}}\left({\sf{Tr}}\left(\rho_{\Lambda} \frac{d x^{H}}{d t}\right)\right)^{2} d t \leq 2\left(\frac{c_{L}  {\sf a}  k_{B} T}{\hbar}\right)^{2}.
\end{equation} 
As it must, the bound of Eq. (\ref{speedharmonic}) for this single
harmonic mode is consistent with the general velocity bounds of
Eqs. (\ref{vbound12},\ref{longABABA''}, \ref{eq:speed_limit})
\cite{bound}. In a system with a higher number of modes, identical
results follow. More generally, in an arbitrary number of spatial
dimensions $d$, summing Eq. (\ref{longABABA''}) over all Cartesian
components $\ell = 1,2, \ldots d$, irrespective of the nature of the
interactions (whether harmonic or not), the average squared speed per
particle $\frac{1}{N_{\Lambda} } \sum_{i=1}^{N_{\Lambda}} \Big\langle
\Big ( \frac{d {\bf r}_{i}^{H}}{dt} \Big)^2  \Big\rangle  \le \frac{2
  (k_B T)^2}{\hbar^2} {\sf Tr} \Big( \rho_{\Lambda}^{\sf classical~
  canonical} ({\bf \Delta r}^H_{i})^2 \Big)$. In a solid satisfying
the Lindemann criterion, we have ${\sf Tr} \Big( \rho_{\Lambda}^{\sf
  classical~ canonical} ({\bf \Delta r}^H_{i})^2 \Big) \le (c_{L} a)^2$, and a trivial extension of Eq. (\ref{speedharmonic}) rigorously follows for general solids irrespective of the nature of the interaction between their constituents. In \cite{melting-speed}, a related bound was advanced for the melting speed (one in which the temperature $T$ is set to be the melting temperature of the harmonic solid). This bound was empirically tested across a broad array of solids \cite{melting-speed}. 

We conclude this Appendix by mentioning that the bounds that we just derived are not restricted to the canonical ensemble. 
Rather, these inequalities hold more broadly when the reduced one body density matrix has the same diagonal entries in the energy eigenbasis as the canonical density matrix,
\begin{eqnarray}
\left(\rho_{\Lambda}\right)_{n n}=\frac{e^{-\beta E_{n}}}{Z_{\Lambda}}=\frac{e^{-\beta\left(n+\frac{1}{2}\right) \hbar \omega}}{Z_{\Lambda}},
\end{eqnarray}
with the partition function of Eq. (\ref{Zosc}). The off-diagonal matrix elements (in the eigenbasis of $H_{\Lambda}$)
of this single body density matrix can be completely arbitrary in the calculations in this Appendix. 
This is so since (see also Section \ref{quantum-thermalization} and Eq. (\ref{lllong}) in particular) 
the long time average of any off-diagonal contributions will trivially vanish since
$\lim _{{\cal{T}} \rightarrow \infty} \frac{1}{{\cal{T}}} \int_{0}^{{\cal{T}}} d t^{\prime} e^{i \omega_{n m} t^{\prime}} =0$ when $\omega_{n m}  \neq 0$. Thus, arbitrary off-diagonal matrix elements will not influence any of bounds that we derived on the long time averages. 

\subsection{An XY spin model}
\label{sec:XY}

As a second toy model, we next consider a spin $S=\frac{1}{2}$ nearest neighbor XY system on a general lattice,
\begin{eqnarray}
H_{\Lambda}=-J \sum_{\langle i j\rangle}\left(S_{i}^{x} S_{j}^{x}+S_{i}^{y} S_{j}^{y}\right).
\end{eqnarray}
The time derivative of $Q^H_i=S^{xH}_{i}$ is generated by terms of the form
\begin{equation}
\label{XYHH}
\tilde{H}_{i}^{H}=-J S_{i}^{y} \sum_{|j-i|=1} S_{j}^{y},
\end{equation}
where, to avoid cumbersome notation, we omit the Heisenberg picture ($^H$) superscript on all spin operators. Henceforth, these will be implicitly assumed. Squaring Eq. (\ref{XYHH}), 
$\left(\tilde{H}_{i}^{H}\right)^{2}=J^{2}\left(S_{i}^{y} \sum_{|j-i|=1} S_{j}^{y}\right)^{2}=J^{2} \sum_{j, j^{\prime}} S_{j}^{y} S_{j^{\prime}}^{y}$ with the sum over $j$ and $j^{\prime}$ being that over all sites $j, j^{\prime}$ that are nearest neighbor sites of $i$. In this sum, if $j=j^{\prime}$ then, trivially, $S_{j}^{y} S_{j^{\prime}}^{y}=\left(S_{j}^{y}\right)^{2}=\frac{\hbar^{2}}{4}$.
If $j \neq j^{\prime}$ the corresponding expectation values $\left({\sf{Tr}} (\rho_{\Lambda}^{\sf canonical}
S_{j}^{y} S_{j^{\prime}}^{y})\right)$ is the next nearest neighbor correlator of the y components of the spin. For the computation of ${\sf{Tr}} (\rho_{\Lambda}^{\sf canonical} \left(\tilde{H}_{i}^{H}\right)^{2} )$  we need to further determine ${\sf{Tr}}( \rho_{\Lambda}^{\sf canonical} \left(S_{i}^{y} \sum_{|j-i|=1} S_{j}^{y}\right)^{2})$. The average ${\sf{Tr}} (\rho_{\Lambda}^{\sf canonical} S_{i}^{y} \sum_{|j-i|=1} S_{j}^{y})$ is half the of the internal energy $\left(-{\sf{Tr}}( \rho_{\Lambda}^{\sf canonical} H_{\Lambda}) / J\right)$ per bond of the XY system. The origin of the latter factor of a half is that both $x$ and $y$ components have equal contributions to the internal energy. Generally not 
all next nearest neighbor correlators ${\sf{Tr}} (\rho_{\Lambda}^{\sf canonical} S_{j}^{y} 
S_{j^{\prime}}^{y})$ assume the same value.

 We next turn to a hypercubic lattice rendition of the model. We set the lattice constant to be one. If on such a lattice, ${\sf{Tr}} (\rho_{\Lambda}^{\sf canonical} S_{j}^{y} S_{j^{\prime}}^{y})=G(\sqrt{2})$ when $\left|j-j^{\prime}\right|=\sqrt{2}$ and ${\sf{Tr}}( \rho_{\Lambda}^{\sf canonical} S_{j}^{y} S_{j^{\prime}}^{y})=G(2)$ for $\left|j-j^{\prime}\right|=2$ 
 then 
\begin{eqnarray}
\label{varXY}
{\sf{Tr}}\Big( \rho_{\Lambda}^{\sf canonical}\left(\tilde{H}_{i}^{H}\right)^{2} \Big)-\left({\sf{Tr}}( \rho_{\Lambda}^{\sf canonical}\tilde{H}_{i}^{H})\right)^{2} \nonumber
\\ ={\sf z}  \frac{\hbar^{4}}{16}+2\left({\sf z}^{\prime} G(\sqrt{2})+{\sf z}^{\prime \prime} G(2)\right)-\left({\sf z} G_{n n}\right)^{2}.
\end{eqnarray}
The origin of the factor of $\hbar^{4} / 16=\left(\hbar^{2} / 4\right)^{2}$ is the square of the spin bilinear arising when when $j=j^{\prime}$. Here, ${\sf z}, {\sf z}^{\prime}$ and ${\sf z}^{\prime \prime}$ are, respectively,
the number of nearest neighbors of site $i,$ the number of sites a distance $\sqrt{2}$ away, and the number of sites a distance 2 away. The origin 
of the prefactor of two in the second term of Eq. (\ref{varXY}) is that both correlation functions ${\sf{Tr}} (\rho_{\Lambda}^{\sf canonical} S_{j}^{y} S_{j^{\prime}}^{y})$ and ${\sf{Tr}}( \rho_{\Lambda}^{\sf canonical} S_{j^{\prime}}^{y} S_{j}^{y})$ appear when computing ${\sf{Tr}} \Big( \rho_{\Lambda}^{\sf canonical}\left(\tilde{H}_{i}^{H}\right)^{2} \Big)$.
On a $d$-dimensional hypercubic lattice, ${\sf z}={\sf z}^{\prime \prime}=2 d$ and ${\sf z}^{\prime}=2 d(d-1)$.
Eq. (\ref{varXY}) forms the righthand side of Eq. (\ref{cv2}). In $d=1$ (i.e., a spin chain), the correlation functions $G(z)$ in Eq. (\ref{varXY}) may be exactly determined via a Jordan-Wigner transformation (see the ``classical'' work of Lieb, Schultz, and Mattis \cite{LSM} for exact expressions). In higher dimensional ($d>1$) realizations, $G(z)$ cannot be computed exactly. 

Putting all of the pieces together, we may now substitute $Q_i=S^{x}_i$ in the bound of Eq. (\ref{cv1}) with $\overline{\mathcal{O}}$ defined by Eqs. (\ref{overdef},\ref{cv3}). Notably, the energy variance of Eq. (\ref{cv2}) is now given by Eq. (\ref{varXY}). As we sketched in Appendix \ref{high-moment-derivative}, higher moments of the time derivative of the spin operator $S^x_i$ may be bounded analogously.

\section{Minimal prethermalization times and metastable states}
\label{minprethermal}
The systems noted in Section \ref{sec:absent} and others may require a very long time to thermalize to the true equilibrium canonical distribution. We now discuss the situation in which the non-equilibrium density matrix $\rho_{\Lambda}^{\sf neq}$ is nearly stationary and the system may appear to thermalize to this state. Such an occurrence appears in prethermalized systems \cite{prethermal1,prethermal2} in which a steady state different from that of true equilibrium is first obtained (followed by true thermalization later on). Even for such a nearly stationary $\rho_{\Lambda}^{\sf neq}$, there trivially is, as is seen from Eq. (\ref{lllong}), a required minimal time window width beyond which the resulting average for any local observable $Q_i$ will not change and seem to approach an effective equilibrium (stationary) value. For prethermalization, similar long time integrals appear that mirror those appearing for the equilibrium state of Section \ref{sec:therm_time_int}. The long time average is equal to the prethermal state average. Repeating the considerations that led to Eq. (\ref{lllong}), we see that this minimal time is given by $\frac{\hbar}{\sqrt{k_{B} T^2 C^{\sf neq}_{v,i}}}$. Here, Eq. (\ref{cv2}) is replaced by the variance of {\it the local Hamiltonian} $\tilde{H}_{i}^{H}$ in the nearly steady non-equilibrium state, 
\begin{eqnarray}
k_{B} T^2 C^{\sf neq}_{v,i} \equiv {\sf{Tr}}\left(\rho_{\Lambda}^{\sf neq}\left(\Delta \tilde{H}_{i}^{H}(t)\right)^{2}\right).
\end{eqnarray}
Similar to the above, metastable systems that appear in numerous arenas may, on measurable time scales, seem to equilibrate to a thermal state associated with an effective Hamiltonian rather than the thermal state associated with the microscopic Hamiltonian $H_{\Lambda}$. For such metastable systems, the thermalization bounds that we derived above and in earlier Appendices and Sections may be extended to the fictive thermal state associated with this effective Hamiltonian.

\section{Lower limit on measurable time windows in macrosopic systems}
\label{sec:clock}

We conclude with a simple comment regarding a universal limit on the possible rates of change of bounded local observables in any system (no matter how large) with bounded interactions. Returning to Eq. (\ref{timel}), in systems with bounded operators $Q_i^H$ (e.g., a spin component), the standard deviation of $\sigma_{Q_{i}^{H}}$ is bounded.
In an analogous fashion, the standard deviation of the local Hamiltonian $\sigma_{\tilde{H}^{H}_{i}}$
is also bounded. This implies that the rate of change of any such local observable is bounded and thus there is a lower (temperature independent) bound on the time scales that may be measured by any such ``clock''.

\bibliography{planckian-biblio}

\begin{thebibliography}{222}%
\makeatletter
\providecommand \@ifxundefined [1]{%
 \@ifx{#1\undefined}
}%
\providecommand \@ifnum [1]{%
 \ifnum #1\expandafter \@firstoftwo
 \else \expandafter \@secondoftwo
 \fi
}%
\providecommand \@ifx [1]{%
 \ifx #1\expandafter \@firstoftwo
 \else \expandafter \@secondoftwo
 \fi
}%
\providecommand \natexlab [1]{#1}%
\providecommand \enquote  [1]{``#1''}%
\providecommand \bibnamefont  [1]{#1}%
\providecommand \bibfnamefont [1]{#1}%
\providecommand \citenamefont [1]{#1}%
\providecommand \href@noop [0]{\@secondoftwo}%
\providecommand \href [0]{\begingroup \@sanitize@url \@href}%
\providecommand \@href[1]{\@@startlink{#1}\@@href}%
\providecommand \@@href[1]{\endgroup#1\@@endlink}%
\providecommand \@sanitize@url [0]{\catcode `\\12\catcode `\$12\catcode
  `\&12\catcode `\#12\catcode `\^12\catcode `\_12\catcode `\%12\relax}%
\providecommand \@@startlink[1]{}%
\providecommand \@@endlink[0]{}%
\providecommand \url  [0]{\begingroup\@sanitize@url \@url }%
\providecommand \@url [1]{\endgroup\@href {#1}{\urlprefix }}%
\providecommand \urlprefix  [0]{URL }%
\providecommand \Eprint [0]{\href }%
\providecommand \doibase [0]{http://dx.doi.org/}%
\providecommand \selectlanguage [0]{\@gobble}%
\providecommand \bibinfo  [0]{\@secondoftwo}%
\providecommand \bibfield  [0]{\@secondoftwo}%
\providecommand \translation [1]{[#1]}%
\providecommand \BibitemOpen [0]{}%
\providecommand \bibitemStop [0]{}%
\providecommand \bibitemNoStop [0]{.\EOS\space}%
\providecommand \EOS [0]{\spacefactor3000\relax}%
\providecommand \BibitemShut  [1]{\csname bibitem#1\endcsname}%
\let\auto@bib@innerbib\@empty
\bibitem [{\citenamefont {Sackur}(1913)}]{Sackur}%
  \BibitemOpen
  \bibfield  {author} {\bibinfo {author} {\bibfnamefont {O.}~\bibnamefont
  {Sackur}},\ }\bibfield  {title} {\enquote {\bibinfo {title} {{Die universelle
  Bedeutung des sog. elementaren Wirkungsquantums}},}\ }\href {\doibase
  https://doi.org/10.1002/andp.19133450103} {\bibfield  {journal} {\bibinfo
  {journal} {Annalen der Physik}\ }\textbf {\bibinfo {volume} {345}},\ \bibinfo
  {pages} {67--86} (\bibinfo {year} {1913})}\BibitemShut {NoStop}%
\bibitem [{\citenamefont {Tetrode}(1912)}]{tetrode}%
  \BibitemOpen
  \bibfield  {author} {\bibinfo {author} {\bibfnamefont {H.}~\bibnamefont
  {Tetrode}},\ }\bibfield  {title} {\enquote {\bibinfo {title} {{Die chemische
  Konstante der Gase und das elementare Wirkungsquantum}},}\ }\href {\doibase
  https://doi.org/10.1002/andp.19123430708} {\bibfield  {journal} {\bibinfo
  {journal} {Annalen der Physik}\ }\textbf {\bibinfo {volume} {343}},\ \bibinfo
  {pages} {434--442} (\bibinfo {year} {1912})}\BibitemShut {NoStop}%
\bibitem [{\citenamefont {Grimus}(2013)}]{experiment}%
  \BibitemOpen
  \bibfield  {author} {\bibinfo {author} {\bibfnamefont {Walter}\ \bibnamefont
  {Grimus}},\ }\bibfield  {title} {\enquote {\bibinfo {title} {{100th
  anniversary of the Sackur-Tetrode equation}},}\ }\href {\doibase
  https://doi.org/10.1002/andp.201300720} {\bibfield  {journal} {\bibinfo
  {journal} {Annalen der Physik}\ }\textbf {\bibinfo {volume} {525}},\ \bibinfo
  {pages} {A32--A35} (\bibinfo {year} {2013})}\BibitemShut {NoStop}%
\bibitem [{\citenamefont {Kovtun}\ \emph {et~al.}(2003)\citenamefont {Kovtun},
  \citenamefont {Son},\ and\ \citenamefont {Starinets}}]{KSS1}%
  \BibitemOpen
  \bibfield  {author} {\bibinfo {author} {\bibfnamefont {Pavel}\ \bibnamefont
  {Kovtun}}, \bibinfo {author} {\bibfnamefont {Dam~T}\ \bibnamefont {Son}}, \
  and\ \bibinfo {author} {\bibfnamefont {Andrei~O}\ \bibnamefont {Starinets}},\
  }\bibfield  {title} {\enquote {\bibinfo {title} {{Holography and
  hydrodynamics: diffusion on stretched horizons}},}\ }\href {\doibase
  10.1088/1126-6708/2003/10/064} {\bibfield  {journal} {\bibinfo  {journal}
  {Journal of High Energy Physics}\ }\textbf {\bibinfo {volume} {2003}},\
  \bibinfo {pages} {064--064} (\bibinfo {year} {2003})}\BibitemShut {NoStop}%
\bibitem [{\citenamefont {Kovtun}\ \emph {et~al.}(2005)\citenamefont {Kovtun},
  \citenamefont {Son},\ and\ \citenamefont {Starinets}}]{KSS2}%
  \BibitemOpen
  \bibfield  {author} {\bibinfo {author} {\bibfnamefont {P.~K.}\ \bibnamefont
  {Kovtun}}, \bibinfo {author} {\bibfnamefont {D.~T.}\ \bibnamefont {Son}}, \
  and\ \bibinfo {author} {\bibfnamefont {A.~O.}\ \bibnamefont {Starinets}},\
  }\bibfield  {title} {\enquote {\bibinfo {title} {{Viscosity in Strongly
  Interacting Quantum Field Theories from Black Hole Physics}},}\ }\href
  {\doibase 10.1103/PhysRevLett.94.111601} {\bibfield  {journal} {\bibinfo
  {journal} {Phys. Rev. Lett.}\ }\textbf {\bibinfo {volume} {94}},\ \bibinfo
  {pages} {111601} (\bibinfo {year} {2005})}\BibitemShut {NoStop}%
\bibitem [{\citenamefont {Shuryak}(2004)}]{Shuryak}%
  \BibitemOpen
  \bibfield  {author} {\bibinfo {author} {\bibfnamefont {Edward}\ \bibnamefont
  {Shuryak}},\ }\bibfield  {title} {\enquote {\bibinfo {title} {Why does the
  quark-gluon plasma at rhic behave as a nearly ideal fluid?}}\ }\href@noop {}
  {\bibfield  {journal} {\bibinfo  {journal} {Prog. Part. Nucl. Phys.}\
  }\textbf {\bibinfo {volume} {53}},\ \bibinfo {pages} {273} (\bibinfo {year}
  {2004})}\BibitemShut {NoStop}%
\bibitem [{\citenamefont {Sch{\"a}fer}\ and\ \citenamefont
  {Teaney}(2009)}]{Tom}%
  \BibitemOpen
  \bibfield  {author} {\bibinfo {author} {\bibfnamefont {Thomas}\ \bibnamefont
  {Sch{\"a}fer}}\ and\ \bibinfo {author} {\bibfnamefont {Derek}\ \bibnamefont
  {Teaney}},\ }\bibfield  {title} {\enquote {\bibinfo {title} {{Nearly perfect
  fluidity: from cold atomic gases to hot quark gluon plasmas}},}\ }\href
  {\doibase 10.1088/0034-4885/72/12/126001} {\bibfield  {journal} {\bibinfo
  {journal} {Reports on Progress in Physics}\ }\textbf {\bibinfo {volume}
  {72}},\ \bibinfo {pages} {126001} (\bibinfo {year} {2009})}\BibitemShut
  {NoStop}%
\bibitem [{\citenamefont {Nussinov}\ \emph
  {et~al.}(2014{\natexlab{a}})\citenamefont {Nussinov}, \citenamefont
  {Nogueira}, \citenamefont {Blodgett},\ and\ \citenamefont {Kelton}}]{nnbk}%
  \BibitemOpen
  \bibfield  {author} {\bibinfo {author} {\bibfnamefont {Z.}~\bibnamefont
  {Nussinov}}, \bibinfo {author} {\bibfnamefont {F.}~\bibnamefont {Nogueira}},
  \bibinfo {author} {\bibfnamefont {M.}~\bibnamefont {Blodgett}}, \ and\
  \bibinfo {author} {\bibfnamefont {K.~F.}\ \bibnamefont {Kelton}},\ }\bibfield
   {title} {\enquote {\bibinfo {title} {{Thermalization and possible quantum
  relaxation times in ``classical'' fluids: theory and experiment}},}\ }\href
  {https://arxiv.org/abs/1409.1915} {\bibfield  {journal} {\bibinfo  {journal}
  {arXiv preprint arXiv:1409.1915}\ } (\bibinfo {year}
  {2014}{\natexlab{a}})}\BibitemShut {NoStop}%
\bibitem [{\citenamefont {Blodgett}\ \emph {et~al.}(2015)\citenamefont
  {Blodgett}, \citenamefont {Egami}, \citenamefont {Nussinov},\ and\
  \citenamefont {Kelton}}]{benk}%
  \BibitemOpen
  \bibfield  {author} {\bibinfo {author} {\bibfnamefont {M.~E.}\ \bibnamefont
  {Blodgett}}, \bibinfo {author} {\bibfnamefont {T.}~\bibnamefont {Egami}},
  \bibinfo {author} {\bibfnamefont {Z.}~\bibnamefont {Nussinov}}, \ and\
  \bibinfo {author} {\bibfnamefont {K.~F.}\ \bibnamefont {Kelton}},\ }\bibfield
   {title} {\enquote {\bibinfo {title} {Proposal for universality in the
  viscosity of metallic liquids},}\ }\href {\doibase 10.1038/srep13837}
  {\bibfield  {journal} {\bibinfo  {journal} {Scientific Reports}\ }\textbf
  {\bibinfo {volume} {5}},\ \bibinfo {pages} {13837} (\bibinfo {year}
  {2015})}\BibitemShut {NoStop}%
\bibitem [{\citenamefont {Nussinov}(2020)}]{bound}%
  \BibitemOpen
  \bibfield  {author} {\bibinfo {author} {\bibfnamefont {Z.}~\bibnamefont
  {Nussinov}},\ }\bibfield  {title} {\enquote {\bibinfo {title} {{Macroscopic
  length correlations in non-equilibrium systems and their possible
  realizations}},}\ }\href {\doibase
  https://doi.org/10.1016/j.nuclphysb.2020.114948} {\bibfield  {journal}
  {\bibinfo  {journal} {Nuclear Physics B}\ }\textbf {\bibinfo {volume}
  {953}},\ \bibinfo {pages} {114948} (\bibinfo {year} {2020})}\BibitemShut
  {NoStop}%
\bibitem [{\citenamefont {Zaanen}\ \emph {et~al.}(2015)\citenamefont {Zaanen},
  \citenamefont {Liu}, \citenamefont {Sun},\ and\ \citenamefont
  {Schalm}}]{jan-book}%
  \BibitemOpen
  \bibfield  {author} {\bibinfo {author} {\bibfnamefont {Jan}\ \bibnamefont
  {Zaanen}}, \bibinfo {author} {\bibfnamefont {Yan}\ \bibnamefont {Liu}},
  \bibinfo {author} {\bibfnamefont {Ya-Wen}\ \bibnamefont {Sun}}, \ and\
  \bibinfo {author} {\bibfnamefont {Koenraad}\ \bibnamefont {Schalm}},\
  }\href@noop {} {\emph {\bibinfo {title} {{Holographic duality in condensed
  matter physics}}}}\ (\bibinfo  {publisher} {Cambridge University Press},\
  \bibinfo {year} {2015})\BibitemShut {NoStop}%
\bibitem [{\citenamefont {Hartnoll}\ \emph {et~al.}(2018)\citenamefont
  {Hartnoll}, \citenamefont {Lucas},\ and\ \citenamefont {Sachdev}}]{HLS-book}%
  \BibitemOpen
  \bibfield  {author} {\bibinfo {author} {\bibfnamefont {S.~A.}\ \bibnamefont
  {Hartnoll}}, \bibinfo {author} {\bibfnamefont {A.}~\bibnamefont {Lucas}}, \
  and\ \bibinfo {author} {\bibfnamefont {S.}~\bibnamefont {Sachdev}},\
  }\href@noop {} {\emph {\bibinfo {title} {{Holographic Quantum Matter}}}}\
  (\bibinfo  {publisher} {MIT Press},\ \bibinfo {year} {2018})\BibitemShut
  {NoStop}%
\bibitem [{\citenamefont {Baggioli}(2019)}]{Matteo2019}%
  \BibitemOpen
  \bibfield  {author} {\bibinfo {author} {\bibfnamefont {Matteo}\ \bibnamefont
  {Baggioli}},\ }\href@noop {} {\emph {\bibinfo {title} {{A Practical
  Mini-Course on Applied Holography}}}}\ (\bibinfo  {publisher} {Springer
  Briefs in Physics},\ \bibinfo {year} {2019})\BibitemShut {NoStop}%
\bibitem [{\citenamefont {Zaanen}(2019)}]{jan-planck}%
  \BibitemOpen
  \bibfield  {author} {\bibinfo {author} {\bibfnamefont {Jan}\ \bibnamefont
  {Zaanen}},\ }\bibfield  {title} {\enquote {\bibinfo {title} {{{Planckian
  dissipation, minimal viscosity and the transport in cuprate strange
  metals}}},}\ }\href {\doibase 10.21468/SciPostPhys.6.5.061} {\bibfield
  {journal} {\bibinfo  {journal} {SciPost Phys.}\ }\textbf {\bibinfo {volume}
  {6}},\ \bibinfo {pages} {61} (\bibinfo {year} {2019})}\BibitemShut {NoStop}%
\bibitem [{\citenamefont {Trachenko}\ and\ \citenamefont
  {Brazhkin}(2020)}]{KT}%
  \BibitemOpen
  \bibfield  {author} {\bibinfo {author} {\bibfnamefont {K.}~\bibnamefont
  {Trachenko}}\ and\ \bibinfo {author} {\bibfnamefont {V.~V.}\ \bibnamefont
  {Brazhkin}},\ }\bibfield  {title} {\enquote {\bibinfo {title} {{Minimal
  quantum viscosity from fundamental physical constants}},}\ }\href {\doibase
  10.1126/sciadv.aba3747} {\bibfield  {journal} {\bibinfo  {journal} {Science
  Advances}\ }\textbf {\bibinfo {volume} {6}} (\bibinfo {year} {2020}),\
  10.1126/sciadv.aba3747}\BibitemShut {NoStop}%
\bibitem [{\citenamefont {Bruin}\ \emph {et~al.}(2013)\citenamefont {Bruin},
  \citenamefont {Sakai}, \citenamefont {Perry},\ and\ \citenamefont
  {Mackenzie}}]{planck1}%
  \BibitemOpen
  \bibfield  {author} {\bibinfo {author} {\bibfnamefont {J.~A.~N.}\
  \bibnamefont {Bruin}}, \bibinfo {author} {\bibfnamefont {H.}~\bibnamefont
  {Sakai}}, \bibinfo {author} {\bibfnamefont {R.~S.}\ \bibnamefont {Perry}}, \
  and\ \bibinfo {author} {\bibfnamefont {A.~P.}\ \bibnamefont {Mackenzie}},\
  }\bibfield  {title} {\enquote {\bibinfo {title} {{Similarity of Scattering
  Rates in Metals Showing T-Linear Resistivity}},}\ }\href {\doibase
  10.1126/science.1227612} {\bibfield  {journal} {\bibinfo  {journal}
  {Science}\ }\textbf {\bibinfo {volume} {339}},\ \bibinfo {pages} {804--807}
  (\bibinfo {year} {2013})}\BibitemShut {NoStop}%
\bibitem [{\citenamefont {Zaanen}(2004)}]{planck3}%
  \BibitemOpen
  \bibfield  {author} {\bibinfo {author} {\bibfnamefont {Jan}\ \bibnamefont
  {Zaanen}},\ }\bibfield  {title} {\enquote {\bibinfo {title} {{Why the
  temperature is high}},}\ }\href {\doibase 10.1038/430512a} {\bibfield
  {journal} {\bibinfo  {journal} {Nature}\ }\textbf {\bibinfo {volume} {430}},\
  \bibinfo {pages} {512--513} (\bibinfo {year} {2004})}\BibitemShut {NoStop}%
\bibitem [{\citenamefont {Hartnoll}(2015)}]{planck4}%
  \BibitemOpen
  \bibfield  {author} {\bibinfo {author} {\bibfnamefont {Sean~A.}\ \bibnamefont
  {Hartnoll}},\ }\bibfield  {title} {\enquote {\bibinfo {title} {{Theory of
  universal incoherent metallic transport}},}\ }\href {\doibase
  10.1038/nphys3174} {\bibfield  {journal} {\bibinfo  {journal} {Nature
  Physics}\ }\textbf {\bibinfo {volume} {11}},\ \bibinfo {pages} {54--61}
  (\bibinfo {year} {2015})}\BibitemShut {NoStop}%
\bibitem [{\citenamefont {Maldacena}\ \emph {et~al.}(2016)\citenamefont
  {Maldacena}, \citenamefont {Shenker},\ and\ \citenamefont
  {Stanford}}]{juan-martin}%
  \BibitemOpen
  \bibfield  {author} {\bibinfo {author} {\bibfnamefont {Juan}\ \bibnamefont
  {Maldacena}}, \bibinfo {author} {\bibfnamefont {Stephen~H.}\ \bibnamefont
  {Shenker}}, \ and\ \bibinfo {author} {\bibfnamefont {Douglas}\ \bibnamefont
  {Stanford}},\ }\bibfield  {title} {\enquote {\bibinfo {title} {{A bound on
  chaos}},}\ }\href {\doibase 10.1007/JHEP08(2016)106} {\bibfield  {journal}
  {\bibinfo  {journal} {Journal of High Energy Physics}\ }\textbf {\bibinfo
  {volume} {2016}},\ \bibinfo {pages} {106} (\bibinfo {year}
  {2016})}\BibitemShut {NoStop}%
\bibitem [{\citenamefont {Mousatov}\ and\ \citenamefont
  {Hartnoll}(2020)}]{melting-speed}%
  \BibitemOpen
  \bibfield  {author} {\bibinfo {author} {\bibfnamefont {Connie~H.}\
  \bibnamefont {Mousatov}}\ and\ \bibinfo {author} {\bibfnamefont {Sean~A.}\
  \bibnamefont {Hartnoll}},\ }\bibfield  {title} {\enquote {\bibinfo {title}
  {{On the Planckian bound for heat diffusion in insulators}},}\ }\href
  {\doibase 10.1038/s41567-020-0828-6} {\bibfield  {journal} {\bibinfo
  {journal} {Nature Physics}\ }\textbf {\bibinfo {volume} {16}},\ \bibinfo
  {pages} {579--584} (\bibinfo {year} {2020})}\BibitemShut {NoStop}%
\bibitem [{\citenamefont {Chakravarty}(2019)}]{sudip}%
  \BibitemOpen
  \bibfield  {author} {\bibinfo {author} {\bibfnamefont {S.}~\bibnamefont
  {Chakravarty}},\ }\bibfield  {title} {\enquote {\bibinfo {title} {{Quantum
  critical fluctuations, Planckian dissipation, and compactification scale}},}\
  }\href {https://arxiv.org/pdf/1907.12163.pdf} {\bibfield  {journal} {\bibinfo
   {journal} {arXiv preprint arXiv:1907.12163}\ } (\bibinfo {year}
  {2019})}\BibitemShut {NoStop}%
\bibitem [{\citenamefont {Zhang}\ \emph {et~al.}(2019)\citenamefont {Zhang},
  \citenamefont {Kountz}, \citenamefont {Behnia},\ and\ \citenamefont
  {Kapitulnik}}]{kapitulnik}%
  \BibitemOpen
  \bibfield  {author} {\bibinfo {author} {\bibfnamefont {Jiecheng}\
  \bibnamefont {Zhang}}, \bibinfo {author} {\bibfnamefont {Erik~D.}\
  \bibnamefont {Kountz}}, \bibinfo {author} {\bibfnamefont {Kamran}\
  \bibnamefont {Behnia}}, \ and\ \bibinfo {author} {\bibfnamefont {Aharon}\
  \bibnamefont {Kapitulnik}},\ }\bibfield  {title} {\enquote {\bibinfo {title}
  {Thermalization and possible signatures of quantum chaos in complex
  crystalline materials},}\ }\href {\doibase 10.1073/pnas.1910131116}
  {\bibfield  {journal} {\bibinfo  {journal} {Proceedings of the National
  Academy of Sciences}\ }\textbf {\bibinfo {volume} {116}},\ \bibinfo {pages}
  {19869--19874} (\bibinfo {year} {2019})}\BibitemShut {NoStop}%
\bibitem [{\citenamefont {Hartnoll}\ and\ \citenamefont
  {Mackenzie}(2021)}]{HM}%
  \BibitemOpen
  \bibfield  {author} {\bibinfo {author} {\bibfnamefont {S.~A.}\ \bibnamefont
  {Hartnoll}}\ and\ \bibinfo {author} {\bibfnamefont {A.~P.}\ \bibnamefont
  {Mackenzie}},\ }\bibfield  {title} {\enquote {\bibinfo {title} {Planckian
  dissipation in metals},}\ }\href {https://arxiv.org/pdf/2107.07802.pdf}
  {\bibfield  {journal} {\bibinfo  {journal} {arXiv preprint arXiv:2107.07802}\
  } (\bibinfo {year} {2021})}\BibitemShut {NoStop}%
\bibitem [{\citenamefont {Lucas}(2019)}]{Lucas1}%
  \BibitemOpen
  \bibfield  {author} {\bibinfo {author} {\bibfnamefont {A.}~\bibnamefont
  {Lucas}},\ }\bibfield  {title} {\enquote {\bibinfo {title} {Operator size at
  finite temperature and planckian bounds on quantum dynamics},}\ }\href
  {\doibase 10.1103/PhysRevLett.122.216601} {\bibfield  {journal} {\bibinfo
  {journal} {Phys. Rev. Lett.}\ }\textbf {\bibinfo {volume} {122}},\ \bibinfo
  {pages} {216601} (\bibinfo {year} {2019})}\BibitemShut {NoStop}%
\bibitem [{\citenamefont {Grissonnanche}\ \emph {et~al.}(2021)\citenamefont
  {Grissonnanche}, \citenamefont {Fang}, \citenamefont {Legros}, \citenamefont
  {Verret}, \citenamefont {Laliberte}, \citenamefont {Collignon}, \citenamefont
  {Ataei}, \citenamefont {Dion}, \citenamefont {Zhou}, \citenamefont {Graf},
  \citenamefont {Lawler}, \citenamefont {Goddard}, \citenamefont {Taillefer},\
  and\ \citenamefont {Ramshaw}}]{Ramshaw}%
  \BibitemOpen
  \bibfield  {author} {\bibinfo {author} {\bibfnamefont {G.}~\bibnamefont
  {Grissonnanche}}, \bibinfo {author} {\bibfnamefont {Y.}~\bibnamefont {Fang}},
  \bibinfo {author} {\bibfnamefont {A.}~\bibnamefont {Legros}}, \bibinfo
  {author} {\bibfnamefont {S.}~\bibnamefont {Verret}}, \bibinfo {author}
  {\bibfnamefont {F.}~\bibnamefont {Laliberte}}, \bibinfo {author}
  {\bibfnamefont {C.}~\bibnamefont {Collignon}}, \bibinfo {author}
  {\bibfnamefont {A.}~\bibnamefont {Ataei}}, \bibinfo {author} {\bibfnamefont
  {M.}~\bibnamefont {Dion}}, \bibinfo {author} {\bibfnamefont {J.}~\bibnamefont
  {Zhou}}, \bibinfo {author} {\bibfnamefont {D.}~\bibnamefont {Graf}}, \bibinfo
  {author} {\bibfnamefont {M.~J.}\ \bibnamefont {Lawler}}, \bibinfo {author}
  {\bibfnamefont {P.}~\bibnamefont {Goddard}}, \bibinfo {author} {\bibfnamefont
  {L.}~\bibnamefont {Taillefer}}, \ and\ \bibinfo {author} {\bibfnamefont
  {B.~J.}\ \bibnamefont {Ramshaw}},\ }\bibfield  {title} {\enquote {\bibinfo
  {title} {Linear-in temperature resistivity from an isotropic planckian
  scattering rate},}\ }\href {\doibase 10.1038/s41586-021-03697-8} {\bibfield
  {journal} {\bibinfo  {journal} {Nature}\ }\textbf {\bibinfo {volume} {595}},\
  \bibinfo {pages} {667} (\bibinfo {year} {2021})}\BibitemShut {NoStop}%
\bibitem [{\citenamefont {Saso}\ \emph {et~al.}(2016)\citenamefont {Saso},
  \citenamefont {Kaplis},\ and\ \citenamefont {Starinets}}]{sazo}%
  \BibitemOpen
  \bibfield  {author} {\bibinfo {author} {\bibfnamefont {Grozdanov}\
  \bibnamefont {Saso}}, \bibinfo {author} {\bibfnamefont {Nikolaos}\
  \bibnamefont {Kaplis}}, \ and\ \bibinfo {author} {\bibfnamefont {Andrei~O.}\
  \bibnamefont {Starinets}},\ }\bibfield  {title} {\enquote {\bibinfo {title}
  {From strong to weak coupling in holographic models of thermalization},}\
  }\href {\doibase 10.1007/JHEP07(2016)151} {\bibfield  {journal} {\bibinfo
  {journal} {JHEP}\ }\textbf {\bibinfo {volume} {151}},\ \bibinfo {pages}
  {1607} (\bibinfo {year} {2016})}\BibitemShut {NoStop}%
\bibitem [{\citenamefont {Maldacena}(1999)}]{ads1}%
  \BibitemOpen
  \bibfield  {author} {\bibinfo {author} {\bibfnamefont {Juan}\ \bibnamefont
  {Maldacena}},\ }\bibfield  {title} {\enquote {\bibinfo {title} {{The Large-N
  Limit of Superconformal Field Theories and Supergravity}},}\ }\href {\doibase
  10.1023/A:1026654312961} {\bibfield  {journal} {\bibinfo  {journal}
  {International Journal of Theoretical Physics}\ }\textbf {\bibinfo {volume}
  {38}},\ \bibinfo {pages} {1113--1133} (\bibinfo {year} {1999})}\BibitemShut
  {NoStop}%
\bibitem [{\citenamefont {Witten}(1998)}]{Witten}%
  \BibitemOpen
  \bibfield  {author} {\bibinfo {author} {\bibfnamefont {Edward}\ \bibnamefont
  {Witten}},\ }\bibfield  {title} {\enquote {\bibinfo {title} {{Anti de Sitter
  space and holography}},}\ }\href
  {https://dx.doi.org/10.4310/ATMP.1998.v2.n2.a2} {\bibfield  {journal}
  {\bibinfo  {journal} {Advances in Theoretical and Mathematical Physics}\
  }\textbf {\bibinfo {volume} {2}},\ \bibinfo {pages} {253--291} (\bibinfo
  {year} {1998})}\BibitemShut {NoStop}%
\bibitem [{\citenamefont {Planck}(1900)}]{Planck1900}%
  \BibitemOpen
  \bibfield  {author} {\bibinfo {author} {\bibfnamefont {Max Karl
  Ernst~Ludwig}\ \bibnamefont {Planck}},\ }\bibfield  {title} {\enquote
  {\bibinfo {title} {{{Zur Theorie des Gesetzes der Energieverteilung im
  Normalspectrum}}},}\ }\href {https://cds.cern.ch/record/262745} {\bibfield
  {journal} {\bibinfo  {journal} {Verhandl. Dtsc. Phys. Ges.}\ }\textbf
  {\bibinfo {volume} {2}},\ \bibinfo {pages} {237} (\bibinfo {year}
  {1900})}\BibitemShut {NoStop}%
\bibitem [{\citenamefont {Einstein}(1907)}]{Einstein}%
  \BibitemOpen
  \bibfield  {author} {\bibinfo {author} {\bibfnamefont {A.}~\bibnamefont
  {Einstein}},\ }\bibfield  {title} {\enquote {\bibinfo {title} {{Die
  Plancksche Theorie der Strahlung und die Theorie der spezifischen
  W{\"a}rme}},}\ }\href {\doibase https://doi.org/10.1002/andp.19063270110}
  {\bibfield  {journal} {\bibinfo  {journal} {Annalen der Physik}\ }\textbf
  {\bibinfo {volume} {327}},\ \bibinfo {pages} {180--190} (\bibinfo {year}
  {1907})}\BibitemShut {NoStop}%
\bibitem [{\citenamefont {Debye}(1912)}]{Debye}%
  \BibitemOpen
  \bibfield  {author} {\bibinfo {author} {\bibfnamefont {P.}~\bibnamefont
  {Debye}},\ }\bibfield  {title} {\enquote {\bibinfo {title} {{Zur Theorie der
  spezifischen W{\"a}rmen}},}\ }\href {\doibase
  https://doi.org/10.1002/andp.19123441404} {\bibfield  {journal} {\bibinfo
  {journal} {Annalen der Physik}\ }\textbf {\bibinfo {volume} {344}},\ \bibinfo
  {pages} {789--839} (\bibinfo {year} {1912})}\BibitemShut {NoStop}%
\bibitem [{Note1()}]{Note1}%
  \BibitemOpen
  \bibinfo {note} {Notwithstanding similarity in name, these ubiquitous thermal
  time scales set by $\tau _{Planck}$ are, of course, not to be confused with
  the (possibly experimentally unattainable \cite {aba1}) Planck time ($t_{P} =
  \protect \sqrt {\hbar G/c^5}$ with $G$ the gravitational constant and $c$ the
  speed of light) below which non-renormalizable quantum gravity effects
  appears. Both the thermal ``Planck scale'' $\tau _{Planck}$ (a term coined,
  in the context of thermal dissipation, by Zaanen \cite {planck3}) and the
  Planck time $t_P$ are associated with hitherto conjectured shortest possible
  timescales. In the current work, we rigorously establish the thermal Plackian
  time $\tau _{Planck}$ to be a minimal possible scale for the variation of
  local observables in typical thermal systems. As we will explain, our exact
  inequalities involve effective local heat capacities whose scale is typically
  set by the Boltzmann constant $k_B$ yet not exactly equal to it (especially
  at low temperatures where there may be a strong quantum suppression of these
  heat capacities leading to yet stronger inequalities on the allowed rates of
  change of local observables). This, in turn, gives rise to bounds with
  dimensionless prefactors amending $\tau _{Planck}$.}\BibitemShut {Stop}%
\bibitem [{\citenamefont {Eyring}(1935)}]{eyring}%
  \BibitemOpen
  \bibfield  {author} {\bibinfo {author} {\bibfnamefont {Henry}\ \bibnamefont
  {Eyring}},\ }\bibfield  {title} {\enquote {\bibinfo {title} {The activated
  complex in chemical reactions},}\ }\href {\doibase 10.1063/1.1749604}
  {\bibfield  {journal} {\bibinfo  {journal} {The Journal of Chemical Physics}\
  }\textbf {\bibinfo {volume} {3}},\ \bibinfo {pages} {107--115} (\bibinfo
  {year} {1935})}\BibitemShut {NoStop}%
\bibitem [{\citenamefont {Eyring}(1936)}]{eyringViscosity}%
  \BibitemOpen
  \bibfield  {author} {\bibinfo {author} {\bibfnamefont {Henry}\ \bibnamefont
  {Eyring}},\ }\bibfield  {title} {\enquote {\bibinfo {title} {Viscosity,
  plasticity, and diffusion as examples of absolute reaction rates},}\ }\href
  {\doibase 10.1063/1.1749836} {\bibfield  {journal} {\bibinfo  {journal} {The
  Journal of Chemical Physics}\ }\textbf {\bibinfo {volume} {4}},\ \bibinfo
  {pages} {283--291} (\bibinfo {year} {1936})}\BibitemShut {NoStop}%
\bibitem [{\citenamefont {Wigner}(1932)}]{Wigner}%
  \BibitemOpen
  \bibfield  {author} {\bibinfo {author} {\bibfnamefont {E.}~\bibnamefont
  {Wigner}},\ }\bibfield  {title} {\enquote {\bibinfo {title} {{On the Quantum
  Correction For Thermodynamic Equilibrium}},}\ }\href {\doibase
  10.1103/PhysRev.40.749} {\bibfield  {journal} {\bibinfo  {journal} {Physical
  Review}\ }\textbf {\bibinfo {volume} {40}},\ \bibinfo {pages} {749} (\bibinfo
  {year} {1932})}\BibitemShut {NoStop}%
\bibitem [{\citenamefont {Abrikosov}\ \emph {et~al.}(1975)\citenamefont
  {Abrikosov}, \citenamefont {Gor'kov},\ and\ \citenamefont
  {Dzyaloshinshkii}}]{AGD}%
  \BibitemOpen
  \bibfield  {author} {\bibinfo {author} {\bibfnamefont {A.~A.}\ \bibnamefont
  {Abrikosov}}, \bibinfo {author} {\bibfnamefont {L.~P.}\ \bibnamefont
  {Gor'kov}}, \ and\ \bibinfo {author} {\bibfnamefont {I.~E.}\ \bibnamefont
  {Dzyaloshinshkii}},\ }\href@noop {} {\emph {\bibinfo {title} {{Methods of
  Quantum Field Theory in Statistical Physics}}}}\ (\bibinfo  {publisher}
  {Dover Publications},\ \bibinfo {year} {1975})\BibitemShut {NoStop}%
\bibitem [{\citenamefont {Coleman}(2015)}]{coleman}%
  \BibitemOpen
  \bibfield  {author} {\bibinfo {author} {\bibfnamefont {Piers}\ \bibnamefont
  {Coleman}},\ }\href@noop {} {\emph {\bibinfo {title} {{Introduction to
  many-body physics}}}}\ (\bibinfo  {publisher} {Cambridge University Press},\
  \bibinfo {year} {2015})\BibitemShut {NoStop}%
\bibitem [{\citenamefont {Das}(1997)}]{das}%
  \BibitemOpen
  \bibfield  {author} {\bibinfo {author} {\bibfnamefont {Ashok}\ \bibnamefont
  {Das}},\ }\href@noop {} {\emph {\bibinfo {title} {{Finite temperature field
  theory}}}}\ (\bibinfo  {publisher} {World scientific},\ \bibinfo {year}
  {1997})\BibitemShut {NoStop}%
\bibitem [{\citenamefont {Hawking}(1975)}]{Hawking}%
  \BibitemOpen
  \bibfield  {author} {\bibinfo {author} {\bibfnamefont {S.~W.}\ \bibnamefont
  {Hawking}},\ }\bibfield  {title} {\enquote {\bibinfo {title} {Particle
  creation by black holes},}\ }\href {\doibase 10.1007/BF02345020} {\bibfield
  {journal} {\bibinfo  {journal} {Comm. Math. Phys.}\ }\textbf {\bibinfo
  {volume} {43}},\ \bibinfo {pages} {199} (\bibinfo {year} {1975})}\BibitemShut
  {NoStop}%
\bibitem [{\citenamefont {Davies}(1975)}]{Un1}%
  \BibitemOpen
  \bibfield  {author} {\bibinfo {author} {\bibfnamefont {P.~C.~W.}\
  \bibnamefont {Davies}},\ }\bibfield  {title} {\enquote {\bibinfo {title}
  {Scalar production in schwarzschild and rindler metrics},}\ }\href
  {https://doi.org/10.1088/0305-4470/8/4/022} {\bibfield  {journal} {\bibinfo
  {journal} {Journal of Physics A}\ }\textbf {\bibinfo {volume} {8}} (\bibinfo
  {year} {1975})}\BibitemShut {NoStop}%
\bibitem [{\citenamefont {Unruh}(1976)}]{Un-2}%
  \BibitemOpen
  \bibfield  {author} {\bibinfo {author} {\bibfnamefont {W.~G.}\ \bibnamefont
  {Unruh}},\ }\bibfield  {title} {\enquote {\bibinfo {title} {Notes on
  black-hole evaporation},}\ }\href
  {https://link.aps.org/doi/10.1103/PhysRevD.14.870} {\bibfield  {journal}
  {\bibinfo  {journal} {Physical Review D}\ }\textbf {\bibinfo {volume} {14}}
  (\bibinfo {year} {1976})}\BibitemShut {NoStop}%
\bibitem [{\citenamefont {van~der Marel}\ \emph {et~al.}(2003)\citenamefont
  {van~der Marel}, \citenamefont {Molegraaf}, \citenamefont {Zannen},
  \citenamefont {Nussinov}, \citenamefont {Carbone}, \citenamefont
  {Damascelli}, \citenamefont {Eisaki}, \citenamefont {Greven}, \citenamefont
  {Kes},\ and\ \citenamefont {Li}}]{Marel}%
  \BibitemOpen
  \bibfield  {author} {\bibinfo {author} {\bibfnamefont {D.}~\bibnamefont
  {van~der Marel}}, \bibinfo {author} {\bibfnamefont {H.~J.~A.}\ \bibnamefont
  {Molegraaf}}, \bibinfo {author} {\bibfnamefont {J.}~\bibnamefont {Zannen}},
  \bibinfo {author} {\bibfnamefont {Z.}~\bibnamefont {Nussinov}}, \bibinfo
  {author} {\bibfnamefont {F.}~\bibnamefont {Carbone}}, \bibinfo {author}
  {\bibfnamefont {A.}~\bibnamefont {Damascelli}}, \bibinfo {author}
  {\bibfnamefont {H.}~\bibnamefont {Eisaki}}, \bibinfo {author} {\bibfnamefont
  {M.}~\bibnamefont {Greven}}, \bibinfo {author} {\bibfnamefont {P.~H.}\
  \bibnamefont {Kes}}, \ and\ \bibinfo {author} {\bibfnamefont
  {M.}~\bibnamefont {Li}},\ }\bibfield  {title} {\enquote {\bibinfo {title}
  {Quantum critical behaviour in a high-$t_c$ superconductor},}\ }\href
  {\doibase 10.1038/nature01978} {\bibfield  {journal} {\bibinfo  {journal}
  {Nature}\ }\textbf {\bibinfo {volume} {425}},\ \bibinfo {pages} {271}
  (\bibinfo {year} {2003})}\BibitemShut {NoStop}%
\bibitem [{\citenamefont {Hertz}(1976)}]{Hertz}%
  \BibitemOpen
  \bibfield  {author} {\bibinfo {author} {\bibfnamefont {John~A.}\ \bibnamefont
  {Hertz}},\ }\bibfield  {title} {\enquote {\bibinfo {title} {{Quantum critical
  phenomena}},}\ }\href {\doibase 10.1103/PhysRevB.14.1165} {\bibfield
  {journal} {\bibinfo  {journal} {Phys. Rev. B}\ }\textbf {\bibinfo {volume}
  {14}},\ \bibinfo {pages} {1165--1184} (\bibinfo {year} {1976})}\BibitemShut
  {NoStop}%
\bibitem [{\citenamefont {Sachdev}(2011)}]{subir}%
  \BibitemOpen
  \bibfield  {author} {\bibinfo {author} {\bibfnamefont {Subir}\ \bibnamefont
  {Sachdev}},\ }\href {\doibase 10.1017/CBO9780511973765} {\emph {\bibinfo
  {title} {{Quantum Phase Transitions}}}},\ \bibinfo {edition} {2nd}\ ed.\
  (\bibinfo  {publisher} {Cambridge University Press},\ \bibinfo {year}
  {2011})\BibitemShut {NoStop}%
\bibitem [{\citenamefont {Varma}\ \emph {et~al.}(2002)\citenamefont {Varma},
  \citenamefont {Nussinov},\ and\ \citenamefont {{van Saarloos}}}]{NFL}%
  \BibitemOpen
  \bibfield  {author} {\bibinfo {author} {\bibfnamefont {C.M.}\ \bibnamefont
  {Varma}}, \bibinfo {author} {\bibfnamefont {Z.}~\bibnamefont {Nussinov}}, \
  and\ \bibinfo {author} {\bibfnamefont {Wim}\ \bibnamefont {{van Saarloos}}},\
  }\bibfield  {title} {\enquote {\bibinfo {title} {{Singular or non-Fermi
  liquids}},}\ }\href {\doibase https://doi.org/10.1016/S0370-1573(01)00060-6}
  {\bibfield  {journal} {\bibinfo  {journal} {Physics Reports}\ }\textbf
  {\bibinfo {volume} {361}},\ \bibinfo {pages} {267--417} (\bibinfo {year}
  {2002})}\BibitemShut {NoStop}%
\bibitem [{\citenamefont {Nussinov}\ \emph
  {et~al.}(2014{\natexlab{b}})\citenamefont {Nussinov}, \citenamefont
  {Madziwa-Nussinov},\ and\ \citenamefont {Nussinov}}]{musketeers}%
  \BibitemOpen
  \bibfield  {author} {\bibinfo {author} {\bibfnamefont {S}~\bibnamefont
  {Nussinov}}, \bibinfo {author} {\bibfnamefont {T}~\bibnamefont
  {Madziwa-Nussinov}}, \ and\ \bibinfo {author} {\bibfnamefont {Z}~\bibnamefont
  {Nussinov}},\ }\bibfield  {title} {\enquote {\bibinfo {title} {{Decoherence
  due to thermal effects in two quintessential quantum systems}},}\ }\href
  {\doibase 10.1007/s40509-014-0004-8} {\bibfield  {journal} {\bibinfo
  {journal} {Quantum Studies: Mathematics and Foundations}\ }\textbf {\bibinfo
  {volume} {1}},\ \bibinfo {pages} {155--164} (\bibinfo {year}
  {2014}{\natexlab{b}})}\BibitemShut {NoStop}%
\bibitem [{\citenamefont {Sekino}\ and\ \citenamefont
  {Susskind}(2008)}]{op-scram1}%
  \BibitemOpen
  \bibfield  {author} {\bibinfo {author} {\bibfnamefont {Yasuhiro}\
  \bibnamefont {Sekino}}\ and\ \bibinfo {author} {\bibfnamefont
  {L}~\bibnamefont {Susskind}},\ }\bibfield  {title} {\enquote {\bibinfo
  {title} {{Fast scramblers}},}\ }\href {\doibase
  10.1088/1126-6708/2008/10/065} {\bibfield  {journal} {\bibinfo  {journal}
  {Journal of High Energy Physics}\ }\textbf {\bibinfo {volume} {2008}},\
  \bibinfo {pages} {065--065} (\bibinfo {year} {2008})}\BibitemShut {NoStop}%
\bibitem [{\citenamefont {Chen}\ and\ \citenamefont {Zhou}(2018)}]{op-scram2}%
  \BibitemOpen
  \bibfield  {author} {\bibinfo {author} {\bibfnamefont {Xiao}\ \bibnamefont
  {Chen}}\ and\ \bibinfo {author} {\bibfnamefont {Tianci}\ \bibnamefont
  {Zhou}},\ }\href {https://arxiv.org/pdf/1804.08655} {\enquote {\bibinfo
  {title} {{Operator scrambling and quantum chaos}},}\ } (\bibinfo {year}
  {2018})\BibitemShut {NoStop}%
\bibitem [{\citenamefont {Liu}\ and\ \citenamefont {Suh}(2014)}]{ent-growth}%
  \BibitemOpen
  \bibfield  {author} {\bibinfo {author} {\bibfnamefont {Hong}\ \bibnamefont
  {Liu}}\ and\ \bibinfo {author} {\bibfnamefont {S.~Josephine}\ \bibnamefont
  {Suh}},\ }\bibfield  {title} {\enquote {\bibinfo {title} {{Entanglement
  growth during thermalization in holographic systems}},}\ }\href {\doibase
  10.1103/PhysRevD.89.066012} {\bibfield  {journal} {\bibinfo  {journal} {Phys.
  Rev. D}\ }\textbf {\bibinfo {volume} {89}},\ \bibinfo {pages} {066012}
  (\bibinfo {year} {2014})}\BibitemShut {NoStop}%
\bibitem [{\citenamefont {Goldstein}\ \emph {et~al.}(2015)\citenamefont
  {Goldstein}, \citenamefont {Hara},\ and\ \citenamefont {Tasaki}}]{typical0}%
  \BibitemOpen
  \bibfield  {author} {\bibinfo {author} {\bibfnamefont {Sheldon}\ \bibnamefont
  {Goldstein}}, \bibinfo {author} {\bibfnamefont {Takashi}\ \bibnamefont
  {Hara}}, \ and\ \bibinfo {author} {\bibfnamefont {Hal}\ \bibnamefont
  {Tasaki}},\ }\bibfield  {title} {\enquote {\bibinfo {title} {{Extremely quick
  thermalization in a macroscopic quantum system for a typical nonequilibrium
  subspace}},}\ }\href {\doibase 10.1088/1367-2630/17/4/045002} {\bibfield
  {journal} {\bibinfo  {journal} {New Journal of Physics}\ }\textbf {\bibinfo
  {volume} {17}},\ \bibinfo {pages} {045002} (\bibinfo {year}
  {2015})}\BibitemShut {NoStop}%
\bibitem [{\citenamefont {Goldstein}\ \emph {et~al.}(2013)\citenamefont
  {Goldstein}, \citenamefont {Hara},\ and\ \citenamefont {Tasaki}}]{typical}%
  \BibitemOpen
  \bibfield  {author} {\bibinfo {author} {\bibfnamefont {Sheldon}\ \bibnamefont
  {Goldstein}}, \bibinfo {author} {\bibfnamefont {Takashi}\ \bibnamefont
  {Hara}}, \ and\ \bibinfo {author} {\bibfnamefont {Hal}\ \bibnamefont
  {Tasaki}},\ }\bibfield  {title} {\enquote {\bibinfo {title} {{Time Scales in
  the Approach to Equilibrium of Macroscopic Quantum Systems}},}\ }\href
  {\doibase 10.1103/PhysRevLett.111.140401} {\bibfield  {journal} {\bibinfo
  {journal} {Phys. Rev. Lett.}\ }\textbf {\bibinfo {volume} {111}},\ \bibinfo
  {pages} {140401} (\bibinfo {year} {2013})}\BibitemShut {NoStop}%
\bibitem [{\citenamefont {Sachdev}\ and\ \citenamefont {Ye}(1993)}]{SY-SYK}%
  \BibitemOpen
  \bibfield  {author} {\bibinfo {author} {\bibfnamefont {S.}~\bibnamefont
  {Sachdev}}\ and\ \bibinfo {author} {\bibfnamefont {J.}~\bibnamefont {Ye}},\
  }\bibfield  {title} {\enquote {\bibinfo {title} {{Gapless Spin-Fluid Ground
  State in a Random Quantum Heisenberg Magnet}},}\ }\href {\doibase
  10.1103/PhysRevLett.70.3339} {\bibfield  {journal} {\bibinfo  {journal}
  {Phys. Rev. Lett.}\ }\textbf {\bibinfo {volume} {70}},\ \bibinfo {pages}
  {3339} (\bibinfo {year} {1993})}\BibitemShut {NoStop}%
\bibitem [{K-S()}]{K-SYK}%
  \BibitemOpen
  \href@noop {} {\ }\bibinfo {note} {A. Kitaev, A simple model of quantum
  holography,
  \href{http://online.kitp.ucsb.edu/online/entangled15/kitaev/}{http://online.kitp.ucsb.edu/online/entangled15/kitaev/},
  \href{http://online.kitp.ucsb.edu/online/entangled15/kitaev2/}{http://online.kitp.ucsb.edu/online/entangled15/kitaev2/}}\BibitemShut
  {NoStop}%
\bibitem [{\citenamefont {Maldacena}\ and\ \citenamefont
  {Stanford}(2016)}]{Juan-SYK}%
  \BibitemOpen
  \bibfield  {author} {\bibinfo {author} {\bibfnamefont {J.}~\bibnamefont
  {Maldacena}}\ and\ \bibinfo {author} {\bibfnamefont {D.}~\bibnamefont
  {Stanford}},\ }\bibfield  {title} {\enquote {\bibinfo {title} {{Remarks on
  the Sachdev-Ye-Kitaev model}},}\ }\href {\doibase 10.1103/PhysRevD.94.106002}
  {\bibfield  {journal} {\bibinfo  {journal} {Phys. Rev. D}\ }\textbf {\bibinfo
  {volume} {94}},\ \bibinfo {pages} {106002} (\bibinfo {year}
  {2016})}\BibitemShut {NoStop}%
\bibitem [{\citenamefont {Gross}\ and\ \citenamefont
  {Rosenhaus}(2017)}]{0-SYK}%
  \BibitemOpen
  \bibfield  {author} {\bibinfo {author} {\bibfnamefont {David~J.}\
  \bibnamefont {Gross}}\ and\ \bibinfo {author} {\bibfnamefont {Vladimir}\
  \bibnamefont {Rosenhaus}},\ }\bibfield  {title} {\enquote {\bibinfo {title}
  {{A Generalization of Sachdev-Ye-Kitaev Model}},}\ }\href
  {https://doi.org/10.1007/JHEP02(2017)093} {\bibfield  {journal} {\bibinfo
  {journal} {J. High Energy Phys}\ }\textbf {\bibinfo {volume} {2017}},\
  \bibinfo {pages} {93} (\bibinfo {year} {2017})}\BibitemShut {NoStop}%
\bibitem [{\citenamefont {Davison}\ \emph {et~al.}(2017)\citenamefont
  {Davison}, \citenamefont {Fu}, \citenamefont {Georges}, \citenamefont {Gu},
  \citenamefont {Jensen},\ and\ \citenamefont {Sachdev}}]{1-SYK}%
  \BibitemOpen
  \bibfield  {author} {\bibinfo {author} {\bibfnamefont {Richard~A.}\
  \bibnamefont {Davison}}, \bibinfo {author} {\bibfnamefont {Wenbo}\
  \bibnamefont {Fu}}, \bibinfo {author} {\bibfnamefont {Antoine}\ \bibnamefont
  {Georges}}, \bibinfo {author} {\bibfnamefont {Yingfei}\ \bibnamefont {Gu}},
  \bibinfo {author} {\bibfnamefont {Kristan}\ \bibnamefont {Jensen}}, \ and\
  \bibinfo {author} {\bibfnamefont {Subir}\ \bibnamefont {Sachdev}},\
  }\bibfield  {title} {\enquote {\bibinfo {title} {Thermoelectric transport in
  disordered metals without quasiparticles: The sachdev-ye-kitaev models and
  holography},}\ }\href {\doibase 10.1103/PhysRevB.95.155131} {\bibfield
  {journal} {\bibinfo  {journal} {Phys. Rev. B}\ }\textbf {\bibinfo {volume}
  {95}},\ \bibinfo {pages} {155131} (\bibinfo {year} {2017})}\BibitemShut
  {NoStop}%
\bibitem [{\citenamefont {Kitaev}\ and\ \citenamefont {Suh}(2018)}]{2-SYK}%
  \BibitemOpen
  \bibfield  {author} {\bibinfo {author} {\bibfnamefont {A.}~\bibnamefont
  {Kitaev}}\ and\ \bibinfo {author} {\bibfnamefont {S.~Josephine}\ \bibnamefont
  {Suh}},\ }\bibfield  {title} {\enquote {\bibinfo {title} {{The soft mode in
  the Sachdev-Ye-Kitaev model and its gravity dual}},}\ }\href {\doibase
  10.1007/JHEP05(2018)183} {\bibfield  {journal} {\bibinfo  {journal} {J. High
  Energy Phys.}\ }\textbf {\bibinfo {volume} {2018}},\ \bibinfo {pages} {183}
  (\bibinfo {year} {2018})}\BibitemShut {NoStop}%
\bibitem [{\citenamefont {Song}\ \emph {et~al.}(2017)\citenamefont {Song},
  \citenamefont {Jian},\ and\ \citenamefont {Balents}}]{3-SYK}%
  \BibitemOpen
  \bibfield  {author} {\bibinfo {author} {\bibfnamefont {Xue-Yang}\
  \bibnamefont {Song}}, \bibinfo {author} {\bibfnamefont {Chao-Ming}\
  \bibnamefont {Jian}}, \ and\ \bibinfo {author} {\bibfnamefont {Leon}\
  \bibnamefont {Balents}},\ }\bibfield  {title} {\enquote {\bibinfo {title}
  {Strongly correlated metal built from sachdev-ye-kitaev models},}\ }\href
  {\doibase 10.1103/PhysRevLett.119.216601} {\bibfield  {journal} {\bibinfo
  {journal} {Phys. Rev. Lett.}\ }\textbf {\bibinfo {volume} {119}},\ \bibinfo
  {pages} {216601} (\bibinfo {year} {2017})}\BibitemShut {NoStop}%
\bibitem [{\citenamefont {Patel}\ \emph {et~al.}(2018)\citenamefont {Patel},
  \citenamefont {McGreevy}, \citenamefont {Arovas},\ and\ \citenamefont
  {Sachdev}}]{4-SYK}%
  \BibitemOpen
  \bibfield  {author} {\bibinfo {author} {\bibfnamefont {Aavishkar~A.}\
  \bibnamefont {Patel}}, \bibinfo {author} {\bibfnamefont {John}\ \bibnamefont
  {McGreevy}}, \bibinfo {author} {\bibfnamefont {Daniel~P.}\ \bibnamefont
  {Arovas}}, \ and\ \bibinfo {author} {\bibfnamefont {Subir}\ \bibnamefont
  {Sachdev}},\ }\bibfield  {title} {\enquote {\bibinfo {title}
  {Magnetotransport in a model of a disordered strange metal},}\ }\href
  {\doibase 10.1103/PhysRevX.8.021049} {\bibfield  {journal} {\bibinfo
  {journal} {Phys. Rev. X}\ }\textbf {\bibinfo {volume} {8}},\ \bibinfo {pages}
  {021049} (\bibinfo {year} {2018})}\BibitemShut {NoStop}%
\bibitem [{\citenamefont {Tulipman}\ and\ \citenamefont {Berg}(2020)}]{5-SYK}%
  \BibitemOpen
  \bibfield  {author} {\bibinfo {author} {\bibfnamefont {Evyatar}\ \bibnamefont
  {Tulipman}}\ and\ \bibinfo {author} {\bibfnamefont {Erez}\ \bibnamefont
  {Berg}},\ }\bibfield  {title} {\enquote {\bibinfo {title} {{Strongly coupled
  quantum phonon fluid in a solvable model}},}\ }\href {\doibase
  10.1103/PhysRevResearch.2.033431} {\bibfield  {journal} {\bibinfo  {journal}
  {Physical Review Research}\ }\textbf {\bibinfo {volume} {2}},\ \bibinfo
  {pages} {033431} (\bibinfo {year} {2020})}\BibitemShut {NoStop}%
\bibitem [{\citenamefont {Araki}\ and\ \citenamefont {Lieb}(2002)}]{lieb-pure}%
  \BibitemOpen
  \bibfield  {author} {\bibinfo {author} {\bibfnamefont {Huzihiro}\
  \bibnamefont {Araki}}\ and\ \bibinfo {author} {\bibfnamefont {Elliott~H}\
  \bibnamefont {Lieb}},\ }\bibfield  {title} {\enquote {\bibinfo {title}
  {{Entropy inequalities}},}\ }in\ \href {\doibase 10.1007/978-3-642-55925-9_4}
  {\emph {\bibinfo {booktitle} {Inequalities}}}\ (\bibinfo  {publisher}
  {Springer},\ \bibinfo {year} {2002})\ pp.\ \bibinfo {pages}
  {47--57}\BibitemShut {NoStop}%
\bibitem [{\citenamefont {Heisenberg}(1927)}]{uncertain}%
  \BibitemOpen
  \bibfield  {author} {\bibinfo {author} {\bibfnamefont {W.}~\bibnamefont
  {Heisenberg}},\ }\bibfield  {title} {\enquote {\bibinfo {title} {{{\"U}}ber
  den anschaulichen inhalt der quantentheoretischen kinematik und mechanik},}\
  }\href {\doibase 10.1007/BF01397280} {\bibfield  {journal} {\bibinfo
  {journal} {Zeitschrift f{\"u}r Physik}\ }\textbf {\bibinfo {volume} {43}},\
  \bibinfo {pages} {172--198} (\bibinfo {year} {1927})}\BibitemShut {NoStop}%
\bibitem [{\citenamefont {Kennard}(1927)}]{Konrad}%
  \BibitemOpen
  \bibfield  {author} {\bibinfo {author} {\bibfnamefont {E.~H.}\ \bibnamefont
  {Kennard}},\ }\bibfield  {title} {\enquote {\bibinfo {title} {{Zur
  Quantenmechanik einfacher Bewegungstypen}},}\ }\href {\doibase
  10.1007/BF01391200} {\bibfield  {journal} {\bibinfo  {journal} {Zeitschrift
  f{\"u}r Physik}\ }\textbf {\bibinfo {volume} {44}},\ \bibinfo {pages}
  {326--352} (\bibinfo {year} {1927})}\BibitemShut {NoStop}%
\bibitem [{\citenamefont {Robertson}(1929)}]{Robertson}%
  \BibitemOpen
  \bibfield  {author} {\bibinfo {author} {\bibfnamefont {H.~P.}\ \bibnamefont
  {Robertson}},\ }\bibfield  {title} {\enquote {\bibinfo {title} {{The
  Uncertainty Principle}},}\ }\href {\doibase 10.1103/PhysRev.34.163}
  {\bibfield  {journal} {\bibinfo  {journal} {Phys. Rev.}\ }\textbf {\bibinfo
  {volume} {34}},\ \bibinfo {pages} {163--164} (\bibinfo {year}
  {1929})}\BibitemShut {NoStop}%
\bibitem [{\citenamefont {L.}\ and\ \citenamefont
  {I.}(1945)}]{Mandelshtam-Tamm}%
  \BibitemOpen
  \bibfield  {author} {\bibinfo {author} {\bibfnamefont {Mandelstam}\
  \bibnamefont {L.}}\ and\ \bibinfo {author} {\bibfnamefont {Tamm}\
  \bibnamefont {I.}},\ }\bibfield  {title} {\enquote {\bibinfo {title} {{The
  Uncertainty Principle}},}\ }\href {\doibase 10.1007/978-3-642-74626-0_8}
  {\bibfield  {journal} {\bibinfo  {journal} {J. Phys. (USSR)}\ }\textbf
  {\bibinfo {volume} {9}},\ \bibinfo {pages} {249} (\bibinfo {year}
  {1945})}\BibitemShut {NoStop}%
\bibitem [{\citenamefont {Susskind}\ and\ \citenamefont
  {Glogower}(1964)}]{lenny}%
  \BibitemOpen
  \bibfield  {author} {\bibinfo {author} {\bibfnamefont {Leonard}\ \bibnamefont
  {Susskind}}\ and\ \bibinfo {author} {\bibfnamefont {Jonathan}\ \bibnamefont
  {Glogower}},\ }\bibfield  {title} {\enquote {\bibinfo {title} {{Quantum
  mechanical phase and time operator}},}\ }\href {\doibase
  10.1103/PhysicsPhysiqueFizika.1.49} {\bibfield  {journal} {\bibinfo
  {journal} {Physics Physique Fizika}\ }\textbf {\bibinfo {volume} {1}},\
  \bibinfo {pages} {49--61} (\bibinfo {year} {1964})}\BibitemShut {NoStop}%
\bibitem [{\citenamefont {Davidson}(1965)}]{Davidson}%
  \BibitemOpen
  \bibfield  {author} {\bibinfo {author} {\bibfnamefont {Ernest~R.}\
  \bibnamefont {Davidson}},\ }\bibfield  {title} {\enquote {\bibinfo {title}
  {{On Derivations of the Uncertainty Principle}},}\ }\href {\doibase
  10.1063/1.1696139} {\bibfield  {journal} {\bibinfo  {journal} {The Journal of
  Chemical Physics}\ }\textbf {\bibinfo {volume} {42}},\ \bibinfo {pages}
  {1461--1462} (\bibinfo {year} {1965})}\BibitemShut {NoStop}%
\bibitem [{\citenamefont {Anandan}\ and\ \citenamefont
  {Aharonov}(1990)}]{Anandan-Aharonov}%
  \BibitemOpen
  \bibfield  {author} {\bibinfo {author} {\bibfnamefont {J.}~\bibnamefont
  {Anandan}}\ and\ \bibinfo {author} {\bibfnamefont {Y.}~\bibnamefont
  {Aharonov}},\ }\bibfield  {title} {\enquote {\bibinfo {title} {{Geometry of
  quantum evolution}},}\ }\href {\doibase 10.1103/PhysRevLett.65.1697}
  {\bibfield  {journal} {\bibinfo  {journal} {Phys. Rev. Lett.}\ }\textbf
  {\bibinfo {volume} {65}},\ \bibinfo {pages} {1697--1700} (\bibinfo {year}
  {1990})}\BibitemShut {NoStop}%
\bibitem [{\citenamefont {Bogoliubov}(1962)}]{Bogol}%
  \BibitemOpen
  \bibfield  {author} {\bibinfo {author} {\bibfnamefont {N.~N.}\ \bibnamefont
  {Bogoliubov}},\ }\href@noop {} {\bibfield  {journal} {\bibinfo  {journal}
  {Physik. Abhandl. Sowjetunion}\ }\textbf {\bibinfo {volume} {6}},\ \bibinfo
  {pages} {113} (\bibinfo {year} {1962})}\BibitemShut {NoStop}%
\bibitem [{\citenamefont {Mermin}\ and\ \citenamefont {Wagner}(1966)}]{MW}%
  \BibitemOpen
  \bibfield  {author} {\bibinfo {author} {\bibfnamefont {N.~D.}\ \bibnamefont
  {Mermin}}\ and\ \bibinfo {author} {\bibfnamefont {H.}~\bibnamefont
  {Wagner}},\ }\bibfield  {title} {\enquote {\bibinfo {title} {{Absence of
  Ferromagnetism or Antiferromagnetism in One- or Two-Dimensional Isotropic
  Heisenberg Models}},}\ }\href {\doibase 10.1103/PhysRevLett.17.1133}
  {\bibfield  {journal} {\bibinfo  {journal} {Phys. Rev. Lett.}\ }\textbf
  {\bibinfo {volume} {17}},\ \bibinfo {pages} {1133--1136} (\bibinfo {year}
  {1966})}\BibitemShut {NoStop}%
\bibitem [{\citenamefont {Nussinov}(2004)}]{naglass}%
  \BibitemOpen
  \bibfield  {author} {\bibinfo {author} {\bibfnamefont {Zohar}\ \bibnamefont
  {Nussinov}},\ }\bibfield  {title} {\enquote {\bibinfo {title} {{Avoided Phase
  Transitions and Glassy Dynamics In Geometrically Frustrated Systems and
  Non-Abelian Theories}},}\ }\href {\doibase 10.1103/PhysRevB.69.014208}
  {\bibfield  {journal} {\bibinfo  {journal} {Physical Review B}\ }\textbf
  {\bibinfo {volume} {69}},\ \bibinfo {pages} {014208} (\bibinfo {year}
  {2004})}\BibitemShut {NoStop}%
\bibitem [{\citenamefont {Harris}\ \emph {et~al.}(2003)\citenamefont {Harris},
  \citenamefont {Yildirim}, \citenamefont {Aharony}, \citenamefont
  {Entin-Wohlman},\ and\ \citenamefont {Korenblit}}]{harrisa}%
  \BibitemOpen
  \bibfield  {author} {\bibinfo {author} {\bibfnamefont {A.~B.}\ \bibnamefont
  {Harris}}, \bibinfo {author} {\bibfnamefont {Taner}\ \bibnamefont
  {Yildirim}}, \bibinfo {author} {\bibfnamefont {Amnon}\ \bibnamefont
  {Aharony}}, \bibinfo {author} {\bibfnamefont {Ora}\ \bibnamefont
  {Entin-Wohlman}}, \ and\ \bibinfo {author} {\bibfnamefont {I.~Ya.}\
  \bibnamefont {Korenblit}},\ }\bibfield  {title} {\enquote {\bibinfo {title}
  {{Unusual Symmetries in the Kugel-Khomskii Hamiltonian}},}\ }\href {\doibase
  10.1103/PhysRevLett.91.087206} {\bibfield  {journal} {\bibinfo  {journal}
  {Phys. Rev. Lett.}\ }\textbf {\bibinfo {volume} {91}},\ \bibinfo {pages}
  {087206} (\bibinfo {year} {2003})}\BibitemShut {NoStop}%
\bibitem [{\citenamefont {Batista}\ and\ \citenamefont
  {Nussinov}(2005)}]{boundary-bounds1}%
  \BibitemOpen
  \bibfield  {author} {\bibinfo {author} {\bibfnamefont {C.~D.}\ \bibnamefont
  {Batista}}\ and\ \bibinfo {author} {\bibfnamefont {Zohar}\ \bibnamefont
  {Nussinov}},\ }\bibfield  {title} {\enquote {\bibinfo {title} {{Generalized
  Elitzur's theorem and dimensional reductions}},}\ }\href {\doibase
  10.1103/PhysRevB.72.045137} {\bibfield  {journal} {\bibinfo  {journal} {Phys.
  Rev. B}\ }\textbf {\bibinfo {volume} {72}},\ \bibinfo {pages} {045137}
  (\bibinfo {year} {2005})}\BibitemShut {NoStop}%
\bibitem [{\citenamefont {Nussinov}\ and\ \citenamefont
  {Ortiz}(2009)}]{topohigh}%
  \BibitemOpen
  \bibfield  {author} {\bibinfo {author} {\bibfnamefont {Z.}~\bibnamefont
  {Nussinov}}\ and\ \bibinfo {author} {\bibfnamefont {G.}~\bibnamefont
  {Ortiz}},\ }\bibfield  {title} {\enquote {\bibinfo {title} {{A symmetry
  principle for topological quantum order}},}\ }\href {\doibase
  https://doi.org/10.1016/j.aop.2008.11.002} {\bibfield  {journal} {\bibinfo
  {journal} {Annals of Physics}\ }\textbf {\bibinfo {volume} {324}},\ \bibinfo
  {pages} {977} (\bibinfo {year} {2009})}\BibitemShut {NoStop}%
\bibitem [{\citenamefont {Everett}(1957)}]{info0}%
  \BibitemOpen
  \bibfield  {author} {\bibinfo {author} {\bibfnamefont {Hugh}\ \bibnamefont
  {Everett}},\ }\bibfield  {title} {\enquote {\bibinfo {title} {{"Relative
  State" Formulation of Quantum Mechanics}},}\ }\href {\doibase
  10.1103/RevModPhys.29.454} {\bibfield  {journal} {\bibinfo  {journal} {Rev.
  Mod. Phys.}\ }\textbf {\bibinfo {volume} {29}},\ \bibinfo {pages} {454--462}
  (\bibinfo {year} {1957})}\BibitemShut {NoStop}%
\bibitem [{\citenamefont {Hirschman}(1957)}]{info1}%
  \BibitemOpen
  \bibfield  {author} {\bibinfo {author} {\bibfnamefont {I.~I.}\ \bibnamefont
  {Hirschman}},\ }\bibfield  {title} {\enquote {\bibinfo {title} {{A Note on
  Entropy}},}\ }\href {http://www.jstor.org/stable/2372390} {\bibfield
  {journal} {\bibinfo  {journal} {American Journal of Mathematics}\ }\textbf
  {\bibinfo {volume} {79}},\ \bibinfo {pages} {152--156} (\bibinfo {year}
  {1957})}\BibitemShut {NoStop}%
\bibitem [{\citenamefont {Beckner}(1975)}]{info2}%
  \BibitemOpen
  \bibfield  {author} {\bibinfo {author} {\bibfnamefont {William}\ \bibnamefont
  {Beckner}},\ }\bibfield  {title} {\enquote {\bibinfo {title} {{Inequalities
  in Fourier Analysis}},}\ }\href {http://www.jstor.org/stable/1970980}
  {\bibfield  {journal} {\bibinfo  {journal} {Annals of Mathematics}\ }\textbf
  {\bibinfo {volume} {102}},\ \bibinfo {pages} {159--182} (\bibinfo {year}
  {1975})}\BibitemShut {NoStop}%
\bibitem [{\citenamefont {Bia{\l}ynicki-Birula}\ and\ \citenamefont
  {Mycielski}(1975)}]{info3}%
  \BibitemOpen
  \bibfield  {author} {\bibinfo {author} {\bibfnamefont {Iwo}\ \bibnamefont
  {Bia{\l}ynicki-Birula}}\ and\ \bibinfo {author} {\bibfnamefont {Jerzy}\
  \bibnamefont {Mycielski}},\ }\bibfield  {title} {\enquote {\bibinfo {title}
  {{Uncertainty relations for information entropy in wave mechanics}},}\ }\href
  {\doibase 10.1007/BF01608825} {\bibfield  {journal} {\bibinfo  {journal}
  {Communications in Mathematical Physics}\ }\textbf {\bibinfo {volume} {44}},\
  \bibinfo {pages} {129--132} (\bibinfo {year} {1975})}\BibitemShut {NoStop}%
\bibitem [{\citenamefont {Deutsch}(1983)}]{info4}%
  \BibitemOpen
  \bibfield  {author} {\bibinfo {author} {\bibfnamefont {David}\ \bibnamefont
  {Deutsch}},\ }\bibfield  {title} {\enquote {\bibinfo {title} {{Uncertainty in
  Quantum Measurements}},}\ }\href {\doibase 10.1103/PhysRevLett.50.631}
  {\bibfield  {journal} {\bibinfo  {journal} {Phys. Rev. Lett.}\ }\textbf
  {\bibinfo {volume} {50}},\ \bibinfo {pages} {631--633} (\bibinfo {year}
  {1983})}\BibitemShut {NoStop}%
\bibitem [{\citenamefont {Kraus}(1987)}]{info5}%
  \BibitemOpen
  \bibfield  {author} {\bibinfo {author} {\bibfnamefont {K.}~\bibnamefont
  {Kraus}},\ }\bibfield  {title} {\enquote {\bibinfo {title} {{Complementary
  observables and uncertainty relations}},}\ }\href {\doibase
  10.1103/PhysRevD.35.3070} {\bibfield  {journal} {\bibinfo  {journal} {Phys.
  Rev. D}\ }\textbf {\bibinfo {volume} {35}},\ \bibinfo {pages} {3070--3075}
  (\bibinfo {year} {1987})}\BibitemShut {NoStop}%
\bibitem [{\citenamefont {Maassen}\ and\ \citenamefont {Uffink}(1988)}]{info6}%
  \BibitemOpen
  \bibfield  {author} {\bibinfo {author} {\bibfnamefont {Hans}\ \bibnamefont
  {Maassen}}\ and\ \bibinfo {author} {\bibfnamefont {J.~B.~M.}\ \bibnamefont
  {Uffink}},\ }\bibfield  {title} {\enquote {\bibinfo {title} {{Generalized
  entropic uncertainty relations}},}\ }\href {\doibase
  10.1103/PhysRevLett.60.1103} {\bibfield  {journal} {\bibinfo  {journal}
  {Phys. Rev. Lett.}\ }\textbf {\bibinfo {volume} {60}},\ \bibinfo {pages}
  {1103--1106} (\bibinfo {year} {1988})}\BibitemShut {NoStop}%
\bibitem [{\citenamefont {Ghirardi}\ \emph {et~al.}(2003)\citenamefont
  {Ghirardi}, \citenamefont {Marinatto},\ and\ \citenamefont {Romano}}]{info7}%
  \BibitemOpen
  \bibfield  {author} {\bibinfo {author} {\bibfnamefont {GianCarlo}\
  \bibnamefont {Ghirardi}}, \bibinfo {author} {\bibfnamefont {Luca}\
  \bibnamefont {Marinatto}}, \ and\ \bibinfo {author} {\bibfnamefont
  {Raffaele}\ \bibnamefont {Romano}},\ }\bibfield  {title} {\enquote {\bibinfo
  {title} {{An optimal entropic uncertainty relation in a two-dimensional
  Hilbert space}},}\ }\href {\doibase
  https://doi.org/10.1016/j.physleta.2003.08.029} {\bibfield  {journal}
  {\bibinfo  {journal} {Physics Letters A}\ }\textbf {\bibinfo {volume}
  {317}},\ \bibinfo {pages} {32--36} (\bibinfo {year} {2003})}\BibitemShut
  {NoStop}%
\bibitem [{\citenamefont {Christandl}\ and\ \citenamefont
  {Winter}(2005)}]{info8}%
  \BibitemOpen
  \bibfield  {author} {\bibinfo {author} {\bibfnamefont {M.}~\bibnamefont
  {Christandl}}\ and\ \bibinfo {author} {\bibfnamefont {A.}~\bibnamefont
  {Winter}},\ }\bibfield  {title} {\enquote {\bibinfo {title} {{Uncertainty,
  monogamy, and locking of quantum correlations}},}\ }\href {\doibase
  10.1109/TIT.2005.853338} {\bibfield  {journal} {\bibinfo  {journal} {IEEE
  Transactions on Information Theory}\ }\textbf {\bibinfo {volume} {51}},\
  \bibinfo {pages} {3159--3165} (\bibinfo {year} {2005})}\BibitemShut {NoStop}%
\bibitem [{\citenamefont {de~Vicente}\ and\ \citenamefont
  {S\'anchez-Ruiz}(2008)}]{info9}%
  \BibitemOpen
  \bibfield  {author} {\bibinfo {author} {\bibfnamefont {Julio~I.}\
  \bibnamefont {de~Vicente}}\ and\ \bibinfo {author} {\bibfnamefont {Jorge}\
  \bibnamefont {S\'anchez-Ruiz}},\ }\bibfield  {title} {\enquote {\bibinfo
  {title} {{Improved bounds on entropic uncertainty relations}},}\ }\href
  {\doibase 10.1103/PhysRevA.77.042110} {\bibfield  {journal} {\bibinfo
  {journal} {Phys. Rev. A}\ }\textbf {\bibinfo {volume} {77}},\ \bibinfo
  {pages} {042110} (\bibinfo {year} {2008})}\BibitemShut {NoStop}%
\bibitem [{\citenamefont {Renes}\ and\ \citenamefont {Boileau}(2009)}]{info10}%
  \BibitemOpen
  \bibfield  {author} {\bibinfo {author} {\bibfnamefont {Joseph~M.}\
  \bibnamefont {Renes}}\ and\ \bibinfo {author} {\bibfnamefont
  {Jean-Christian}\ \bibnamefont {Boileau}},\ }\bibfield  {title} {\enquote
  {\bibinfo {title} {{Conjectured Strong Complementary Information
  Tradeoff}},}\ }\href {\doibase 10.1103/PhysRevLett.103.020402} {\bibfield
  {journal} {\bibinfo  {journal} {Phys. Rev. Lett.}\ }\textbf {\bibinfo
  {volume} {103}},\ \bibinfo {pages} {020402} (\bibinfo {year}
  {2009})}\BibitemShut {NoStop}%
\bibitem [{\citenamefont {Berta}\ \emph {et~al.}(2010)\citenamefont {Berta},
  \citenamefont {Christandl}, \citenamefont {Colbeck}, \citenamefont {Renes},\
  and\ \citenamefont {Renner}}]{info11}%
  \BibitemOpen
  \bibfield  {author} {\bibinfo {author} {\bibfnamefont {Mario}\ \bibnamefont
  {Berta}}, \bibinfo {author} {\bibfnamefont {Matthias}\ \bibnamefont
  {Christandl}}, \bibinfo {author} {\bibfnamefont {Roger}\ \bibnamefont
  {Colbeck}}, \bibinfo {author} {\bibfnamefont {Joseph~M.}\ \bibnamefont
  {Renes}}, \ and\ \bibinfo {author} {\bibfnamefont {Renato}\ \bibnamefont
  {Renner}},\ }\bibfield  {title} {\enquote {\bibinfo {title} {{The uncertainty
  principle in the presence of quantum memory}},}\ }\href {\doibase
  10.1038/nphys1734} {\bibfield  {journal} {\bibinfo  {journal} {Nature
  Physics}\ }\textbf {\bibinfo {volume} {6}},\ \bibinfo {pages} {659--662}
  (\bibinfo {year} {2010})}\BibitemShut {NoStop}%
\bibitem [{\citenamefont {Coles}\ \emph {et~al.}(2012)\citenamefont {Coles},
  \citenamefont {Colbeck}, \citenamefont {Yu},\ and\ \citenamefont
  {Zwolak}}]{info12}%
  \BibitemOpen
  \bibfield  {author} {\bibinfo {author} {\bibfnamefont {Patrick~J.}\
  \bibnamefont {Coles}}, \bibinfo {author} {\bibfnamefont {Roger}\ \bibnamefont
  {Colbeck}}, \bibinfo {author} {\bibfnamefont {Li}~\bibnamefont {Yu}}, \ and\
  \bibinfo {author} {\bibfnamefont {Michael}\ \bibnamefont {Zwolak}},\
  }\bibfield  {title} {\enquote {\bibinfo {title} {{Uncertainty Relations from
  Simple Entropic Properties}},}\ }\href {\doibase
  10.1103/PhysRevLett.108.210405} {\bibfield  {journal} {\bibinfo  {journal}
  {Phys. Rev. Lett.}\ }\textbf {\bibinfo {volume} {108}},\ \bibinfo {pages}
  {210405} (\bibinfo {year} {2012})}\BibitemShut {NoStop}%
\bibitem [{\citenamefont {Yunger~Halpern}\ \emph {et~al.}(2019)\citenamefont
  {Yunger~Halpern}, \citenamefont {Bartolotta},\ and\ \citenamefont
  {Pollack}}]{info13}%
  \BibitemOpen
  \bibfield  {author} {\bibinfo {author} {\bibfnamefont {Nicole}\ \bibnamefont
  {Yunger~Halpern}}, \bibinfo {author} {\bibfnamefont {Anthony}\ \bibnamefont
  {Bartolotta}}, \ and\ \bibinfo {author} {\bibfnamefont {Jason}\ \bibnamefont
  {Pollack}},\ }\bibfield  {title} {\enquote {\bibinfo {title} {{Entropic
  uncertainty relations for quantum information scrambling}},}\ }\href
  {\doibase 10.1038/s42005-019-0179-8} {\bibfield  {journal} {\bibinfo
  {journal} {Communications Physics}\ }\textbf {\bibinfo {volume} {2}},\
  \bibinfo {pages} {92} (\bibinfo {year} {2019})}\BibitemShut {NoStop}%
\bibitem [{\citenamefont {Barato}\ and\ \citenamefont
  {Seifert}(2015)}]{thermo1}%
  \BibitemOpen
  \bibfield  {author} {\bibinfo {author} {\bibfnamefont {Andre~C.}\
  \bibnamefont {Barato}}\ and\ \bibinfo {author} {\bibfnamefont {Udo}\
  \bibnamefont {Seifert}},\ }\bibfield  {title} {\enquote {\bibinfo {title}
  {{Thermodynamic Uncertainty Relation for Biomolecular Processes}},}\ }\href
  {\doibase 10.1103/PhysRevLett.114.158101} {\bibfield  {journal} {\bibinfo
  {journal} {Phys. Rev. Lett.}\ }\textbf {\bibinfo {volume} {114}},\ \bibinfo
  {pages} {158101} (\bibinfo {year} {2015})}\BibitemShut {NoStop}%
\bibitem [{\citenamefont {Pietzonka}\ \emph {et~al.}(2016)\citenamefont
  {Pietzonka}, \citenamefont {Barato},\ and\ \citenamefont
  {Seifert}}]{thermo2}%
  \BibitemOpen
  \bibfield  {author} {\bibinfo {author} {\bibfnamefont {Patrick}\ \bibnamefont
  {Pietzonka}}, \bibinfo {author} {\bibfnamefont {Andre~C.}\ \bibnamefont
  {Barato}}, \ and\ \bibinfo {author} {\bibfnamefont {Udo}\ \bibnamefont
  {Seifert}},\ }\bibfield  {title} {\enquote {\bibinfo {title} {{Universal
  bounds on current fluctuations}},}\ }\href {\doibase
  10.1103/PhysRevE.93.052145} {\bibfield  {journal} {\bibinfo  {journal} {Phys.
  Rev. E}\ }\textbf {\bibinfo {volume} {93}},\ \bibinfo {pages} {052145}
  (\bibinfo {year} {2016})}\BibitemShut {NoStop}%
\bibitem [{\citenamefont {Nicholson}\ \emph {et~al.}(2020)\citenamefont
  {Nicholson}, \citenamefont {Garc{\'\i}a-Pintos}, \citenamefont {del Campo},\
  and\ \citenamefont {Green}}]{time-information}%
  \BibitemOpen
  \bibfield  {author} {\bibinfo {author} {\bibfnamefont {Schuyler~B.}\
  \bibnamefont {Nicholson}}, \bibinfo {author} {\bibfnamefont {Luis~Pedro}\
  \bibnamefont {Garc{\'\i}a-Pintos}}, \bibinfo {author} {\bibfnamefont
  {Adolfo}\ \bibnamefont {del Campo}}, \ and\ \bibinfo {author} {\bibfnamefont
  {Jason~R.}\ \bibnamefont {Green}},\ }\bibfield  {title} {\enquote {\bibinfo
  {title} {{Time--information uncertainty relations in thermodynamics}},}\
  }\href {\doibase 10.1038/s41567-020-0981-y} {\bibfield  {journal} {\bibinfo
  {journal} {Nature Physics}\ }\textbf {\bibinfo {volume} {16}},\ \bibinfo
  {pages} {1211--1215} (\bibinfo {year} {2020})}\BibitemShut {NoStop}%
\bibitem [{\citenamefont {Miller}\ and\ \citenamefont {Anders}(2018)}]{E-T}%
  \BibitemOpen
  \bibfield  {author} {\bibinfo {author} {\bibfnamefont {H.~J.~D.}\
  \bibnamefont {Miller}}\ and\ \bibinfo {author} {\bibfnamefont
  {J.}~\bibnamefont {Anders}},\ }\bibfield  {title} {\enquote {\bibinfo {title}
  {{Energy-temperature uncertainty relation in quantum thermodynamics}},}\
  }\href {\doibase 10.1038/s41467-018-04536-7} {\bibfield  {journal} {\bibinfo
  {journal} {Nature Communications}\ }\textbf {\bibinfo {volume} {9}},\
  \bibinfo {pages} {2203} (\bibinfo {year} {2018})}\BibitemShut {NoStop}%
\bibitem [{\citenamefont {Margolus}\ and\ \citenamefont
  {Levitin}(1998)}]{speed2}%
  \BibitemOpen
  \bibfield  {author} {\bibinfo {author} {\bibfnamefont {N.}~\bibnamefont
  {Margolus}}\ and\ \bibinfo {author} {\bibfnamefont {L.~B.}\ \bibnamefont
  {Levitin}},\ }\bibfield  {title} {\enquote {\bibinfo {title} {The maximum
  speed of dynamical evolution},}\ }\href {\doibase
  10.1016/S0167-2789(98)00054-2} {\bibfield  {journal} {\bibinfo  {journal}
  {Physica D: Nonlinear Phenomena}\ }\textbf {\bibinfo {volume} {120}},\
  \bibinfo {pages} {188} (\bibinfo {year} {1998})}\BibitemShut {NoStop}%
\bibitem [{\citenamefont {Deffner}\ and\ \citenamefont
  {Campbell}(2017)}]{speed1}%
  \BibitemOpen
  \bibfield  {author} {\bibinfo {author} {\bibfnamefont {S.}~\bibnamefont
  {Deffner}}\ and\ \bibinfo {author} {\bibfnamefont {S.}~\bibnamefont
  {Campbell}},\ }\bibfield  {title} {\enquote {\bibinfo {title} {{Quantum speed
  limits: from Heisenberg's uncertainty principle to optimal quantum
  control}},}\ }\href {\doibase 10.1088/1751-8121/aa86c6} {\bibfield  {journal}
  {\bibinfo  {journal} {J. Phys. A: Math. Theor.}\ }\textbf {\bibinfo {volume}
  {50}},\ \bibinfo {pages} {453001} (\bibinfo {year} {2017})}\BibitemShut
  {NoStop}%
\bibitem [{\citenamefont {del Campo}\ \emph {et~al.}(2013)\citenamefont {del
  Campo}, \citenamefont {Egusquiza}, \citenamefont {Plenio},\ and\
  \citenamefont {Huelga}}]{campo}%
  \BibitemOpen
  \bibfield  {author} {\bibinfo {author} {\bibfnamefont {A.}~\bibnamefont {del
  Campo}}, \bibinfo {author} {\bibfnamefont {I.~L.}\ \bibnamefont {Egusquiza}},
  \bibinfo {author} {\bibfnamefont {M.~B.}\ \bibnamefont {Plenio}}, \ and\
  \bibinfo {author} {\bibfnamefont {S.~F.}\ \bibnamefont {Huelga}},\ }\bibfield
   {title} {\enquote {\bibinfo {title} {{Quantum Speed Limits in Open System
  Dynamics}},}\ }\href {\doibase 10.1103/PhysRevLett.110.050403} {\bibfield
  {journal} {\bibinfo  {journal} {Phys. Rev. Lett.}\ }\textbf {\bibinfo
  {volume} {110}},\ \bibinfo {pages} {050403} (\bibinfo {year}
  {2013})}\BibitemShut {NoStop}%
\bibitem [{\citenamefont {Lieb}\ and\ \citenamefont
  {Robinson}(1972)}]{Lieb_Robinson}%
  \BibitemOpen
  \bibfield  {author} {\bibinfo {author} {\bibfnamefont {Elliott~H.}\
  \bibnamefont {Lieb}}\ and\ \bibinfo {author} {\bibfnamefont {Derek~W.}\
  \bibnamefont {Robinson}},\ }\bibfield  {title} {\enquote {\bibinfo {title}
  {{The finite group velocity of quantum spin systems}},}\ }\href {\doibase
  10.1007/BF01645779} {\bibfield  {journal} {\bibinfo  {journal}
  {Communications in Mathematical Physics}\ }\textbf {\bibinfo {volume} {28}},\
  \bibinfo {pages} {251--257} (\bibinfo {year} {1972})}\BibitemShut {NoStop}%
\bibitem [{\citenamefont {Nachtergaele}\ and\ \citenamefont
  {Sims}(2006)}]{Bruno}%
  \BibitemOpen
  \bibfield  {author} {\bibinfo {author} {\bibfnamefont {Bruno}\ \bibnamefont
  {Nachtergaele}}\ and\ \bibinfo {author} {\bibfnamefont {Robert}\ \bibnamefont
  {Sims}},\ }\bibfield  {title} {\enquote {\bibinfo {title} {{Lieb-Robinson
  Bounds and the Exponential Clustering Theorem}},}\ }\href {\doibase
  10.1007/s00220-006-1556-1} {\bibfield  {journal} {\bibinfo  {journal}
  {Communications in Mathematical Physics}\ }\textbf {\bibinfo {volume}
  {265}},\ \bibinfo {pages} {119--130} (\bibinfo {year} {2006})}\BibitemShut
  {NoStop}%
\bibitem [{\citenamefont {Bravyi}\ \emph {et~al.}(2006)\citenamefont {Bravyi},
  \citenamefont {Hastings},\ and\ \citenamefont {Verstraete}}]{Sergey}%
  \BibitemOpen
  \bibfield  {author} {\bibinfo {author} {\bibfnamefont {S.}~\bibnamefont
  {Bravyi}}, \bibinfo {author} {\bibfnamefont {M.~B.}\ \bibnamefont
  {Hastings}}, \ and\ \bibinfo {author} {\bibfnamefont {F.}~\bibnamefont
  {Verstraete}},\ }\bibfield  {title} {\enquote {\bibinfo {title}
  {{Lieb-Robinson Bounds and the Generation of Correlations and Topological
  Quantum Order}},}\ }\href {\doibase 10.1103/PhysRevLett.97.050401} {\bibfield
   {journal} {\bibinfo  {journal} {Phys. Rev. Lett.}\ }\textbf {\bibinfo
  {volume} {97}},\ \bibinfo {pages} {050401} (\bibinfo {year}
  {2006})}\BibitemShut {NoStop}%
\bibitem [{\citenamefont {Hamma}\ \emph {et~al.}(2009)\citenamefont {Hamma},
  \citenamefont {Markopoulou}, \citenamefont {Pr\'emont-Schwarz},\ and\
  \citenamefont {Severini}}]{alioscia}%
  \BibitemOpen
  \bibfield  {author} {\bibinfo {author} {\bibfnamefont {Alioscia}\
  \bibnamefont {Hamma}}, \bibinfo {author} {\bibfnamefont {Fotini}\
  \bibnamefont {Markopoulou}}, \bibinfo {author} {\bibfnamefont {Isabeau}\
  \bibnamefont {Pr\'emont-Schwarz}}, \ and\ \bibinfo {author} {\bibfnamefont
  {Simone}\ \bibnamefont {Severini}},\ }\bibfield  {title} {\enquote {\bibinfo
  {title} {{Lieb-Robinson Bounds and the Speed of Light from Topological
  Order}},}\ }\href {\doibase 10.1103/PhysRevLett.102.017204} {\bibfield
  {journal} {\bibinfo  {journal} {Phys. Rev. Lett.}\ }\textbf {\bibinfo
  {volume} {102}},\ \bibinfo {pages} {017204} (\bibinfo {year}
  {2009})}\BibitemShut {NoStop}%
\bibitem [{\citenamefont {Wang}\ and\ \citenamefont {Hazzard}(2020)}]{kaden}%
  \BibitemOpen
  \bibfield  {author} {\bibinfo {author} {\bibfnamefont {Zhiyuan}\ \bibnamefont
  {Wang}}\ and\ \bibinfo {author} {\bibfnamefont {Kaden~R.A.}\ \bibnamefont
  {Hazzard}},\ }\bibfield  {title} {\enquote {\bibinfo {title} {{Tightening the
  Lieb-Robinson Bound in Locally Interacting Systems}},}\ }\href {\doibase
  10.1103/PRXQuantum.1.010303} {\bibfield  {journal} {\bibinfo  {journal} {PRX
  Quantum}\ }\textbf {\bibinfo {volume} {1}},\ \bibinfo {pages} {010303}
  (\bibinfo {year} {2020})}\BibitemShut {NoStop}%
\bibitem [{\citenamefont {Else}\ \emph {et~al.}(2020)\citenamefont {Else},
  \citenamefont {Machado},\ and\ \citenamefont {Yao}}]{Else20}%
  \BibitemOpen
  \bibfield  {author} {\bibinfo {author} {\bibfnamefont {V.}~\bibnamefont
  {Else}}, \bibinfo {author} {\bibfnamefont {C.}~\bibnamefont {Machado},
  \bibfnamefont {F.~an~Nayak}}, \ and\ \bibinfo {author} {\bibfnamefont
  {N.~Y.}\ \bibnamefont {Yao}},\ }\bibfield  {title} {\enquote {\bibinfo
  {title} {{Improved Lieb-Robinson bound for many-body hamiltonians with
  power-law interactions}},}\ }\href {\doibase 10.1103/PhysRevA.101.022333}
  {\bibfield  {journal} {\bibinfo  {journal} {Physical Review A}\ }\textbf
  {\bibinfo {volume} {101}},\ \bibinfo {pages} {022333} (\bibinfo {year}
  {2020})}\BibitemShut {NoStop}%
\bibitem [{\citenamefont {Ioffe}\ and\ \citenamefont
  {Regel}(1960)}]{Ioffe-Regel}%
  \BibitemOpen
  \bibfield  {author} {\bibinfo {author} {\bibfnamefont {AF}~\bibnamefont
  {Ioffe}}\ and\ \bibinfo {author} {\bibfnamefont {AR}~\bibnamefont {Regel}},\
  }\bibfield  {title} {\enquote {\bibinfo {title} {{Non-crystalline, amorphous
  and liquid electronic semiconductors}},}\ }\href@noop {} {\bibfield
  {journal} {\bibinfo  {journal} {Prog. Semicond}\ }\textbf {\bibinfo {volume}
  {4}},\ \bibinfo {pages} {237--291} (\bibinfo {year} {1960})}\BibitemShut
  {NoStop}%
\bibitem [{\citenamefont {Lifshitz}\ and\ \citenamefont
  {Pitaevskii}(2013)}]{LL}%
  \BibitemOpen
  \bibfield  {author} {\bibinfo {author} {\bibfnamefont {Evgenii~Mikhailovich}\
  \bibnamefont {Lifshitz}}\ and\ \bibinfo {author} {\bibfnamefont
  {Lev~Petrovich}\ \bibnamefont {Pitaevskii}},\ }\href@noop {} {\emph {\bibinfo
  {title} {{Statistical physics: theory of the condensed state}}}},\
  Vol.~\bibinfo {volume} {9}\ (\bibinfo  {publisher} {Elsevier},\ \bibinfo
  {year} {2013})\BibitemShut {NoStop}%
\bibitem [{\citenamefont {Hwang}\ and\ \citenamefont
  {Das~Sarma}(2019)}]{sarma}%
  \BibitemOpen
  \bibfield  {author} {\bibinfo {author} {\bibfnamefont {E.~H.}\ \bibnamefont
  {Hwang}}\ and\ \bibinfo {author} {\bibfnamefont {S.}~\bibnamefont
  {Das~Sarma}},\ }\bibfield  {title} {\enquote {\bibinfo {title} {{Linear-in-T
  resistivity in dilute metals: A Fermi liquid perspective}},}\ }\href
  {\doibase 10.1103/PhysRevB.99.085105} {\bibfield  {journal} {\bibinfo
  {journal} {Phys. Rev. B}\ }\textbf {\bibinfo {volume} {99}},\ \bibinfo
  {pages} {085105} (\bibinfo {year} {2019})}\BibitemShut {NoStop}%
\bibitem [{\citenamefont {Wu}\ \emph {et~al.}(2019)\citenamefont {Wu},
  \citenamefont {Hwang},\ and\ \citenamefont {Das~Sarma}}]{dassarmaPh}%
  \BibitemOpen
  \bibfield  {author} {\bibinfo {author} {\bibfnamefont {Fengcheng}\
  \bibnamefont {Wu}}, \bibinfo {author} {\bibfnamefont {Euyheon}\ \bibnamefont
  {Hwang}}, \ and\ \bibinfo {author} {\bibfnamefont {Sankar}\ \bibnamefont
  {Das~Sarma}},\ }\bibfield  {title} {\enquote {\bibinfo {title}
  {{Phonon-induced giant linear-in-T resistivity in magic angle twisted bilayer
  graphene: Ordinary strangeness and exotic superconductivity}},}\ }\href
  {\doibase 10.1103/PhysRevB.99.165112} {\bibfield  {journal} {\bibinfo
  {journal} {Phys. Rev. B}\ }\textbf {\bibinfo {volume} {99}},\ \bibinfo
  {pages} {165112} (\bibinfo {year} {2019})}\BibitemShut {NoStop}%
\bibitem [{\citenamefont {Deutsch}(1991)}]{eth1}%
  \BibitemOpen
  \bibfield  {author} {\bibinfo {author} {\bibfnamefont {J.~M.}\ \bibnamefont
  {Deutsch}},\ }\bibfield  {title} {\enquote {\bibinfo {title} {{Quantum
  statistical mechanics in a closed system}},}\ }\href {\doibase
  10.1103/PhysRevA.43.2046} {\bibfield  {journal} {\bibinfo  {journal} {Phys.
  Rev. A}\ }\textbf {\bibinfo {volume} {43}},\ \bibinfo {pages} {2046--2049}
  (\bibinfo {year} {1991})}\BibitemShut {NoStop}%
\bibitem [{\citenamefont {Srednicki}(1994)}]{eth2}%
  \BibitemOpen
  \bibfield  {author} {\bibinfo {author} {\bibfnamefont {Mark}\ \bibnamefont
  {Srednicki}},\ }\bibfield  {title} {\enquote {\bibinfo {title} {{Chaos and
  quantum thermalization}},}\ }\href {\doibase 10.1103/PhysRevE.50.888}
  {\bibfield  {journal} {\bibinfo  {journal} {Phys. Rev. E}\ }\textbf {\bibinfo
  {volume} {50}},\ \bibinfo {pages} {888--901} (\bibinfo {year}
  {1994})}\BibitemShut {NoStop}%
\bibitem [{\citenamefont {Rigol}\ \emph {et~al.}(2008)\citenamefont {Rigol},
  \citenamefont {Dunjko},\ and\ \citenamefont {Olshanii}}]{eth3}%
  \BibitemOpen
  \bibfield  {author} {\bibinfo {author} {\bibfnamefont {Marcos}\ \bibnamefont
  {Rigol}}, \bibinfo {author} {\bibfnamefont {Vanja}\ \bibnamefont {Dunjko}}, \
  and\ \bibinfo {author} {\bibfnamefont {Maxim}\ \bibnamefont {Olshanii}},\
  }\bibfield  {title} {\enquote {\bibinfo {title} {{Thermalization and its
  mechanism for generic isolated quantum systems}},}\ }\href {\doibase
  10.1038/nature06838} {\bibfield  {journal} {\bibinfo  {journal} {Nature}\
  }\textbf {\bibinfo {volume} {452}},\ \bibinfo {pages} {854--858} (\bibinfo
  {year} {2008})}\BibitemShut {NoStop}%
\bibitem [{\citenamefont {Borgonovi}\ \emph {et~al.}(2016)\citenamefont
  {Borgonovi}, \citenamefont {Izrailev}, \citenamefont {Santos},\ and\
  \citenamefont {Zelevinsky}}]{eth4}%
  \BibitemOpen
  \bibfield  {author} {\bibinfo {author} {\bibfnamefont {F.}~\bibnamefont
  {Borgonovi}}, \bibinfo {author} {\bibfnamefont {F.M.}\ \bibnamefont
  {Izrailev}}, \bibinfo {author} {\bibfnamefont {L.F.}\ \bibnamefont {Santos}},
  \ and\ \bibinfo {author} {\bibfnamefont {V.G.}\ \bibnamefont {Zelevinsky}},\
  }\bibfield  {title} {\enquote {\bibinfo {title} {{Quantum chaos and
  thermalization in isolated systems of interacting particles}},}\ }\href
  {\doibase https://doi.org/10.1016/j.physrep.2016.02.005} {\bibfield
  {journal} {\bibinfo  {journal} {Physics Reports}\ }\textbf {\bibinfo {volume}
  {626}},\ \bibinfo {pages} {1--58} (\bibinfo {year} {2016})},\ \bibinfo {note}
  {quantum chaos and thermalization in isolated systems of interacting
  particles}\BibitemShut {NoStop}%
\bibitem [{\citenamefont {Rigol}(2009)}]{rigol}%
  \BibitemOpen
  \bibfield  {author} {\bibinfo {author} {\bibfnamefont {Marcos}\ \bibnamefont
  {Rigol}},\ }\bibfield  {title} {\enquote {\bibinfo {title} {{Breakdown of
  Thermalization in Finite One-Dimensional Systems}},}\ }\href {\doibase
  10.1103/PhysRevLett.103.100403} {\bibfield  {journal} {\bibinfo  {journal}
  {Phys. Rev. Lett.}\ }\textbf {\bibinfo {volume} {103}},\ \bibinfo {pages}
  {100403} (\bibinfo {year} {2009})}\BibitemShut {NoStop}%
\bibitem [{\citenamefont {Polkovnikov}\ \emph {et~al.}(2011)\citenamefont
  {Polkovnikov}, \citenamefont {Sengupta}, \citenamefont {Silva},\ and\
  \citenamefont {Vengalattore}}]{pol}%
  \BibitemOpen
  \bibfield  {author} {\bibinfo {author} {\bibfnamefont {Anatoli}\ \bibnamefont
  {Polkovnikov}}, \bibinfo {author} {\bibfnamefont {Krishnendu}\ \bibnamefont
  {Sengupta}}, \bibinfo {author} {\bibfnamefont {Alessandro}\ \bibnamefont
  {Silva}}, \ and\ \bibinfo {author} {\bibfnamefont {Mukund}\ \bibnamefont
  {Vengalattore}},\ }\bibfield  {title} {\enquote {\bibinfo {title}
  {{Colloquium: Nonequilibrium dynamics of closed interacting quantum
  systems}},}\ }\href {\doibase 10.1103/RevModPhys.83.863} {\bibfield
  {journal} {\bibinfo  {journal} {Rev. Mod. Phys.}\ }\textbf {\bibinfo {volume}
  {83}},\ \bibinfo {pages} {863--883} (\bibinfo {year} {2011})}\BibitemShut
  {NoStop}%
\bibitem [{\citenamefont {Santos}\ \emph {et~al.}(2011)\citenamefont {Santos},
  \citenamefont {Polkovnikov},\ and\ \citenamefont {Rigol}}]{polkovnikov1}%
  \BibitemOpen
  \bibfield  {author} {\bibinfo {author} {\bibfnamefont {Lea~F.}\ \bibnamefont
  {Santos}}, \bibinfo {author} {\bibfnamefont {Anatoli}\ \bibnamefont
  {Polkovnikov}}, \ and\ \bibinfo {author} {\bibfnamefont {Marcos}\
  \bibnamefont {Rigol}},\ }\bibfield  {title} {\enquote {\bibinfo {title}
  {{Entropy of Isolated Quantum Systems after a Quench}},}\ }\href {\doibase
  10.1103/PhysRevLett.107.040601} {\bibfield  {journal} {\bibinfo  {journal}
  {Phys. Rev. Lett.}\ }\textbf {\bibinfo {volume} {107}},\ \bibinfo {pages}
  {040601} (\bibinfo {year} {2011})}\BibitemShut {NoStop}%
\bibitem [{\citenamefont {D'Alessio}\ \emph {et~al.}(2016)\citenamefont
  {D'Alessio}, \citenamefont {Kafri}, \citenamefont {Polkovnikov},\ and\
  \citenamefont {Rigol}}]{polkovnikov2}%
  \BibitemOpen
  \bibfield  {author} {\bibinfo {author} {\bibfnamefont {Luca}\ \bibnamefont
  {D'Alessio}}, \bibinfo {author} {\bibfnamefont {Yariv}\ \bibnamefont
  {Kafri}}, \bibinfo {author} {\bibfnamefont {Anatoli}\ \bibnamefont
  {Polkovnikov}}, \ and\ \bibinfo {author} {\bibfnamefont {Marcos}\
  \bibnamefont {Rigol}},\ }\bibfield  {title} {\enquote {\bibinfo {title}
  {{From quantum chaos and eigenstate thermalization to statistical mechanics
  and thermodynamics}},}\ }\href {\doibase 10.1080/00018732.2016.1198134}
  {\bibfield  {journal} {\bibinfo  {journal} {Advances in Physics}\ }\textbf
  {\bibinfo {volume} {65}},\ \bibinfo {pages} {239--362} (\bibinfo {year}
  {2016})}\BibitemShut {NoStop}%
\bibitem [{\citenamefont {Srednicki}(1996)}]{srednicki-95}%
  \BibitemOpen
  \bibfield  {author} {\bibinfo {author} {\bibfnamefont {Mark}\ \bibnamefont
  {Srednicki}},\ }\bibfield  {title} {\enquote {\bibinfo {title} {{Thermal
  fluctuations in quantized chaotic systems}},}\ }\href {\doibase
  10.1088/0305-4470/29/4/003} {\bibfield  {journal} {\bibinfo  {journal}
  {Journal of Physics A: Mathematical and General}\ }\textbf {\bibinfo {volume}
  {29}},\ \bibinfo {pages} {L75--L79} (\bibinfo {year} {1996})}\BibitemShut
  {NoStop}%
\bibitem [{alS()}]{alSpeed}%
  \BibitemOpen
  \href {https://www.rshydro.co.uk/sound-speeds/} {\bibinfo  {journal}
  {https://www.rshydro.co.uk/sound-speeds/}\ }\BibitemShut {NoStop}%
\bibitem [{alu()}]{aluminum}%
  \BibitemOpen
\bibfield  {journal} {  }\href
  {http://www.chemspider.com/Chemical-Structure.4514248.html} {\bibinfo
  {journal} {http://www.chemspider.com/Chemical-Structure.4514248.html}\
  }\BibitemShut {NoStop}%
\bibitem [{\citenamefont {Grimvall}(1999)}]{lindemannAl}%
  \BibitemOpen
\bibfield  {journal} {  }\bibfield  {author} {\bibinfo {author} {\bibfnamefont
  {G\"{o}ran}\ \bibnamefont {Grimvall}},\ }\bibfield  {title} {\enquote
  {\bibinfo {title} {Chapter 19 - estimations and correlations},}\ }in\ \href
  {\doibase https://doi.org/10.1016/B978-044482794-4/50020-6} {\emph {\bibinfo
  {booktitle} {Thermophysical Properties of Materials}}},\ \bibinfo {editor}
  {edited by\ \bibinfo {editor} {\bibfnamefont {G\"{o}ran}\ \bibnamefont
  {Grimvall}}}\ (\bibinfo  {publisher} {Elsevier Science B.V.},\ \bibinfo
  {address} {Amsterdam},\ \bibinfo {year} {1999})\ pp.\ \bibinfo {pages}
  {331--352}\BibitemShut {NoStop}%
\bibitem [{\citenamefont {Easteal}\ \emph {et~al.}(1989)\citenamefont
  {Easteal}, \citenamefont {Price},\ and\ \citenamefont {Woolf}}]{waterD}%
  \BibitemOpen
  \bibfield  {author} {\bibinfo {author} {\bibfnamefont {Allan~J.}\
  \bibnamefont {Easteal}}, \bibinfo {author} {\bibfnamefont {William~E.}\
  \bibnamefont {Price}}, \ and\ \bibinfo {author} {\bibfnamefont {Lawrence~A.}\
  \bibnamefont {Woolf}},\ }\bibfield  {title} {\enquote {\bibinfo {title}
  {{Diaphragm cell for high-temperature diffusion measurements. Tracer
  Diffusion coefficients for water to 363 K}},}\ }\href {\doibase
  10.1039/F19898501091} {\bibfield  {journal} {\bibinfo  {journal} {J. Chem.
  Soc.{,} Faraday Trans. 1}\ }\textbf {\bibinfo {volume} {85}},\ \bibinfo
  {pages} {1091--1097} (\bibinfo {year} {1989})}\BibitemShut {NoStop}%
\bibitem [{\citenamefont {Skinner}\ \emph {et~al.}(2013)\citenamefont
  {Skinner}, \citenamefont {Huang}, \citenamefont {Schlesinger}, \citenamefont
  {Pettersson},\ and\ \citenamefont {Nilsson}}]{watercoord1}%
  \BibitemOpen
  \bibfield  {author} {\bibinfo {author} {\bibfnamefont {Lawrie~B.}\
  \bibnamefont {Skinner}}, \bibinfo {author} {\bibfnamefont {Congcong}\
  \bibnamefont {Huang}}, \bibinfo {author} {\bibfnamefont {Daniel}\
  \bibnamefont {Schlesinger}}, \bibinfo {author} {\bibfnamefont {Lars G.~M.}\
  \bibnamefont {Pettersson}}, \ and\ \bibinfo {author} {\bibfnamefont
  {Chris~J.}\ \bibnamefont {Nilsson}, \bibfnamefont {Anders an~Benmore}},\
  }\bibfield  {title} {\enquote {\bibinfo {title} {{Benchmark oxygen-oxygen
  pair-distribution function of ambient water from x-ray diffraction
  measurements with a wide Q-range}},}\ }\href {\doibase 10.1063/1.4790861}
  {\bibfield  {journal} {\bibinfo  {journal} {The Journal of Chemical Physics}\
  }\textbf {\bibinfo {volume} {138}},\ \bibinfo {pages} {074506} (\bibinfo
  {year} {2013})}\BibitemShut {NoStop}%
\bibitem [{\citenamefont {Willow}\ \emph {et~al.}(2015)\citenamefont {Willow},
  \citenamefont {Salim}, \citenamefont {Kim},\ and\ \citenamefont
  {Hirara}}]{watercoord2}%
  \BibitemOpen
  \bibfield  {author} {\bibinfo {author} {\bibfnamefont {Soohaeng~Yoo}\
  \bibnamefont {Willow}}, \bibinfo {author} {\bibfnamefont {Michael~A.}\
  \bibnamefont {Salim}}, \bibinfo {author} {\bibfnamefont {Kwang~S.}\
  \bibnamefont {Kim}}, \ and\ \bibinfo {author} {\bibfnamefont
  {So}~\bibnamefont {Hirara}},\ }\bibfield  {title} {\enquote {\bibinfo {title}
  {Ab initio molecular dynamics of liquid water using embedded-fragment
  second-order many-body perturbation theory towards its accurate property
  prediction},}\ }\href {\doibase 10.1038/srep14358} {\bibfield  {journal}
  {\bibinfo  {journal} {Scientific Reports}\ }\textbf {\bibinfo {volume} {5}},\
  \bibinfo {pages} {14358} (\bibinfo {year} {2015})}\BibitemShut {NoStop}%
\bibitem [{\citenamefont {Xue}\ \emph {et~al.}(2021)\citenamefont {Xue},
  \citenamefont {Nogueira}, \citenamefont {Kelton},\ and\ \citenamefont
  {Nussinov}}]{jing}%
  \BibitemOpen
  \bibfield  {author} {\bibinfo {author} {\bibfnamefont {Jing}\ \bibnamefont
  {Xue}}, \bibinfo {author} {\bibfnamefont {Flavio~S.}\ \bibnamefont
  {Nogueira}}, \bibinfo {author} {\bibfnamefont {Ken~F.}\ \bibnamefont
  {Kelton}}, \ and\ \bibinfo {author} {\bibfnamefont {Zohar}\ \bibnamefont
  {Nussinov}},\ }\bibfield  {title} {\enquote {\bibinfo {title} {{Deviations
  from Arrhenius dynamics in high temperature liquids, a possible collapse, and
  a viscosity bound}},}\ }\href {https://arxiv.org/pdf/2105.13397.pdf}
  {\bibfield  {journal} {\bibinfo  {journal} {arXiv preprint arXiv:2105.13397}\
  } (\bibinfo {year} {2021})}\BibitemShut {NoStop}%
\bibitem [{NIS()}]{NIST}%
  \BibitemOpen
  \bibfield  {title} {\enquote {\bibinfo {title} {{National Institute of
  Standards and Technology database}},}\ }\href
  {https://webbook.nist.gov/chemistry/fluid} {\bibinfo  {journal}
  {https://webbook.nist.gov/chemistry/fluid}\ }\BibitemShut {NoStop}%
\bibitem [{\citenamefont {Nussinov}(2017)}]{glass}%
  \BibitemOpen
\bibfield  {journal} {  }\bibfield  {author} {\bibinfo {author} {\bibfnamefont
  {Zohar}\ \bibnamefont {Nussinov}},\ }\bibfield  {title} {\enquote {\bibinfo
  {title} {{A one parameter fit for glassy dynamics as a quantum corollary of
  the liquid to solid transition}},}\ }\href {\doibase
  10.1080/14786435.2016.1274837} {\bibfield  {journal} {\bibinfo  {journal}
  {Philosophical Magazine}\ }\textbf {\bibinfo {volume} {97}},\ \bibinfo
  {pages} {1509--1566} (\bibinfo {year} {2017})}\BibitemShut {NoStop}%
\bibitem [{\citenamefont {Rigol}\ \emph {et~al.}(2007)\citenamefont {Rigol},
  \citenamefont {Dunjko}, \citenamefont {Yurovsky},\ and\ \citenamefont
  {Olshanii}}]{GGE1}%
  \BibitemOpen
  \bibfield  {author} {\bibinfo {author} {\bibfnamefont {Marcos}\ \bibnamefont
  {Rigol}}, \bibinfo {author} {\bibfnamefont {Vanja}\ \bibnamefont {Dunjko}},
  \bibinfo {author} {\bibfnamefont {Vladimir}\ \bibnamefont {Yurovsky}}, \ and\
  \bibinfo {author} {\bibfnamefont {Maxim}\ \bibnamefont {Olshanii}},\
  }\bibfield  {title} {\enquote {\bibinfo {title} {{Relaxation in a Completely
  Integrable Many-Body Quantum System: An Ab Initio Study of the Dynamics of
  the Highly Excited States of 1D Lattice Hard-Core Bosons}},}\ }\href
  {\doibase 10.1103/PhysRevLett.98.050405} {\bibfield  {journal} {\bibinfo
  {journal} {Phys. Rev. Lett.}\ }\textbf {\bibinfo {volume} {98}},\ \bibinfo
  {pages} {050405} (\bibinfo {year} {2007})}\BibitemShut {NoStop}%
\bibitem [{\citenamefont {Foini}\ \emph {et~al.}(2017)\citenamefont {Foini},
  \citenamefont {Gambassi}, \citenamefont {Konik},\ and\ \citenamefont
  {Cugliandolo}}]{GGE3}%
  \BibitemOpen
  \bibfield  {author} {\bibinfo {author} {\bibfnamefont {Laura}\ \bibnamefont
  {Foini}}, \bibinfo {author} {\bibfnamefont {Andrea}\ \bibnamefont
  {Gambassi}}, \bibinfo {author} {\bibfnamefont {Robert}\ \bibnamefont
  {Konik}}, \ and\ \bibinfo {author} {\bibfnamefont {Leticia~F.}\ \bibnamefont
  {Cugliandolo}},\ }\bibfield  {title} {\enquote {\bibinfo {title} {{Measuring
  effective temperatures in a generalized Gibbs ensemble}},}\ }\href {\doibase
  10.1103/PhysRevE.95.052116} {\bibfield  {journal} {\bibinfo  {journal} {Phys.
  Rev. E}\ }\textbf {\bibinfo {volume} {95}},\ \bibinfo {pages} {052116}
  (\bibinfo {year} {2017})}\BibitemShut {NoStop}%
\bibitem [{\citenamefont {Goldenfeld}(2018)}]{Nigel}%
  \BibitemOpen
  \bibfield  {author} {\bibinfo {author} {\bibfnamefont {Nigel}\ \bibnamefont
  {Goldenfeld}},\ }\href@noop {} {\emph {\bibinfo {title} {{Lectures on phase
  transitions and the renormalization group}}}}\ (\bibinfo  {publisher} {CRC
  Press},\ \bibinfo {year} {2018})\BibitemShut {NoStop}%
\bibitem [{Note2()}]{Note2}%
  \BibitemOpen
  \bibinfo {note} {If the transition matrix $W$ is special in that it has a
  microstate (a unit vector in the basis of micro states in which $W$ is
  defined) as an eigenstate and if the initial state is such a microstate then
  the system will remain trivially stationary.}\BibitemShut {Stop}%
\bibitem [{\citenamefont {Haag}\ \emph {et~al.}(1967)\citenamefont {Haag},
  \citenamefont {Hugenholtz},\ and\ \citenamefont {Winnink}}]{KMS1}%
  \BibitemOpen
  \bibfield  {author} {\bibinfo {author} {\bibfnamefont {R.}~\bibnamefont
  {Haag}}, \bibinfo {author} {\bibfnamefont {N.~M.}\ \bibnamefont
  {Hugenholtz}}, \ and\ \bibinfo {author} {\bibfnamefont {M.}~\bibnamefont
  {Winnink}},\ }\bibfield  {title} {\enquote {\bibinfo {title} {{On the
  equilibrium states in quantum statistical mechanics}},}\ }\href {\doibase
  10.1007/BF01646342} {\bibfield  {journal} {\bibinfo  {journal}
  {Communications in Mathematical Physics}\ }\textbf {\bibinfo {volume} {5}},\
  \bibinfo {pages} {215--236} (\bibinfo {year} {1967})}\BibitemShut {NoStop}%
\bibitem [{\citenamefont {Kubo}(1957)}]{KMS2}%
  \BibitemOpen
  \bibfield  {author} {\bibinfo {author} {\bibfnamefont {Ryogo}\ \bibnamefont
  {Kubo}},\ }\bibfield  {title} {\enquote {\bibinfo {title}
  {{Statistical-Mechanical Theory of Irreversible Processes. I. General Theory
  and Simple Applications to Magnetic and Conduction Problems}},}\ }\href
  {\doibase 10.1143/JPSJ.12.570} {\bibfield  {journal} {\bibinfo  {journal}
  {Journal of the Physical Society of Japan}\ }\textbf {\bibinfo {volume}
  {12}},\ \bibinfo {pages} {570--586} (\bibinfo {year} {1957})}\BibitemShut
  {NoStop}%
\bibitem [{\citenamefont {Martin}\ and\ \citenamefont
  {Schwinger}(1959)}]{KMS3}%
  \BibitemOpen
  \bibfield  {author} {\bibinfo {author} {\bibfnamefont {Paul~C.}\ \bibnamefont
  {Martin}}\ and\ \bibinfo {author} {\bibfnamefont {Julian}\ \bibnamefont
  {Schwinger}},\ }\bibfield  {title} {\enquote {\bibinfo {title} {{Theory of
  Many-Particle Systems. I}},}\ }\href {\doibase 10.1103/PhysRev.115.1342}
  {\bibfield  {journal} {\bibinfo  {journal} {Phys. Rev.}\ }\textbf {\bibinfo
  {volume} {115}},\ \bibinfo {pages} {1342--1373} (\bibinfo {year}
  {1959})}\BibitemShut {NoStop}%
\bibitem [{Note3()}]{Note3}%
  \BibitemOpen
  \bibinfo {note} {If particle (or site) $i \equiv i_0$ is decoupled from a
  distant similar particle $i_{1}$ which in turn is decoupled from $i_{2},
  \protect \cdots i_{n'}$, then we may view $i, i_{1}, i_{2}, \protect \ldots ,
  i_{n'}$ as belonging to different realizations of a system $\Lambda $. The
  ensemble average over these realizations of $\Lambda $ will coincide with the
  global average of the expectation value of $Q^H$ at sites $i, i_{1}i_{2},
  \protect \ldots $ . In such situations in which the correlation length is
  finite and particles (or sites) $i, i_{1}, i_{2}, \protect \ldots , i_{n'}$
  are decoupled, we can see how relations such as \begin {equation} {\protect
  \sf {Tr}}\left (\rho _{\Lambda }^{\protect \sf {canonical }} Q^H_{i}\right
  )=\protect \qopname \relax m{lim}_{n' \rightarrow \infty } \protect \frac
  {1}{n'} \DOTSB \sum@ \slimits@ _{i=0}^{n'} {\protect \sf {Tr}}\left (\rho
  _{\Lambda } Q^H_{i_{n'}}\right ) \end {equation} are obtained.}\BibitemShut
  {Stop}%
\bibitem [{\citenamefont {Pearl}(1988)}]{Markovblanket}%
  \BibitemOpen
  \bibfield  {author} {\bibinfo {author} {\bibfnamefont {Judea}\ \bibnamefont
  {Pearl}},\ }\href@noop {} {\emph {\bibinfo {title} {{Probabilistic reasoning
  in intelligent systems: networks of plausible inference}}}}\ (\bibinfo
  {publisher} {San Mateo, California: Morgan Kaufmann Publishers},\ \bibinfo
  {year} {1988})\BibitemShut {NoStop}%
\bibitem [{Note4()}]{Note4}%
  \BibitemOpen
  \bibinfo {note} {Specifically, in \cite {bound} it was demonstrated that the
  relative rate of change at which general global observables $Q$ may vary is
  bounded is set by the thermodynamic heat capacity, \begin {eqnarray} \label
  {Qsee} \tau ^{-1}_{Q} \equiv \protect \frac {|\langle \protect \frac {dQ}{dt}
  \rangle |}{\sigma _{Q}} \le \protect \frac {2 T \protect \sqrt {k_{B}
  C_{v}^{(\Lambda )}}} {\hbar }. \end {eqnarray} This global relaxation rate is
  the counterpart of the local relaxation rate of Eq. (\ref {cv3}). Since the
  global constant volume heat capacity $C_{v}^{(\Lambda )} = T (\protect \frac
  {\partial S^{(\Lambda )}}{\partial T})_{v}$, this bound is set by the
  temperature derivative of the system entropy $S^{(\Lambda )}$ (or information
  content of $\Lambda $ as measured by the Shannon entropy that differs from
  $S^{(\Lambda )}$ by a trivial constant multiplicative factor (of $1/(k_{B}
  \protect \qopname \relax o{ln}2)$)). Similarly, in equilibrated systems, the
  uniform susceptibility of the global quantity $Q$ is bounded by a value that
  is further set by the global heat capacity, \begin {eqnarray} \label
  {fluct-eqq} \chi _{Q} \ge \protect \frac {\hbar ^{2}}{4 k_{B}^{2} T^{3}
  C_{v}^{(\Lambda )}} \leavevmode@ifvmode {\setbox \z@ \hbox {\mathsurround \z@
  $\nulldelimiterspace \z@ \left |\vcenter to1.5\big@size {}\right .$}\box \z@
  } \protect \frac {dQ}{dt} \leavevmode@ifvmode {\setbox \z@ \hbox
  {\mathsurround \z@ $\nulldelimiterspace \z@ \left |\vcenter to1.5\big@size
  {}\right .$}\box \z@ }^{2}. \end {eqnarray}}\BibitemShut {NoStop}%
\bibitem [{\citenamefont {Chakravarty}\ \emph {et~al.}(2007)\citenamefont
  {Chakravarty}, \citenamefont {Debenedetti},\ and\ \citenamefont
  {Stillinger}}]{Lindemann}%
  \BibitemOpen
  \bibfield  {author} {\bibinfo {author} {\bibfnamefont {Charusita}\
  \bibnamefont {Chakravarty}}, \bibinfo {author} {\bibfnamefont {Pablo~G.}\
  \bibnamefont {Debenedetti}}, \ and\ \bibinfo {author} {\bibfnamefont
  {Frank~H.}\ \bibnamefont {Stillinger}},\ }\bibfield  {title} {\enquote
  {\bibinfo {title} {Lindemann measures for the solid-liquid phase
  transition},}\ }\href {\doibase 10.1063/1.2737054} {\bibfield  {journal}
  {\bibinfo  {journal} {The Journal of Chemical Physics}\ }\textbf {\bibinfo
  {volume} {126}},\ \bibinfo {pages} {204508} (\bibinfo {year}
  {2007})}\BibitemShut {NoStop}%
\bibitem [{\citenamefont {Green}(1954)}]{Green1954}%
  \BibitemOpen
  \bibfield  {author} {\bibinfo {author} {\bibfnamefont {M.~S.}\ \bibnamefont
  {Green}},\ }\bibfield  {title} {\enquote {\bibinfo {title} {{Markoff Random
  Processes and the Statistical Mechanics of Time-Dependent Phenomena. II.
  Irreversible Processes in Fluid}},}\ }\href {\doibase 10.1063/1.1740082}
  {\bibfield  {journal} {\bibinfo  {journal} {J. Chem. Phys.}\ }\textbf
  {\bibinfo {volume} {22}},\ \bibinfo {pages} {398} (\bibinfo {year}
  {1954})}\BibitemShut {NoStop}%
\bibitem [{\citenamefont {Zwanzig}(1965)}]{liqGv1}%
  \BibitemOpen
  \bibfield  {author} {\bibinfo {author} {\bibfnamefont {R.}~\bibnamefont
  {Zwanzig}},\ }\bibfield  {title} {\enquote {\bibinfo {title}
  {{Time-Correlation Functions and Transport Coefficients in Statistical
  Mechanics}},}\ }\href {\doibase 10.1146/annurev.pc.16.100165.000435}
  {\bibfield  {journal} {\bibinfo  {journal} {Annual Review of Physical
  Chemistry}\ }\textbf {\bibinfo {volume} {16}},\ \bibinfo {pages} {67}
  (\bibinfo {year} {1965})}\BibitemShut {NoStop}%
\bibitem [{\citenamefont {Zwanzig}\ and\ \citenamefont {Bixon}(1970)}]{liqGv2}%
  \BibitemOpen
  \bibfield  {author} {\bibinfo {author} {\bibfnamefont {Robert}\ \bibnamefont
  {Zwanzig}}\ and\ \bibinfo {author} {\bibfnamefont {Mordechai}\ \bibnamefont
  {Bixon}},\ }\bibfield  {title} {\enquote {\bibinfo {title} {Hydrodynamic
  theory of the velocity correlation function},}\ }\href {\doibase
  10.1103/PhysRevA.2.2005} {\bibfield  {journal} {\bibinfo  {journal} {Phys.
  Rev. A}\ }\textbf {\bibinfo {volume} {2}},\ \bibinfo {pages} {2005--2012}
  (\bibinfo {year} {1970})}\BibitemShut {NoStop}%
\bibitem [{\citenamefont {Harp}\ and\ \citenamefont {Berne}(1970)}]{liqGv3}%
  \BibitemOpen
  \bibfield  {author} {\bibinfo {author} {\bibfnamefont {G.~D.}\ \bibnamefont
  {Harp}}\ and\ \bibinfo {author} {\bibfnamefont {B.~J.}\ \bibnamefont
  {Berne}},\ }\bibfield  {title} {\enquote {\bibinfo {title} {Time-correlation
  functions, memory functions, and molecular dynamics},}\ }\href {\doibase
  10.1103/PhysRevA.2.975} {\bibfield  {journal} {\bibinfo  {journal} {Phys.
  Rev. A}\ }\textbf {\bibinfo {volume} {2}},\ \bibinfo {pages} {975--996}
  (\bibinfo {year} {1970})}\BibitemShut {NoStop}%
\bibitem [{\citenamefont {Alder}\ and\ \citenamefont
  {Wainwright}(1970)}]{liqGv4}%
  \BibitemOpen
  \bibfield  {author} {\bibinfo {author} {\bibfnamefont {B.~J.}\ \bibnamefont
  {Alder}}\ and\ \bibinfo {author} {\bibfnamefont {T.~E.}\ \bibnamefont
  {Wainwright}},\ }\bibfield  {title} {\enquote {\bibinfo {title} {Decay of the
  velocity autocorrelation function},}\ }\href {\doibase 10.1103/PhysRevA.1.18}
  {\bibfield  {journal} {\bibinfo  {journal} {Phys. Rev. A}\ }\textbf {\bibinfo
  {volume} {1}},\ \bibinfo {pages} {18--21} (\bibinfo {year}
  {1970})}\BibitemShut {NoStop}%
\bibitem [{\citenamefont {Ernst}\ \emph {et~al.}(1970)\citenamefont {Ernst},
  \citenamefont {Hauge},\ and\ \citenamefont {van Leeuwen}}]{liqGv5}%
  \BibitemOpen
  \bibfield  {author} {\bibinfo {author} {\bibfnamefont {M.~H.}\ \bibnamefont
  {Ernst}}, \bibinfo {author} {\bibfnamefont {E.~H.}\ \bibnamefont {Hauge}}, \
  and\ \bibinfo {author} {\bibfnamefont {J.~M.~J.}\ \bibnamefont {van
  Leeuwen}},\ }\bibfield  {title} {\enquote {\bibinfo {title} {Asymptotic time
  behavior of correlation functions},}\ }\href {\doibase
  10.1103/PhysRevLett.25.1254} {\bibfield  {journal} {\bibinfo  {journal}
  {Phys. Rev. Lett.}\ }\textbf {\bibinfo {volume} {25}},\ \bibinfo {pages}
  {1254--1256} (\bibinfo {year} {1970})}\BibitemShut {NoStop}%
\bibitem [{\citenamefont {Ernst}\ \emph {et~al.}(1971)\citenamefont {Ernst},
  \citenamefont {Hauge},\ and\ \citenamefont {van Leeuwen}}]{liqGv6}%
  \BibitemOpen
  \bibfield  {author} {\bibinfo {author} {\bibfnamefont {M.~H.}\ \bibnamefont
  {Ernst}}, \bibinfo {author} {\bibfnamefont {E.~H.}\ \bibnamefont {Hauge}}, \
  and\ \bibinfo {author} {\bibfnamefont {J.~M.~J.}\ \bibnamefont {van
  Leeuwen}},\ }\bibfield  {title} {\enquote {\bibinfo {title} {Asymptotic time
  behavior of correlation functions. i. kinetic terms},}\ }\href {\doibase
  10.1103/PhysRevA.4.2055} {\bibfield  {journal} {\bibinfo  {journal} {Phys.
  Rev. A}\ }\textbf {\bibinfo {volume} {4}},\ \bibinfo {pages} {2055--2065}
  (\bibinfo {year} {1971})}\BibitemShut {NoStop}%
\bibitem [{\citenamefont {Zwanzig}\ and\ \citenamefont {Bixon}(1975)}]{liqGv7}%
  \BibitemOpen
  \bibfield  {author} {\bibinfo {author} {\bibfnamefont {Robert}\ \bibnamefont
  {Zwanzig}}\ and\ \bibinfo {author} {\bibfnamefont {Mordechai}\ \bibnamefont
  {Bixon}},\ }\bibfield  {title} {\enquote {\bibinfo {title} {Compressibility
  effects in the hydrodynamic theory of brownian motion},}\ }\href {\doibase
  10.1017/S0022112075001280} {\bibfield  {journal} {\bibinfo  {journal}
  {Journal of Fluid Mechanics}\ }\textbf {\bibinfo {volume} {69}},\ \bibinfo
  {pages} {2125} (\bibinfo {year} {1975})}\BibitemShut {NoStop}%
\bibitem [{\citenamefont {Chakraborty}(2011)}]{liqGv8}%
  \BibitemOpen
  \bibfield  {author} {\bibinfo {author} {\bibfnamefont {D.}~\bibnamefont
  {Chakraborty}},\ }\bibfield  {title} {\enquote {\bibinfo {title} {{Velocity
  autocorrelation function of a Brownian particle}},}\ }\href {\doibase
  10.1140/epjb/e2011-20395-3} {\bibfield  {journal} {\bibinfo  {journal} {Eur.
  Phys. J. B}\ }\textbf {\bibinfo {volume} {83}},\ \bibinfo {pages} {375}
  (\bibinfo {year} {2011})}\BibitemShut {NoStop}%
\bibitem [{Note5()}]{Note5}%
  \BibitemOpen
  \bibinfo {note} {In certain circles, the time $t^v$ in Eq. (\ref {eqnD+}) is
  referred to as to onset time of the so-called ``correlation hole'' \cite
  {Lev86} in which the autocorrelation function drops below its asymptotic long
  time value (which, in the current case, amounts to the first zero of the
  velocity autocorrelation function).}\BibitemShut {Stop}%
\bibitem [{Note6()}]{Note6}%
  \BibitemOpen
  \bibinfo {note} {We emphasize that $\sigma _{\protect \dot {Y}}$ so defined
  is the standard deviation of $\protect \dot {Y}^{H}$ since the corresponding
  (classical) canonical ensemble average of any time derivative trivially
  vanishes, ${\protect \sf Tr}\leavevmode@ifvmode {\setbox \z@ \hbox
  {\mathsurround \z@ $\nulldelimiterspace \z@ \left (\vcenter to1.5\big@size
  {}\right .$}\box \z@ }\rho _{\Lambda }^{\protect \sf classical~canonical}
  ~\protect \dot {Y}^{H} \leavevmode@ifvmode {\setbox \z@ \hbox {\mathsurround
  \z@ $\nulldelimiterspace \z@ \left )\vcenter to1.5\big@size {}\right .$}\box
  \z@ } =0$.}\BibitemShut {Stop}%
\bibitem [{\citenamefont {Hoover}\ \emph {et~al.}(1980)\citenamefont {Hoover},
  \citenamefont {Evans}, \citenamefont {Hickman}, \citenamefont {Ladd},
  \citenamefont {Ashurst},\ and\ \citenamefont {Moran}}]{Hoover1980}%
  \BibitemOpen
  \bibfield  {author} {\bibinfo {author} {\bibfnamefont {William~G.}\
  \bibnamefont {Hoover}}, \bibinfo {author} {\bibfnamefont {Denis~J.}\
  \bibnamefont {Evans}}, \bibinfo {author} {\bibfnamefont {Richard~B.}\
  \bibnamefont {Hickman}}, \bibinfo {author} {\bibfnamefont {Anthony J.~C.}\
  \bibnamefont {Ladd}}, \bibinfo {author} {\bibfnamefont {William~T.}\
  \bibnamefont {Ashurst}}, \ and\ \bibinfo {author} {\bibfnamefont {Bill}\
  \bibnamefont {Moran}},\ }\bibfield  {title} {\enquote {\bibinfo {title}
  {{Lennard-Jones triple-point bulk and shear viscosities. Green-Kubo theory,
  Hamiltonian mechanics, and nonequilibrium molecular dynamics}},}\ }\href
  {\doibase 10.1103/PhysRevA.22.1690} {\bibfield  {journal} {\bibinfo
  {journal} {Phys. Rev. A}\ }\textbf {\bibinfo {volume} {22}},\ \bibinfo
  {pages} {1690--1697} (\bibinfo {year} {1980})}\BibitemShut {NoStop}%
\bibitem [{\citenamefont {Hiura}\ and\ \citenamefont {Sasa}(2018)}]{Hiura}%
  \BibitemOpen
  \bibfield  {author} {\bibinfo {author} {\bibfnamefont {Ken}\ \bibnamefont
  {Hiura}}\ and\ \bibinfo {author} {\bibfnamefont {Shin-ichi}\ \bibnamefont
  {Sasa}},\ }\bibfield  {title} {\enquote {\bibinfo {title} {{How Does Pressure
  Fluctuate in Equilibrium?}}}\ }\href {\doibase 10.1007/s10955-018-2134-6}
  {\bibfield  {journal} {\bibinfo  {journal} {Journal of Statistical Physics}\
  }\textbf {\bibinfo {volume} {173}},\ \bibinfo {pages} {285--294} (\bibinfo
  {year} {2018})}\BibitemShut {NoStop}%
\bibitem [{Note7()}]{Note7}%
  \BibitemOpen
  \bibinfo {note} {For a function $f$ of general Heisenberg picture position
  operators, we may, in all of the equations of this Section, substitute the
  Schrodinger picture operators by their corresponding form in the Heisenberg
  picture, $p_{i\ell } \to p^{H}_{i \ell }$ and $r_{i \ell } \to r_{i \ell
  }^{H}$, to obtain identical inequalities.}\BibitemShut {Stop}%
\bibitem [{\citenamefont {Larkin}\ and\ \citenamefont
  {Ovchinnikov}(1969)}]{Larkin}%
  \BibitemOpen
  \bibfield  {author} {\bibinfo {author} {\bibfnamefont {A.~I.}\ \bibnamefont
  {Larkin}}\ and\ \bibinfo {author} {\bibfnamefont {Yu~N.}\ \bibnamefont
  {Ovchinnikov}},\ }\bibfield  {title} {\enquote {\bibinfo {title}
  {{Quasiclassical method in the theory of superconductivity}},}\ }\href@noop
  {} {\bibfield  {journal} {\bibinfo  {journal} {Sov. Phys. JETP}\ }\textbf
  {\bibinfo {volume} {28}},\ \bibinfo {pages} {1200--1205} (\bibinfo {year}
  {1969})}\BibitemShut {NoStop}%
\bibitem [{\citenamefont {Roberts}\ \emph {et~al.}(2015)\citenamefont
  {Roberts}, \citenamefont {Stanford},\ and\ \citenamefont {Susskind}}]{RSL}%
  \BibitemOpen
  \bibfield  {author} {\bibinfo {author} {\bibfnamefont {Daniel~A.}\
  \bibnamefont {Roberts}}, \bibinfo {author} {\bibfnamefont {Douglas}\
  \bibnamefont {Stanford}}, \ and\ \bibinfo {author} {\bibfnamefont {Leonard}\
  \bibnamefont {Susskind}},\ }\bibfield  {title} {\enquote {\bibinfo {title}
  {{Localized shocks}},}\ }\href {\doibase 10.1007/JHEP03(2015)051} {\bibfield
  {journal} {\bibinfo  {journal} {Journal of High Energy Physics}\ }\textbf
  {\bibinfo {volume} {2015}},\ \bibinfo {pages} {51} (\bibinfo {year}
  {2015})}\BibitemShut {NoStop}%
\bibitem [{\citenamefont {Kundu}(2021)}]{subleading}%
  \BibitemOpen
  \bibfield  {author} {\bibinfo {author} {\bibfnamefont {Sandipan}\
  \bibnamefont {Kundu}},\ }\bibfield  {title} {\enquote {\bibinfo {title}
  {Subleading bounds on chaos},}\ }\href {https://arxiv.org/pdf/2109.03826.pdf}
  {\bibfield  {journal} {\bibinfo  {journal} {arXiv preprint arXiv:2109.03826}\
  } (\bibinfo {year} {2021})}\BibitemShut {NoStop}%
\bibitem [{\citenamefont {Kukuljan}\ \emph {et~al.}(2017)\citenamefont
  {Kukuljan}, \citenamefont {Saso},\ and\ \citenamefont {Prosen}}]{saso2}%
  \BibitemOpen
  \bibfield  {author} {\bibinfo {author} {\bibfnamefont {Ivan}\ \bibnamefont
  {Kukuljan}}, \bibinfo {author} {\bibfnamefont {Grozdanov}\ \bibnamefont
  {Saso}}, \ and\ \bibinfo {author} {\bibfnamefont {Tomaz}\ \bibnamefont
  {Prosen}},\ }\bibfield  {title} {\enquote {\bibinfo {title} {Weak quantum
  chaos},}\ }\href {\doibase 10.1103/PhysRevB.96.060301} {\bibfield  {journal}
  {\bibinfo  {journal} {Phys. Rev. B}\ }\textbf {\bibinfo {volume} {96}},\
  \bibinfo {pages} {060301(R)} (\bibinfo {year} {2017})}\BibitemShut {NoStop}%
\bibitem [{\citenamefont {Shenker}\ and\ \citenamefont
  {Stanford}(2014)}]{Shenker}%
  \BibitemOpen
  \bibfield  {author} {\bibinfo {author} {\bibfnamefont {Stephen~H.}\
  \bibnamefont {Shenker}}\ and\ \bibinfo {author} {\bibfnamefont {Douglas}\
  \bibnamefont {Stanford}},\ }\bibfield  {title} {\enquote {\bibinfo {title}
  {{Black holes and the butterfly effect}},}\ }\href {\doibase
  10.1007/JHEP03(2014)067} {\bibfield  {journal} {\bibinfo  {journal} {Journal
  of High Energy Physics}\ }\textbf {\bibinfo {volume} {2014}},\ \bibinfo
  {pages} {67} (\bibinfo {year} {2014})}\BibitemShut {NoStop}%
\bibitem [{\citenamefont {Patel}\ and\ \citenamefont
  {Sachdev}(2017)}]{PatelSachdev}%
  \BibitemOpen
  \bibfield  {author} {\bibinfo {author} {\bibfnamefont {Aavishkar~A.}\
  \bibnamefont {Patel}}\ and\ \bibinfo {author} {\bibfnamefont {Subir}\
  \bibnamefont {Sachdev}},\ }\bibfield  {title} {\enquote {\bibinfo {title}
  {Quantum chaos on a critical fermi surface},}\ }\href {\doibase
  10.1073/pnas.1618185114} {\bibfield  {journal} {\bibinfo  {journal}
  {Proceedings of the National Academy of Sciences}\ }\textbf {\bibinfo
  {volume} {114}},\ \bibinfo {pages} {1844--1849} (\bibinfo {year}
  {2017})}\BibitemShut {NoStop}%
\bibitem [{\citenamefont {Zanardi}\ and\ \citenamefont
  {Anand}(2021)}]{Zanardi}%
  \BibitemOpen
  \bibfield  {author} {\bibinfo {author} {\bibfnamefont {Paolo}\ \bibnamefont
  {Zanardi}}\ and\ \bibinfo {author} {\bibfnamefont {Namit}\ \bibnamefont
  {Anand}},\ }\bibfield  {title} {\enquote {\bibinfo {title} {{Information
  Scrambling and Chaos in Open Quantum Systems}},}\ }\href
  {https://arxiv.org/pdf/2012.13172.pdf} {\bibfield  {journal} {\bibinfo
  {journal} {Physical Review A, to appear}\ } (\bibinfo {year}
  {2021})}\BibitemShut {NoStop}%
\bibitem [{\citenamefont {Foini}\ and\ \citenamefont {Kurchan}(2019)}]{Foini}%
  \BibitemOpen
  \bibfield  {author} {\bibinfo {author} {\bibfnamefont {L.}~\bibnamefont
  {Foini}}\ and\ \bibinfo {author} {\bibfnamefont {J.}~\bibnamefont
  {Kurchan}},\ }\bibfield  {title} {\enquote {\bibinfo {title} {The eigenstate
  thermalization hypothesis and out of time order correlators},}\ }\href
  {\doibase 10.1103/PhysRevE.99.042139} {\bibfield  {journal} {\bibinfo
  {journal} {Phys. Rev. E}\ }\textbf {\bibinfo {volume} {99}},\ \bibinfo
  {pages} {042139} (\bibinfo {year} {2019})}\BibitemShut {NoStop}%
\bibitem [{\citenamefont {Murthy}\ and\ \citenamefont
  {Srednicki}(2019)}]{MM19}%
  \BibitemOpen
  \bibfield  {author} {\bibinfo {author} {\bibfnamefont {Chaitanya}\
  \bibnamefont {Murthy}}\ and\ \bibinfo {author} {\bibfnamefont {Mark}\
  \bibnamefont {Srednicki}},\ }\bibfield  {title} {\enquote {\bibinfo {title}
  {Bounds on chaos from the eigenstate thermalization hypothesis},}\ }\href
  {\doibase 10.1103/PhysRevLett.123.230606} {\bibfield  {journal} {\bibinfo
  {journal} {Phys. Rev. Lett.}\ }\textbf {\bibinfo {volume} {123}},\ \bibinfo
  {pages} {230606} (\bibinfo {year} {2019})}\BibitemShut {NoStop}%
\bibitem [{\citenamefont {Kurchan}(2018)}]{Kurchan}%
  \BibitemOpen
  \bibfield  {author} {\bibinfo {author} {\bibfnamefont {J.}~\bibnamefont
  {Kurchan}},\ }\bibfield  {title} {\enquote {\bibinfo {title} {{Quantum Bound
  to Chaos and the Semiclassical Limit}},}\ }\href {\doibase
  10.1007/s10955-018-2052-7} {\bibfield  {journal} {\bibinfo  {journal}
  {Journal of Statistical Physics}\ }\textbf {\bibinfo {volume} {171}},\
  \bibinfo {pages} {965} (\bibinfo {year} {2018})}\BibitemShut {NoStop}%
\bibitem [{Note8()}]{Note8}%
  \BibitemOpen
  \bibinfo {note} {Our results are actually stronger than those suggested by
  the semiclassical analysis alone that led to (\ref {longobvi1111}). This
  equation and all that followed from Eq. (\ref {longobvi}) are applicable to
  semiclassical systems (since Eq. (\ref {longobvi}) is a consequence of the
  semiclassical Eq. (\ref {squared-ineq})). We work within the semiclassical
  regime since the notions of chaos (including the definition of the Lyapunov
  exponent) are better described classically. Fundamentally, however, in both
  the real basic quantum as well as the semiclassical setting, the very
  existence of different initial conditions allowing for the common textbook
  definition of Lyapunov exponents does allow one to simply set both initial
  conditions (or associated thermal probability densities $\rho _{\Lambda
  }^{(1)}$ and $\rho _{\Lambda }^{(2)}$ (with these defined as in Section \ref
  {setup})) to be the same (i.e., with both of these probability densities
  being equal to $\rho _{\Lambda }^{\protect \sf canonical}$). Thus, in order
  to define the problem so as to address what is typically measured classically
  when making small changes to the initial conditions, one cannot readily set
  the initial probability densities to be the canonical probability density
  since this will yield a vanishing result for the measured ${\protect \sf
  Tr}\leavevmode@ifvmode {\setbox \z@ \hbox {\mathsurround \z@
  $\nulldelimiterspace \z@ \left (\vcenter to1.5\big@size {}\right .$}\box \z@
  }(\rho _{\Lambda }^{(1)} - \rho _{\Lambda }^{(2)} ) Q_{i}^{H}(t)
  \leavevmode@ifvmode {\setbox \z@ \hbox {\mathsurround \z@
  $\nulldelimiterspace \z@ \left )\vcenter to1.5\big@size {}\right .$}\box \z@
  }$. Our construct of $\rho _{\Lambda }^{1 \cup 2}$ and proof in Section \ref
  {sec:chaos} circumvents these problems. With all of the above cautionary
  remarks now made, we note that along lines identical to those pursued in the
  current Section \ref {sec:chaos}, we may alternatively derive Eq. (\ref
  {longobvi1111}) as an exact inequality at all times by directly invoking Eqs.
  (\ref {central1}, \ref {a-central-1}) within a more general (i.e., not
  necessarily only semiclassical) framework.}\BibitemShut {Stop}%
\bibitem [{Note9()}]{Note9}%
  \BibitemOpen
  \bibinfo {note} {Violations of Eq. (\ref {DOTOC}) are found in crystalline
  systems. As we remarked earlier, during the long time oscillatory motion of
  ions in a crystal, the velocity autocorrelation function does not exhibit a
  simple exponential decay in time and notable negative contributions to the
  Green-Kubo integral can arise. The behavior of the long time velocity
  autocorrelation function is not universal. Depending on system details,
  oscillatory long time tails of the autocorrelation function may also lead to
  positive contributions to the Green-Kubo integral.}\BibitemShut {Stop}%
\bibitem [{\citenamefont {Leviandier}\ \emph {et~al.}(1986)\citenamefont
  {Leviandier}, \citenamefont {Jost},\ and\ \citenamefont {Pique}}]{Lev86}%
  \BibitemOpen
  \bibfield  {author} {\bibinfo {author} {\bibfnamefont {L.}~\bibnamefont
  {Leviandier}}, \bibinfo {author} {\bibfnamefont {R.}~\bibnamefont {Jost}}, \
  and\ \bibinfo {author} {\bibfnamefont {J.~P.}\ \bibnamefont {Pique}},\
  }\bibfield  {title} {\enquote {\bibinfo {title} {Fourier transform: A tool to
  measure statistical level properties in very complex spectra},}\ }\href
  {\doibase 10.1103/PhysRevLett.56.2449} {\bibfield  {journal} {\bibinfo
  {journal} {Phys. Rev. Lett.}\ }\textbf {\bibinfo {volume} {56}},\ \bibinfo
  {pages} {2449} (\bibinfo {year} {1986})}\BibitemShut {NoStop}%
\bibitem [{\citenamefont {Gunnarsson}\ \emph {et~al.}(2003)\citenamefont
  {Gunnarsson}, \citenamefont {Calandra},\ and\ \citenamefont
  {Han}}]{saturate1}%
  \BibitemOpen
  \bibfield  {author} {\bibinfo {author} {\bibfnamefont {O.}~\bibnamefont
  {Gunnarsson}}, \bibinfo {author} {\bibfnamefont {M.}~\bibnamefont
  {Calandra}}, \ and\ \bibinfo {author} {\bibfnamefont {J.~E.}\ \bibnamefont
  {Han}},\ }\bibfield  {title} {\enquote {\bibinfo {title} {{Colloquium:
  Saturation of electrical resistivity}},}\ }\href {\doibase
  10.1103/RevModPhys.75.1085} {\bibfield  {journal} {\bibinfo  {journal} {Rev.
  Mod. Phys.}\ }\textbf {\bibinfo {volume} {75}},\ \bibinfo {pages} {1085}
  (\bibinfo {year} {2003})}\BibitemShut {NoStop}%
\bibitem [{\citenamefont {Emery}\ and\ \citenamefont {Kivelson}(1995)}]{bad}%
  \BibitemOpen
  \bibfield  {author} {\bibinfo {author} {\bibfnamefont {V.~J.}\ \bibnamefont
  {Emery}}\ and\ \bibinfo {author} {\bibfnamefont {S.~A.}\ \bibnamefont
  {Kivelson}},\ }\bibfield  {title} {\enquote {\bibinfo {title}
  {{Superconductivity in bad metals }},}\ }\href {\doibase
  10.1103/PhysRevLett.74.3253} {\bibfield  {journal} {\bibinfo  {journal}
  {Phys. Rev. Lett.}\ }\textbf {\bibinfo {volume} {74}},\ \bibinfo {pages}
  {3253} (\bibinfo {year} {1995})}\BibitemShut {NoStop}%
\bibitem [{\citenamefont {Hussey}\ \emph {et~al.}(2004)\citenamefont {Hussey},
  \citenamefont {Takenaka},\ and\ \citenamefont {Takagi}}]{badmetal}%
  \BibitemOpen
  \bibfield  {author} {\bibinfo {author} {\bibfnamefont {N.~E.}\ \bibnamefont
  {Hussey}}, \bibinfo {author} {\bibfnamefont {K.}~\bibnamefont {Takenaka}}, \
  and\ \bibinfo {author} {\bibfnamefont {H.}~\bibnamefont {Takagi}},\
  }\bibfield  {title} {\enquote {\bibinfo {title} {Universality of the
  mott–ioffe–regel limit in metals},}\ }\href {\doibase
  10.1080/14786430410001716944} {\bibfield  {journal} {\bibinfo  {journal}
  {Philosophical Magazine}\ }\textbf {\bibinfo {volume} {84}},\ \bibinfo
  {pages} {2847--2864} (\bibinfo {year} {2004})}\BibitemShut {NoStop}%
\bibitem [{\citenamefont {Legros}\ \emph {et~al.}(2018)\citenamefont {Legros},
  \citenamefont {Benhabib}, \citenamefont {Lalibert´e}, \citenamefont {Dion},
  \citenamefont {Lizaire}, \citenamefont {Vignolle}, \citenamefont {Vignolles},
  \citenamefont {Raffy}, \citenamefont {Li}, \citenamefont {Auban-Senzier},
  \citenamefont {Doiron-Leyraud}, \citenamefont {Fournier}, \citenamefont
  {Colson}, \citenamefont {Taillefer}, ,\ and\ \citenamefont
  {Proust}}]{Legros18A}%
  \BibitemOpen
  \bibfield  {author} {\bibinfo {author} {\bibfnamefont {A.}~\bibnamefont
  {Legros}}, \bibinfo {author} {\bibfnamefont {W.}~\bibnamefont {Benhabib},
  \bibfnamefont {S.~an~Tabis}}, \bibinfo {author} {\bibfnamefont
  {F.}~\bibnamefont {Lalibert´e}}, \bibinfo {author} {\bibfnamefont
  {M.}~\bibnamefont {Dion}}, \bibinfo {author} {\bibfnamefont {M.}~\bibnamefont
  {Lizaire}}, \bibinfo {author} {\bibfnamefont {B.}~\bibnamefont {Vignolle}},
  \bibinfo {author} {\bibfnamefont {D.}~\bibnamefont {Vignolles}}, \bibinfo
  {author} {\bibfnamefont {H.}~\bibnamefont {Raffy}}, \bibinfo {author}
  {\bibfnamefont {Z.~Z.}\ \bibnamefont {Li}}, \bibinfo {author} {\bibfnamefont
  {P.}~\bibnamefont {Auban-Senzier}}, \bibinfo {author} {\bibfnamefont
  {N.}~\bibnamefont {Doiron-Leyraud}}, \bibinfo {author} {\bibfnamefont
  {P.}~\bibnamefont {Fournier}}, \bibinfo {author} {\bibfnamefont
  {.}~\bibnamefont {Colson}}, \bibinfo {author} {\bibfnamefont
  {L.}~\bibnamefont {Taillefer}}, , \ and\ \bibinfo {author} {\bibfnamefont
  {C.}~\bibnamefont {Proust}},\ }\bibfield  {title} {\enquote {\bibinfo {title}
  {{Universal T-linear resistivity and Planckian dissipation in overdoped
  cuprates}},}\ }\href@noop {} {\bibfield  {journal} {\bibinfo  {journal}
  {Nature Physics}\ }\textbf {\bibinfo {volume} {15}},\ \bibinfo {pages} {142}
  (\bibinfo {year} {2018})}\BibitemShut {NoStop}%
\bibitem [{\citenamefont {Lindner}\ and\ \citenamefont
  {Auerbach}(2010)}]{AssaBose}%
  \BibitemOpen
  \bibfield  {author} {\bibinfo {author} {\bibfnamefont {Netanel~H.}\
  \bibnamefont {Lindner}}\ and\ \bibinfo {author} {\bibfnamefont {Assa}\
  \bibnamefont {Auerbach}},\ }\bibfield  {title} {\enquote {\bibinfo {title}
  {{Conductivity of hard core bosons: A paradigm of a bad metal}},}\ }\href
  {\doibase 10.1103/PhysRevB.81.054512} {\bibfield  {journal} {\bibinfo
  {journal} {Phys. Rev. B}\ }\textbf {\bibinfo {volume} {81}},\ \bibinfo
  {pages} {054512} (\bibinfo {year} {2010})}\BibitemShut {NoStop}%
\bibitem [{\citenamefont {Mukerjee}\ \emph {et~al.}(2006)\citenamefont
  {Mukerjee}, \citenamefont {Oganesyan},\ and\ \citenamefont {Huse}}]{VadimT}%
  \BibitemOpen
  \bibfield  {author} {\bibinfo {author} {\bibfnamefont {Subroto}\ \bibnamefont
  {Mukerjee}}, \bibinfo {author} {\bibfnamefont {Vadim}\ \bibnamefont
  {Oganesyan}}, \ and\ \bibinfo {author} {\bibfnamefont {David}\ \bibnamefont
  {Huse}},\ }\bibfield  {title} {\enquote {\bibinfo {title} {Statistical theory
  of transport by strongly interacting lattice fermions},}\ }\href {\doibase
  10.1103/PhysRevB.73.035113} {\bibfield  {journal} {\bibinfo  {journal} {Phys.
  Rev. B}\ }\textbf {\bibinfo {volume} {73}},\ \bibinfo {pages} {035113}
  (\bibinfo {year} {2006})}\BibitemShut {NoStop}%
\bibitem [{\citenamefont {Mott}(1929)}]{NMott}%
  \BibitemOpen
  \bibfield  {author} {\bibinfo {author} {\bibfnamefont {Nevill}\ \bibnamefont
  {Mott}},\ }\bibfield  {title} {\enquote {\bibinfo {title} {{The Wave
  Mechanics of $\alpha$-Ray Tracks}},}\ }\href {\doibase
  10.1098/rspa.1929.0205} {\bibfield  {journal} {\bibinfo  {journal}
  {Proceedings of the Royal Society}\ }\textbf {\bibinfo {volume} {A126}},\
  \bibinfo {pages} {79} (\bibinfo {year} {1929})}\BibitemShut {NoStop}%
\bibitem [{\citenamefont {von Neumann}(1955)}]{johnvN}%
  \BibitemOpen
  \bibfield  {author} {\bibinfo {author} {\bibfnamefont {John}\ \bibnamefont
  {von Neumann}},\ }\href@noop {} {\emph {\bibinfo {title} {{Mathematical
  Foundations of Quantum Mechanics}}}}\ (\bibinfo  {publisher} {Princeton
  University Press},\ \bibinfo {year} {1955})\BibitemShut {NoStop}%
\bibitem [{\citenamefont {Allahverdyan}\ \emph {et~al.}(2013)\citenamefont
  {Allahverdyan}, \citenamefont {Balian},\ and\ \citenamefont
  {Nieuwenhuizen}}]{measure_dyn}%
  \BibitemOpen
  \bibfield  {author} {\bibinfo {author} {\bibfnamefont {Armen~E.}\
  \bibnamefont {Allahverdyan}}, \bibinfo {author} {\bibfnamefont {Roger}\
  \bibnamefont {Balian}}, \ and\ \bibinfo {author} {\bibfnamefont {Theo~M.}\
  \bibnamefont {Nieuwenhuizen}},\ }\bibfield  {title} {\enquote {\bibinfo
  {title} {{Understanding quantum measurement from the solution of dynamical
  models}},}\ }\href {\doibase 10.1016/j.physrep.2012.11.001} {\bibfield
  {journal} {\bibinfo  {journal} {Physics Reports}\ }\textbf {\bibinfo {volume}
  {525}},\ \bibinfo {pages} {1} (\bibinfo {year} {2013})}\BibitemShut {NoStop}%
\bibitem [{Note10()}]{Note10}%
  \BibitemOpen
  \bibinfo {note} {Strictly speaking, we should look at the frequency variance
  of the integrand appearing in Eq. (\ref {finite_time_average}) and recall
  that the Hamiltonian $\protect \tilde {H}^{H}_{i}(t)$ defining the time
  ordered exponential ${\protect \mathbbm {U}}_{i}(t)$ is time dependent. We
  approximate the standard deviation by the time independent average of
  $(k_{B}T^{2} C _{v,i})$ as defined by Eq. (\ref {cv2}) over all
  $i$.}\BibitemShut {Stop}%
\bibitem [{Note11()}]{Note11}%
  \BibitemOpen
  \bibinfo {note} {Whenever $H_{\Lambda }$ has degeneracies, we may construct a
  common diagonal eigenbasis of this long time average operator and $H_{\Lambda
  }$.}\BibitemShut {Stop}%
\bibitem [{\citenamefont {Farquhar}(1964)}]{book-far}%
  \BibitemOpen
  \bibfield  {author} {\bibinfo {author} {\bibfnamefont {I.~E.}\ \bibnamefont
  {Farquhar}},\ }\href@noop {} {\emph {\bibinfo {title} {{Ergodic theory in
  statistical mechanics}}}}\ (\bibinfo  {publisher} {Interscience},\ \bibinfo
  {year} {1964})\BibitemShut {NoStop}%
\bibitem [{\citenamefont {Jaynes}(1957{\natexlab{a}})}]{Jaynes1}%
  \BibitemOpen
  \bibfield  {author} {\bibinfo {author} {\bibfnamefont {E.~T.}\ \bibnamefont
  {Jaynes}},\ }\bibfield  {title} {\enquote {\bibinfo {title} {{Information
  Theory and Statistical Mechanics}},}\ }\href {\doibase
  10.1103/PhysRev.106.620} {\bibfield  {journal} {\bibinfo  {journal} {Physical
  Review}\ }\textbf {\bibinfo {volume} {106}},\ \bibinfo {pages} {620}
  (\bibinfo {year} {1957}{\natexlab{a}})}\BibitemShut {NoStop}%
\bibitem [{\citenamefont {Jaynes}(1957{\natexlab{b}})}]{Jaynes2}%
  \BibitemOpen
  \bibfield  {author} {\bibinfo {author} {\bibfnamefont {E.~T.}\ \bibnamefont
  {Jaynes}},\ }\bibfield  {title} {\enquote {\bibinfo {title} {{Information
  Theory and Statistical Mechanics II}},}\ }\href {\doibase
  10.1103/PhysRev.108.171} {\bibfield  {journal} {\bibinfo  {journal} {Physical
  Review}\ }\textbf {\bibinfo {volume} {108}},\ \bibinfo {pages} {171}
  (\bibinfo {year} {1957}{\natexlab{b}})}\BibitemShut {NoStop}%
\bibitem [{\citenamefont {Villain}\ \emph {et~al.}(1980)\citenamefont
  {Villain}, \citenamefont {Bidaux}, \citenamefont {Carton},\ and\
  \citenamefont {Conte}}]{OBD}%
  \BibitemOpen
  \bibfield  {author} {\bibinfo {author} {\bibfnamefont {J.}~\bibnamefont
  {Villain}}, \bibinfo {author} {\bibfnamefont {R.}~\bibnamefont {Bidaux}},
  \bibinfo {author} {\bibfnamefont {J.~P.}\ \bibnamefont {Carton}}, \ and\
  \bibinfo {author} {\bibfnamefont {R.}~\bibnamefont {Conte}},\ }\bibfield
  {title} {\enquote {\bibinfo {title} {Order as an effect of disorder},}\
  }\href {\doibase 10.1051/jphys:0198000410110126300} {\bibfield  {journal}
  {\bibinfo  {journal} {J. Physique}\ }\textbf {\bibinfo {volume} {41}},\
  \bibinfo {pages} {1263} (\bibinfo {year} {1980})}\BibitemShut {NoStop}%
\bibitem [{\citenamefont {Henley}(1989)}]{OBD1}%
  \BibitemOpen
  \bibfield  {author} {\bibinfo {author} {\bibfnamefont {C.~L.}\ \bibnamefont
  {Henley}},\ }\bibfield  {title} {\enquote {\bibinfo {title} {Ordering due to
  disorder in a frustrated vector antiferromagnet},}\ }\href {\doibase
  10.1103/PhysRevLett.62.2056} {\bibfield  {journal} {\bibinfo  {journal}
  {Phys. Rev. Lett.}\ }\textbf {\bibinfo {volume} {62}},\ \bibinfo {pages}
  {2056} (\bibinfo {year} {1989})}\BibitemShut {NoStop}%
\bibitem [{\citenamefont {Nussinov}\ \emph {et~al.}(2004)\citenamefont
  {Nussinov}, \citenamefont {Biskup}, \citenamefont {Chayes},\ and\
  \citenamefont {van~den Brink}}]{NBCB}%
  \BibitemOpen
  \bibfield  {author} {\bibinfo {author} {\bibfnamefont {Z.}~\bibnamefont
  {Nussinov}}, \bibinfo {author} {\bibfnamefont {M.}~\bibnamefont {Biskup}},
  \bibinfo {author} {\bibfnamefont {L.}~\bibnamefont {Chayes}}, \ and\ \bibinfo
  {author} {\bibfnamefont {J.}~\bibnamefont {van~den Brink}},\ }\bibfield
  {title} {\enquote {\bibinfo {title} {Orbital order in classical models of
  transition-metal compounds},}\ }\href {\doibase 10.1209/epl/i2004-10134-5}
  {\bibfield  {journal} {\bibinfo  {journal} {Europhysics Letters}\ }\textbf
  {\bibinfo {volume} {67}},\ \bibinfo {pages} {990} (\bibinfo {year}
  {2004})}\BibitemShut {NoStop}%
\bibitem [{\citenamefont {Biskup}\ \emph {et~al.}(2005)\citenamefont {Biskup},
  \citenamefont {Chayes},\ and\ \citenamefont {Nussinov}}]{BCN}%
  \BibitemOpen
  \bibfield  {author} {\bibinfo {author} {\bibfnamefont {M.}~\bibnamefont
  {Biskup}}, \bibinfo {author} {\bibfnamefont {L.}~\bibnamefont {Chayes}}, \
  and\ \bibinfo {author} {\bibfnamefont {Z.}~\bibnamefont {Nussinov}},\
  }\bibfield  {title} {\enquote {\bibinfo {title} {{Orbital Ordering in
  Transition Metal Compounds: I. The 120-Degree Model}},}\ }\href {\doibase
  10.1007/s00220-004-1272-7} {\bibfield  {journal} {\bibinfo  {journal}
  {Communications in Mathematical Physics}\ }\textbf {\bibinfo {volume}
  {255}},\ \bibinfo {pages} {253} (\bibinfo {year} {2005})}\BibitemShut
  {NoStop}%
\bibitem [{\citenamefont {Jackeli}\ and\ \citenamefont {Avella}(2015)}]{QOBDK}%
  \BibitemOpen
  \bibfield  {author} {\bibinfo {author} {\bibfnamefont {George}\ \bibnamefont
  {Jackeli}}\ and\ \bibinfo {author} {\bibfnamefont {Adolfo}\ \bibnamefont
  {Avella}},\ }\bibfield  {title} {\enquote {\bibinfo {title} {{Quantum order
  by disorder in the Kitaev model on a triangular lattice}},}\ }\href {\doibase
  10.1103/PhysRevB.92.184416} {\bibfield  {journal} {\bibinfo  {journal} {Phys.
  Rev. B}\ }\textbf {\bibinfo {volume} {92}},\ \bibinfo {pages} {184416}
  (\bibinfo {year} {2015})}\BibitemShut {NoStop}%
\bibitem [{Note12()}]{Note12}%
  \BibitemOpen
  \bibinfo {note} {This a trivial extension of the variational principle. For
  fluctuations about the $m$-th eigenstate of the Hamiltonian, the energy of
  the state $| \psi \rangle = \DOTSB \sum@ \slimits@ _{n \protect \neq m}
  (c_{n}^{(0)} + \delta c_{n} )| \phi _{n} \rangle + \protect \sqrt {1- \DOTSB
  \sum@ \slimits@ _{n \protect \neq m} |(c_{n}^{(0)} + \delta c_{n})|^{2}}
  |\phi _{m} \rangle $ will vary only quadratically in $\{\delta c_{n}\}$ about
  that of $|\psi ^{0} \rangle $. That is, for such states, the energy $E =
  \DOTSB \sum@ \slimits@ _{n} |\delta c_{n}|^{2} E_{n} + (1- \DOTSB \sum@
  \slimits@ _{n \protect \neq m} |\delta c_{n}|^{2}) E_{m}$. By contrast,
  fluctuations about states that are not eigenstates of the Hamiltonian will
  generally yield contributions to the energy that are linear in $\{\delta
  c_{n}\}$.}\BibitemShut {Stop}%
\bibitem [{Note13()}]{Note13}%
  \BibitemOpen
  \bibinfo {note} {Here we consider a non-relativistic system given by Eq.
  (\ref {eq:HLAMBDA}). As usual for such Hamiltonians with a potential energy
  that depends only on particle positions, the classical canonical probability
  density $\rho _{\Lambda }^{\protect \sf classical~canonical}$ factorizes into
  real and momentum space distributions. The expectation value of any single
  particle ratio $(hm/\protect \bm {p}^2)$ involves only the single particle
  momentum space average displayed in Eq. (\ref {classicalh}).}\BibitemShut
  {Stop}%
\bibitem [{\citenamefont {Reif}(1965)}]{Reif}%
  \BibitemOpen
  \bibfield  {author} {\bibinfo {author} {\bibfnamefont {F.}~\bibnamefont
  {Reif}},\ }\href@noop {} {\emph {\bibinfo {title} {{Fundamentals of
  statistical and thermal physics}}}}\ (\bibinfo  {publisher} {McGraw-Hill, New
  York NY},\ \bibinfo {year} {1965})\BibitemShut {NoStop}%
\bibitem [{\citenamefont {Tisza}(1942)}]{Tisza1941}%
  \BibitemOpen
  \bibfield  {author} {\bibinfo {author} {\bibfnamefont {L.}~\bibnamefont
  {Tisza}},\ }\bibfield  {title} {\enquote {\bibinfo {title} {{Supersonic
  Absorption and Stokes' Viscosity Relation}},}\ }\href {\doibase
  10.1103/PhysRev.61.531} {\bibfield  {journal} {\bibinfo  {journal} {Phys.
  Rev.}\ }\textbf {\bibinfo {volume} {61}},\ \bibinfo {pages} {531--536}
  (\bibinfo {year} {1942})}\BibitemShut {NoStop}%
\bibitem [{\citenamefont {Jaeger}\ \emph {et~al.}(2018)\citenamefont {Jaeger},
  \citenamefont {Matar},\ and\ \citenamefont {Muller}}]{Muller2018}%
  \BibitemOpen
  \bibfield  {author} {\bibinfo {author} {\bibfnamefont {F.}~\bibnamefont
  {Jaeger}}, \bibinfo {author} {\bibfnamefont {O.~M.}\ \bibnamefont {Matar}}, \
  and\ \bibinfo {author} {\bibfnamefont {E.~A.}\ \bibnamefont {Muller}},\
  }\bibfield  {title} {\enquote {\bibinfo {title} {Bulk viscosity of molecular
  fluids},}\ }\href {10.1063/1.5022752} {\bibfield  {journal} {\bibinfo
  {journal} {The Journal of Physical Chemistry}\ }\textbf {\bibinfo {volume}
  {148}},\ \bibinfo {pages} {174504} (\bibinfo {year} {2018})}\BibitemShut
  {NoStop}%
\bibitem [{Note14()}]{Note14}%
  \BibitemOpen
  \bibinfo {note} {The mean of Eq (\ref {relativistich}) is a classical
  canonical expectation value associated with a general position dependent
  potential energy augmenting the sum of single particle ultra-relativistic
  kinetic energies $c \DOTSB \sum@ \slimits@ _{i=1}^{N_{\Lambda }} |\protect
  \bm {p}_{i}|$. For such classical canonical averages (as in the
  non-relativistic system of Eq. (\ref {classicalh})), the probability density
  $\rho _{\Lambda }^{\protect \sf classical~canonical}$ factorizes into real
  and momentum space parts with the average of the single particle
  $h/(c|\protect \bm {p}|)$ set by the momentum space integral ratio of Eq.
  (\ref {relativistich}).}\BibitemShut {Stop}%
\bibitem [{Note15()}]{Note15}%
  \BibitemOpen
  \bibinfo {note} {Among other approximations, apart from performing classical
  phase (momentum) space averages (Eqs. (\ref {classicalh}, \ref
  {relativistich})), we wish to highlight that we assumed that the kinetic
  energy per particle following the collisions (interactions) with the other
  particles remains of the same order of magnitude as the average value set by
  the thermal average at temperature $T$ after equilibration}\BibitemShut
  {NoStop}%
\bibitem [{Note16()}]{Note16}%
  \BibitemOpen
  \bibinfo {note} {Our focus in deriving Eqs. (\ref {classicalh}, \ref
  {relativistich}) has been on simplified (semi-)classical systems. These
  considerations do not apply to numerous quantum extensions of classical
  systems, including, e.g, the dilute classical gases that we discussed. For
  instance, as is well known, in dilute degenerate gases at ultra cold
  temperatures, cross-sections can become large (as in the divergent scattering
  length at the Feshbach resonance marking the BCS-BEC crossover e.g., \cite
  {BEC-BCS-R},\cite {BEC-BCS}) and the mean-free path may then be very
  small.}\BibitemShut {Stop}%
\bibitem [{\citenamefont {Kelton}\ and\ \citenamefont
  {Greer}(2010)}]{Ken-book}%
  \BibitemOpen
  \bibfield  {author} {\bibinfo {author} {\bibfnamefont {K.~F.}\ \bibnamefont
  {Kelton}}\ and\ \bibinfo {author} {\bibfnamefont {A.~L.}\ \bibnamefont
  {Greer}},\ }\href@noop {} {\emph {\bibinfo {title} {{Nucleation in Condensed
  Matter, Applications in Materials and Biology}}}}\ (\bibinfo  {publisher}
  {Elsevier (Pergamon Materials Series)},\ \bibinfo {year} {2010})\BibitemShut
  {NoStop}%
\bibitem [{\citenamefont {Weingartner}\ \emph {et~al.}(2016)\citenamefont
  {Weingartner}, \citenamefont {Pubelo}, \citenamefont {Nogueira},
  \citenamefont {Kelton},\ and\ \citenamefont {Nussinov}}]{Nick2}%
  \BibitemOpen
  \bibfield  {author} {\bibinfo {author} {\bibfnamefont {N.~B.}\ \bibnamefont
  {Weingartner}}, \bibinfo {author} {\bibfnamefont {C.}~\bibnamefont {Pubelo}},
  \bibinfo {author} {\bibfnamefont {F.~S.}\ \bibnamefont {Nogueira}}, \bibinfo
  {author} {\bibfnamefont {K.~F.}\ \bibnamefont {Kelton}}, \ and\ \bibinfo
  {author} {\bibfnamefont {Z.}~\bibnamefont {Nussinov}},\ }\bibfield  {title}
  {\enquote {\bibinfo {title} {A phase space approach to supercooled liquids
  and a universal collapse of their viscosity},}\ }\href {\doibase
  10.3389/fmats.2016.00050} {\bibfield  {journal} {\bibinfo  {journal}
  {Frontiers in Materials}\ }\textbf {\bibinfo {volume} {3}},\ \bibinfo {pages}
  {50} (\bibinfo {year} {2016})}\BibitemShut {NoStop}%
\bibitem [{\citenamefont {Berthier}\ and\ \citenamefont {Biroli}(2011)}]{Ludo}%
  \BibitemOpen
  \bibfield  {author} {\bibinfo {author} {\bibfnamefont {Ludovic}\ \bibnamefont
  {Berthier}}\ and\ \bibinfo {author} {\bibfnamefont {Giulio}\ \bibnamefont
  {Biroli}},\ }\bibfield  {title} {\enquote {\bibinfo {title} {{Theoretical
  perspective on the glass transition and amorphous materials}},}\ }\href
  {\doibase 10.1103/RevModPhys.83.587} {\bibfield  {journal} {\bibinfo
  {journal} {Rev. Mod. Phys.}\ }\textbf {\bibinfo {volume} {83}},\ \bibinfo
  {pages} {587} (\bibinfo {year} {2011})}\BibitemShut {NoStop}%
\bibitem [{\citenamefont {Stillinger}\ and\ \citenamefont
  {Debenedetti}(2013)}]{Glass2}%
  \BibitemOpen
  \bibfield  {author} {\bibinfo {author} {\bibfnamefont {Frank~H.}\
  \bibnamefont {Stillinger}}\ and\ \bibinfo {author} {\bibfnamefont {Pablo~G.}\
  \bibnamefont {Debenedetti}},\ }\bibfield  {title} {\enquote {\bibinfo {title}
  {Glass transition thermodynamics and kinetics},}\ }\href {\doibase
  10.1146/annurev-conmatphys-030212-184329} {\bibfield  {journal} {\bibinfo
  {journal} {Annual Review of Condensed Matter Physics}\ }\textbf {\bibinfo
  {volume} {4}},\ \bibinfo {pages} {263} (\bibinfo {year} {2013})}\BibitemShut
  {NoStop}%
\bibitem [{\citenamefont {Richter}\ \emph {et~al.}(2020)\citenamefont
  {Richter}, \citenamefont {Dymarsky}, \citenamefont {Steinigeweg},\ and\
  \citenamefont {Gemmer}}]{Tolya1}%
  \BibitemOpen
  \bibfield  {author} {\bibinfo {author} {\bibfnamefont {J.}~\bibnamefont
  {Richter}}, \bibinfo {author} {\bibfnamefont {A.}~\bibnamefont {Dymarsky}},
  \bibinfo {author} {\bibfnamefont {R.}~\bibnamefont {Steinigeweg}}, \ and\
  \bibinfo {author} {\bibfnamefont {J.}~\bibnamefont {Gemmer}},\ }\bibfield
  {title} {\enquote {\bibinfo {title} {Eigenstate thermalization hypothesis
  beyond standard indicators: Emergence of random-matrix behavior at small
  frequencies},}\ }\href {\doibase 10.1103/PhysRevE.102.042127} {\bibfield
  {journal} {\bibinfo  {journal} {Phys. Rev. E}\ }\textbf {\bibinfo {volume}
  {102}},\ \bibinfo {pages} {042127} (\bibinfo {year} {2020})}\BibitemShut
  {NoStop}%
\bibitem [{\citenamefont {Dymarsky}(2018)}]{Tolya2}%
  \BibitemOpen
  \bibfield  {author} {\bibinfo {author} {\bibfnamefont {A.}~\bibnamefont
  {Dymarsky}},\ }\bibfield  {title} {\enquote {\bibinfo {title} {Bound on
  eigenstate thermalization from transport},}\ }\href
  {https://arxiv.org/pdf/1804.08626.pdf} {\bibfield  {journal} {\bibinfo
  {journal} {arXiv preprint arXiv:1804.08626}\ } (\bibinfo {year}
  {2018})}\BibitemShut {NoStop}%
\bibitem [{Note17()}]{Note17}%
  \BibitemOpen
  \bibinfo {note} {The average $ [\protect \dot {Q}^{ETH}_{mn}]_{\protect \cal
  {P}} =0$ since $[R_{mn}]_{\protect \cal {P}} =0$. Thus, the variance is
  indeed $[|\protect \dot {Q}^{ETH}_{mn}|^2]_{\protect \cal {P}} - ([\protect
  \dot {Q}^{ETH}_{mn}]_{\protect \cal {P}})^2 = [|\protect \dot
  {Q}^{ETH}_{mn}|^2]_{\protect \cal {P}}$.}\BibitemShut {Stop}%
\bibitem [{Note18()}]{Note18}%
  \BibitemOpen
  \bibinfo {note} {For the particular case of an exponential decay of
  correlations with distance in ${\protect \sf G}_1^{i'j'}$, the variance of
  $(H_1/N)$ scales as $1/N$ while for an algebraic decay of ${\protect \sf
  G}_1^{i'j'} \propto |i'-j'|^{p}$ for a system in $d$ spatial dimensions, the
  variance of $(H_1/N)$ scales as $N^{-d/p}$. In both cases, the variance tends
  to zero as $N \to \infty $. More generally, this variance will vanish
  whenever the correlations decay to zero in the limit of large
  distances.}\BibitemShut {Stop}%
\bibitem [{Note19()}]{Note19}%
  \BibitemOpen
  \bibinfo {note} {For completeness, we remark that there are numerous
  situations involving open systems in which intensive state variables are not
  sharp and divergent long range connected correlations ${\protect \sf
  G}^{i'j'}$ exist. These appear, e.g., in systems that are coupled to an
  external drive \cite {bound} as well as the condensate number fluctuations in
  a Bose condensate within the grand canonical ensemble (where the chemical
  potential term plays the role of an external bath that couples to all
  particles in the system). For such an open system with a chemical potential,
  below the Bose-Einstein condensate temperature, the standard deviation
  $\sigma _{N_{0}}$ of the total number of particles in the condensate and the
  number of particles in the condensate are the of the scale of the number of
  particles in the system, $\sigma _{N_{0}} = {\protect \cal {O}}(N)$. Thus,
  the condensate fraction $N_{0}/N$ is not sharp and the associated connected
  condensate number correlation functions do not decay to zero at long
  distances. Similar physics also appears in the classical mean spherical model
  below its critical temperature in which the mean spherical constraint is
  implemented by a Lagrange multiplier (serving the role of an effective
  chemical potential).}\BibitemShut {Stop}%
\bibitem [{\citenamefont {Beugeling}\ \emph {et~al.}(2015)\citenamefont
  {Beugeling}, \citenamefont {Moessner},\ and\ \citenamefont
  {Haque}}]{offdiag}%
  \BibitemOpen
  \bibfield  {author} {\bibinfo {author} {\bibfnamefont {Wouter}\ \bibnamefont
  {Beugeling}}, \bibinfo {author} {\bibfnamefont {Roderich}\ \bibnamefont
  {Moessner}}, \ and\ \bibinfo {author} {\bibfnamefont {Masudul}\ \bibnamefont
  {Haque}},\ }\bibfield  {title} {\enquote {\bibinfo {title} {Off-diagonal
  matrix elements of local operators in many-body quantum systems},}\ }\href
  {\doibase 10.1103/PhysRevE.91.012144} {\bibfield  {journal} {\bibinfo
  {journal} {Phys. Rev. E}\ }\textbf {\bibinfo {volume} {91}},\ \bibinfo
  {pages} {012144} (\bibinfo {year} {2015})}\BibitemShut {NoStop}%
\bibitem [{Note20()}]{Note20}%
  \BibitemOpen
  \bibinfo {note} {We stress that here $f$ is a general function of the
  coordinates and their derivatives. The derivative (as well as finite higher
  order derivatives) of the Gaussian $\protect \EuScript {F}$ is non-vanishing
  at a distance $\sigma _{F}$ where the density matrix vanishes. It follows
  that the expectation value of the commutator $[f, \protect \EuScript {F}$]
  will vanish. This is so since (i) where the derivative of the Gaussian is
  finite the density matrix vanishes and (ii) where the density matrix is
  finite the derivative of the Gaussian is vanishingly small.}\BibitemShut
  {Stop}%
\bibitem [{Note21()}]{Note21}%
  \BibitemOpen
  \bibinfo {note} {In those special cases in which we will seek a specific
  decomposition of the form of Eq. (\ref {eq:decoupled+sum}), we will attempt
  to find a transformation $\protect \tilde {H}_{i'}^{H} \to \protect \tilde
  {H}_{i'}^{H} + W^H_{i'}$ minimizing ${\protect \sf {Tr}}\left (\rho
  ^{\protect \sf {canonical}}_{\Lambda }\left (\protect \tilde
  {H}^H_{i'}+W^H_{i'}\right )^{2}\right )-\left ({\protect \sf {Tr}}\left (\rho
  ^{\protect \sf {canonical}}_{\Lambda }\left (\protect \tilde
  {H}^H_{i'}+W^H_{i'}\right )\right )^{2}\right )$ and also further satisfying
  the constraint $\DOTSB \sum@ \slimits@ _{i'} W^H_{i'} =0$.}\BibitemShut
  {Stop}%
\bibitem [{\citenamefont {Nussinov}\ \emph {et~al.}(2012)\citenamefont
  {Nussinov}, \citenamefont {Ortiz},\ and\ \citenamefont
  {Cobanera}}]{boundary-bounds2}%
  \BibitemOpen
  \bibfield  {author} {\bibinfo {author} {\bibfnamefont {Zohar}\ \bibnamefont
  {Nussinov}}, \bibinfo {author} {\bibfnamefont {Gerardo}\ \bibnamefont
  {Ortiz}}, \ and\ \bibinfo {author} {\bibfnamefont {Emilio}\ \bibnamefont
  {Cobanera}},\ }\bibfield  {title} {\enquote {\bibinfo {title} {{Effective and
  exact holographies from symmetries and dualities}},}\ }\href {\doibase
  https://doi.org/10.1016/j.aop.2012.07.001} {\bibfield  {journal} {\bibinfo
  {journal} {Annals of Physics}\ }\textbf {\bibinfo {volume} {327}},\ \bibinfo
  {pages} {2491--2521} (\bibinfo {year} {2012})}\BibitemShut {NoStop}%
\bibitem [{\citenamefont {Wehling}\ \emph {et~al.}(2014)\citenamefont
  {Wehling}, \citenamefont {Black-Schaffer},\ and\ \citenamefont
  {Balatsky}}]{Dirac_Sasha}%
  \BibitemOpen
  \bibfield  {author} {\bibinfo {author} {\bibfnamefont {T.O.}\ \bibnamefont
  {Wehling}}, \bibinfo {author} {\bibfnamefont {A.M.}\ \bibnamefont
  {Black-Schaffer}}, \ and\ \bibinfo {author} {\bibfnamefont {A.V.}\
  \bibnamefont {Balatsky}},\ }\bibfield  {title} {\enquote {\bibinfo {title}
  {Dirac materials},}\ }\href {\doibase 10.1080/00018732.2014.927109}
  {\bibfield  {journal} {\bibinfo  {journal} {Advances in Physics}\ }\textbf
  {\bibinfo {volume} {63}},\ \bibinfo {pages} {1--76} (\bibinfo {year}
  {2014})}\BibitemShut {NoStop}%
\bibitem [{\citenamefont {Ashcroft}\ and\ \citenamefont
  {Mermin}(1976)}]{AMbook}%
  \BibitemOpen
  \bibfield  {author} {\bibinfo {author} {\bibfnamefont {Neil~W.}\ \bibnamefont
  {Ashcroft}}\ and\ \bibinfo {author} {\bibfnamefont {N.~David}\ \bibnamefont
  {Mermin}},\ }\href@noop {} {\emph {\bibinfo {title} {{Solid State
  Physics}}}}\ (\bibinfo  {publisher} {Saunders College Publishing},\ \bibinfo
  {year} {1976})\BibitemShut {NoStop}%
\bibitem [{\citenamefont {Girvin}\ and\ \citenamefont {Yang}(2019)}]{GYbook}%
  \BibitemOpen
  \bibfield  {author} {\bibinfo {author} {\bibfnamefont {Steven~M.}\
  \bibnamefont {Girvin}}\ and\ \bibinfo {author} {\bibfnamefont {Kun}\
  \bibnamefont {Yang}},\ }\href@noop {} {\emph {\bibinfo {title} {{Modern
  Condensed Matter Physics 1st Edition}}}}\ (\bibinfo  {publisher} {Cambridge
  University Press},\ \bibinfo {year} {2019})\BibitemShut {NoStop}%
\bibitem [{\citenamefont {Simons}(2013)}]{stevensimons}%
  \BibitemOpen
  \bibfield  {author} {\bibinfo {author} {\bibfnamefont {Steven~H.}\
  \bibnamefont {Simons}},\ }\href@noop {} {\emph {\bibinfo {title} {{The Oxford
  Solid State Basics}}}}\ (\bibinfo  {publisher} {Oxford University Press},\
  \bibinfo {year} {2013})\BibitemShut {NoStop}%
\bibitem [{\citenamefont {Bloch}(1930)}]{B1}%
  \BibitemOpen
  \bibfield  {author} {\bibinfo {author} {\bibfnamefont {F.}~\bibnamefont
  {Bloch}},\ }\bibfield  {title} {\enquote {\bibinfo {title} {Zum elektrischen
  widerstandsgesetz bei tiefen temperaturen},}\ }\href@noop {} {\bibfield
  {journal} {\bibinfo  {journal} {Zeitschrift fur Physik}\ }\textbf {\bibinfo
  {volume} {59}},\ \bibinfo {pages} {208} (\bibinfo {year} {1930})}\BibitemShut
  {NoStop}%
\bibitem [{\citenamefont {Gruneisen}(1933)}]{G1}%
  \BibitemOpen
  \bibfield  {author} {\bibinfo {author} {\bibfnamefont {E.}~\bibnamefont
  {Gruneisen}},\ }\bibfield  {title} {\enquote {\bibinfo {title} {Die
  abhangigkeit des elektrischen widerstandes reiner metalle von der
  temperatur},}\ }\href@noop {} {\bibfield  {journal} {\bibinfo  {journal}
  {Annalen der Physik}\ }\textbf {\bibinfo {volume} {408}},\ \bibinfo {pages}
  {530} (\bibinfo {year} {1933})}\BibitemShut {NoStop}%
\bibitem [{\citenamefont {Glimm}\ and\ \citenamefont
  {Jaffe}(2012)}]{Reflection-Positivity1}%
  \BibitemOpen
  \bibfield  {author} {\bibinfo {author} {\bibfnamefont {James}\ \bibnamefont
  {Glimm}}\ and\ \bibinfo {author} {\bibfnamefont {Arthur}\ \bibnamefont
  {Jaffe}},\ }\href@noop {} {\emph {\bibinfo {title} {{Quantum physics: a
  functional integral point of view}}}}\ (\bibinfo  {publisher} {Springer
  Science \& Business Media},\ \bibinfo {year} {2012})\BibitemShut {NoStop}%
\bibitem [{\citenamefont {Osterwalder}\ and\ \citenamefont
  {Schrader}(1973)}]{Reflection-Positivity2-1}%
  \BibitemOpen
  \bibfield  {author} {\bibinfo {author} {\bibfnamefont {Konrad}\ \bibnamefont
  {Osterwalder}}\ and\ \bibinfo {author} {\bibfnamefont {Robert}\ \bibnamefont
  {Schrader}},\ }\bibfield  {title} {\enquote {\bibinfo {title} {{Axioms for
  Euclidean Green's functions}},}\ }\href {\doibase 10.1007/BF01645738}
  {\bibfield  {journal} {\bibinfo  {journal} {Communications in Mathematical
  Physics}\ }\textbf {\bibinfo {volume} {31}},\ \bibinfo {pages} {83--112}
  (\bibinfo {year} {1973})}\BibitemShut {NoStop}%
\bibitem [{\citenamefont {Osterwalder}\ and\ \citenamefont
  {Schrader}(1975)}]{Reflection-Positivity2-2}%
  \BibitemOpen
  \bibfield  {author} {\bibinfo {author} {\bibfnamefont {Konrad}\ \bibnamefont
  {Osterwalder}}\ and\ \bibinfo {author} {\bibfnamefont {Robert}\ \bibnamefont
  {Schrader}},\ }\bibfield  {title} {\enquote {\bibinfo {title} {{Axioms for
  Euclidean Green's functions II}},}\ }\href {\doibase 10.1007/BF01608978}
  {\bibfield  {journal} {\bibinfo  {journal} {Communications in Mathematical
  Physics}\ }\textbf {\bibinfo {volume} {42}},\ \bibinfo {pages} {281--305}
  (\bibinfo {year} {1975})}\BibitemShut {NoStop}%
\bibitem [{\citenamefont {Fr{\"o}hlich}\ \emph {et~al.}(1978)\citenamefont
  {Fr{\"o}hlich}, \citenamefont {Israel}, \citenamefont {Lieb},\ and\
  \citenamefont {Simon}}]{Reflection-Positivity3}%
  \BibitemOpen
  \bibfield  {author} {\bibinfo {author} {\bibfnamefont {J{\"u}rg}\
  \bibnamefont {Fr{\"o}hlich}}, \bibinfo {author} {\bibfnamefont {Robert}\
  \bibnamefont {Israel}}, \bibinfo {author} {\bibfnamefont {Elliot~H.}\
  \bibnamefont {Lieb}}, \ and\ \bibinfo {author} {\bibfnamefont {Barry}\
  \bibnamefont {Simon}},\ }\bibfield  {title} {\enquote {\bibinfo {title}
  {{Phase transitions and reflection positivity. I. General theory and long
  range lattice models}},}\ }\href {\doibase 10.1007/BF01940327} {\bibfield
  {journal} {\bibinfo  {journal} {Communications in Mathematical Physics}\
  }\textbf {\bibinfo {volume} {62}},\ \bibinfo {pages} {1--34} (\bibinfo {year}
  {1978})}\BibitemShut {NoStop}%
\bibitem [{\citenamefont {Griffiths}(1967)}]{Grif1}%
  \BibitemOpen
  \bibfield  {author} {\bibinfo {author} {\bibfnamefont {Robert~B.}\
  \bibnamefont {Griffiths}},\ }\bibfield  {title} {\enquote {\bibinfo {title}
  {{Correlations in Ising Ferromagnets. I}},}\ }\href {\doibase
  10.1063/1.1705219} {\bibfield  {journal} {\bibinfo  {journal} {Journal of
  Mathematical Physics}\ }\textbf {\bibinfo {volume} {8}},\ \bibinfo {pages}
  {478--483} (\bibinfo {year} {1967})}\BibitemShut {NoStop}%
\bibitem [{\citenamefont {Kelly}\ and\ \citenamefont {Sherman}(1968)}]{Grif2}%
  \BibitemOpen
  \bibfield  {author} {\bibinfo {author} {\bibfnamefont {D.~G.}\ \bibnamefont
  {Kelly}}\ and\ \bibinfo {author} {\bibfnamefont {S.}~\bibnamefont
  {Sherman}},\ }\bibfield  {title} {\enquote {\bibinfo {title} {{General
  Griffiths' Inequalities on Correlations in Ising Ferromagnets}},}\ }\href
  {\doibase 10.1063/1.1664600} {\bibfield  {journal} {\bibinfo  {journal}
  {Journal of Mathematical Physics}\ }\textbf {\bibinfo {volume} {9}},\
  \bibinfo {pages} {466--484} (\bibinfo {year} {1968})}\BibitemShut {NoStop}%
\bibitem [{\citenamefont {Griffiths}(1969)}]{Grif3}%
  \BibitemOpen
  \bibfield  {author} {\bibinfo {author} {\bibfnamefont {Robert~B.}\
  \bibnamefont {Griffiths}},\ }\bibfield  {title} {\enquote {\bibinfo {title}
  {{Rigorous Results for Ising Ferromagnets of Arbitrary Spin}},}\ }\href
  {\doibase 10.1063/1.1665005} {\bibfield  {journal} {\bibinfo  {journal}
  {Journal of Mathematical Physics}\ }\textbf {\bibinfo {volume} {10}},\
  \bibinfo {pages} {1559--1565} (\bibinfo {year} {1969})}\BibitemShut {NoStop}%
\bibitem [{\citenamefont {Ginibre}(1970)}]{Grif4}%
  \BibitemOpen
  \bibfield  {author} {\bibinfo {author} {\bibfnamefont {J.}~\bibnamefont
  {Ginibre}},\ }\bibfield  {title} {\enquote {\bibinfo {title} {{General
  formulation of Griffiths' inequalities}},}\ }\href {\doibase
  10.1007/BF01646537} {\bibfield  {journal} {\bibinfo  {journal}
  {Communications in Mathematical Physics}\ }\textbf {\bibinfo {volume} {16}},\
  \bibinfo {pages} {310--328} (\bibinfo {year} {1970})}\BibitemShut {NoStop}%
\bibitem [{Note22()}]{Note22}%
  \BibitemOpen
  \bibinfo {note} {In the classical high temperature regime, the probability
  density $\rho _{\Lambda }^{\protect \sf {classical~canonical }}=\protect
  \frac {e^{-\beta H_{\Lambda }}}{Z_{\Lambda }}$, with the harmonic oscillator
  Hamiltonian $H_{\Lambda }$, factorizes into a Gaussian in the momenta
  multiplying a Gaussian in the spatial coordinates and Eq. (\ref {osc:var:2})
  also immediately follows from a trivial application of Wick's
  theorem.}\BibitemShut {Stop}%
\bibitem [{\citenamefont {Lieb}\ \emph {et~al.}(1961)\citenamefont {Lieb},
  \citenamefont {Schultz},\ and\ \citenamefont {Mattis}}]{LSM}%
  \BibitemOpen
  \bibfield  {author} {\bibinfo {author} {\bibfnamefont {Elliot}\ \bibnamefont
  {Lieb}}, \bibinfo {author} {\bibfnamefont {Theodore}\ \bibnamefont
  {Schultz}}, \ and\ \bibinfo {author} {\bibfnamefont {Daniel}\ \bibnamefont
  {Mattis}},\ }\bibfield  {title} {\enquote {\bibinfo {title} {{Two Soluble
  Models of an Antiferromagnetic Chain}},}\ }\href {\doibase
  10.1016/0003-4916(61)90115-4} {\bibfield  {journal} {\bibinfo  {journal}
  {Annals of Physics}\ }\textbf {\bibinfo {volume} {126}},\ \bibinfo {pages}
  {407} (\bibinfo {year} {1961})}\BibitemShut {NoStop}%
\bibitem [{\citenamefont {Berges}\ \emph {et~al.}(2004)\citenamefont {Berges},
  \citenamefont {Borsanyi},\ and\ \citenamefont {Wetterich}}]{prethermal1}%
  \BibitemOpen
  \bibfield  {author} {\bibinfo {author} {\bibfnamefont {J.}~\bibnamefont
  {Berges}}, \bibinfo {author} {\bibfnamefont {Sz.}\ \bibnamefont {Borsanyi}},
  \ and\ \bibinfo {author} {\bibfnamefont {C.}~\bibnamefont {Wetterich}},\
  }\bibfield  {title} {\enquote {\bibinfo {title} {Prethermalization},}\ }\href
  {\doibase 10.1103/PhysRevLett.93.142002} {\bibfield  {journal} {\bibinfo
  {journal} {Phys. Rev. Lett.}\ }\textbf {\bibinfo {volume} {93}},\ \bibinfo
  {pages} {142002} (\bibinfo {year} {2004})}\BibitemShut {NoStop}%
\bibitem [{\citenamefont {Gring}\ \emph {et~al.}(2012)\citenamefont {Gring},
  \citenamefont {Kuhnert}, \citenamefont {Langen}, \citenamefont {Kitagawa},
  \citenamefont {Rauer}, \citenamefont {Schreitl}, \citenamefont {Mazets},
  \citenamefont {Adu~Smith}, \citenamefont {Demler},\ and\ \citenamefont
  {Schmiedmayer}}]{prethermal2}%
  \BibitemOpen
  \bibfield  {author} {\bibinfo {author} {\bibfnamefont {M.}~\bibnamefont
  {Gring}}, \bibinfo {author} {\bibfnamefont {M.}~\bibnamefont {Kuhnert}},
  \bibinfo {author} {\bibfnamefont {T.}~\bibnamefont {Langen}}, \bibinfo
  {author} {\bibfnamefont {T.}~\bibnamefont {Kitagawa}}, \bibinfo {author}
  {\bibfnamefont {B.}~\bibnamefont {Rauer}}, \bibinfo {author} {\bibfnamefont
  {M.}~\bibnamefont {Schreitl}}, \bibinfo {author} {\bibfnamefont
  {I.}~\bibnamefont {Mazets}}, \bibinfo {author} {\bibfnamefont
  {D.}~\bibnamefont {Adu~Smith}}, \bibinfo {author} {\bibfnamefont
  {E.}~\bibnamefont {Demler}}, \ and\ \bibinfo {author} {\bibfnamefont
  {J.}~\bibnamefont {Schmiedmayer}},\ }\bibfield  {title} {\enquote {\bibinfo
  {title} {Relaxation and pre-thermalization in an isolated quantum system},}\
  }\href {\doibase 10.1126/science.1224953} {\bibfield  {journal} {\bibinfo
  {journal} {Science}\ }\textbf {\bibinfo {volume} {337}},\ \bibinfo {pages}
  {1318} (\bibinfo {year} {2012})}\BibitemShut {NoStop}%
\bibitem [{\citenamefont {Casher}\ and\ \citenamefont {Nussinov}(1995)}]{aba1}%
  \BibitemOpen
  \bibfield  {author} {\bibinfo {author} {\bibfnamefont {A.}~\bibnamefont
  {Casher}}\ and\ \bibinfo {author} {\bibfnamefont {S.}~\bibnamefont
  {Nussinov}},\ }\bibfield  {title} {\enquote {\bibinfo {title} {Some
  speculations on the ultimate planck energy accelerators},}\ }\href
  {https://arxiv.org/pdf/hep-ph/9510364.pdf} {\bibfield  {journal} {\bibinfo
  {journal} {arXiv preprint arXiv:hep-ph/9510364}\ } (\bibinfo {year}
  {1995})}\BibitemShut {NoStop}%
\bibitem [{\citenamefont {Strinati}\ \emph {et~al.}(2018)\citenamefont
  {Strinati}, \citenamefont {Pieri}, \citenamefont {Roepke}, \citenamefont
  {Schuck},\ and\ \citenamefont {Urban}}]{BEC-BCS-R}%
  \BibitemOpen
  \bibfield  {author} {\bibinfo {author} {\bibfnamefont {Giancarlo~Calvanese}\
  \bibnamefont {Strinati}}, \bibinfo {author} {\bibfnamefont {Pierbiagio}\
  \bibnamefont {Pieri}}, \bibinfo {author} {\bibfnamefont {Gerd}\ \bibnamefont
  {Roepke}}, \bibinfo {author} {\bibfnamefont {Peter}\ \bibnamefont {Schuck}},
  \ and\ \bibinfo {author} {\bibfnamefont {Michael}\ \bibnamefont {Urban}},\
  }\bibfield  {title} {\enquote {\bibinfo {title} {{The BCS-BEC crossover: From
  ultra-cold Fermi gases to nuclear systems}},}\ }\href {\doibase
  10.1016/j.physrep.2018.02.004} {\bibfield  {journal} {\bibinfo  {journal}
  {Physics Reports}\ }\textbf {\bibinfo {volume} {738}},\ \bibinfo {pages} {1}
  (\bibinfo {year} {2018})}\BibitemShut {NoStop}%
\bibitem [{\citenamefont {Nussinov}\ and\ \citenamefont
  {Nussinov}(2006)}]{BEC-BCS}%
  \BibitemOpen
  \bibfield  {author} {\bibinfo {author} {\bibfnamefont {Z.}~\bibnamefont
  {Nussinov}}\ and\ \bibinfo {author} {\bibfnamefont {S.}~\bibnamefont
  {Nussinov}},\ }\bibfield  {title} {\enquote {\bibinfo {title} {Triviality of
  the bcs-bec crossover in extended dimensions: implications for the ground
  state energy},}\ }\href {\doibase 10.1103/PhysRevA.74.053622} {\bibfield
  {journal} {\bibinfo  {journal} {Phys. Rev. A}\ }\textbf {\bibinfo {volume}
  {74}},\ \bibinfo {pages} {053622} (\bibinfo {year} {2006})}\BibitemShut
  {NoStop}%
\end{thebibliography}%

\end{document}